\documentclass[sigplan,10pt]{acmart}
\renewcommand\footnotetextcopyrightpermission[1]{} 

\usepackage{enumitem}
\usepackage{makecell}
\usepackage{wrapfig}
\usepackage{lipsum}
\usepackage{float}
\usepackage{algorithm}
\usepackage{algpseudocode}
\usepackage{amsmath,amsfonts}

\usepackage{microtype}
\setlist[enumerate]{itemsep=0mm}

\newcommand{\name}{{D-STACK}}
\newcommand{\ie}{{\em i.e., \/}}
\newcommand{\eg}{{\em e.g., \/}}
\newcommand{\etc}{{\em etc.\/}}

\newcommand{\sknote}[1]{{\color{brown}[SK: #1]}}
\newcommand{\anote}[1]{{\color{magenta}[AD: #1]}}

\newcommand{\knote}[1]{\todo[color=yellow,inline]{KK: #1}}
\newcommand{\Scut}[1] {}
\newcommand{\update}[1]{{\color{blue}#1}}
\newcommand{\cut}[1]{{\color{gray}\sout{[CUT: #1]}}}

\newcommand{\revise}[1]{{\hl{#1}}}
\renewcommand{\revise}[1]{{#1}}
\renewcommand{\cut}[1]{}
\renewcommand{\update}[1]{{#1}}

\newcommand{\newcut}[1] {}

\usepackage{subcaption}

\AtBeginDocument{%
  \providecommand\BibTeX{{%
    \normalfont B\kern-0.5em{\scshape i\kern-0.25em b}\kern-0.8em\TeX}}}

\setcopyright{none}

\begin{document}

\title{\name: High Throughput DNN Inference by Effective Multiplexing and Spatio-Temporal Scheduling of GPUs}

\author{Aditya Dhakal}
\email{adhak001@ucr.edu}
\affiliation{%
  \institution{University of California, Riverside}
  \country{USA}
}

\author{Sameer G. Kulkarni}
\email{sameergk@iitgn.ac.in}
\affiliation{%
  \institution{IIT Gandhinagar}
  \country{India}}

\author{K. K. Ramakrishnan}
\email{kk@cs.ucr.edu}
\affiliation{%
  \institution{University of California, Riverside}
  \country{USA}
}
\renewcommand{\shortauthors}{Dhakal, et al.}

\begin{abstract}
  
 
{
Hardware accelerators such as GPUs are required for real-time, low latency inference with Deep Neural Networks (DNN). However, due to the inherent limits to the parallelism they can exploit, DNNs often under-utilize the capacity of today's high-end accelerators\Scut{GPUs}. Although spatial multiplexing of the GPU, while limiting the GPU resources (GPU\%) to each DNN to the right amount, leads to higher GPU utilization and higher inference throughput, 
there remain a number of challenges. 
Finding the GPU\% for right-sizing the GPU for each DNN through profiling, determining an optimal batching of requests to balance throughput improvement while meeting  application-specific deadlines and service level objectives (SLOs), and maximizing throughput by appropriately scheduling DNNs are still significant challenges.\looseness-1

This paper, introduces a dynamic and fair spatio-temporal scheduler (\name{}) that enables multiple DNNs to run in the GPU concurrently. To help allocate the appropriate GPU\% (we call it the "Knee"), we develop and validate a model that estimates the parallelism each DNN can utilize.
We also develop a lightweight optimization formulation to find an efficient batch size for each DNN operating with \name{}. 
We bring together our optimizations and our spatio-temporal scheduler to provide a holistic inference framework. We demonstrate its ability to provide high throughput while meeting application SLOs. We compare \name{} with an ideal scheduler that can allocate the right GPU\% for every DNN kernel. \name{} gets higher than 90\% throughput and GPU utilization compared to the ideal scheduler. We also compare \name{} with other GPU multiplexing and scheduling methods (e.g., NVIDIA Triton, Clipper, Nexus), using popular DNN models.\Scut{Finally, we evaluate our approach's ability to adjust the GPU\% dynamically with minimal downtime.} Our controlled experiments with multiplexing several popular DNN models achieve up to $1.6\times$ improvement in GPU utilization and up to $4\times$ improvement in inference throughput.\looseness-1

\Scut{
Deep Neural Networks (DNNs) have become the algorithms of choice for machine learning (ML) applications. The high compute requirement of DNNs makes GPUs an indispensable hardware accelerator for low-latency ML training and inference. 
\Scut{The development of libraries such as CUDA, cuDNN, RoCM, etc., and ML frameworks such as PyTorch, TensorFlow, MxNet, etc., has also helped fuel the rapid development and deployment of these DNN applications. }
However, DNNs have inherent limits to the parallelism that they can exploit of the massive parallelism offered by current, high-end GPUs. This often results in an under-utilization of GPU resources. 
Spatial multiplexing enables concurrent execution of multiple models in a GPU and improves GPU utilization. But, it needs to be complemented by an intelligent scheduler to ensure that application specific deadlines and service level objectives (SLOs) are met. We also need to understand the GPU resource requirements of individual DNN models to help allocate the GPU resources appropriately. 
    
We develop a model to estimate the parallelism that can be utilized by DNNs. We validate this model based on detailed evaluations of popularly used DNN types such as feed-forward convolutional neural networks (CNN) and recurrent neural networks (RNN). We use the model and insights gained to propose the estimation of an appropriate percentage of GPU resources necessary for efficient operation of the DNN. 
\Scut{This allows us to efficiently run multiple DNN applications concurrently, enhancing current state-of-the-art GPU multiplexing.} 
We leverage\Scut{\knote{leverage??}} controlled spatial sharing (CSS) to judiciously adjust the GPU resource allocation for each DNNs and run multiple DNN applications concurrently, thus improving the utilization of GPU resources. 
We also propose a lightweight optimization that right-sizes the batching of DNN tasks on the GPU, to maximize  throughput and GPU utilization, while being SLO-aware.
Further, our spatio-temporal scheduler accounts for the DNN model's SLO per inference, GPU resource allocation and the batch size to provide a schedule that improves the likelihood of satisfying the SLOs across multiple DNN models. 
We demonstrate the benefits of our framework by bringing together CSS, the optimal batch size, 
and the spatio-temporal scheduler to run multiple real DNN models. Our controlled experiments with multiplexing a number of popular DNN models achieve upto $1.6\times$ improvement in GPU utilization and upto $4\times$ improvement in inference throughput.
}
\Scut{
We then present a set of software primitives needed by ML platforms, and an overall solution to support CSS, to effectively utilize the GPU hardware. These primitives include efficient coordination of the CPU-GPU interaction while running asynchronous tasks, data transfer offloading to the GPU’s DMA while processing high volumes of DNN requests, and timely notification to reduce GPU idle time and efficient sharing of GPU resources with GPU resource isolation.
\knote{talk about overlap for multiplexing and scheduling.}
These help increase hardware utilization as well as improve the latency and throughput of inference. 
We also propose a lightweight optimization that right-sizes the batching of tasks on the GPU along with controlled spatial multiplexing, while respecting service level objectives.
\knote{say how you bring all together to show that this overall approach works for multiple real models.}
}
}
\Scut{
  A clear and well-documented \LaTeX\ document is presented as an
  article formatted for publication by ACM in a conference proceedings
  or journal publication. Based on the ``acmart'' document class, this
  article presents and explains many of the common variations, as well
  as many of the formatting elements an author may use in the
  preparation of the documentation of their work.
  }
\end{abstract}

\Scut{



\begin{teaserfigure}
  \includegraphics[width=\textwidth]{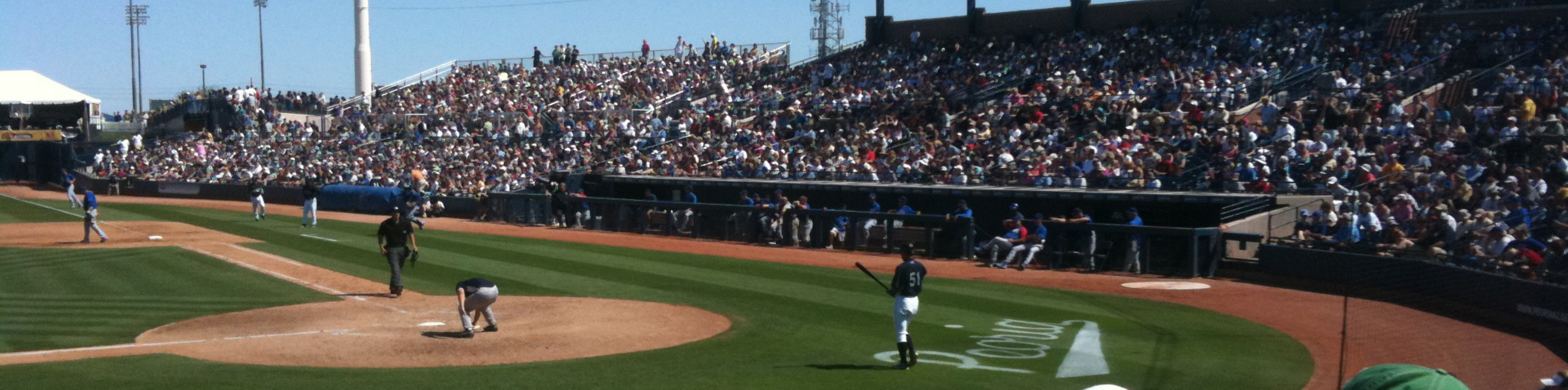}
  \caption{Seattle Mariners at Spring Training, 2010.}
  \Description{Enjoying the baseball game from the third-base
  seats. Ichiro Suzuki preparing to bat.}
  \label{fig:teaser}
\end{teaserfigure}
}
\maketitle
\section{Introduction}
Deep Neural Networks (DNNs) are widely used for many applications, including image recognition, natural language processing, \etc{} 
Accelerators have become indispensable for DNN learning and inference. Accelerators such as GPUs, TensorCores~\cite{markidis2018nvidialong}, and TPU~\cite{jouppi2017datacenterlong} reduce the DNN inference times, often by 2-3 orders of magnitude compared to even using a high-end CPU cluster.  These accelerators are widely used by cloud services as a part of their \textit{inference-as-a-service} (IaaS) offerings, where trained DNN models are hosted in a Cloud or an Edge Cloud (especially for low-latency operation). User requests are inferred using the GPUs deployed in the cloud.
\color{black}
\looseness-1

Most DNN models running in inference frameworks (PyTorch~\cite{NEURIPS2019_9015}, TensorFlow Serving~\cite{Tensorflowserving}, NVIDIA's Triton~\cite{tritonserver} \etc) 
often execute far fewer floating-point operations per second (FLOPS) than the capacity of these high-end GPUs~\cite{GSLICE,zhang2019laius,architectural-impli}, TPUs~\cite{wang2020systematic} and other accelerators~\cite{kong2021edlab}. We observed that DNN models, when performing inference even using a single GPU, do not significantly reduce the DNN's processing latency when provided with additional GPU resources (\ie{} number of Streaming Multiprocessors (SMs) -  GPU compute units analogous to CPU cores) beyond a certain point. We call this point as a \textbf{"Knee"} for the DNN (expressed as a percentage of the total SMs available in the GPU, \eg 50\% of a V100 GPU (which has 80 SMs in total) is 40 SMs.). Running applications with resources matching the Knee is desirable for a cloud operator providing Inference as a Service, since multiplexing a GPU (or similar accelerator) across as many applications as possible keeps costs low. Operating at the Knee also keeps the latency low for the user. When more GPU resources are provided for a DNN (e.g., by giving the full GPU to an application, possibly using temporal sharing), it is wasteful as the GPU is not fully utilized.

We see two fundamental reasons for this under-utilization of multi-core accelerators such as GPUs by DNNs when given more than the Knee's resources: i) Amount of parallelism over the entirety of DNN's execution is not uniform, \ie many DNN functions (\eg convolution, ReLU \etc) are unable to fully utilize the parallelism offered by the accelerator; ii) DNN operations also involve other overheads\Scut{number of sequential tasks that cause accelerator to be idle for significant periods} (\eg kernel launches, memory read-write, \etc).   We study the execution of a variety of DNN models\Scut{with real GPU systems} to understand the root causes of under-utilization of such accelerators, particularly GPUs, and develop methods to improve the overall system utilization, thus improving throughput and reducing inference latency.\Scut{The spatio-temporal multiplexing and scheduling methods and algorithms we develop seek to balance the processing resource demand (space) and time-based demand (temporal) on a multi-processor accelerator. We extend our techniques to utilize DNN inference primitives such as batching, and demonstrate our technique's effectiveness in real GPUs using several popular DNN models.}\looseness-1

\Scut{
\begin{figure}[t]
    \centering
    \vspace{-2mm}
    \includegraphics[width=\linewidth]{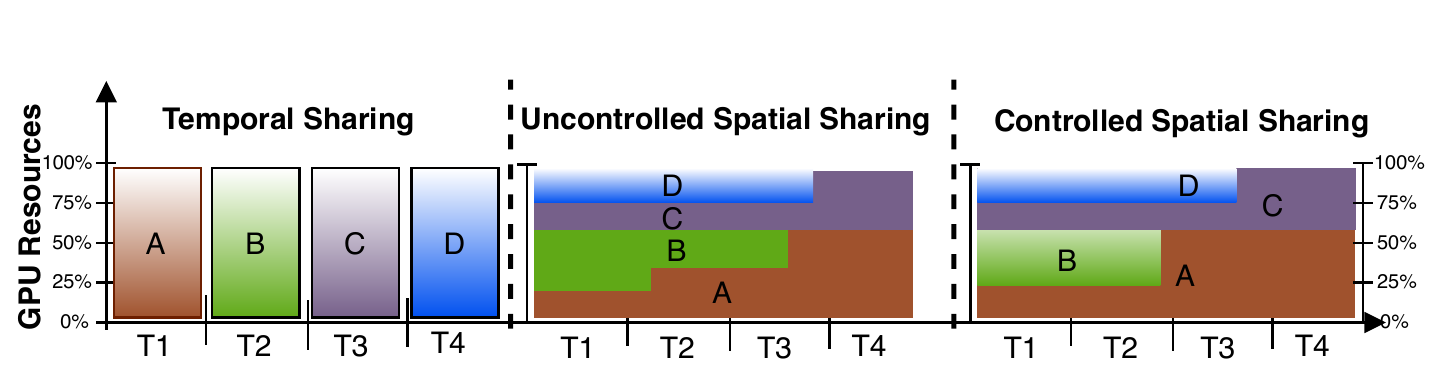}\vspace{-4mm}
    \caption{GPU multiplexing methods for apps. A,B,C and D}
    \label{fig:GPU_Sharing_methods}
    \vspace{-7mm}
\end{figure}
}
\textbf{\textit{Multiplexing GPUs in the Edge Cloud}}:

\Scut{
We consider the increasingly popular case of DNN-based inference as a service, with user inference requests being processed in a cloud server with GPUs. The goal is to achieve high inference throughput while processing each request within a fixed deadline (\eg $<$100ms). This potentially reduces the operating cost for the cloud operator.}

DNN inference requests for applications such as autonomous driving, augmented reality, \etc{,} have stringent deadlines (\eg $<$100ms). A cloud providing IaaS also has to account for the network latency. Edge Clouds offer a sweet spot reducing both latency and offering the necessary processing resources, although more constrained than centralized cloud services. Multiplexing the expensive hardware accelerator is therefore very desirable. 
Current GPU virtualization and inference service frameworks such as Nexus~\cite{shen2019nexuslong}, NVIDIA's Triton Inference Server (Triton)~\cite{tritonserver}, gPipe~\cite{huang2019gpipe}, and PipeDream~\cite{narayanan2019pipedream} either use a 'single GPU per DNN' model or time-share the GPU across multiple DNN models. These current state-of-the-art frameworks for DNNs allocate the full GPU (\ie{} 100\% of GPU) for the time quantum as shown in Fig.~\ref{fig:spatio-temporal-scheduling} (left). 
However, dedicating an entire GPU to run a single DNN model at a time can be  wasteful. Furthermore, interleaving execution of tenant applications by temporally sharing increases inference latency for all of them, because of the significant cost of frequent switching between applications. Multiplexing several applications on the GPU to run concurrently, through spatial as well as temporal multiplexing, helps to better utilize the GPU and achieve much higher aggregate inference throughput.\looseness-1
\color{black}


 Our approach utilizes the CUDA Multi-process Service (MPS)~\cite{nvidiamps2019} to spatially share the GPU across several applications, similar to 
 GSLICE~\cite{GSLICE}.
But, existing approaches of spatial multiplexing with the GPU either only statically partition the GPU for each application or does not guarantee computing resource isolation while multiplexing. This has the potential to allocate fewer resources than necessary for an application. It also causes interference among the multiplexed applications when too many models share the GPU, thus, increasing the inference latency. 

We illustrate with an example when four different models have to be run on a V100 GPU (three are already executing and a fourth is added). Temporal sharing allocates the GPU to each model for a time slice. Static spatial sharing with CUDA-MPS will allow all 4 models to run in an uncontrolled manner, causing interference as noted in ~\cite{GSLICE}.  GSLICE will initially spatially share the 3 models, and allocate GPU resources according to their Knee GPU\% capacities. 
When the fourth model is added (in Fig.~\ref{fig:spatio-temporal-scheduling}(middle)), the VGG-19 model's GPU\% is reduced from 50\% to 25\%, causing increased inference latency for that more complex VGG-19 model, which also is undesirable.\looseness-1
\begin{figure}
    \includegraphics[width=\linewidth]{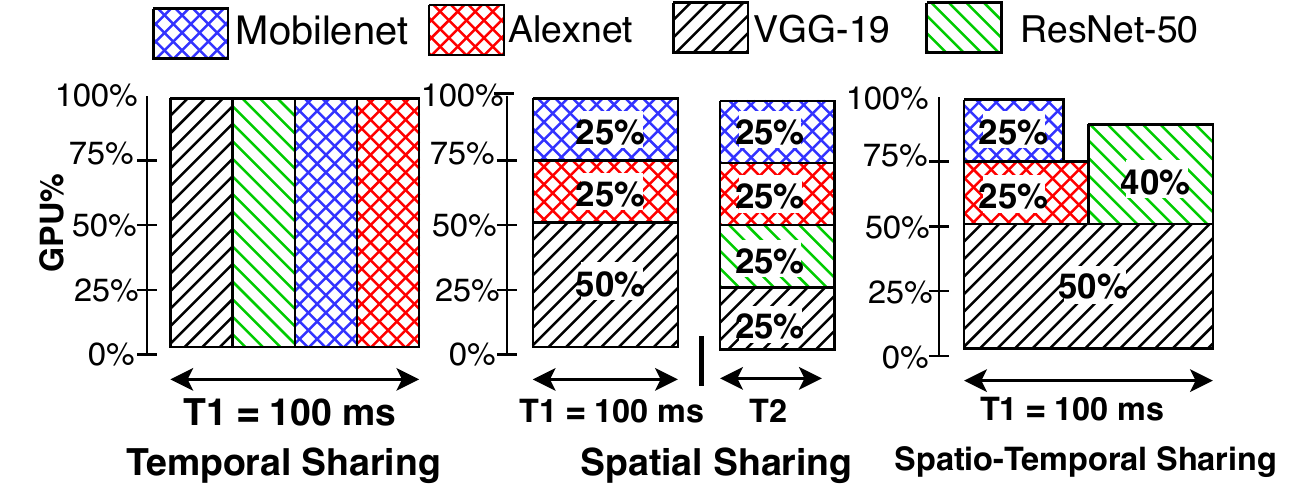}
    \caption{GPU multiplexing scenarios\Scut{\\ with different DNNs}}
    \label{fig:spatio-temporal-scheduling}
\end{figure}

On the other hand, our GPU virtualization framework, with our spatio-temporal scheduler, \textbf{Dynamic Spatio-Temporal pACK} (\name{}), can run on multiple NVIDIA GPU-based systems (single GPU or GPU clusters). \textbf{\name{}} schedules DNNs based on spatial resources (Knee GPU\%, number of SMs), and the appropriate time slice. Combining spatial and temporal scheduling, \name{} is designed to meet the inference deadline for each DNN model. \name{} goes well beyond the basic idea of simple temporal or static spatial multiplexing of a GPU presented in earlier works~\cite{GSLICE,architectural-impli,tritonserver}. The example of Spatio-Temporal scheduling in Fig.~\ref{fig:spatio-temporal-scheduling} (right), has all 4 models getting their Knee GPU\%.  When a model completes its inference, another model utilizes the GPU resources, thus, sharing the GPU resources both temporally and spatially. \name{}'s scheduler further utilizes the idle processing resource of the GPU by dynamically running any 'ready' models, thus maximizing GPU utilization. 

\textbf{\textit{\name{}'s Innovations}}: 

\textit{i.) Understanding a DNN's demand}: For efficient utilization of the GPU, \name{} requires information about the resource requirements of each DNN model. Providing the right resources for the DNN is not just a challenge for the GPU, but is fundamental for all such accelerators that utilize a multitude of compute engines for parallel processing. 
In this paper, along with our analytical models of DNN execution and scheduling, we estimate what would be theoretically possible for a DNN to exploit available parallelism by knowing exactly how much computational capacity is required,  assuming that instantaneous switching between multiplexed tasks is possible. We then show how close we come to that theoretical optimal by implementing our GPU virtualization framework using our \name{} scheduler on a GPU cluster.\looseness-1


\textit{ii.) Dynamic Resource Allocation in GPU: }
Currently, dynamic resource allocation of the GPU requires reloading of applications with their new desired GPU\%. For typical 
DNN models, this reloading time can be 10s of seconds, during which the GPU is idle, lowering the overall system utilization and throughput. In \name{}, we address the dynamic allocation of GPU resources by overlapping the loading of a DNN model with the new resource allocation, by continuing to execute the existing DNN model, thus effectively masking the loading latency. We thus reduce the time the GPU is idle to less than 100 micro-seconds with \name{}. \looseness-1


\textit{iii.) Multi-GPU Cluster: }
Understanding the use of a single GPU and increasing its utilization translates to improving overall throughput of a GPU cluster. \name{}'s optimization can be easily extended to a multi-GPU cluster. In this paper we present the implementation of \name{}'s Spatio-temporal scheduler across multi-GPU cluster to increase the system througput by 200\%.

\Scut{\textbf{\textit{Fundamental Nature of Spatio-Temporal Scheduling in GPUs: }}We believe the problem we address here is fundamental. Hardware accelerators for parallel processing, such as GPUs and TPUs are getting more powerful, with new iterations capable of performing multiple Tera-FLOPS. Similar to the trend with CPUs, most of the improvement in computational power comes from adding more parallel processing compute engines rather than increased clock speed (which usually has physical limits) of an individual processor.}

\Scut{DNNs are emblematic of the class of applications that have a lot of parallelism. Thus, they can benefit from accelerators which possesses more parallel processing compute engines \eg{} GPUs. Effective utilization of the accelerator requires utilizing all of its compute units as efficiently and completely as possible. DNNs however have uneven processing demand across its multiple kernels. Multiplexing several different tasks can be key to improving that utilization. However, switching from one task to another for these parallel processing engines is not instantaneous, often having to wait for synchronization among processors, completion of work by stragglers, aggregation of results from different processors \etc{} 
When using an accelerator for DNN processing, we seek a framework that can very effectively multiplex and batch requests dynamically, and fairly schedule the tasks being multiplexed.
However, the amount of parallelism over the entirety of a DNN's execution is not uniform. It 
is also limited due to the fixed dimensions of the matrices used in the DNN computation. Therefore, we see a need to use accelerators judiciously.
}


\Scut{
\textbf{\name{}} schedules DNNs based on space (Knee GPU\%), and time slice, meeting their inference deadlines. We show an example of Spatio-Temporal scheduling in Fig.~\ref{fig:spatio-temporal-scheduling} (right), where all 4 models get their Knee GPU\%. Rather than a static allocation\Scut{ of the GPU}, the scheduler dynamically re-allocates GPU resources once a model's inference is completed, so other models can effectively utilize that set of GPU SMs.
For efficient utilization of the GPU, \Scut{, while, still meeting the inference deadline, }\name{} requires information about the resource requirements of each DNN model. Finding this right GPU\% for each application is a crucial problem that we tackle in this paper, thus going well beyond the basic idea of spatial multiplexing of GPU presented in earlier work~\cite{GSLICE,architectural-impli,tritonserver}.\looseness-1}
\Scut{ CSS in ~\cite{GSLICE}.}
\Scut{
For this, we derive a theoretical model 
for a DNN which demonstrates why there are inherent limits in the parallelism of a DNN model in a single GPU, and how that leads to under-utilization of the GPU. We also profile DNN models in the MLPerf benchmark~\cite{mattson2020mlperf} as well as additional popular DNN models on actual GPU hardware to show that these limits exist.}\looseness-1
\Scut{Further, batching incoming DNN requests together is another important feature that allows DNN models to exploit the parallelism offered by the GPU even more, thus increasing the inference throughput. State-of-the-art works, like~\cite{crankshaw2017clipperlong,shen2019nexuslong,GSLICE}, incorporate adaptive batching\Scut{techniques} to improve\Scut{GPU utilization and} inference throughput but fail to consider the change in inference latency of a specific batch, with a change in allocated GPU\% due to spatial sharing. We observe that for a given DNN model, a particular GPU\% and batch size combination provides the optimal latency. This sweet spot 
also leads to more efficient GPU utilization and improves aggregate inference throughput. We present an optimization framework in this paper, to find this optimal
\Scut{operating point } 
share of GPU for a DNN, for different batch sizes. 
\name{} uses the knowledge of this optimal operating point of a DNN model to schedule and run multiple DNN models concurrently. Thus, it attains higher throughput than existing state-of-the-art DNN inference platforms such as NVIDIA's Triton Server.\looseness-1} 

\begin{table}[h]
\captionof{table}{\Scut{Task completion time in} Triton and \name{} with 4 DNN models}
    \vspace{-2mm}
    \resizebox{\linewidth}{!}{
    \begin{tabular}{|c|c|c|c|}\hline
         & \makecell{Triton\\Server} &
         \makecell{\name{}} & \makecell{Latency\\Reduction(\%)} \\\hline
         Task completion (sec.)& 58.61\Scut{& 55.55 }& 35.59 & 37\%\\\hline
    \end{tabular}}
    \label{tab:triton-4-models}
\end{table}

\textbf{\textit{Comparing with State-of-the-art: }}We present a comparison of \name{} with NVIDIA's Triton Inference Server. We evaluate the total time taken to infer with 4 different DNN models, Alexnet, Mobilenet, ResNet-50, and VGG-19 being multiplexed on one V100 GPU, each concurrently inferring 10000 images each.\Scut{For the \name{} scheduler, we also sent concurrent requests to all 4 models.} The results in Table.~\mbox{\ref{tab:triton-4-models}} show that\Scut{both the modes of execution in} the Triton server\Scut{(temporally interleaved and concurrent execution)} takes about 58 seconds to finish inference. The \name{} scheduler completes inference on all requests more than 37\% faster (only \textbf{36} seconds). \name{}'s spatial multiplexing, providing just the right amount of GPU\% and its dynamic spatio-temporal scheduling results in more effective use of the GPU and achieving higher DNN inference throughput than NVIDIA's Triton server, while also lowering task completion time. Based on these experiments, we see that implementation of Spatio-temporal scheduling can further enhance throughput when inferring with multiple different models concurrently.\Scut{This is a strong motivation for the mathematical models we develop to gain insight for designing our \name{} scheduler (\S~\ref{sec:scheduling}).\looseness-1}
 
\Scut{
and sees about 1.6$\times$ improvement in utilization of a GPU's SMs compared to a purely temporal scheduler.\Scut{Higher GPU utilization means higher throughput. More SMs being utilized for computation, thus, higher throughput.With the spatio-temporal scheduler,} Due to higher GPU utilization, DNN inference throughput increases by up to 4$\times$ compared to just temporal 
scheduling, while avoiding any deadline (SLO) violations. Our Spatio-Temporal scheduler can be implemented without modifications to the GPU architecture or runtime, thus, it can help enhance the performance of existing inference frameworks (e.g., Triton Server).\looseness-1
}

\Scut{
Going beyond providing the desired GPU\% for a DNN (\eg~\cite{GSLICE}), the comprehensive GPU virtualization framework we propose here utilizes a spatio-temporal scheduler that uses the knowledge of the optimal operating point of a DNN model to schedule and run multiple DNN models concurrently.\looseness-1  
The scheduler optimizes spatial sharing of the GPU with as many DNN models as possible, while meeting the deadline of each model. With the spatio-temporal scheduler, we see about 1.6$\times$ improvement in utilization of a GPU's SMs compared to a purely temporal scheduler. \Scut{Higher GPU utilization means higher throughput. More SMs being utilized for computation, thus, higher throughput.With the spatio-temporal scheduler,} Due to higher GPU utilization, DNN inference throughput is increased by up to 4$\times$ compared to plain temporal 
scheduling, 
while avoiding any deadline (SLO) violation.\looseness-1}\Scut{ due to maximizing the total GPU\% utilized while avoiding any deadline violation.\looseness-1 }


\Scut{
Batching incoming DNN requests together is another important feature that allows DNN models to exploit the parallelism offered by the GPU even more, thus increasing the inference throughput.\Scut{ However, batching a larger number of tasks together also increases the overall DNN inference latency. While this might be acceptable for training, it is likely not as acceptable for inference tasks conducted in real-time, usually with fixed deadlines and service level objectives (SLOs).}
State-of-the-art works, like~\cite{crankshaw2017clipperlong,shen2019nexuslong,GSLICE}, incorporate adaptive batching\Scut{techniques} to improve\Scut{GPU utilization and} inference throughput but fail to consider the change in inference latency of a specific batch, with a change in allocated GPU\% due to spatial sharing. We observe that for a given DNN model, a particular GPU\% and batch size combination provides the optimal latency. 
This sweet spot 
also leads to more efficient GPU utilization and improves aggregate inference throughput. We present an optimization framework in this paper, to find this optimal
\Scut{operating point } 
share of GPU for a DNN, for different batch sizes.} \Scut{We quantify the optimal operating point (batch, GPU\%), 
for several DNNs commonly used in production.Utilizing our profiling approach} 

\textbf{\textit{Contributions: }}\name{} improves GPU utilization by 60\% and increases in DNN inference throughput by 4$\times$ compared to a pure temporal scheduler, while still avoiding any deadline (SLO) violations. 
\Scut{We propose a GPU virtualization framework that utilizes a spatio-temporal scheduler.  We build this framework upon GSLICE~\cite{GSLICE} platform. \knote{why do you bring this up on the last para, rather than earlier?? Isn't this one of the major pieces of the paper?}
With the knowledge of optimal operating point of a DNN model, we schedule DNNs with optimal resource on spatial and temporal basis. }Our key contributions are:\looseness-1
\vspace{-1mm}
\begin{itemize}[itemsep=-0.50mm,leftmargin=*]
 \item We investigate the extent to which a DNN can exploit parallelism (\S\mbox{\ref{sec:background}}), and devise an analytical model to demonstrate this limitation of typical DNNs \Scut{inherent }when performing inference with GPUs(\S~\ref{sec:model_dnn_parallelism}).
 
    \item \Scut{Finally, }We develop a Spatio-Temporal scheduler for DNNs, using the GPU\% and batch size derived from our analytical models, to maximize inference throughput while allocating GPU resources fairly (\S\mbox{\ref{sec:scheduling}}).
   
    \item We develop an optimization framework to determine the optimal \Scut{combination of }DNN Batch size and GPU\%. We evaluate the efficacy of GPU usage when choosing the optimal batch size and Knee GPU\%.\Scut{allocation} (\S~\ref{sec:batching_in_dnn}).
    
    \item We compare \name's approach with the Triton server and other state-of-the-art scheduling algorithms.\looseness-1 
\end{itemize}

\Scut{
We use measurements to understand the extent of parallelism (\S~\mbox{\ref{sec:background}}) and then use those insights to create a model that establishes the existence of a Knee for DNNs (\S~\mbox{\ref{sec:model_dnn_parallelism}}). Since batching is critical to throughput but has an impact on inference delay, we build an optimization formulation to find the right balance(\S~\mbox{\ref{sec:batching_in_dnn}}). Armed with all this, we build a fair spatio-temporal scheduler(\S~\mbox{\ref{sec:scheduling}}).}
\Scut{We first describe advantages of spatial sharing of GPU with CSS and necessity of spatio-temporal sharing of GPU in \S~\mbox{\ref{sec:background}}. Based on our observations for DNN inference, we develop a model for DNNs to establish the existence of a Knee\% in \S~\mbox{\ref{sec:model_dnn_parallelism}}.  Knowing the Knee\% of a DNN model, we formulate an optimization that will strike a balance between GPU\% and inference batch size such that a DNN model utilizes the GPU efficiently while meeting its inference SLO in \S~\mbox{\ref{sec:batching_in_dnn}}. We utilize the knowledge of the proper batch size and Knee GPU\% as the 'demand' to be used in a fair, spatio-temporal scheduler described in \S~\mbox{\ref{sec:scheduling}}. We evaluate our framework with popular DNN models, on a server with a single NVIDIA V100 GPU. Our framework can multiplex several DNNs while providing high throughput and controlled, low latency for those DNN applications, enabling them to meet their inference time SLOs as shown in \S~\mbox{\ref{sec:validation}}. }

\Scut{
We create a proper GPU virtualization framework upon GSLICE~\cite{GSLICE} platform that obtains the "right-amount" of GPU resource and batch size for multiplexing DNNs on GPU. This framework schedules DNN models on spatial a

Towards that goal, we simulate a DNN, demonstrating why there are inherent limits in parallelism of a DNN model, and how that leads to under-utilization of the GPU. We also profile popular DNN (CNN and RNN) models on real GPU hardware to show that these limits exist in these popularly used DNNs. With the 
}
\Scut{
\anote{Need to move next statement later}Further, we augment CSS with temporal scheduling of the DNN applications to accommodate running a large number of applications within the specified SLO constraints.

}

\Scut{
 \update{However, these adaptive batching algorithms only consider batching to meet a deadline. However, a change in GPU resource of a DNN model (due to spatial sharing of GPU) can change the inference latency of a certain batch. Moreover, we also see that with very large or small batch size, the DNN inference is not optimal with respect to GPU utilization.

These challenges necessitates a proper GPU virtualization framework that allocates GPU resources to multiplexing DNNs on a spatial \emph{and} temporal basis such that the DNN applications can fulfill their inference deadline while also providing high throughput.\Scut{ \ie big batch size.}\knote{you want a proper GPU virtualization framework} \Scut{a GPU based DNN framework that can schedule multiple applications based on their inference deadline batch size and spatial resource requirement.} \knote{badly written sentence. rewrite: AD Please check again.}
}
\update{The GPU virtualization framework should also provide high throughput for each individual application running in the GPU. Batching the inference requests together allows DNN models to exploit even more GPU parallelism, therefore, increases the inference throughput.} However, Batching \Scut{a larger number of tasks together} also increases the latency for the overall DNN operation. While this might be acceptable for training, it is likely not as acceptable for inference tasks which are conducted in real-time, usually with fixed deadlines and SLOs. Nonetheless, state-of-the-art works like~\cite{crankshaw2017clipperlong,shen2019nexuslong} incorporate adaptive batching techniques to maximize inference throughput.\Scut{ by trying to batch as many requests that can be completed within the specified latency budget.} \update{However, these adaptive batching algorithms only consider batching to meet a deadline. However, a change in GPU resource of a DNN model (due to spatial sharing of GPU) can change the inference latency of a certain batch. Moreover, we also see that with very large or small batch size, the DNN inference is not optimal with respect to GPU utilization. A GPU virtualization framework requires a adaptive batching technique which batches with respect to spatial resources of a model as well as with GPU utilization.}
}

\Scut{
We then use the solution of the optimization formulation in a spatio-temporal scheduler to schedule the DNN model inference. with a max-min fair spatial allocation with the optimal batch size to meet the deadline set by applications.}

\Scut{
In this paper we create and describe a proper GPU virtualization framework that obtains the "right-amount" of GPU resources and batch size for multiplexing DNNs on a GPU. This virtualization framework, allocates right GPU resources, schedules the DNN models on spatial \emph{and} temporal basis such that the DNN applications can fulfill their inference deadline while also providing high throughput. We demonstrate the capabilities of this virtualization framework by multiplexing GPU with real and popular DNN models.
}

\update{\cut{In this paper, we present a framework which utilizes GPU most efficiently while multiplexing multiple DNNs. We first explore the reasons why DNNs do not fully utilize the GPU and seek to scheduling of DNN applications. }
\cut{First, we explore the reasons why DNNs do not fully utilize the GPU resources.} 
} 
\cut{We further present the key CPU-GPU primitives to overcome the challenges of low GPU utilization and provide high inference throughput while meeting SLO specified by the programmer. }

\Scut{
Another application to benefit from offloading DNN computation to edge cloud could be autonomous driving. Autonomous vehicles depend on processing the streaming input from multiple sensors such as cameras, RADAR etc. in real-time to safely navigate the roads. Leading autonomous vehicles manufacturers such as Tesla and Waymo use Deep Neural Networks (DNN) to perform image segmentation, object detection, depth estimation and other safety critical systems~\cite{tesla_autopilot_ai,bojarski2017explaining,bansal2018chauffeurnet}. As the self-driving systems makes further improvements in autonomy, they will require more ML computation. Upgrading and putting new hardware with more compute power could help. However, upgrading hardware for already existing vehicles will be expensive and might not be possible in some cases. An alternative to upgrading hardware is to offload the processing to the edge cloud which consists of hardware such as GPU which is necessary to process data with low latency. Edge cloud servers in road side units (RSUs) and base station (BT) have been proposed to facilitate low latency offloading~\cite{feng2018mobile}.}
\section{Related Work}


\Scut{
Previous works~\cite{yu2017scalpel,kim2018improving,hauswald2015djinn,xiao2018gandiva,shen2019nexus,201468} have explored methods to increase the GPU utilization and to lower the latency and improve the overall throughput of DNN tasks. First is the whitebox approach that tries to improve the performance of a DNN by having algorithmic optimizations of the DNN. These optimizations enable  
DNN operations to take full advantage of GPU resources, \ie use the maximum possible GPU computational power (hardware threads) at any time. 
For example, ML frameworks include features such as graph optimizers which merge compatible DNN layers to increase the amount of computation performed in a single layer~\cite{chen2018tvm}, and \Scut{fusing multiple kernels}~\cite{jia2019optimizing,du2017fused,song2017towards}. 
Algorithmic improvements usually require whitebox models that can be changed~\cite{he2014reshaping,ahn2020chameleon}. 
An alternative approach is to provide system enhancements that effectively utilize the GPU \eg by multiplexing the execution of multiple DNNs on the GPU~\cite{xiao2018gandiva,shen2019nexus,201468}.
Similarly, batching requests together increase the compute requirement for a DNN kernel while also amortizing the overhead of the sequential tasks in the GPU. However, both these solutions which seek to increase GPU utilization have substantial challenges. }
\noindent\textbf{GPU Multiplexing:} Multiplexing GPU to increase the GPU utilization and system throughput has been discussed in many studies. Proprietary products such as Nutanix~\mbox{\cite{nvidianutanix}}, vGPU~\mbox{\cite{nvidiavgpu}} utilize GPU virtualization to multiplex GPU across VMs.
Many 
consider temporal multiplexing and seek increased GPU utilization through batching and better scheduling \revise{\mbox{\cite{crankshaw2017clipperlong,227623long,shen2019nexuslong,gujarati2020serving,gao2018low,sagemaker,10.1145/3369583.3392679}}}.
Gandiva~\cite{xiao2018gandivalong} and Mystic~\cite{ukidave2016mystic} address multiplexing the GPU while observing but not solving the interference caused while multiplexing DNNs in the GPU.\Scut{ Gandiva estimates an 18\% slowdown with two DNNs running in the GPU concurrently.} Unlike these, our work \Scut{develops a solution that }can concurrently run multiple applications in GPU, improve GPU utilization \emph{and} reduce or eliminate the interference through 
controlled spatial multiplexing.\looseness-1 

\Scut{NVIDIA's Triton Inference Server (Triton)~\cite{tritonserver} supports spatial sharing, especially for concurrent inference by multiple instances of the same DNN model, attaining higher inference throughput. However, Triton temporally shares the GPU using the default hardware scheduler~\mbox{\cite{triton-temporal}}  when serving inference requests for different models.
\knote{I wonder if the above sentences regarding Triton should go up into the intro, introducing it before we talk about comparison with \name?}
We find that this results in lower throughput than spatial sharing. With our proposed Spatio-Temporal scheduling, frameworks such as Triton can also get higher throughput, while multiplexing multiple different models.
}

\textbf{Spatial Multiplexing of GPU:} GSLICE~\cite{GSLICE} utilizes CUDA MPS to spatially share the GPU among multiple DNN applications. 
However, it partitions the GPU statically and does not schedule the execution of DNNs. With GSLICE, executing a large number of models potentially cause each model get a small GPU slice (less than the Knee), leading to higher inference latency and lower throughput. Moreover, the lack of a scheduler means it is insufficient for deadline-driven inference scenarios. We compare \name{} with GSLICE in \S\ref{sec:validation}.\looseness-1

Laius~\mbox{\cite{zhang2019laius}}, G-Net~\mbox{\cite{zhang2018glong}}, Gost~\cite{gost} and Baymax \mbox{\cite{chen2016baymax}} spatially multiplex GPU kernels. Unlike these works, our platform focuses on the spatially multiplex entire DNNs consisting of multiple kernels. \Scut{our platform also avoids more intrusive algorithm modifications to DNN execution sequences.} Moreover, we run DNN applications in their native DNN framework (e.g., PyTorch, TensorFlow) without any algorithmic modifications, unlike the whitebox approach of Laius and Baymax.\Scut{
Laius~\cite{zhang2019laius}\Scut{utilizes NVIDIA's MPS and spatially multiplexes the GPU among different applications to improve the throughput.} 
\Scut{Our spatial multiplexing also uses the CUDA MPS similar to Laius's approach.Laius}\Scut{
does not consider SLO, 
nor does it support adaptive batching or variation in batch size while running GPU applications. } maintains a pool of processes with different GPU\% so a kernel can be executed in GPU with desired GPU\%.
Our work differs from Laius as we consider spatio-temporal scheduling for running an entire DNNs made of multiple kernels\Scut{instead of individual kernels as Laius}.
\revise{Baymax\mbox{\cite{chen2016baymax}} uses ML to predict a kernel's runtime to reorder the execution sequence to meet deadlines. Unlike Baymax that focuses on individual GPU kernels, our approach focuses on the DNNs, avoiding the more intrusive algorithm modification to the kernel execution sequence. 
Moreover, we run DNN applications in their native DNN framework (e.g., PyTorch, TensorFlow) without any modification compared to Laius and 
Baymax's whitebox approach. We also consider batching during DNN inference, and  support multiplexing applications with varying SLOs in this paper.}\looseness-1\Scut{and changing spatial demand due to batching in DNN.\anote{Need to include follow on work for Laius}}}
\Scut{Some earlier work has considered spatial sharing of the GPU using CUDA streams and CUDA Multi-Process Service (MPS)\Scut{to increase GPU utilization}}\Scut{ \cite{zhang2018glong} used multiple virtual network functions 
to spatially share the GPU with streams.}\Scut{\mbox{\cite{deadline-aware}} utilizes streams to schedule multiple tasks, prioritizing them based on how much head-room each task has.} S3DNN\mbox{\cite{zhou2018s3dnn}} (uses Streams) and Prophet\mbox{~\cite{chen2017prophet}} (uses MPS) and CuMAS~\cite{cuMas}
profile each kernel and use a shim to 
capture kernel launches and reorder kernel executions for proper spatial sharing.\Scut{utilizes a similar kernel launch capture method. } In contrast, our approach does not require a shim or reordering of kernels and works in a black box manner, without requiring an application's individual kernel profile (which may not be available).\looseness-1

\Scut{
Furthermore, utilizing streams requires users to have in-depth understanding of workload of each kernel, so that kernels in different streams can execute concurrently without interference.}
\Scut{While these works understand the limits of temporal sharing of GPU and propose spatial sharing, }\Scut{ they 
use techniques such as limiting the number of threads per kernel, scheduling kernels appropriately so as to fit while 
running concurrently, and launching the kernels from a single process utilizing CUDA streams. they require a more in-depth characterization and modification of the kernels before launching them to execute in the GPU.  In contrast, our approach does not need to modify the application's kernel and programming logic.}\Scut{ and provides resource isolation. Moreover, user-driven dynamic allocation of GPU resources still remains a challenge, which we address in this paper.}\looseness-1\Scut{ CSS avoids interference seen by sharing the GPU temporally or spatially with default MPS.} 

\textbf{DNN's limits on Utilizing GPUs:} Several works~\cite{jeonanalysis,yeung2020towards2, yeung2020horuslong} have discussed the under-utilization of GPU by DNNs, and have proposed algorithmic optimizations that make DNN kernel computation more efficient\Scut{such as fusing multiple layers}~\cite{jia2019optimizing,du2017fused,song2017towards,chen2018tvmlong}\Scut{ and making individual GPU kernel's computation more efficient~\cite{}}. These solutions require whitebox models that can be changed. There have been works analyzing how DNN's exploit parallelism. \cite{DBLP:journals/corr/abs-1901-10008, jain2018dynamiclong} show that DNNs attain a much smaller number of FLOPS than what a GPU can provide. Poise~\cite{dublish2019poise} and~\cite{kayiran2013neitherlong} shows that the high data load latency from the GPU memory to the processing unit is also a reason for the limit in parallelism.
\cite{liang2022modeldriven} creates an analytical model to predict the inference latency and mainly utilize temporal queuing solution to meet deadlines.~\cite{liang2022modeldriven}'s model uses default MPS, and due to interference causing increased latency, they limit the number of models spatially sharing the GPU at a time. On the other hand, \name{} provides fine-grained spatial and temporal control of resources of the GPU and thus is able to run far more models with larger batch sizes without interference. With a spatio-temporal scheduler \name{} utilizes resources both spatially and temporally to meet the inference deadline. \cite{architectural-impli} shows lack of resources in CPU and GPU spatial resources will greatly slowdown  GPU execution. Our work complements~\cite{architectural-impli} by demonstrating a method to find the Knee beyond which applications fail to utilize GPU efficiently. We utilize understanding from these related work to create an analytical DNN model that helps deriving the Knee\% necessary for inference without slowdowns. Furthermore, we evaluate our methods in a real system.\looseness-1

\textbf{Multi-Instance GPUs (MIGs)} such as the NVIDIA A100 are hardware-based approaches for coarser-grained, spatial multiplexing. MIGs allow static partitioning of a GPU into multiple smaller GPU instances (up to 7 instances with the A100). However, MIGs require the GPU to be reset or VMs to be restarted to change the resource allocation. This causes significant downtimes as all the processing using the GPU has to also be restarted. \name{}'s spatio-temporal scheduling avoids the GPU reset and quickly allocates the desired GPU resources.  
Moreover, note that MIG GPUs are also able to run as a single GPU (similar to V100). Thus, they can benefit from \name{} without any modification.

\Scut{
\noindent\textbf{Studies of GPU Architectures:}
Pagoda~\mbox{\cite{yeh2017pagoda}} and \mbox{\cite{pourghassemi2020brief}}, show that launching small kernels incurs high latency.\Scut{ and low GPU utilization. Therefore, they launch small kernels by using a kernel that is already running in the GPU, lowering the overhead and this increases GPU utilization. In our work, we provide lower GPU resources to tasks that have limited parallelism using CSS, and therefore can utilize more of the GPU for other tasks.} HSM\mbox{\cite{zhao2020hsm}} presents a model to predict the extent of slowdown for kernels while multiplexing the GPU in a simulation. Wrapped-Slicer~\cite{warped-slicer} and~\cite{simultaneous-multikernel} explore GPU architecture to spatially multiplex individual SMs in a GPU.\Scut{ HSM's model can predict the extent of slowdown accurately in a simulation.} 
These works focus on increasing the throughput of individual SMs and requires\Scut{additional memory counters and} architectural changes to existing hardware. 
In contrast, our work focuses on increasing the overall throughput of the inference system while working with existing real hardware. Additionally, we spatially share the GPU at the granularity of an SM, so that it is easily implementable in current GPU runtime environments with a software enhancement.\looseness-1 }\Scut{And in~\cite{pourghassemi2020brief}, the authors show that the time to load multiple sequentially-run GPU kernels as being responsible for a limit on the parallelism a DNN can exploit. In~\cite{kayiran2013neitherlong}, the authors show that using more GPU threads does not always result in better performance because memory latency becomes a limiting factor.} \looseness-1
\Scut{Equalizer~\cite{sethia2014equalizer} notes that heavy demand of resource by a thread in GPU can interfere with other concurrently running threads. This limits the performance of GPU applications running in GPU. Equalizer proposes a 
runtime system that monitors the number of waiting warps launched by kernels to check for possible contention for memory or another critical resource. On detecting contention, Equalizer determines the appropriate dynamic allocation of resources, resizing of threadblock, as well as change of GPU SM's memory or core frequency, such that bottleneck resource contention is reduced. The authors' demonstrate that their technique speeds up the benchmark kernels on GPGPU-sim. However, Equalizer's solution uses counters and primitives such as changing memory and core frequency, which are not available in real GPU hardware. Thus, user-driven dynamic allocation of GPU resource still remains a challenge, which we address in this paper. }

\Scut{We further analyze how DNN's utilize GPU resources at run time for inference and demonstrate that the amount of parallelism they exploit is limited. We also present a formulation to find the best operating point for a DNN in terms of batch size and the share of GPU resources. This allows for multiplexing the GPU across concurrent models, thus increasing the utilization/efficacy of the GPU.\looseness-1   }

\Scut{
\skote{We need to first make the case for why Modelling the Parallelism of DNN (CNN and RNN) models is important. Then point the learning towards developing the primitives for aiding better GPU utilization and improved application level performance. Finally show the benefits of proposed modelling in terms of GPU utilization and performance for one or more multiplexed applications}
}
\section{Understanding DNN Parallelism through Measurement}
\label{sec:background}

\Scut{
\begin{figure}
    \centering
    \subfloat[V100 GPU]{
    \includegraphics[width=\linewidth]{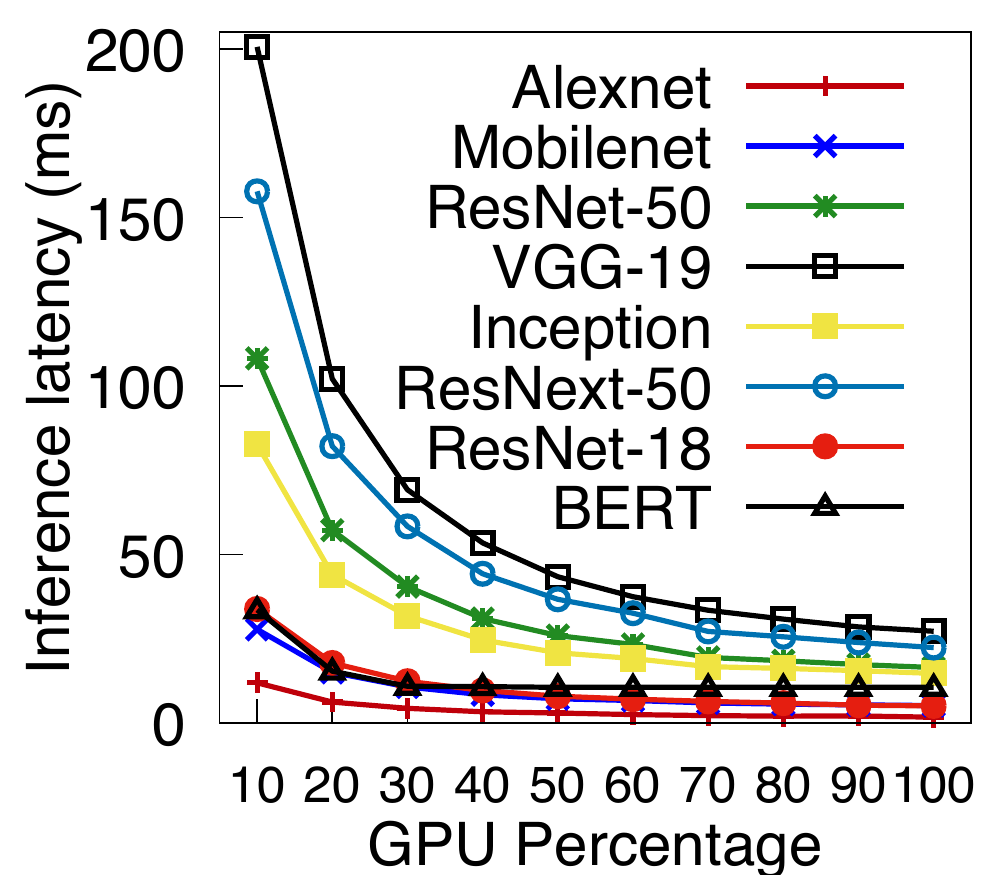}
    \label{fig:inference_latency_batch}
    }
    \Scut{
    \subfloat[P100 and T4 GPUs]{
    \includegraphics[width=.5\linewidth]{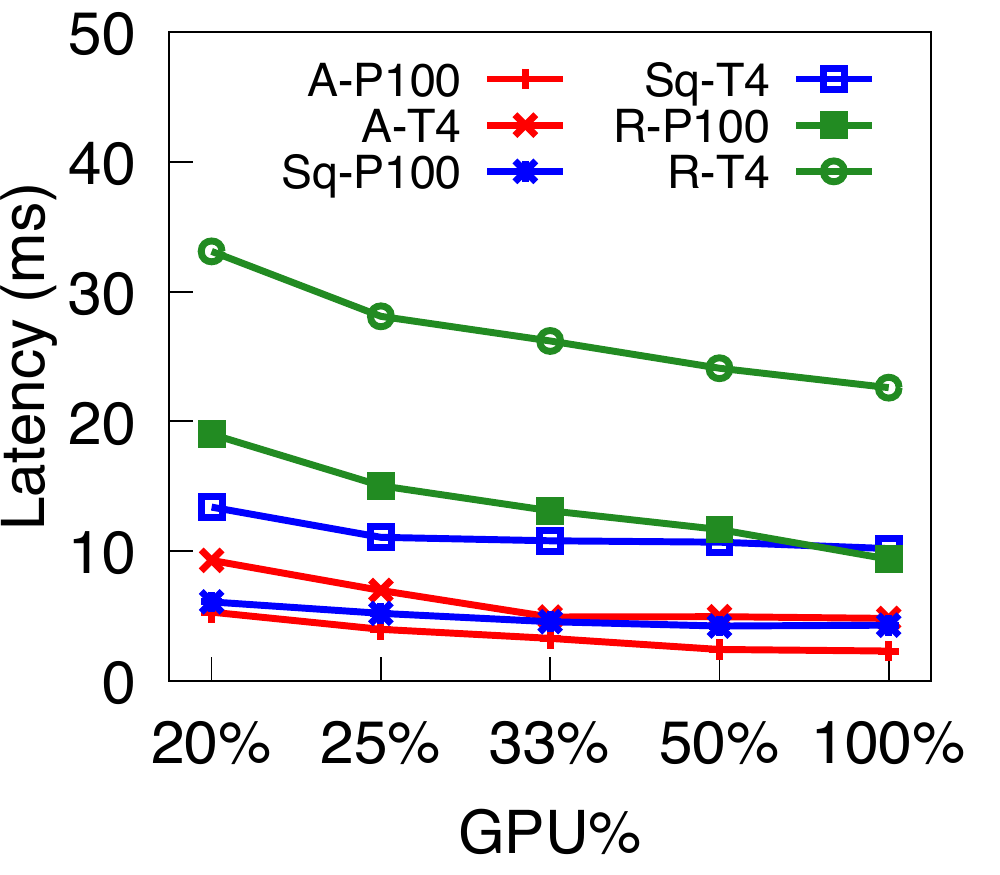}
    \label{fig:p100-t4}
    }}
    \vspace{-2mm}
    \caption{Inference latency of DNN models across different GPU\% (b) A=Alexnet, Sq=Squeezenet, R=ResNet-50}
\end{figure}

}
\Scut{
\begin{figure*}
    \centering
    \begin{minipage}{.24\linewidth}
    \includegraphics[width=\linewidth]{figures/model_latencies/latency_of_batch16.pdf}
    \vspace{-6mm}
    \caption{\Scut{GPU\% vs Inference latency} V100 GPU }
    \label{fig:inference_latency_batch}
    \end{minipage}
    \begin{minipage}{.24\linewidth}
    \includegraphics[width=\linewidth]{figures/t4-p100-results/p100-t4-knee.pdf}
\vspace{-6mm}
\caption{\Scut{Alexnet (A), Squeezenet (Sq) and ResNet-50(R) Latency vs. GPU\% in} \revise{P100 and T4 GPUs\Scut{profile}}}
\label{fig:p100-t4}
    \end{minipage}
\Scut{
    \begin{minipage}{.25\linewidth}
    \vspace{-18mm}
    \centering
        \captionof{table}{\revise{GPU Spatial Sharing options: 3 Alexnet instances}}
        \vspace{-1mm}
        \resizebox{\columnwidth}{!}{%
    \begin{tabular}{|c|c|c|c|}\hline
         & Streams & MPS & CSS  \\\hline
         Mean Lat. & 2.2 ms & 1.78 & 1.55 \\
         P-99 Lat. & 4.2 ms & 2.94 & 1.58\\
       Throughput & 1242 ips. & 1675 & 1836\\\hline
    \end{tabular}}
    \label{tab:streams_vs_css}
    \vspace{-5mm}
\end{minipage}
}
\Scut{
 \begin{minipage}{.44\linewidth}
 \vspace{0pt}
 \captionsetup{justification=centering}
    \includegraphics[width=\linewidth]{figures/GPU_sharing_diagram/gpu_sharing_new.pdf}
    \vspace{-6mm}
    \caption{GPU multiplexing scenarios\Scut{\\ with different DNNs}}
    \label{fig:spatio-temporal-scheduling}
    \end{minipage}
}
 \begin{minipage}{.25\textwidth}
 \includegraphics[width=\linewidth]{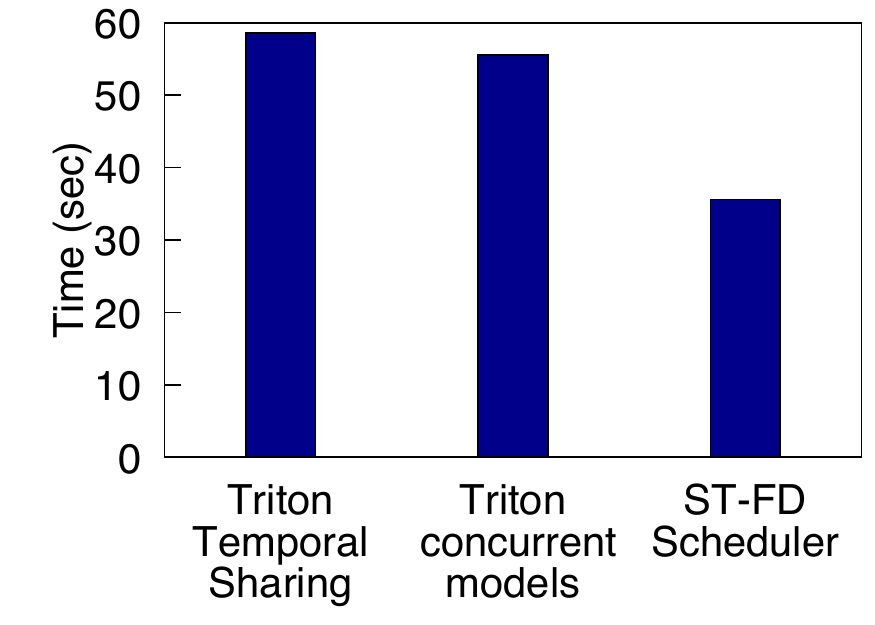}
 \vspace{-6mm}
    \caption{Task completion time: Triton \& S-T scheduler\looseness-1}
    \label{fig:triton-comparison}
 \end{minipage}
    \vspace{-3mm}
\end{figure*}

}

\Scut{It is helpful to judiciously utilize cloud computing resources, especially in environments such as edge clouds, which may have limited resources, in contrast to the ’almost infinite scalability’ of resources in centralized clouds. 
This is true for GPU resources used in the edge to provide low latency services that require significant compute power, such as DNNs. However, sharing the GPU hardware among multiple DNN applications, as a way of improving utilization, presents challenges. Various works have proposed techniques to multiplexing the GPU of a cloud service~\cite{hauswald2015djinn,xiao2018gandiva,shen2019nexus,201468}. However, they consider the GPU as a monolithic entity, and schedule individual applications to run one at a time on the GPU. Multiplexing in the GPU is done by temporally sharing the GPU among multiple applications as seen in Fig.~\ref{fig:GPU_Sharing_methods}. However, modern GPUs 
provide a large number of Streaming Microprocessors (SMs).
Each SM, akin to a CPU core, is capable of executing GPU kernels independent of others. Temporal sharing of a GPU can be wasteful. An application with low GPU compute requirement, needing only a limited number of SMs, is given all of the GPU for its time slice. 

NVIDIA GPUs do allow spatial sharing among multiple applications, with the use of Hyper-Q, CUDA Streams, and CUDA Multi-Process Server (MPS). SMs not used by one application might be used by other concurrently running applications~\cite{zhang2018g,liang2014efficient}. While the default MPS and CUDA streams might be helpful in increasing GPU utilization with applications having a lower computing requirement, compute heavy DNNs often interfere with each other while spatially sharing a GPU in this manner. This interference leads to unpredictable and higher latency during DNN inference~\cite{shen2019nexus}. Newer NVIDIA GPU architectures (Volta, Turing, Ampere) allow GPU compute resource to be provisioned for each application. By providing a GPU\% in an environmental variable associated with a process, we can provision the maximum number of SMs an application can use. As resource provisioning is transparent to the application, all applications can be used without any changes. This makes it possible to consider the GPU as a collection of smaller logical GPUs which can be then scheduled to execute different DNN models, independent of each other as seen in Fig.~\ref{fig:GPU_Sharing_methods}, to facilitate a special form of multiplexing termed as~\textit{Controlled Spatial Sharing} (CSS). }     
\noindent\textbf{Experimental Setup and Testbed}:
We used a Dell Server 
with Intel(R) Xeon(R) Gold 6148 CPU with 20 cores, 256 GB of system memory, and one NVIDIA V100 GPU, and an Intel X710 10GbE NIC as our testbed. The V100 has 80 SMs and 16 GB of memory.\Scut{We also use an 8 NVIDIA Tesla T4 GPU cluster as part of our multi-GPU testbed. Each Tesla T4 GPU has 16 Gigabytes of GPU memory and 40 streaming multiprocessors. 
We use 4 quadport Intel XL710 10 GbE NICs (40Gb aggregate capacity) for serving inference requests to the GPU server/cluster. 
}
Our workload for the vision based DNNs (Alexnet~\cite{alexnetpaper}, Mobilenet~\cite{howard2017mobilenets}, ResNets~\cite{he2016deep}, VGG~\cite{simonyan2014very}, Inception~\cite{szegedy2015going}, ResNext~\cite{xie2017aggregated}) consists of color images of resolution 224$\times$224. This resolution choice is inspired by initial work~\cite{krizhevsky2012imagenet,simonyan2014very,torchvision_model_zoo}. For BERT~\cite{devlin2019bert}, a natural language processing DNN, we utilize sentences of 10 words.\looseness-1

We use OpenNetVM~\cite{zhang_opennetvm:_2016} to host our framework that runs multiple DNN models for inference. We use Moongen~\cite{moongen-imc2015} to transmit \textasciitilde1920 images/sec. on a 10Gbps Ethernet link. Our platform can batch input data to the desired batch size. We primarily report the execution time for inference in the GPU for all our experiments and do not consider the additional latency contributed by network protocols.\Scut{` including HTTP, TCP, or UDP.} Therefore, our results are independent of the network transport protocol used.
\Scut{In this paper, we use GPU\% to denote the percentage of the available\Scut{streaming multiprocessors (SM)(compute unit) of the GPU used.} SMs \Scut{\eg NVIDIA V100 GPU has 80 SMs; }\Scut{a process getting}\eg 50\% of V100 GPU \Scut{for a process} means the process is restricted to using at most 40 (out of 80) SMs. } We utilize CUDA Multi-Process Service (MPS) to spatially multiplex the GPU. We use \mbox{CUDA\_MPS\_ACTIVE\_}
\mbox{THREAD\_PERCENTAGE} environmental variable to provide GPU\%\Scut{for a process by setting as an environmental variable}.\Scut{ Once the process performs GPU initialization,} Once set, the GPU\% cannot be changed for a process.\looseness-1

\Scut{

\subsection{Selecting Spatial Sharing Method}
\Scut{We utilize Controlled Spatial Sharing (CSS) utilized in GSLICE~\cite{GSLICE} to spatially multiplex the GPU. CSS allows operator to provide }
Other existing spatial sharing methods\Scut{used in practice for spatially sharing the GPU, } \eg CUDA Streams and CUDA MPS (Multi-Process Service), do not allow the cloud operator to allocate a specified portion of GPU resources for each DNN model. Several studies~\cite{jain2018dynamiclong,xiao2018gandivalong,scrooge-socc}, as well as our study here, show that multiplexing using CUDA streams and default CUDA MPS leads to the execution of different models interfering with each other. This leads to higher, and unpredictable, inference latency. \Scut{We evaluated different options, such as Streams, CUDA MPS and Controlled Spatial Sharing (CSS)~\mbox{\cite{GSLICE}}, for spatial sharing of the GPU.} 
\Scut{CUDA Streams has been available for NVIDIA GPUs for several generations, while CUDA MPS and CSS are more recent (NVIDIA Volta and newer).} 
We spatially shared the GPU among 3 Alexnet
\begin{figure}
\vspace{-3mm}
    \centering
        \captionof{table}{\revise{GPU Spatial Sharing options: 3 Alexnet instances} (ips.=images/sec)}
        \vspace{-4mm}
    \begin{tabular}{|c|c|c|c|}\hline
         & \makecell{CUDA\\Stream} & \makecell{Default\\MPS} & CSS  \\\hline
         Mean Lat. & 2.2 ms & 1.78 & 1.55 \\
         P-99 Lat. & 4.2 ms & 2.94 & 1.58\\
       Throughput & 1242 ips. & 1675 & 1836\\\hline
    \end{tabular}
    \label{tab:streams_vs_css}
\end{figure}
instances to concurrently infer requests using default CUDA MPS, Streams and CSS, and show the models' mean and $99^{th}$ percentile (P-99) latency and average throughput (ips. = images/sec) in Table~\mbox{\ref{tab:streams_vs_css}}.
For both Streams and default MPS, the models interfere with each other, inefficiently utilizing the GPU (the P-99 is much higher than the mean, reflecting unpredictability) even when spatially sharing the GPU with another instance of the same DNN model. 
On the other hand, CSS's isolation of the compute resource and allocation of just the right, Knee\%, for Alexnet yields a lower latency and 1.5$\times$ higher throughput than Streams and 1.1$\times$ higher than default MPS. CSS's mean and P-99 latency are close, indicating very predictable inference runtimes. Thus, we choose CSS to spatially multiplex DNNs on GPUs. }
\Scut{CSS~\cite{GSLICE}, on the other hand, provides resource isolation, thus, avoiding interference and provides predictable latency for each model's inference. Therefore, we chose to utilize CSS for spatial sharing with our \name{} scheduler.}\looseness-1

\Scut{
\begin{figure}
\begin{minipage}{.33\linewidth}
\includegraphics[width=\linewidth]{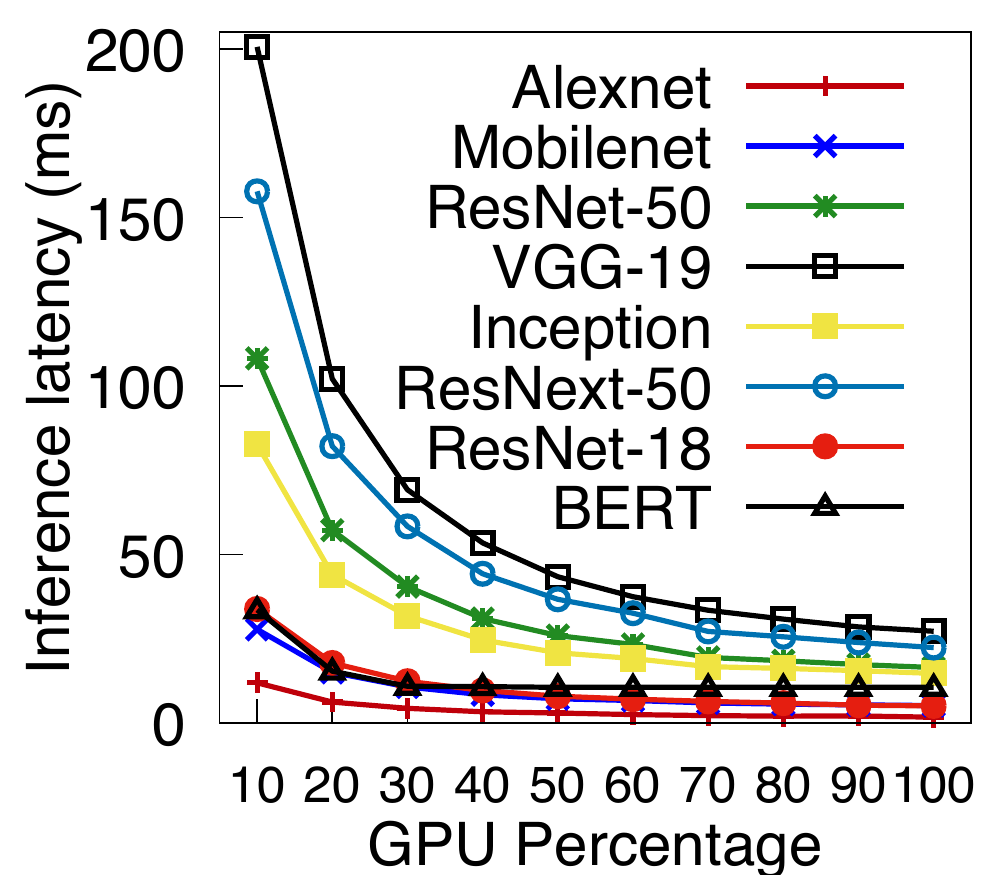}
    \vspace{-8mm}
    \caption{V100 latency vs. GPU\% (Batch = 16)}
    \label{fig:inference_latency_batch}
\end{minipage}
\begin{minipage}{.33\linewidth}
\includegraphics[width=\linewidth]{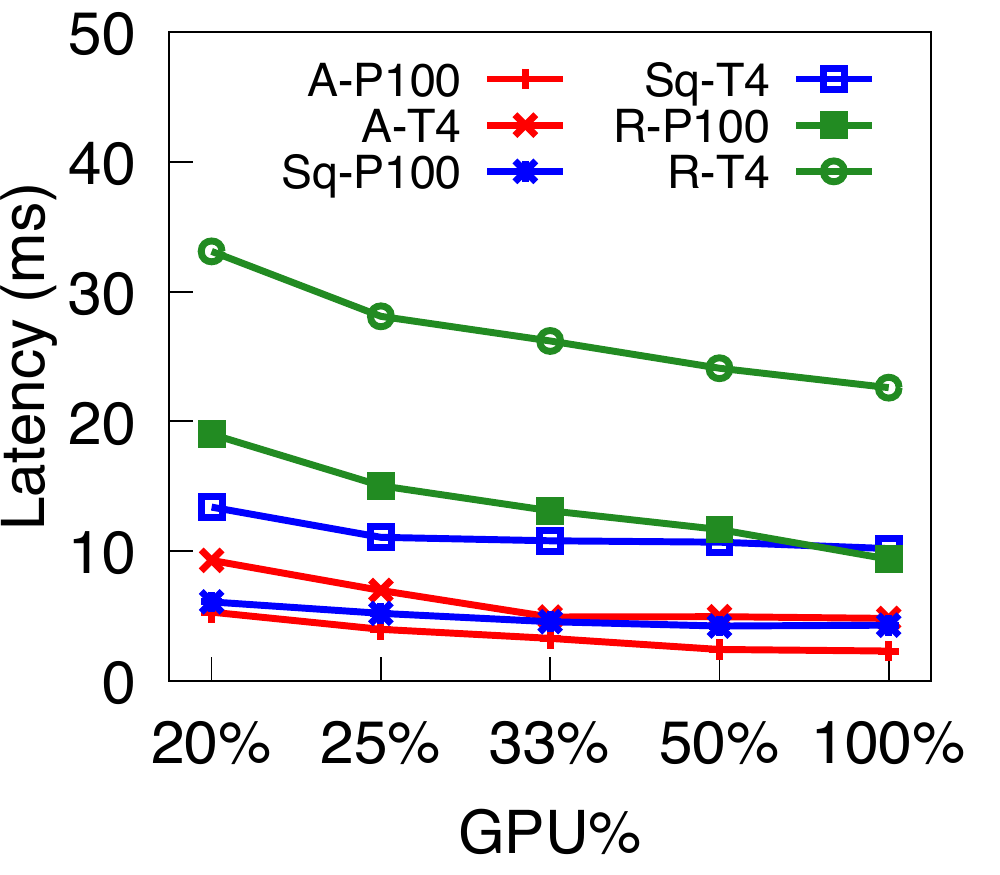}
\vspace{-6mm}
\caption{\Scut{Alexnet (A), Squeezenet (Sq) and ResNet-50(R) Latency vs. GPU\% in} \revise{P100 and T4 GPUs profile}}
\label{fig:p100-t4}
\end{minipage}%
\begin{minipage}{.33\linewidth}
\vspace{-20mm}
    \centering
    \vspace{2mm}
        \captionof{table}{\revise{GPU Spatial Sharing options: 3 Alexnet instances}}
        \vspace{-3mm}
        \resizebox{\columnwidth}{!}{%
    \begin{tabular}{|c|c|c|c|}\hline
         & \makecell{CUDA\\Stream} & \makecell{Default\\MPS} & CSS  \\\hline
         Mean Lat. & 2.2 ms & 1.78 & 1.55 \\
         P-99 Lat. & 4.2 ms & 2.94 & 1.58\\
       Throughput & 1242 ips. & 1675 & 1836\\\hline
    \end{tabular}}
    \label{tab:streams_vs_css}
    
\end{minipage}
\vspace{-7mm}
\end{figure}
}
\subsection{Measurement with ML Models}

We now present measurements performed on our testbed with multiple DNNs, to demonstrate the limits in the parallelism of those DNN models.\Scut{We observe in Fig.~\mbox{\ref{fig:inference_latency_batch}} that some DNN models do not see improved inference latency even when allocated more GPU resources, while others continue to show improvement, but with diminishing returns, \mbox{\ie} increase in GPU resources after a point does not result in a proportional decrease in latency. We call this point a "\textit{Knee}."} We measured the latency for inferring a batch of 16 images/sentences using different GPU\% for several popular DNN models using PyTorch framework. We utilize models with different compute requirements.

\Scut{
\begin{wrapfigure}[10]{l}{.3\columnwidth}
\vspace{-4mm}
\includegraphics[width=\linewidth]{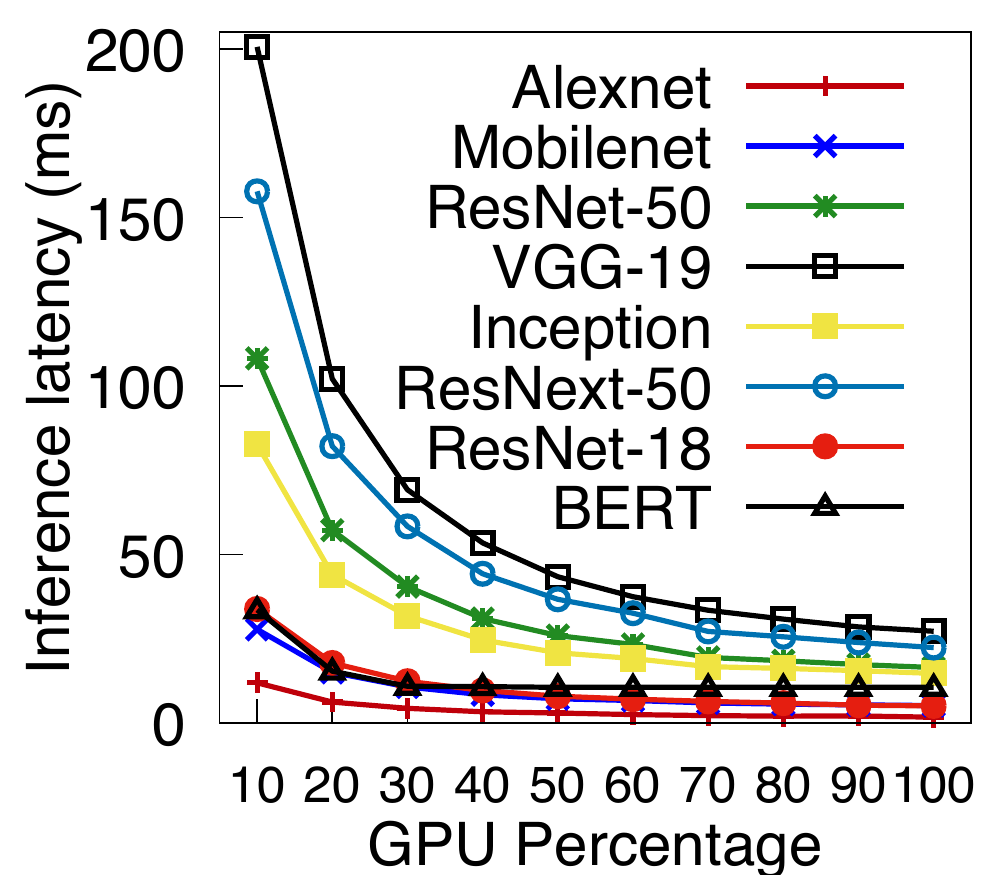}
    \vspace{-6mm}
    \caption{V100 latency vs. GPU\%}
    \label{fig:inference_latency_batch}
    \vspace{-4mm}
\end{wrapfigure}
}
From Fig.~\mbox{\ref{fig:inference_latency_batch}}, we see that the inference latency remains unchanged
above 30-50\% of GPU for most models (Knee point).\Scut{ We call this point a "\textit{Knee}."
\knote{didn't we already define 'knee'? Just say, this is the knee point for this model}}
With a smaller batch size, the Knee\% is lower (20\%-35\%). However, we also observe that using fewer than necessary SMs (low GPU\%) leads to an exponential increase in model latency (also observed in~\mbox{\cite{architectural-impli})}.\Scut{We also repeated the experiment with inference batch size of 16 and see  (Fig.~\mbox{\ref{fig:inference_latency_batch}}) that with higher batch size the knee\% increases to 30-40\% for the low latency models and up to 50-60\% for models that are compute intensive.}\Scut{ Providing 100\% GPU for those models, as done in temporal sharing of the GPU, is wasteful and results in frequent under-utilization of the GPU.} We observed a similar knee with other GPUs as well. We evaluated computationally light models, Alexnet (A-P100 and A-T4) and Squeezenet (Sq-P100 and Sq-T4) on both the P100 and T4 GPUs.\Scut{ and observed similar diminishing return in latency with those GPUs.} The T4 GPU supports CSS, but the P100 only supports default MPS. We present their results in Fig.~\mbox{\ref{fig:p100-t4}}. Even with different GPUs, we see the knee behavior in Alexnet and Squeezenet. Only the computationally dense ResNet-50 (R-P100 and R-T4) does not show an obvious knee. 
Both the P100 and T4 GPUs have lower computational capacity than the V100, therefore, ResNet-50 can fully utilize those GPUs. As the knee for these models exists in other GPUs as well, our platform can be used more generally in other GPUs as well.\Scut{ we ran multiple instance of same models (\eg 4 instances of Alexnet concurrently to provide 25\% GPU to each instance). As the knee exists for these models with other GPUs as well, our platform can be generally used across different GPUs.}\looseness-1

\Scut{
\begin{figure}

    \centering
    \begin{minipage}{.25\linewidth}
    \includegraphics[width=\linewidth]{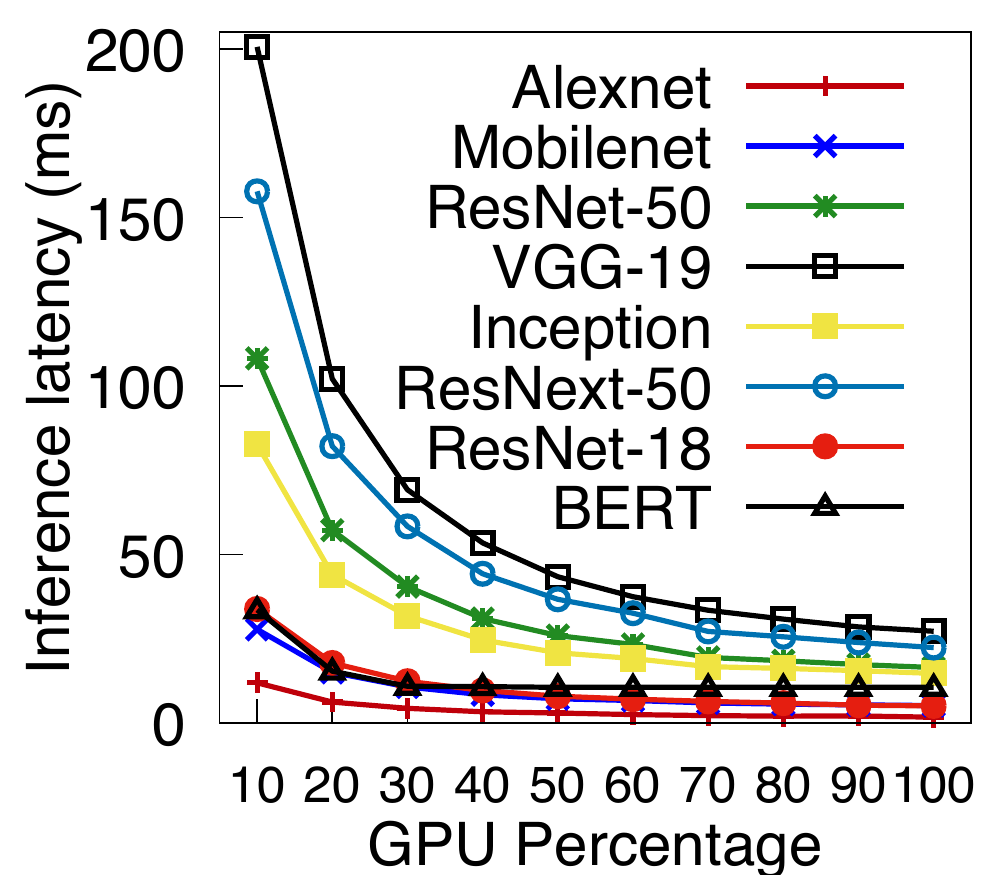}
    \caption{V100 latency vs. GPU\%}
    \label{fig:inference_latency_batch}
    \end{minipage}%
    
    \begin{minipage}{.25\linewidth}
    \captionof{table}{Triton Server Evaluation}
        \resizebox{\columnwidth}{!}{ \begin{tabular}{|c|c|}\hline
         Models & \makecell{Thpt.\\(imgs/s)}  \\\hline
         \makecell{Mobilenet\\\textbf{(2-instances)}}& 1675\\\hline
         \makecell{ResNet-50\\(\textit{R-50})\\\textbf{(2-instances)}}&989\\\hline
         \makecell{Mobilenet \& R-50\\\textbf{Temporal}}&804\\\hline
         \makecell{Mobilenet \& R-50\\\textbf{Concurrent}}&844\\\hline
    \end{tabular}}
    \label{tab:triton-multiple-instance}
    \end{minipage}
\end{figure}

}
\Scut{
We also evaluated a DNN model's limit of parallelism with  NVIDIA P100 (56 SMs) and T4 (40 SMs) GPUs. The T4 GPU supports CSS, but the P100 does not. 
Thus, for the P100 GPU we ran multiple instances of the same model to give each a certain GPU\%. \mbox{\eg} we ran 4 Alexnet model instances to approximate give 25\% of the P100 GPU, and the result is shown in Fig.~\mbox{\ref{fig:p100-t4}}. Even with different GPUs, we see the same knee behavior for Alexnet and Squeezenet. The computationally dense ResNet-50 does not show obvious knee only because this model can fully utilize the lower computational capacity of both the P100 and T4 GPU. But the key is that the knee exists for these DNN models with other GPUs as well. Our platform can be used generally with several different GPUs.\looseness-1}



\Scut{
\begin{figure}
\begin{subfigure}{.5\linewidth}
\includegraphics[width=\linewidth]{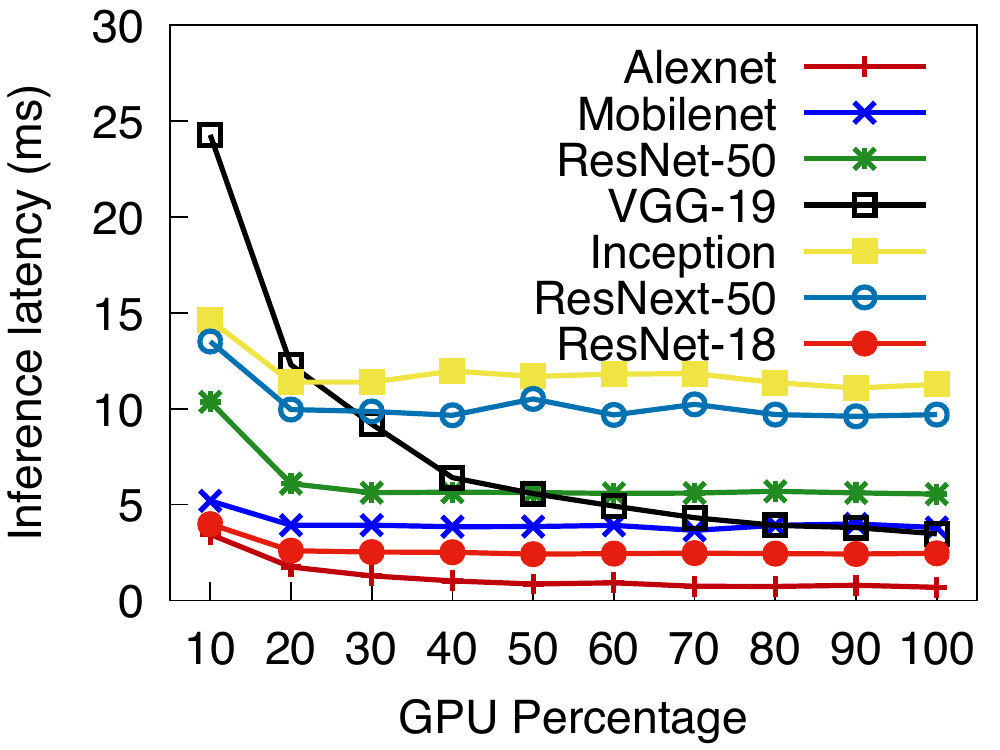}
    \caption{Pytorch}
\end{subfigure}%
\begin{subfigure}{.5\linewidth}
\includegraphics[width=\linewidth]{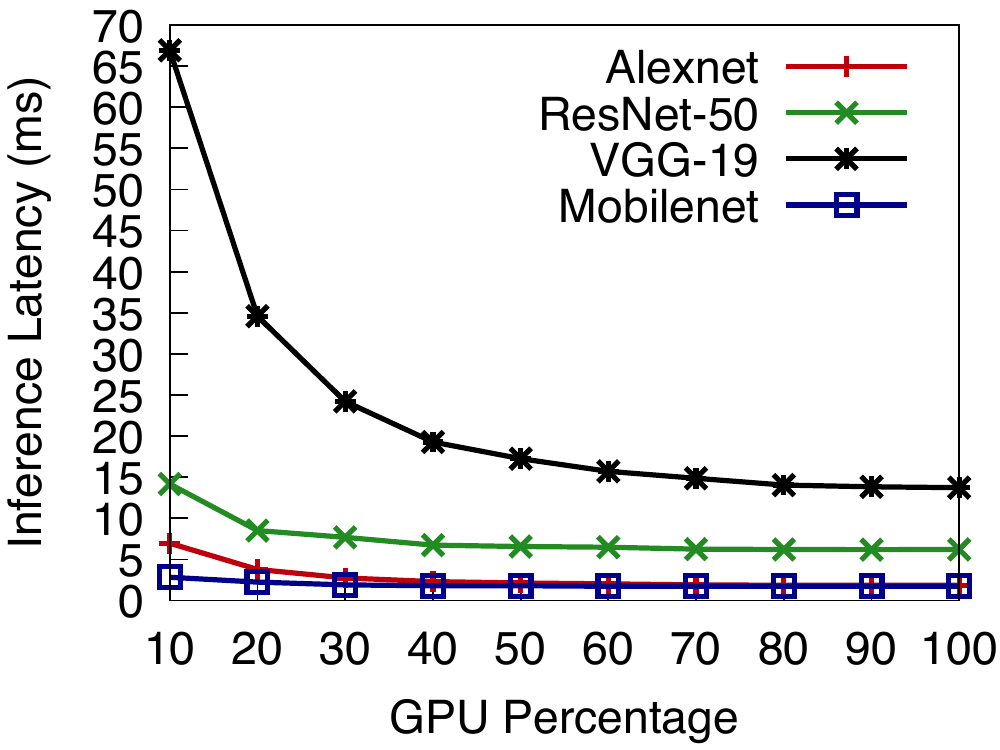}
\caption{TensorRT}
\end{subfigure}
\Scut{
\begin{subfigure}{0.5\linewidth}
\includegraphics[width=\linewidth]{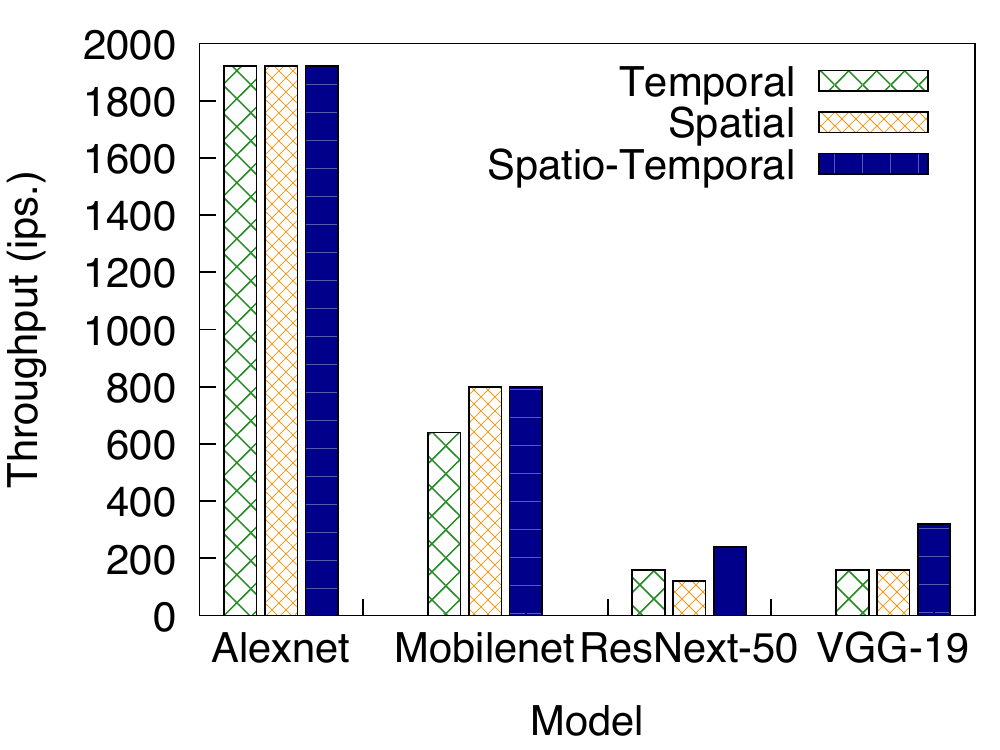}
\caption{Tensor}
\label{fig:spatio-temporal}
\end{subfigure}
}
\caption{Avg.inference latency of models vs. GPU\% \Scut{(b). Avg. throughput of different scheduling scenario}}
\label{fig:inference_latency}
\vspace{-8mm}
\end{figure}
}

\Scut{
\begin{figure}[t]
    \centering
    
\begin{subfigure}{0.25\linewidth}
    \includegraphics[width=\linewidth]{figures/model_latencies/latency_of_different_models.pdf}
    \caption{Pytorch}
    \label{fig:latency_per_gpu_Percentage_pytorch}
    \end{subfigure}%
    \begin{subfigure}{0.25\linewidth}
    \includegraphics[width=\linewidth]{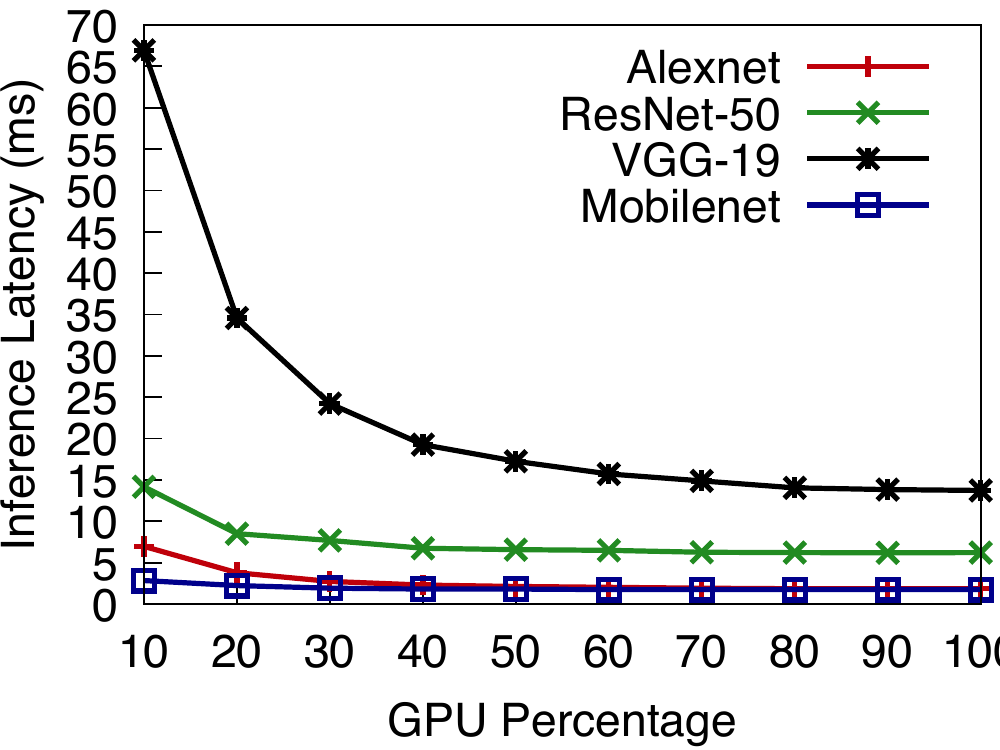}
    \caption{TensorRT}
    \label{fig:latency_per_gpu_Percentage_tensorrt}
    \end{subfigure}
    \caption{Average Inference Latency of Various models with different GPU\% in two different DNN platforms}
    \label{fig:inference_latency}
\end{figure}
}

\Scut{
\begin{figure}
\subfloat[V100 GPU]{
    \includegraphics[width=.49\linewidth]{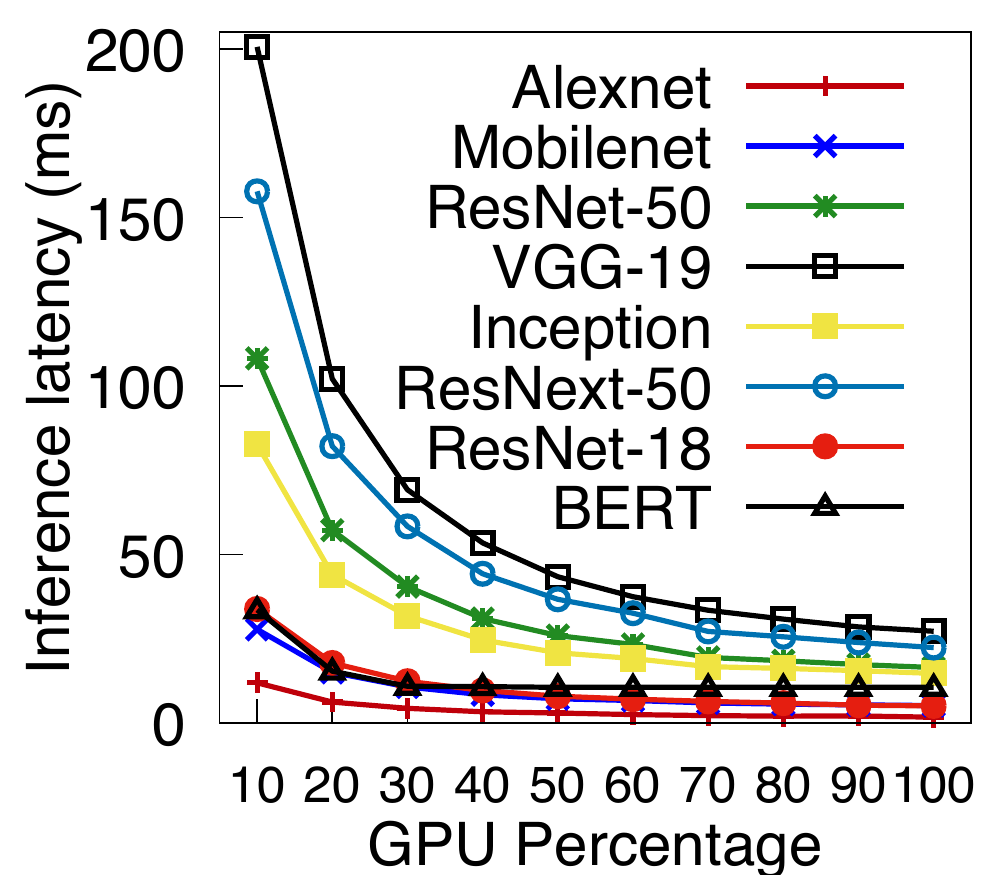}
    \label{fig:inference_latency_batch}
    }
\subfloat[\Scut{Alexnet (A), Squeezenet (Sq) and ResNet-50(R) Latency vs. GPU\% in} \revise{P100 and T4 GPUs profile}]{
\includegraphics[width=.49\linewidth]{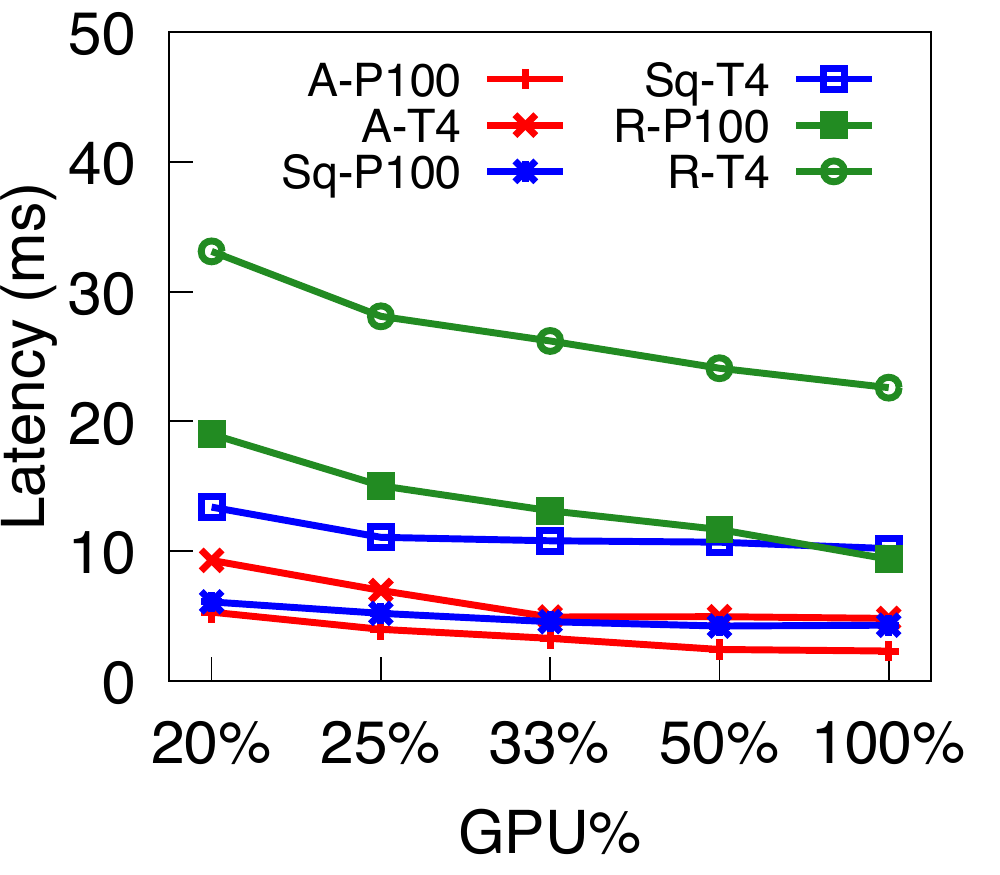}
\label{fig:p100-t4}
}
\caption{(a) Latency vs. GPU\% in V100 GPU (Batch = 16);(b) Alexnet (A), Squeezenet (Sq) and ResNet-50 (R) Latency vs. GPU\% in P100 and T4 GPUs (Batch=1)}
\end{figure}

}

\Scut{
\begin{figure}
    \centering
    \begin{subfigure}{.5\linewidth}
    \includegraphics[width=\linewidth]{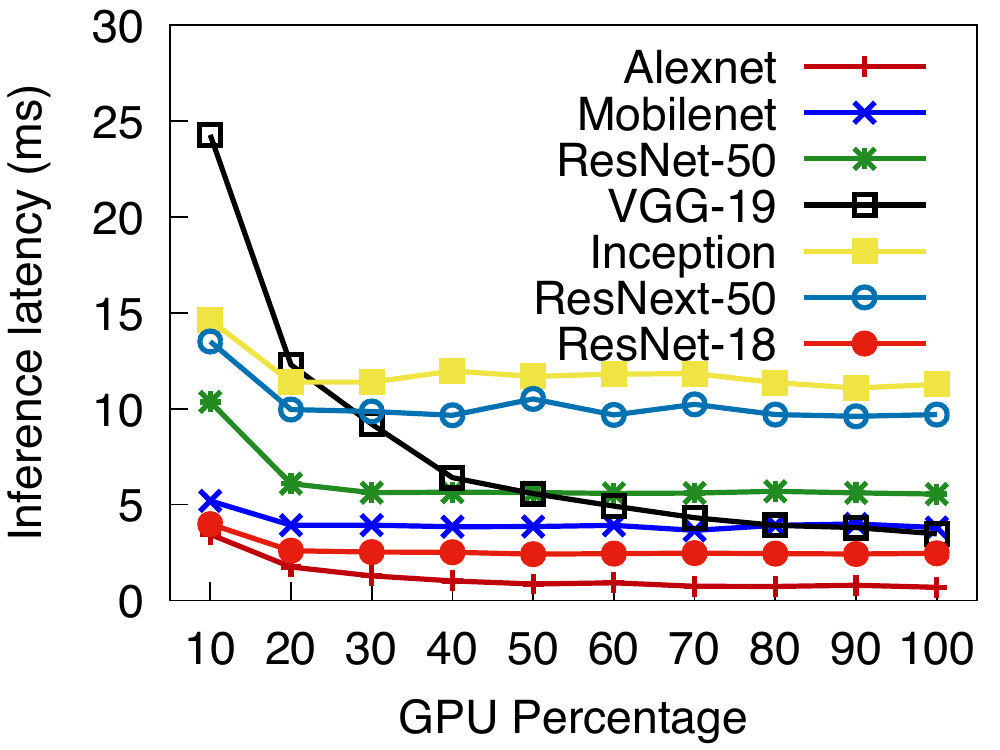}
    \caption{Batch Size = 1}
    \label{fig:inference_latency}
    \end{subfigure}%
    \begin{subfigure}{.5\linewidth}
    \includegraphics[width=\linewidth]{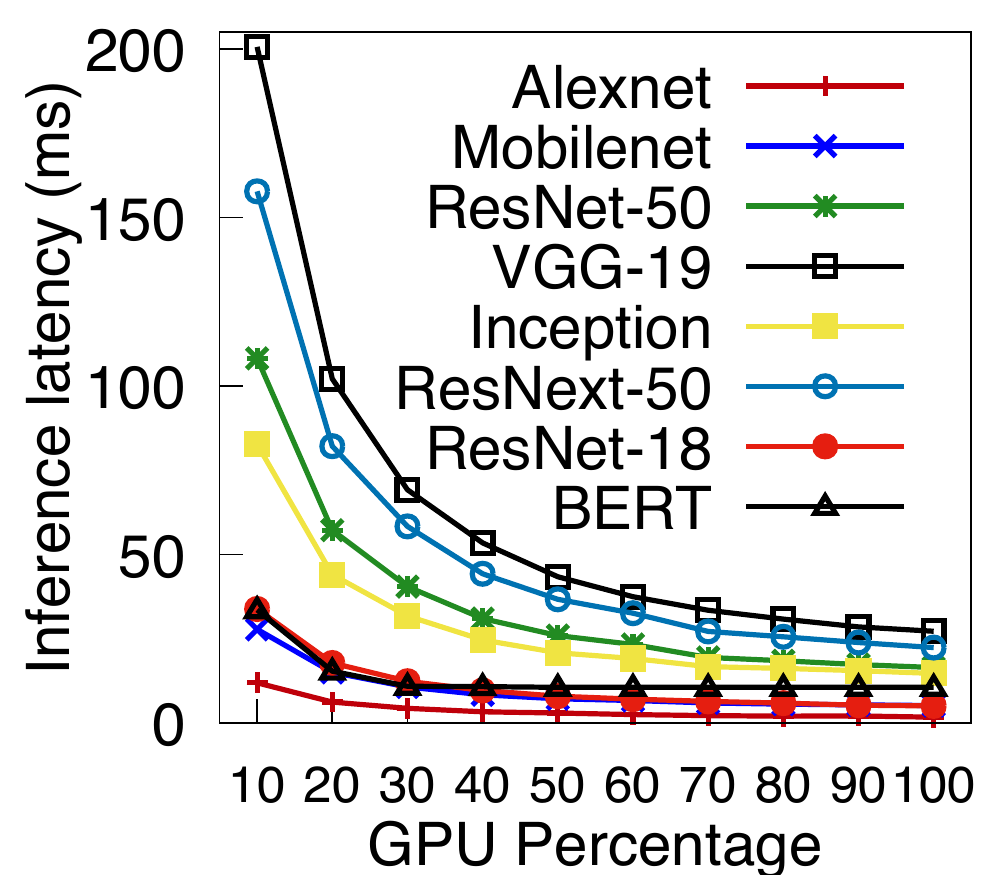}
    \caption{Batch Size = 16}
    \label{fig:inference_latency_batch}
    \end{subfigure}
    \caption{Avg. inference latency (ms) vs GPU\% in PyTorch}
\end{figure}
}
\Scut{
\noindent\textbf{Measurement with ML models}:
\hl{We now present measurements performed in the testbed, using multiple DNNs to demonstrate the limits in the parallelism of those DNN models.}\Scut{ We show that allocating a GPU\% to the DNN (as with CSS in~\cite{GSLICE}) improves over temporal or uncontrolled spatial sharing of the GPU. }These measurements also motivate the need for further improvements for DNNs with inference time constraints.\looseness-1 
}

\Scut{Most pre-trained models available in model-zoos 
\cite{torchvision_model_zoo} are trained with an image resolution of 224$\times$224.
This resolution choice maybe inspired by initial work on Alexnet~\cite{krizhevsky2012imagenet} and VGG~\cite{simonyan2014very}, 
apparently chosen for training. 
We chose this same resolution.}
\Scut{
We note that, increasing the resolution of training image does not increase the accuracy of the model~\cite{simonyan2014very}. \Scut{DNN model's accuracy increases slightly when inferring an image with resolution slightly larger than one where model is trained at. } However, the accuracy of inference decreases when the resolution of inferred image gets larger than the resolution of images used during training~\cite{touvron2019fixing,hoffer2019mix}. 
Most DNN platforms today can dynamically adjust the DNN kernels to have more GPU threads if they receive an image of size bigger than image size used to train DNNs. We have observed that inferring an image of resolution 448$\times$448 requires four times more threads than one of 224$\times$224 in a ResNet-18 model run in PyTorch v1.3. While the dynamic adjusting of kernels provides flexibility on inferring different resolution images, images of different resolutions cannot be batched together in one batch. 
Considering both accuracy of the DNN inference and ease of batching, we chose the image of size 224$\times$224 for our DNN workload. 
}

\Scut{
\hl{Each SM in the recent NVIDIA GPUs (P100, V100) is limited to running 2048 threads concurrently. Therefore, changing the number of SMs also changes the number of GPU threads that can run simultaneously. Allocating a low GPU\%, \mbox{\ie} very few SMs, means only a few thousands of GPU threads can run in parallel. DNN kernels which can utilize tens of thousands of threads during computation, will take longer. }
We can see from Fig.~\ref{fig:inference_latency_batch} that when providing only 10\% GPU, the inference latency is much higher than when provided a higher GPU\%.\looseness-1
}
\Scut{
\hl{However, we observe that some DNN models do not see improved inference latency even when allocated more GPU resources, while others continue to show improvement, but with diminishing returns, \mbox{\ie} increase in GPU resources after a point does not result in a proportional decrease in latency. We call this point a "\textit{Knee}."} \hl{We measured the latency for inferring a batch of 16 images using different GPU\% for several popular DNN models on the PyTorch framework. From Fig.~\mbox{\ref{fig:inference_latency_batch}} we see that for most models, the inference latency remains unchanged
above 30-50\% of GPU. With a smaller batch size, we see even lower knee\% (20\%-35\%).\Scut{We also repeated the experiment with inference batch size of 16 and see  (Fig.~\mbox{\ref{fig:inference_latency_batch}}) that with higher batch size the knee\% increases to 30-40\% for the low latency models and up to 50-60\% for models that are compute intensive.} Providing 100\% GPU for those models, as done in temporal sharing of the GPU, is wasteful and results in frequent under-utilization of the GPU.}\looseness-1
}
\Scut{
We also evaluated spatial sharing with NVIDIA P100 (56 SMs) and T4 (40 SMs). T4 GPU supports CSS, but, P100 does not support providing isolation with GPU\% for each application. For P100 GPU, we ran multiple instance of same model to slice the GPU into certain GPU\%.\mbox{\eg} we ran 4 Alexnet models to approximate 25\% GPU in P100. We present the result in Fig.~\mbox{\ref{fig:p100-t4}}.\hl{ Even with different GPUs, we see knee behavior in Alexnet and Squeezenet,} while, the computationally dense ResNet-50 does not show obvious knee. \hl{Both P100 and T4 GPU have lower computational capacity than V100, therefore, ResNet-50 can fully utilize these GPUs. As the knee for these models exists in other GPUs as well, our platform can be used more generally in other GPUs.}
}
\Scut{Although the latency of inference varies across different platforms for a given model, the trend is similar - the model has limited latency improvement as we increase GPU resources beyond a point. }
\Scut{
\knote{this part seems to be bringing up considerations and CPU/GPU related issues that belong more in Section III.A. It undercuts the contribution of Sec. III by putting it here. If you take this to Sec. III.A, it would not only strengthen that section, where you can talk about time to start a kernel, memory latency, instruction fetch cycles \etc appropriately. Move it away from here. }This limitation on parallelism stems from many factors such as varying compute requirements of different layers of DNN. Sequential nature of DNN means that a layer has to wait for output of previous layer before beginning processing. A layer that requires low amount of GPU resource prevents execution of subsequent layers therefore adding some latency that cannot be lowered even by giving high amount of GPU resources \knote{confusing sentence}. Further factors not related to DNN processing such as time to start a kernel, memory latency, instruction fetch cycles \etc also contribute to the limitation. Therefore, dividing the GPU appropriately and providing appropriate GPU\% to each DNN would allow us to run many DNN in GPU concurrently while also not adversely impacting their latency. \knote{move up to here. But there is probably only a small amount of content here that is really distinct from what is covered in III 1st para and III.A} 

However, we also note that providing a very low GPU\% to a DNN results in a steep increase in inference latency. Therefore, it is appropriate to provide a GPU\% close to knee for that model. Thus, a GPU can only support running certain number of models concurrently, until the sum of all concurrent models' knee reach 100\%. If we need to run more DNN models, we need to share the GPU temporally and schedule the execution of DNN models such that the concurrently running DNN models do not exceed 100\% GPU.
}

\update{



\Scut{Providing 100\% GPU resources to a DNN model when it cannot fully utilize it is wasteful. Therefore, temporal sharing of a GPU is wasteful. In temporal sharing, an application with low GPU compute requirement, needing only a limited number of SMs, is given all of the GPU for its time slice.}

\Scut{
 While the spatial sharing of GPU with default MPS and CUDA streams help in increasing GPU utilization\Scut{with applications having a lower computing requirement}, we have observed that compute heavy DNNs often interfere with each other.
This interference leads to unpredictable and higher latency during DNN inference~\cite{shen2019nexuslong,dhakal2020ecml,jain2018dynamiclong}. To counter the unpredictability due to DNN interference, we adopt CSS from GSLICE~\cite{GSLICE}, which isolates the GPU resource for each multiplexed DNN. CSS virtualizes GPU by providing enough GPU resources to each application to meet the deadline.\Scut{Newer NVIDIA GPU architectures (Volta, Turing, Ampere) allow GPU compute resource to be provisioned for each application. By providing a GPU\% by setting  CUDA\_MPS\_ACTIVE\_THREAD \_PERCENTAGE environmental variable associated with a process, we can provision the maximum number of Streaming multiprocessors (SMs) an application can use. This provisioning should be based on resource contention for the GPU from concurrently active applications.}  We utilize the "knee" information obtained from profiling such as Fig.~\ref{fig:inference_latency} to provide the DNN models with fixed GPU\%.\Scut{ Accordingly, in a CUDA environment we leverage and configure the environment variable CUDA\_MPS\_ACTIVE\_THREAD \_PERCENTAGE for each client process.}
}
\Scut{Modern GPUs provide a large number of Streaming Microprocessors (SMs). Each SM, akin to a CPU core, is capable of executing GPU kernels independent of others. }

\Scut{ NVIDIA GPUs do allow spatial sharing among multiple applications, with the use of Hyper-Q, CUDA Streams, and CUDA Multi-Process Server (MPS). SMs not used by one application might be used by other concurrently running applications~\cite{zhang2018g,liang2014efficient}. }
\Scut{
Unlike the default (uncontrolled) spatial multiplexing with CUDA MPS, \Scut{we make the case that} it is beneficial to provide the ability to explicitly partition GPU resources for each running application. 
}
\Scut{This partition would be based on
their corresponding limit on model parallelism as identified in \S\ref{limits_on_parallelism},\S\ref{sec:modelling_dnn_parallelism}, and the 'knee-point' of the latency-GPU\% curve.}


\Scut{Further, to provide strict isolation on the CPU side, we also pin the Inference applications and the MPS server (MPS control daemon) to different 
CPU cores. We keep track of all the available CPU cores and leverage the ‘taskset’ utility to bind the applications to one or multiple CPU cores.}
\Scut{
\begin{figure}
    \centering
    \subfloat[3 DNNs inferring concurrently ]{
    \includegraphics[width=.5\linewidth]{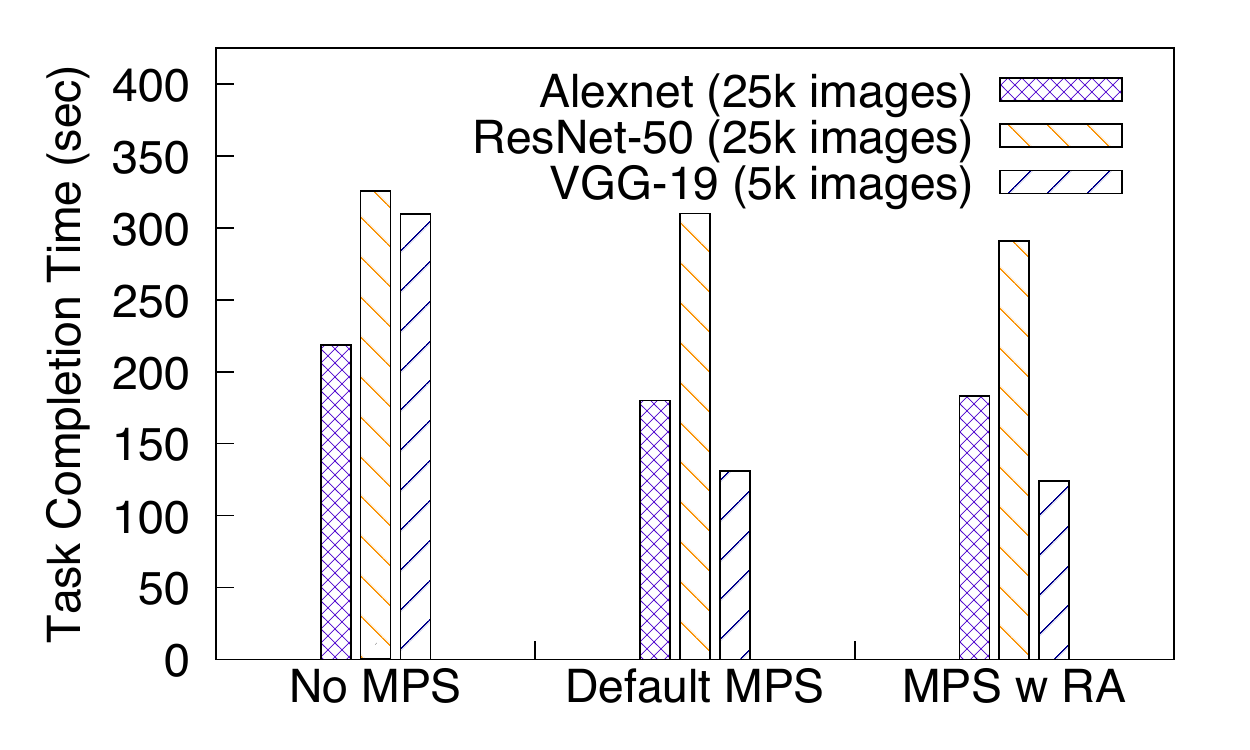}
    \label{fig:3_dnn_concurrent}}
    \subfloat[4 DNNs inferring concurrently ]{
    \includegraphics[width=.5\linewidth]{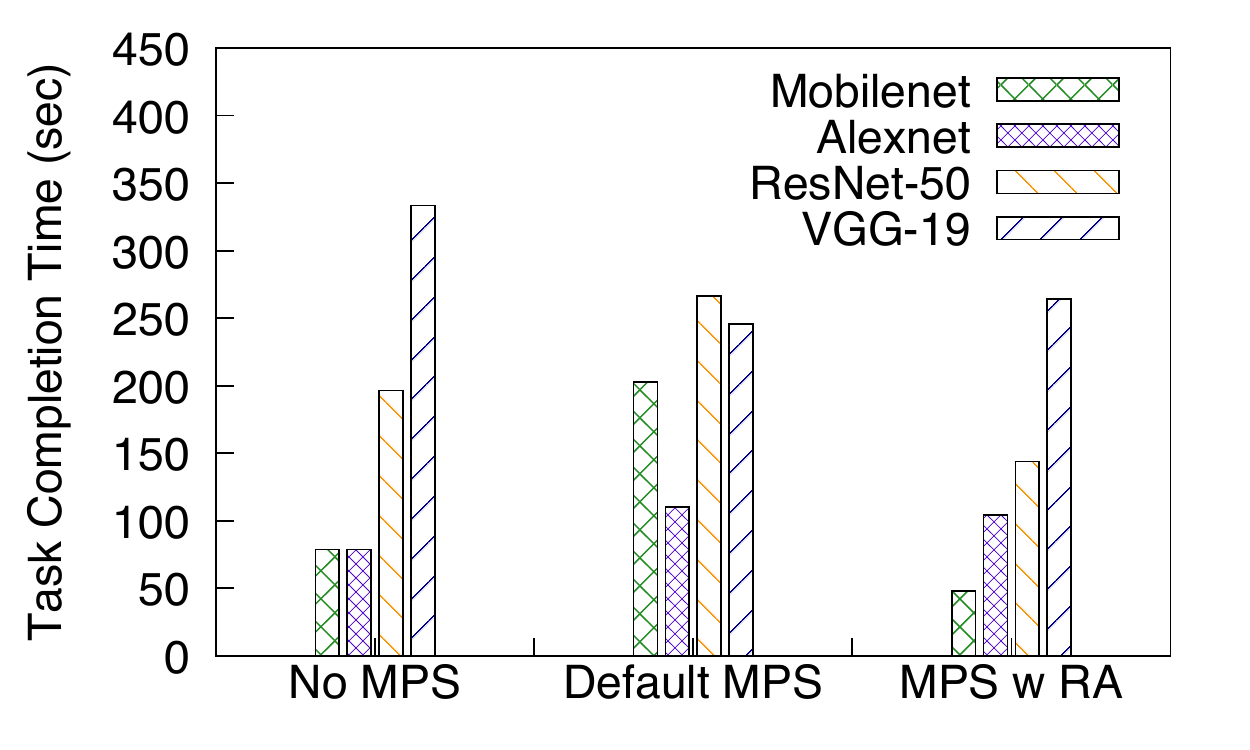}
    \label{fig:4_dnn_concurrent}    
    }
    \caption{Task completion time of multiple DNN running concurrently and sharing GPU with different techniques} \sknote{Rename NOMPS to Temporal sharing instead?}
\end{figure}
}
\Scut{
\begin{table}[]
    \centering
    \caption{Profiles of compute and memory bound kernels}
    \vspace{-3mm}
    \resizebox{\columnwidth}{!}{%
    \begin{tabular}{|c|c|c|c|c|c|}\hline
        Model & Layer & GFLOP & \#Bytes$(10^6)$ & Arit. Int. & Limited by  \\\hline
        Alexnet & Conv. 2 &0.30 &0.22 & 182 & Compute\\
        ResNet-50 & Conv.2 &0.103 &0.121 &393 & Compute\\
        VGG-19 & Conv. 11 &3.7 &9.44 & 391& Compute\\
        GNMT & LSTM &0.016 &8.38  & 2 & Memory \\\hline
    \end{tabular}}
    \label{tab:arithmetic_index_table}
\vspace{-6mm}
\end{table}
}
\Scut{
\noindent\textbf{Selecting Spatial Sharing Method}: We evaluated different options: Streams; CUDA Multi-process Service (MPS) and Controlled Spatial Sharing (CSS)~\mbox{\cite{GSLICE}}, for spatial sharing of a NVIDIA Volta GPU. 
We spatially shared the GPU among 3 Alexnet instances to  concurrently infer requests, and show the models' mean and $99^{th}$ percentile (P-99) latency and average throughput in Table~\mbox{\ref{tab:streams_vs_css}}.
The models interfere with each other, inefficiently utilizing GPU with Streams and MPS resulting in higher latency (P-99 higher than mean, indicating unpredictability) even when spatially sharing GPU with the same DNN model. 
On the other hand, CSS's GPU compute resource isolation and allocation of a Knee\% for Alexnet gives lower latency and 1.5$\times$ higher throughput than Streams and 1.1$\times$ higher than default MPS. CSS's mean and P-99 latency are close, indicating very predictable inference times. Thus, we choose CSS to provide spatial sharing for our ST-FD scheduler.\Scut{ spatially multiplex DNNs on GPUs.}\looseness-1
\Scut{We utilize GPU virtualization techniques of CSS~\mbox{\cite{GSLICE}} for the inference framework described in this paper.\Scut{ 
Spatial sharing of GPU using default API such as CUDA streams, Multi-process service (MPS) do not isolate compute resource across the applications. This results in applications running concurrently and sharing the GPU SMs and interfering with each other causing unpredictable latencies for some models as noted in~\cite{jain2018dynamiclong}.} Unlike default methods of spatially multiplexing GPU such as CUDA streams and CUDA Multi-process service (MPS), CSS isolates GPU compute (SMs) across concurrently running DNN models, while avoiding over-subscription of the GPU (\mbox{\ie} multiplex models whose total demand (each at knee) does not exceed 100\%) to avoid interference between applications. We compare CSS with streams and default MPS while spatially sharing the GPU. We run 3 instances of Alexnet concurrently and we provide 30\% GPU to each instance (Knee\% for Alexnet) for CSS. We report the mean and $99^{th}$ percentile (P99) latency as well as  average throughput obtained by the models in Table~\mbox{\ref{tab:streams_vs_css}}. Using streams and default MPS for spatial sharing leads to higher mean latency and much higher tail latency than CSS. Further, we should also note that P99 latency of CSS is quite close to the mean latency, showing much more predictable latency than other approaches. 
Streams and MPS do not isolate compute resource across the applications. This results in applications sharing the GPU SMs and interfering with each other causing unpredictable latencies for some models. Furthermore, lower mean latency allows CSS to get 1.5$\times$ higher throughput than streams and 1.1$\times$ higher throughput than default MPS. Therefore, we choose CSS to multiplex DNNs on a GPU.}
}
\Scut{
\noindent\textbf{Case for Spatio-temporal Sharing of DNN}:
When multiplexing several DNNs, as we illustrate in a scenario of 4 different models sharing the GPU (Fig.~\ref{fig:spatio-temporal-scheduling}), we still have a challenge as
\emph{CSS does not schedule the DNN models dynamically.} 
Temporal sharing, Fig.~\ref{fig:spatio-temporal-scheduling} (left) provides all models 100\% GPU for a certain amount of GPU processing time. This is wasteful as most DNNs cannot use 100\% GPU. CSS, being static, Fig.~\ref{fig:spatio-temporal-scheduling} (middle) can only run 3 models at their Knee\% so as to not over-subscribe the GPU. Alternatively, CSS could reconfigure (reduce) GPU\% for some models, so they run without oversubscribing the GPU. However, allocating resources below their knee resulting in higher inference latency (and possibly miss their SLO). Fig.~\ref{fig:spatio-temporal-scheduling} (right) shows a spatio-temporal schedule where all DNN run at their Knee\% in GPU and still meet their inference deadline. We design and evaluate spatio-temporal scheduling in \S~\ref{sec:scheduling}.\looseness-1

}
\Scut{ CSS DOES NOT DO SCHEDULING. When multiple kernel run on the same GPU SMs, as with other spatial multiplexing techniques like CUDA Streams and default CUDA MPS, the DNN inference latency becomes highly variable. However, CSS's constraint introduces challenges when multiplexing together many models. We illustrate a scenario of 4 different models sharing the GPU (Fig.~\ref{fig:spatio-temporal-scheduling}). 
Temporal sharing, Fig.~\ref{fig:spatio-temporal-scheduling} (left) provides all models 100\% GPU for a certain amount of GPU processing time. This wastes the GPU. CSS,  Fig.~\ref{fig:spatio-temporal-scheduling} (middle), may also execute models with a high GPU\% demand sequentially, thus introducing additional latency for models with high demand. CSS, since it re-allocates the GPU only after the execution of the high GPU demand DNN, it can cause low GPU utilization as some of the models might end their computation earlier, freeing GPU resources which are not utilized. Spatio-temporal scheduling Fig.~\ref{fig:spatio-temporal-scheduling} (right) on the other hand, can run the models concurrently with controlled sharing of the GPU as well as increase the GPU utilization throughout by running a model whenever GPU resources become available. This motivates us to adopt spatio-temporally scheduling for DNN inference.}

\Scut{

We demonstrate the benefit of CSS with an example workload where the GPU is shared by four inference applications (namely, Mobilenet, Alexnet, ResNet-50, 
and VGG-19), each performing 25,000 inference tasks. We set an SLO of 50 milliseconds to ensure a robust and interactive inference and a adaptive batching scheme that adjusts the batch size.\Scut{ (ranging from 1 to 32 images) while keeping inference latency within 50 ms.} CSS virtualizes the GPU by providing sufficient GPU resources to each application to meet the deadline. We utilize the "\textit{knee}" information obtained from apriori profiling, such as Fig.~\ref{fig:inference_latency}, to provide the DNN models their knee GPU\%.\Scut{We note that oversubscribing the GPU \ie allowing to execute multiple concurrent applications where the aggregate request and corresponding allocation of the GPU resource exceeds 100\% GPU is possible with controlled spatial sharing of GPU. In such a case, some of the SMs of GPU may be shared across applications and isolation of compute resources is not guaranteed. Oversubscribing GPU can lead to interference and increased latency for applications.}
The inference throughput (total number of images inferred per second) is in Fig.~\ref{fig:4_dnn_concurrent}. The overall throughput with CSS is better than the temporal and spatial multiplexing with default CUDA-MPS mode. 
With temporal sharing, the lightweight Mobilenet and Alexnet achieve a similar throughput of $\sim$300 images per second. ResNet-50 and VGG-19, on the other hand, achieve low throughput due to the GPU being shared in equal timeslices with the computationally lighter models. 
We observe that
the default MPS mode 
improves the throughput of VGG-19 considerably. However, the throughput of all other models drop significantly compared to temporal sharing, because of contention from VGG-19 which occupies a major share of the  GPU resources.
In CSS mode, 
the throughput is higher than temporal sharing mode for all the DNN models. Their throughput is higher than default MPS for all the models, except for VGG-19 (which is lower, but close to the Default MPS's VGG-19 throughput).\Scut{Furthermore, the major benefit with CSS is reflected in the latency profile (Fig.~\ref{fig:controlled_sharing_contention}), which is more consistent and guarantees to be 
within the specified SLO budget (50ms). As all the inferences end within their SLO, CSS allows us to repeat their execution and run the models twice in 100 ms time window to greatly improve on the throughput,} 
These results indicate that 
managing and allocating the GPU \% correctly 
across all the DNN applications is critical to ensure high overall system throughput, while completing tasks within their SLO. }
\Scut{
\begin{figure}
\begin{center}
    \includegraphics[width=\linewidth]{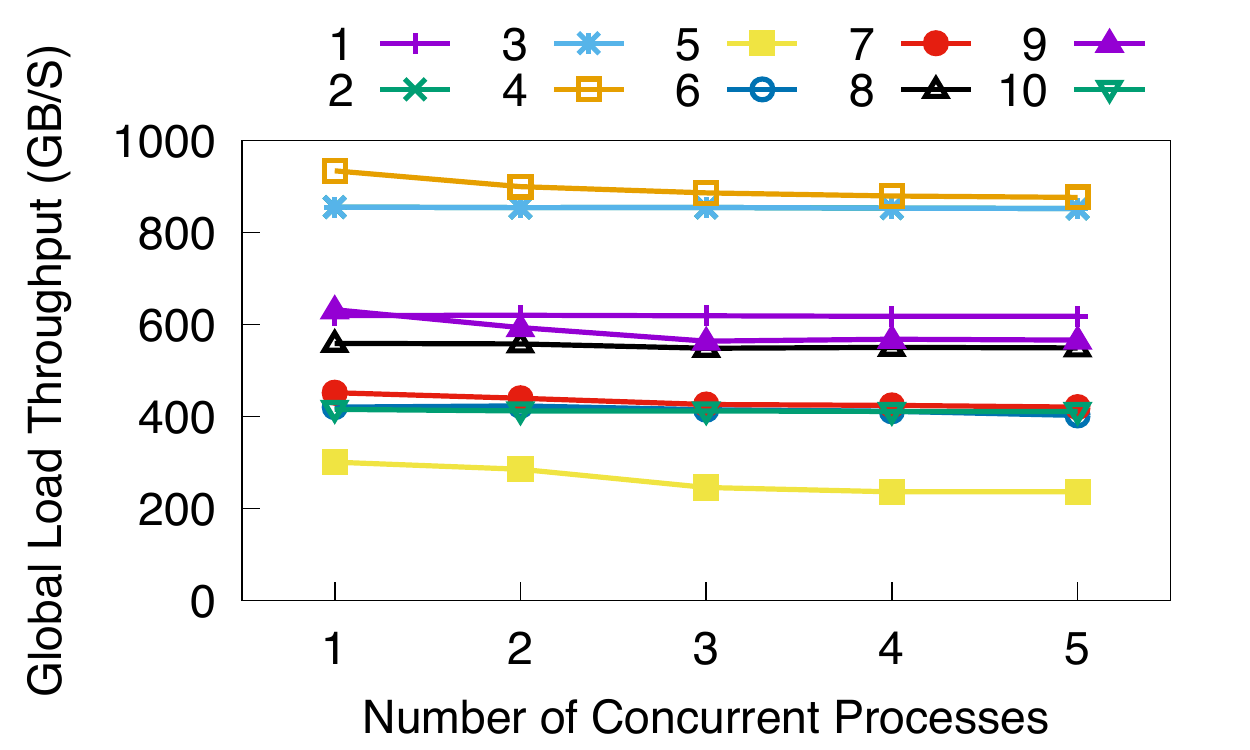}
    \caption{Global Memory Load Throughput of VGG-19 kernels while multiplexing }
    \label{fig:vgg-19-kernels-throughput}
    \end{center}
\end{figure}
}}

\Scut{
\begin{figure}
\includegraphics[width=\linewidth]{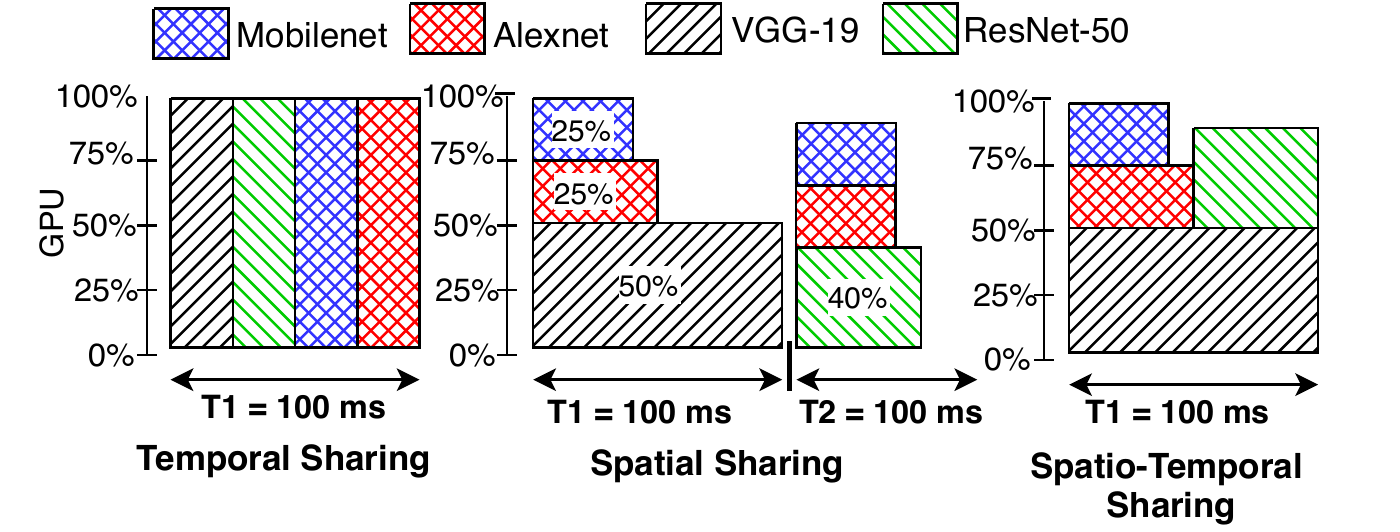}
    \caption{Multiplexing Scenario with 4 different DNNs}
    \label{fig:spatio-temporal-scheduling}
 
\end{figure}
}
\Scut{
To avoid interference caused by spatial multiplexing of the GPU, we avoid oversubscribe the GPU, \ie multiplex models whose total demand (each at knee) exceeds 100\%. This constraint introduces challenges when multiplexing together many models. We illustrate a scenario in Fig.~\ref{fig:spatio-temporal-scheduling} where we try to run 4 different models in the GPU. Temporal sharing Fig.~\ref{fig:spatio-temporal-scheduling} (left) can lead to all models getting certain amount of GPU processing time with 100\% GPU,  resulting in wasting the GPU. CSS, as in Fig.~\ref{fig:spatio-temporal-scheduling} (middle), can cause us to executed models with high GPU\% demand sequentially in different time slices, thus introducing additional latency for models with high demand. Moreover, it can cause low GPU utilization as some of the models might end their computation earlier, freeing GPU resources which are not utilized. Spatio-temporal scheduling Fig.~\ref{fig:spatio-temporal-scheduling} (right) on the other hand, can run the models concurrently with controlled sharing of the GPU as well as increase the GPU utilization throughout by running a model whenever GPU resources become available. This motivates us to adopt spatio-temporally scheduling for DNN inference.
}

\Scut{\noindent\textbf{Fundamental Limitations of Temporal Sharing and Simple Multiplexing of DNNs}} 
\Scut{
\begin{itemize}
    \item DNNs do not use all GPU
    \item provisioning for the biggest kernel is wasteful
    \item finding an amount of GPU\% that can accomodate smaller kernels also can run larger kernel with only a little delay is the best way to approach
    \item we give rest GPU to other DNNs
    \item what temporal sharing platforms do is give all GPU to the applications, does not allow to use the idle resources.
    
\end{itemize}
}
\Scut{DNNs comprise multiple \textit{layers}, computational segments which accept input from a previous layer and pass the computed output to next layer.
\Scut{ output of one layer of DNN fed as input of another layer.} 
Layers such as convolutional, fully-connected, LSTM, ReLU, deconvolutional, \etc, are commonly used in DNNs. \hl{DNN execution in a GPU utilizes \textit{kernels} (GPU functions) from popular libraries such as cuDNN, OpenCL \mbox{\etc} These kernels launch cooperative thread arrays (CTAs) in the GPU to perform the DNN functions of each layer in order.}\Scut{, such as convolution, LSTM, ReLU\etc, and are executed one after another in the GPU with each kernel's output being utilized by the next.}\Scut{GPU implementation consist of multiple kernels performing these layers' computation.} The computational requirement of each GPU kernel depends on the type of computation (heavy convolutions compared to simple ReLU) and the dimensions and size of the matrix.\Scut{varies \eg for the same matrix dimension, a compute heavy convolutional kernel require more compute than kernels such as ReLU, and matrix operations between matrices of larger dimensions requires more compute than one with smaller dimensions.}\Scut{ Furthermore, the dimension of the intermediate feature matrix changes due to varying size of filters used by the kernels, thus, adding to the variability in compute requirement across different kernel within a DNN.} \hl{The DNN design means a subsequent kernel has to wait for the output of an earlier kernel. They cannot run concurrently. This results in a significant part of GPU resources (i.e., SMs) being idle when a kernel 
with a relatively small amount of compute is executing. When multiple DNNs are spatially multiplexed, the idle GPU SMs may be utilized. This is somewhat easier 
with multiple instances of the same DNN model. The computing workload for the SMs can be more evenly distributed, and the instances complete their inferences almost in unison. 
Furthermore, these multiple instances also benefit from pre-loading and caching of weights and parameters of the single DNN model. Thus, we achieve higher throughput. 
However, when different DNN models are multiplexed, the kernels may not be of the same complexity. Compute heavy kernels from one model can occupy SMs while another model's kernels complete execution. But the launch of kernels of another DNN model may need to be delayed until enough SMs a free, thus resulting in SMs being idle and under-utilizing the GPU. Overall, this can increase the overall inference latency for requests.} The increase in latency is also unpredictable as it depends on which models run concurrently. Meanwhile, giving too many SMs to a model is wasteful since only a few compute heavy kernels benefit from the increased GPU resource availability. \hl{Moreover, many recent models, optimized for low latency inference (e.g., Mobilenet, Squeezenet \mbox{\etc}) may not  have even a single kernel that can utilize the entire GPU.} However, we have also observed that using fewer than necessary SMs leads to an exponential increase in model latency \hl{(also observed in ~\mbox{\cite{architectural-impli})}}. Therefore, we seek to give each model right number of GPU SMs where they can run as fast as when they get all the GPU SMs. Further, the SMs that cannot be efficiently utilized by a model can be allocated to other concurrently running models, so that they can be utilized. \hl{This motivates us to create a dynamic and intelligent Spatio-Temporal (ST-FD) scheduling system, that is aware of resource requirement (Knee) of the model and dynamically re-utilizes the GPU SMs that are left idle after a model completes its inference task.  }

\Scut{Isolating SMs for each model can avoid this delay caused during concurrent execution of different model, a technique we utilize in our paper.} \knote{this last sentence DOES NOT FOLLOW from the arguments above. All that the arguments say is that a more intelligent SM allocation (knee) and scheduling(ST-FD) are needed. You need to use this to then say - giving the right amount of GPU will make sure that the model runs as fast as it can; don't give any more, and don't waste those SMs, instead give it to another model, but you need to do it more intelligently, like ST-FD, which is DYNAMIC and Knee aware. The idea of providing entire GPU to these models would be wasteful needs to come out.} \anote{Please see.}
}
\looseness-1\Scut{ Thus, providing entire GPU to these kernels would be wasteful. Some DNN kernels can fully utilize most of the SMs that GPU has to offer, however, we have observed that most kernels in DNNs require much fewer threads than GPU can spawn, inefficiently utilizing the GPU SMs. Providing entire GPU to these kernels would be }
\Scut{
State-of-the-art DNN inference platforms such as the Triton Server can spatially sharing the GPU with multiple instance of the same model. However, it utilizes the default GPU hardware scheduler that time-multiplexes the GPU across kernels from different DNN models~\mbox{\cite{triton-temporal}}.\Scut{This default scheduler allocates the entire GPU, even to kernels which may not fully utilize all the SMs.} Time multiplexing (Temporal sharing) is inefficient for two reasons: first, the GPU is underutilized while running a kernel with limited compute; second, multiplexing multiple DNN models using default scheduler will result in temporal sharing of the GPU, which causes interleaved execution of kernels of different 
models and increases overall latency.\knote{why?? are you implicitly assuming they are time multiplexed?? You leave the reader hanging at this sentence, without saying why}. 
\Scut{We fundamentally differ from other state-of-the-art inference platforms by spatially multiplexing the GPU restricting the number of SMs (GPU\%) using CSS and multiplexing the DNNs with Spatio-Temporal (ST-FD) scheduler, rather than giving exclusive GPU access to all GPU SMs.} \Scut{However, picking the right number of SM is challenging. If too few SMs are allocated for an application, compute heavy kernels which require a lot of CTAs will not have enough SMs to run those CTAs concurrently. We have observed that the increase in overall latency of inference as we reduce number of SMs is exponential. Meanwhile, giving too many SMs to a DNN is wasteful, only few compute heavy kernels benefit from the increased GPU resource. Moreover, many DNN optimized for low latency inference (eg. Mobilenet, Squeezenet \etc) might not even have a single kernel that can utilize entire GPU. With our work, we find the \textit{right} number of SMs (Knee\%) for a DNN. While, running a DNN at the Knee, a few compute heavy kernels will take slightly longer to complete compared to running with entire GPU, however, most of the kernels' runtime will be unaffected as the number of SMs in Knee\% is more than enough for those kernels. This results in overall inferece latency to be same or close to while inferring with entire GPU. Moreover, restricting an application to certain number of SMs means other SMs can be utilized for running kernels from another DNN concurrently. }
}
\Scut{
\hl{We compare Triton server with our ST-FD scheduler approach.\Scut{these approaches to demonstrate the effectiveness of our approach compared to DNN inference platform Triton.} We evaluated the total time taken to infer 10000 images each by 4 different DNN models, Alexnet, Mobilenet, ResNet-50 and VGG-19 (40000 images in total) using one V100 GPU. With the Triton server, we studied two alternatives: first sending inference requests to just 1 model at a time, so that the workload causes interleaving of inference across the 4 models; in the second alternative, we sent requests to all 4 models concurrently. For ST-FD scheduler, we also sent concurrent requests to all 4 models. We present the results in Fig.~\mbox{\ref{fig:triton-comparison}}. The models running in the Triton server complete their inferences in about 55-60 seconds. We see that both modes of execution in Triton server (temporally interleaved and concurrent execution) take about same time to finish inference, while, the ST-FD scheduler completes all the inferences more than 37\% faster. }Triton's utilization of GPU's hardware scheduler leads to temporal sharing of the GPU, thus, utilizing GPU inefficiently, leading to higher tasks completion time. Meanwhile, Spatio-Temporal scheduler provides right amount of GPU as well as resource isolation to concurrently running model, this leads to lower overall task completion time. We use this as the motivation for designing our ST-FD scheduler (\S~\ref{sec:scheduling}) and the mathematical models to provide information required for such a scheduler.\looseness-1
}

\Scut{
\noindent\textbf{Dynamic GPU Resource Reconfiguration without Downtime}:\label{sec:dynamic_gpu_recon1}\Scut{The GPU\% provided to an application in CSS will likely need to change due to the variations in the workload, addition of new DNN,\etc\Scut{ \ie the change in the arrival rate of the tasks (especially with streaming data).}
\Scut{ or the variations in the number of concurrently executing applications. Further, when an application terminates, it is necessary to reclaim and redistribute the GPU resources among the active applications to avoid any resource fragmentation and under-utilization of the GPU.} 
The latest CUDA MPS version\Scut{, R455+}~\cite{nvidiamps2019} limits the configuration of the GPU\% to a one-time (static) allocation set during the GPU initialization for the application. 
This 
}
We observed that the GPU\% of a given model might need to be changed dynamically when working with different workloads of requests to models multiplexed on a GPU. For example, a slight change in GPU\% may allow us to accommodate higher inference batch sizes. However, due to the limitation of CUDA MPS~\mbox{\cite{nvidiamps2019}} to configure the GPU\% \Scut{of an application to a one-time (static) allocation at GPU initialization 
poses a major challenge. }any GPU resource readjustment  requires us to spin up a new CPU process with an updated GPU\%, which results in several seconds of downtime (depending on the ML framework initialization).
\Scut{Some works~\cite{zhang2019laius} maintain a pool 
of (several) instances for each application with different statically assigned GPU\%. This allows the new incoming work to be directed towards an instance with the desired GPU\%. Although effective for smaller GPU applications, we do not believe it is usable for large DNN models which often occupy gigabytes of GPU memory.\Scut{To avoid this idle time of several seconds, we utilize an \textit{overlapped execution} mechanism explained in ~\cite{GSLICE}.} Alternately,~\cite{GSLICE}, maintains a single standby instance of a DNN model that can be loaded into the GPU with dynamically computed GPU\%, while the original instance continues processing the requests. Once the standby instance is ready, it seamlessly takes over the inference processing 
and terminates the original instance.}
We utilize the overlapped execution approach of~\mbox{\cite{GSLICE}}, to mask the DNN loading and setup time while changing the GPU\%. We maintain a \textit{active-standby} pair of process, where the active process performs DNN inference in GPU. The standby process is a hot replica that loads all required software libraries for the DNN model but does not access the GPU (no need to set a GPU\%). When a GPU\% change is desired active process sends a message to standby to start loading the model, while, the active DNN instance continues processing requests until the standby process loads the DNN model on the GPU with a new GPU\% and is ready to infer.\looseness-1\Scut{ Once the standby is ready, it seamlessly takes over the inference and the original instance terminates.}
}

\Scut{
\noindent\textbf{Inefficiencies while running DNNs in State-of-art DNN platforms: }
Latest DNN architecture consist of directed graph with output of one layer of DNN fed as input of another layer. These layers are functions such as convolutions, fully-connected, LSTM, ReLU, etc. DNNs' GPU implementation consist of multiple kernels performing these layers' computation. As the DNN's computation requirement vary across different layers (eg. convolutional layers require more compute than ReLU for same size feature matrices), the kernel's computation also varies. Further, DNN's feature matrix dimension changes as the data get processed by layers due to different layer dimension and filter dimension, thus, adding to the variability in compute requirement across different kernel within a DNN. However, the directed graph design of DNN means even compute heavy kernels have to wait for the output of the earlier kernel which might only require few threads and few SMs (eg. ReLU). Moreover, this results in lot of GPU resources being idle when a kernel with little compute is executing. Thus, providing entire GPU to these kernels would be wasteful.\Scut{ Some DNN kernels can fully utilize most of the SMs that GPU has to offer, however, we have observed that most kernels in DNNs require much fewer threads than GPU can spawn, inefficiently utilizing the GPU SMs. Providing entire GPU to these kernels would be }

Current state-of-the-art DNN inference platform such as Triton server~\cite{triton} and others~\cite{} use default GPU scheduler to schedule different DNN models while performing inference.First, default GPU scheduler interleaves kernels of different applications for fixed time slices. Even the smallest kernel, which cannot even fully utilize single SM is provided with entire GPU, stalling a compute heavy kernel of another DNN. This, is inefficient in two terms. One, GPU is underutilized due to small kernel,two,the latency increases while multiplexing. In this work, we want to find the right number of SMs (Knee) for an entire DNN. For few kernels in some DNNs, the execution time might slightly increase due to running at Knee, as these kernels could potentially use more SMs than at knee, but most of the kernels, which are fairly light in terms of compute, the number of SMs at Knee will be sufficient to run without increasing the execution time of that kernel. Therefore, at the Knee, the latency of execution of entire DNN will be very close to the latency when using 100\% GPU. The remaining SMs in that are not used by a DNN model then can be given to another DNN. Here, we will not be wasting most GPU resource on running small kernels, and also can run other DNNs concurrently without increasing the latency.

}
\Scut{
\noindent\textbf{Limitations of the State-of-the-Art:}
\label{sec:triton_server}
To estimate the overall benefit of our Spatio-Temporal scheduling algorithm (ST-FD), we compare it with the throughput of Triton server~\cite{tritonserver}. 
We evaluated the total time taken to infer 10000 images each by 4 different DNN models, Alexnet, Mobilenet, ResNet-50 and VGG-19 (40000 images in total). With the Triton server, we studied two alternatives: first sending inference requests to just 1 model at a time, so that the workload causes interleaving of inference across the 4 models; in the second alternative, we sent requests to all 4 models concurrently. For ST-FD scheduler, we also sent concurrent requests to all 4 models. We present the results in Fig.~\ref{fig:triton-comparison}. The models running in the Triton server complete their inferences in about 55-60 seconds. We see that both modes of execution in Triton server (temporally interleaved and concurrent execution) take about same time to finish inference. The ST-FD scheduler completes all the inferences more than 37\% faster. These results show that the state-of-the-art NVIDIA Triton server is not efficient at multiplexing different DNN models.\knote{an architectural argument as to WHY Triton is not efficient at multiplexing multiple different models is needed - what is it that prevents Triton from using a lot of parallel processing engines at the same time well? What is the software limitation? What is the application level characteristic (DNN parallelism) that it is not able to recognize well? What are the primitives needed to enable efficient multiplexing so that lot of models can run at high throughput? This is what we are providing with our design. That idea has to come through CLEARLY. We can't just say - we measured, it is no good. We can do better - we have to say WHY we can do better, and that is the point of the paper. } The spatial sharing of GPU utilized by our ST-FD scheduler greatly boosts throughput, hence the lower task completion time. We use this as the motivation for designing our ST-FD scheduler (\S~\ref{sec:scheduling}) and the models to provide information required for such a scheduler.\looseness-1
}

\Scut{
\begin{figure}
    \centering
    \includegraphics[width=\linewidth]{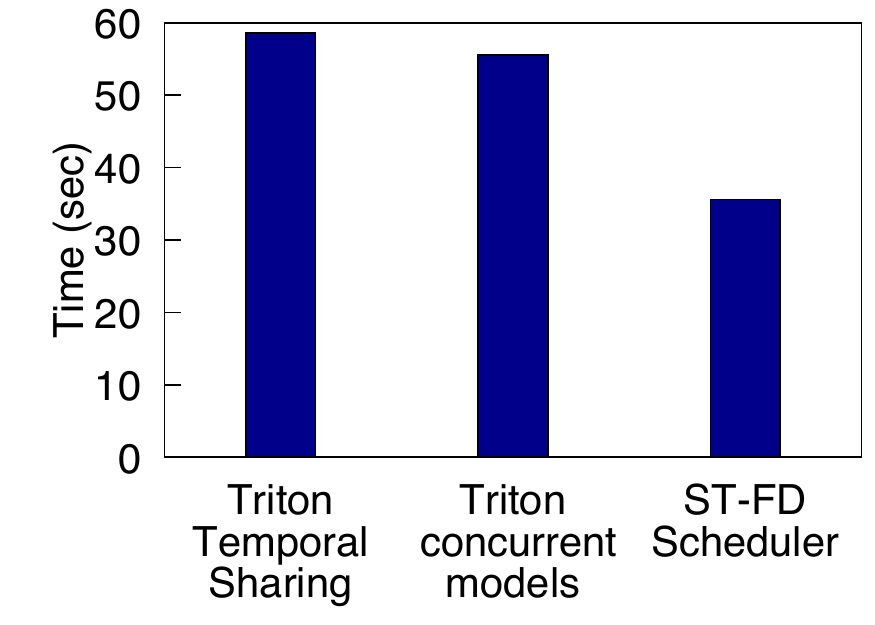}
    \caption{Time taken to infer fixed load with different execution modes of Triton server and our Spatio-Temporal scheduler\looseness-1 }
    \label{fig:triton-comparison}
\end{figure}

}

\Scut{
We now experimentally demonstrate the different scheduling methods for DNN models on a GPU. We run four different image recognition models: Alexnet, Mobilenet, ResNet-50 and VGG-19, in the PyTorch platform. In these experiments we picked the demand (GPU\%) for Alexnet and Mobilenet to be 25\%, the demand for ResNet-50 to be 40\% and VGG-19 to be 50\% of the GPU based on profiling these models as shown in Fig.~\ref{fig:inference_latency}. We further set an inference deadline of 100 ms for all the DNN models to simulate an SLO for each model. \Scut{We set the deadline to be 40ms for Mobilenet, 50 ms for ResNext-50 and Alexnet, and 100 ms for the relatively compute-heavy VGG-19 model.} The deadlines were chosen so that the slowest model can infer a batch of 16 requests several times (2-3 batches). \anote{I have to rethink the experiment a bit. I need to say why the Alexnet and mobilenet quit early. Maybe we should just infer 'n' number of batches. such that the lighter models end early}
We evaluated the performance of temporal sharing of GPU by providing each model a 25 milliseconds time slice as seen in Fig.~\ref{fig:spatio-temporal-scheduling}. We then run the schedule for 2 deadlines,\ie for 200 ms, therefore, each model runs at least twice. \Scut{For controlled spatial sharing, each model was provided 17\% of GPU compute resources and ran for 600 milliseconds}. For the second experiment, we multiplexed the GPU (using controlled spatial sharing, as in ~\cite{GSLICE}) according to the demand above, for each model. However, as the aggregate demand for all 4 models exceeds 100\% of GPU, we cannot schedule all the models at once. Therefore, we infer with VGG-19 in the first time 100 ms time slot, and ResNet-50 in second time 100 ms time slot, along with Alexnet and Mobilenet concurrently in both time slots, as seen in Fig.~\ref{fig:spatio-temporal-scheduling}. Finally, for spatio-temporal scheduling, all the models are again provided a GPU\% based on their demand and deadline. However, rather than the static schedule of spatial sharing, we schedule ResNext-50 to run as soon as enough GPU resource is available, on one or more of the models complete their task. We again run this schedule of models for 200 ms and then note the throughput attained by each model and examine the results. \Scut{and scheduled as follows: Alexnet (25\%), ResNet-50 (35\%), VGG-19 (40\%) running concurrently for  first 300 milliseconds and Mobilenet (33\%), ResNet-18 (33\%) and Inception-V3 (34\%) running concurrently for the second 300 milliseconds timeslice.}

We can see in Fig.~\ref{fig:spatio-temporal} that in most cases pure temporal scheduling provides the lowest throughput while spatial and spatio-temporal execution of DNNs provide much higher throughput. This difference is especially true for the more compute-heavy models (ResNet-50 and VGG-19). Their throughput gains by spatio-temporal scheduling is 1.5$\times$ for ResNet-50 and 2$\times$ for VGG-19. On the other hand, the throughput difference from temporal sharing for the compute-light models is marginal. This demonstrates that giving all the GPU resource for compute-light models is wasteful. Limiting those models to right amount of GPU and sharing the GPU resource in controlled fashion can greatly benefit the concurrently running compute-heavy models. 
\knote{change ResNext to ResNet: Aditya: OK}
We should also note that ResNet-50 gets lower throughput when using only spatial sharing compared to temporal sharing. This is due to the fact that ResNet-50 only gets to run once in 200 ms time window with the spatial sharing schedule, as its high GPU resource demands prohibits placing it together with other models in first time slot (T1 in Fig.~\ref{fig:spatio-temporal-scheduling} (middle)) due to lack of GPU resources. Unfortunately, spatial sharing can also cause the GPU to be idle when some of the models finish their work earlier than deadline. Spatio-temporal scheduling aims to utilize this idle time and run another model to utilize the freed-up GPU space, as seen in Fig~\ref{fig:spatio-temporal-scheduling} (right). Spatio-temporal scheduling combines the best of both temporal and (controlled) spatial sharing. It allows more models to run in the GPU concurrently with controlled sharing of the GPU as well as increasing the GPU utilization throughout by running a model that is waiting for the GPU resources to become available. 

\knote{the throughput improvement is marginal, isn't it? Do you want to emphasize that it is particularly for the large models that you are able to get better throughput? And by how much? AD:Please see now}
}


\Scut{
\subsection{Fundamental difference from other DNN frameworks}
DNNs comprise multiple \textit{layers}, computational segments which accept input from a previous layer and pass the computed output to the next layer.
\Scut{Layers such as convolutional, fully-connected, LSTM, ReLU, deconvolutional, \etc, are commonly used in DNNs.} DNN execution in a GPU utilizes \textit{kernels} from popular libraries such as cuDNN. These kernels launch cooperative thread arrays (CTAs) in the GPU to perform the DNN functions of each layer in order. These kernels vary in compute requirement as well as the number of threads.\Scut{The computational requirement of each GPU kernel depends on the type of computation (heavy convolutions compared to simple ReLU) and the dimensions and size of the matrix.} The DNN design means a subsequent kernel has to wait for the output of an earlier kernel. They cannot run concurrently. Each SM in the recent NVIDIA GPUs (P100, V100) can run up to 2048 threads concurrently.\Scut{Therefore, changing the number of SMs also changes the number of GPU threads that can run simultaneously. Providing \Scut{a low GPU\%, \mbox{\ie} }very few SMs (low GPU\%), means only a few thousands of GPU threads can run in parallel.}\Scut{DNN kernels which require relatively few threads during computation, will not be able to utilize all the SMs.} This results in a significant part of GPU resources (\mbox{\ie} SMs) being idle when a kernel requires a relatively small number of threads.\looseness-1

}
\Scut{
When DNNs are spatially multiplexed, the idle GPU SMs may be utilized. This is somewhat easier with multiple instances of the same DNN model. The computing workload for the SMs can be more evenly distributed, and the instances complete their inferences almost in unison. 
Furthermore, these multiple instances also benefit from pre-loading and caching of weights and parameters of the single DNN model. Thus, we achieve higher throughput. 
However, when different DNN models are multiplexed, the kernels may not be of the same complexity. Compute heavy kernels from one model can occupy enough SMs\Scut{while another model's kernels complete execution. But} to delay the launch of kernels of another model (awaiting required SMs to be free). This can increase the overall inference latency for requests.}\looseness-1\Scut{The increase in latency is also unpredictable as it depends on which models run concurrently. Meanwhile, giving too many SMs to a model is wasteful since only a few compute heavy kernels benefit from the increased GPU resource availability.}

\Scut{
Many recent models optimized for low latency inference (e.g., Mobilenet, Squeezenet \mbox{\etc}) may not  have even a single kernel that can utilize the entire GPU.\Scut{However, we have also observed that using fewer than necessary SMs leads to an exponential increase in model latency (also observed in ~\mbox{\cite{architectural-impli})}} Our ST-SD scheduling fundamentally differs from existing DNN inference frameworks by giving each model right number of GPU SMs where they can run as fast as when they get all the GPU SMs. Further, the SMs that cannot be efficiently utilized by a model can be allocated to other concurrently running models\Scut{, so that they can be utilized}. This motivates us to create a dynamic and intelligent Spatio-Temporal (ST-FD) scheduling system, that is aware of resource requirement (Knee) of the model and dynamically schedules DNN models such that the GPU SMs that are left idle is utilized, and the SMs freed after a model completes its inference task.\looseness-1
}
\Scut{
\subsection{Evaluation of NVIDIA Triton Server} \Scut{State-of-the-art DNN inference platforms such as the Triton Server can spatially sharing the GPU with multiple instance of the same model to greatly increase the inference throughput. However, it utilizes the default GPU hardware scheduler that time-multiplexes the GPU across kernels from different DNN models~\mbox{\cite{triton-temporal}}.} 
\begin{wrapfigure}{l}{.30\columnwidth}
\vspace{-5mm}
\captionof{table}{Triton Server Evaluation}
    \vspace{-2mm}
        \resizebox{\linewidth}{!}{ \begin{tabular}{|c|c|}\hline
         Models & \makecell{Thpt.\\(imgs/s)}  \\\hline
         \makecell{Mobilenet\\\textbf{(2-instances)}}& 1675\\\hline
         \makecell{ResNet-50\\(\textit{R-50})\\\textbf{(2-instances)}}&989\\\hline
         \makecell{Mobilenet \& R-50\\\textbf{Temporal}}&804\\\hline
         \makecell{Mobilenet \& R-50\\\textbf{Concurrent}}&844\\\hline
    \end{tabular}}
    \label{tab:triton-multiple-instance}
    \vspace{-4mm}
\end{wrapfigure}
We experimented in our testbed to observe the performance of running multiple instances of same model and entirely different model concurrently in single GPU using a State-of-the-art Triton server. We utilized Mobilenet model and computed the throughput by spatially sharing GPU with 2 concurrent instance of same model with batch size of 16. For temporal sharing results, we interleaved the inference request to a single instance of Mobilenet model and one instance of ResNet-50 model. Finally, to evaluate Triton's multiplexing of different models, we sent concurrent request to both Mobilenet and ResNet-50 models. We report the combined throughput of each experiment in Table.~\ref{tab:triton-multiple-instance}.We note that the throughput attained during temporal sharing (interleaving) is quite similar to when running different model concurrently. Both temporal sharing and concurrent execution of multiple models in Triton provides much lower throughput than throughput from multiple instance of same model.\looseness-1 \Scut{This experiments demonstrates that spatial sharing could greatly benefit DNN frameworks to achieve higher throughput.}

\Scut{To demonstrate the effectiveness of Spatio-Temporal scheduling while multiplexing more models, we compare the Triton server with our Spatio-Temporal (ST-FD) scheduler.}\Scut{these approaches to demonstrate the effectiveness of our approach compared to DNN inference platform Triton.}

We now compare Triton server with our Spatio-Temporal (ST-FD) scheduler. We evaluate the total time taken to infer 10000 images each by 4 different DNN models, Alexnet, Mobilenet, ResNet-50, and VGG-19 (40000 images in total) using one V100 GPU. With the Triton server, we studied two alternatives: temporal sharing by sending inference requests to just 1 model at a time\Scut{, so that the workload causes interleaving of inference across the 4 models}; in the second alternative, we sent requests to all 4 models concurrently. For the ST-FD scheduler, we also sent concurrent requests to all 4 models.\Scut{We present the results in Table.~\mbox{\ref{tab:triton-4-models}}.}  We observed that\Scut{the models running in the Triton server complete their inferences in about 55-60 seconds. We see that} both modes of execution in the Triton server (temporally interleaved and concurrent execution) take about the same time to finish inference, \textbf{58 and 55 seconds}, respectively. The ST-FD scheduler completes all the inferences more than 37\% faster (\textbf{36} seconds).\Scut{Triton's utilization of GPU's hardware scheduler leads to temporal sharing of the GPU, thus, utilizing GPU inefficiently, leading to higher tasks completion time.} Based on these experiments we see that implementation of Spatio-temporal scheduling can further enhance throughput when inferring with multiple different models concurrently.
\Scut{Spatio-Temporal scheduler spatially share the GPU across multiple model resulting in achieving higher GPU utilization and better throughput compared to purely temporal sharing approaches.}\Scut{ while concurrently running models that spatially share the GPU, this leads to lower overall task completion time.} This is a strong motivation for designing our ST-FD scheduler (\S~\ref{sec:scheduling}) and the mathematical models we develop to gain insight and information for that scheduler.\looseness-1
}
\begin{figure}[h]
\begin{minipage}{.49\linewidth}
\includegraphics[width=\linewidth]{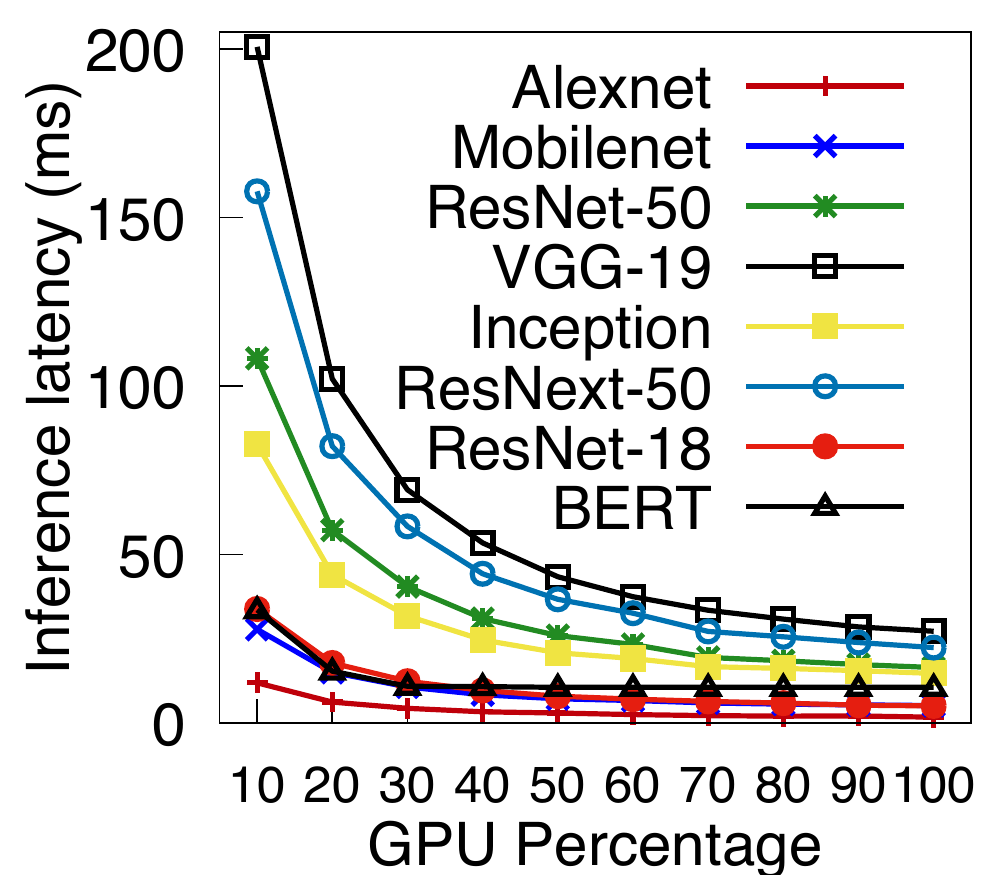}
    \caption{V100 lat. vs. GPU\%(Batch = 16)}
    \label{fig:inference_latency_batch}
\end{minipage}
\begin{minipage}{.49\linewidth}
\includegraphics[width=\linewidth]{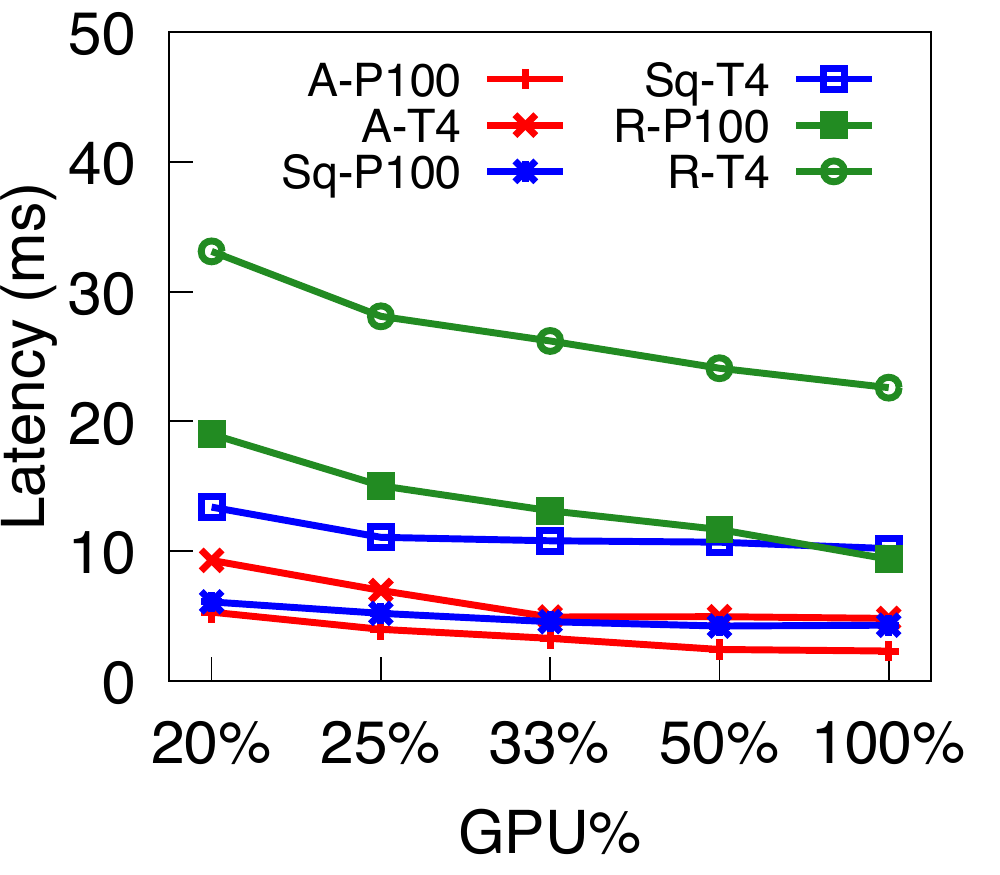}
\caption{\Scut{Alexnet (A), Squeezenet (Sq) and ResNet-50(R) Latency vs. GPU\% in} \revise{P100 and T4 GPUs profile}}
\label{fig:p100-t4}
\end{minipage}%
\end{figure}

\subsection{Dynamic GPU Resource Reconfiguration} \label{sec:dynamic_gpu_recon1} 
\Scut{We observed that the GPU\% of a given model might need to be changed dynamically when working with dynamic workload.}\Scut{ to models multiplexed on a GPU.} Due to the limitation of CUDA MPS~\mbox{\cite{nvidiamps2019}}\Scut{to configure the GPU\%}, any GPU resource readjustment requires us to spin up a new CPU process with an updated GPU\%. This results in several seconds of downtime (depending on the ML framework initialization).
We utilize the overlapped execution approach of GSLICE~\mbox{\cite{GSLICE}}, which maintain an \textit{active-standby} pair of process, where an active process keeps processing incoming requests while a standby process loads the DNN model into the GPU with updated GPU\%. The standby takes over inference when ready, thus, avoiding downtime.\looseness-1

While changing the GPU\%, two instances of the same model, the original and the new model,\Scut{loaded with a new GPU\% are running concurrently} 
occupy the GPU during the brief overlap time. This increases the GPU memory demand. We overcome this drawback through DNN parameter sharing utilized in GSLICE~\cite{GSLICE}.\Scut{by the model in the GPU  to run in the GPU as each instance of the model loads its own model parameters\Scut{ and creates its own buffers }in GPU. We explain }\Scut{ explained
in \S~\ref{sec:dynamic_gpu_recon2}.To counter the increased GPU memory requirement while loading a new model with a different GPU\%
we share DNN parameters and other buffers between the original and new model.} 
\Scut{
As model parameters remain the same in both models, \revise{we separately load the model parameters using an orchestrator process} and create cudaIPC handles for those parameter buffers.\Scut{create cudaIPC handles for all the parameter buffers of the original model and} 
When a new model instance is loaded with a new GPU\%, \Scut{it forgoes copying the model parameters from CPU to GPU. }
it receives cudaIPC handles from the orchestrator and uses them to utilize the model parameters for inference.
Parameter sharing reduces the memory required by the newly loaded DNN model by up to 40\%, depending on size of the model's parameters. Thus,
\Scut{we not only avoid idling the GPU but also} 
we maintain a low GPU memory profile which enables us to run larger DNN models.\looseness-1}
We use cudaIPC to share the weights and parameters loaded by the original model with the new loading model, thus, removing the need of loading the weights again.Parameter sharing reduces the memory required by the newly loaded DNN model by up to 40\%.\looseness-1

\subsection{Loading models without known Knee\%} When a model which is not profiled and whose knee is not known is started, our platform initially provides it a nominal, 30\%, GPU. The GPU\% is then readjusted using Dynamic GPU resource reconfiguration to find the knee based on the inference latency using a simple binary search.\looseness-1 \Scut{The platform profiles the model at different GPU\% using a simple binary search to locate the knee quickly.}

\Scut{
\begin{table}[]
    \centering
    \caption{\Scut{Task completion time in} Triton and ST-FD scheduler with 4 DNN models}
    \vspace{-2mm}
    \resizebox{\columnwidth}{!}{
    \begin{tabular}{|c|c|c|c|}\hline
         & \makecell{Triton\\temporal} & \makecell{Triton\\concurrent} & \makecell{ST-FD\\ Scheduler} \\\hline
         Task completion time (Sec.)& 58.61 & 55.55 & 35.59\\\hline
    \end{tabular}}
    \label{tab:triton-4-models}
\end{table}
}
\Scut{
\begin{table}[]
    \centering
    \begin{tabular}{|c|c|c|c|c|}\hline
         &  \makecell{Alexnet\\2-instances}&\makecell{ResNet-50\\2-instrances}&\makecell{Triton\\temporal}&\makecell{Triton\\concurrent}\\\hline
         Throughput (ips)& 1675 & 989 & 804 & 844\\ \hline
    \end{tabular}
    \caption{Caption}
    \label{tab:my_label}
\end{table}
}
\Scut{
\begin{table}[]
    \centering
    \resizebox{.4\columnwidth}{!}{
    \begin{tabular}{|c|c|}\hline
         & \makecell{Thpt.(ips)}  \\\hline
         \makecell{Alexnet\\(2-instance)}& 1675\\\hline
         \makecell{ResNet-50\\(2-inst)}&989\\\hline
         \makecell{Triton\\temporal}&804\\\hline
         \makecell{Triton\\concurrent}&844\\\hline
    \end{tabular}}
    \caption{Caption}
    \label{tab:my_label}
\end{table}
}
\section{Modeling DNN parallelism}\label{sec:model_dnn_parallelism}

\subsection{Compute Bound vs. Memory Bound Workloads}
\begin{figure}
 \captionof{table}{Compute \& memory bound kernels \Scut{profiles}}
 \vspace{-3mm}
 \resizebox{\columnwidth}{!}{%
    \begin{tabular}{|c|c|c|c|c|c|}\hline
        Model & Layer & GFLOPs & \makecell{Bytes\\$(10^6)$} & \makecell{Arit.\\ Int.} & Limit  \\\hline
        Alexnet & Conv.2 &0.30 &0.22 & 182 & Compute\\
        ResNet50 & Conv.2 &0.103 &0.121 &393 & Compute\\
        VGG-19 & Conv.11 &3.7 &9.44 & 391& Compute\\
        GNMT & LSTM &0.016 &8.38  & 2 & Memory \\\hline
    \end{tabular}}
    \label{tab:arithmetic_index_table}
\end{figure}
\Scut{The latency of execution of a DNN in a GPU does not only depend on compute latency.}The latency of accessing parameters and weights of the DNN layer from the GPU DRAM can be significant. Many studies~\cite{deepcpu} have suggested that\Scut{ performance of } memory-bound \Scut{or memory-limited} DNN kernels may have a small amount of compute and are likely to be limited by GPU memory bandwidth.\Scut{unlike compute bound \Scut{'math-limited' }DNNs} 
NVIDIA has proposed an \textit{arithmetic intensity} (A.int) metric~\mbox{\cite{nvidia_math_limt}} to estimate if a\Scut{DNN layer} kernel is memory or compute bound. The \textit{A. int} of a kernel is computed as a ratio of floating point operations\Scut{a layer performs } to memory (bytes) it fetched\Scut{ from main memory}. \mbox{\ie $A.int=\frac{\#operations}{\#bytes}$}. NVIDIA reports the arithmetic index of V100 GPU (in our testbed) is 139.8 \textit{FLOPS/Byte}~\mbox{\cite{nvidia_math_limt}}. Any kernel lower than the GPU's arithmetic index is memory-bound, while a  kernel with higher index is compute-bound.\looseness-1

We analyzed the most frequently occurring kernels of CNNs Alexnet~\cite{krizhevsky2012imagenet}, ResNet-50~\cite{he2016deep}, VGG-19~\cite{simonyan2014very}, and an RNN, GNMT~\cite{wu2016google}, to illustrate the behavior of compute and memory-bound DNNs. We present the results in Table.~\mbox{\ref{tab:arithmetic_index_table}}. Most convolution layers exceed the GPU's A.int, thus, are compute-bound. These layers can reduce their runtimes if more compute is available. However, kernels like LSTM in GNMT, which operate with large input and output features (1024 features in GNMT), require a lot of data but perform relatively fewer computations compared to convolution. Therefore, they score very low A.int. We should note that DNNs are not entirely constructed of convolution or LSTM layers. However, CNNs, in general, have more convolution kernels.\looseness-1

\subsection{Memory Contention While Multiplexing}
Studies~\cite{gpucache1,jia2018dissecting} of scientific computation workloads have shown that the GPU cache size and occupancy are important factors influencing the latency of kernel execution. We also examine the 
effect of cache contention while running multiple DNN models. However, we observe with DNNs, that the inference latency does not vary significantly \emph{if} SM isolation is maintained. Since we indeed maintain SM isolation with spatial multiplexing using CSS, the impacts of contention in the GPU cache or other memory resources is minimal. We present the 99$^{th}$-percentile inference latency (batch = 16) of DNN models running in isolation (Fig.~\ref{fig:inference_latency_batch}) 
versus the same model multiplexed at its knee GPU\% with 4 other models in Table~\ref{tab:isolationvsmultiplexed}. Inference latency varies less than 3\%, confirming this 
minimal impact. 
Thus, we do not utilize a separate variable for delay caused by the GPU cache. Instead, in the model of a DNN that we discuss in the next subsection, we consider all the memory related delays as a single variable.\looseness-1 
\begin{table}
\begin{minipage}{\linewidth}
    \centering
    \caption{Latency (ms) in isolation and multiplexed }
    \vspace{-4mm}
    \begin{tabular}{|c|c|c|c|}\hline
         Model&Knee\%&Isolation&Multiplexed\\\hline
         Mobilenet&20\%&9.8 (ms)&9.9\\
         ResNet-18&30\%&12.4&12.4\\
         BERT&30\%&9.3 &9.3\\
         ResNet-50&40\%&28.9&28.5\\
         VGG-19&50\%&51.2&52.4\\\hline
    \end{tabular}
    \label{tab:isolationvsmultiplexed}
\end{minipage}%
\end{table}
\begin{table}
\begin{minipage}{\linewidth}
\centering
    \caption{Table of Notations for DNN Model}
    \vspace{-2mm}
    \resizebox{\columnwidth}{!}{
    \begin{tabular}{|c|c|}\hline
        \textbf{Variable} & \textbf{Description} \\\hline
         $b$ & Batch Size\\
         $p$ &  1st kernel's number of concurrent ops. (tasks)\\
         $Kmax$ & Maximum number of kernels\\
         $K_{i}$ & $i^{th}$ kernel\\
         $N_{i}$ & Number of parallelizable operations for $K_{i}$\\
         $R_{i}$ & Number of repetition of $K_{i}$ in DNN\\
         $M$ & Memory Bandwidth per SM\\
         $d_{i}$ & Data for $i^{th}$ kernel (parameters \& input)\\
         $S$ &  Number of allocated SMs  \\\hline
         
    \end{tabular}
    }
    \label{tab:dnn_limit_on_parallelism}
\end{minipage}
\end{table}

\Scut{
Things to focus on:
\begin{enumerate}
    \item Message should be:It would be ideal if the GPU\% can be changed instantaneously but it cannot. And running them in different process with different GPU\% will be too complicated involving many more mechanism like IPC. So we pick a percentage that provides good latency compared to 100\%. 
    Big layers/kernels might run a bit slow with reduced resources, but small layers are not affected at all. But we get free GPU\% that can be utilized by another application
\end{enumerate}
}

\Scut{
\begin{figure*}[t]
\includegraphics[width=\linewidth]{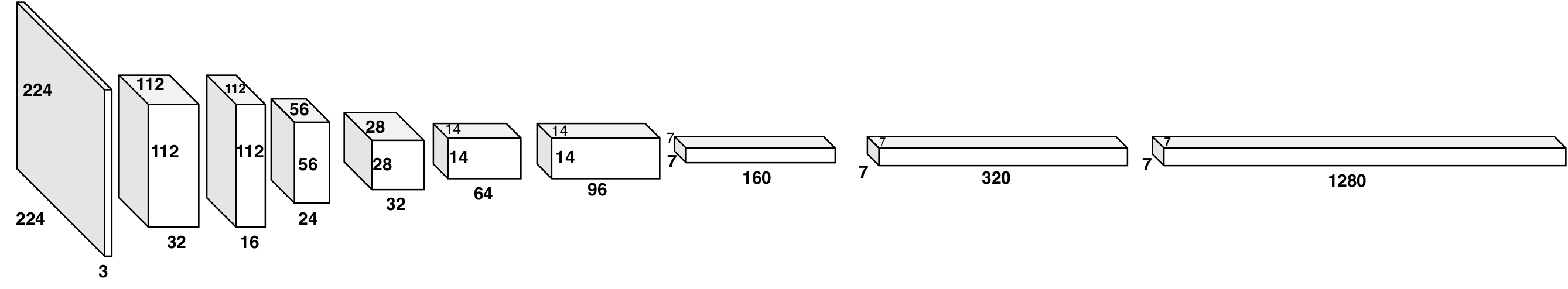}
\label{fig:mobilenet_architecture}
\vspace{-3mm}
\caption{ 
\Scut{(a) DNN Kernels of ResNet-18 running in Pytorch and their average runtime  (total 82 kernel executions) for inferring 1 image of resolution 224$\times$224, with different GPU\% (Kernel 9, 10 and 11 are convolution kernels). (b) }Feature matrix dimensions for image inference in Mobilenet.
}
\vspace{-6mm}
\end{figure*}
}

\cut{DNN inference models typically consist of neural network (NN) with multiple layers between the input and output (\eg Mobilenet, shown in Fig~\ref{fig:mobilenet_architecture}).
These layers vary in their computational requirements based on the size of the input 
and the number of operations, \eg convolutions that the layer needs to perform. These variations often result in under-utilization of the available parallel processing capacity of the GPU. First, we present a model of DNN that illustrates scenarios where the limits to parallelism exist. We then profile and analyze a number of popular DNNs 
to empirically demonstrate such limits 
on real GPU hardware.
Convolution layers often take the major share of the computation~\cite{10.1145/3146347.3146356}(\eg contribute to 90\% computational complexity of a DNN model in ResNet-50).}

\label{sec:modelling_dnn_parallelism}

\Scut{
\begin{figure}[t]
    \includegraphics[width=\linewidth]{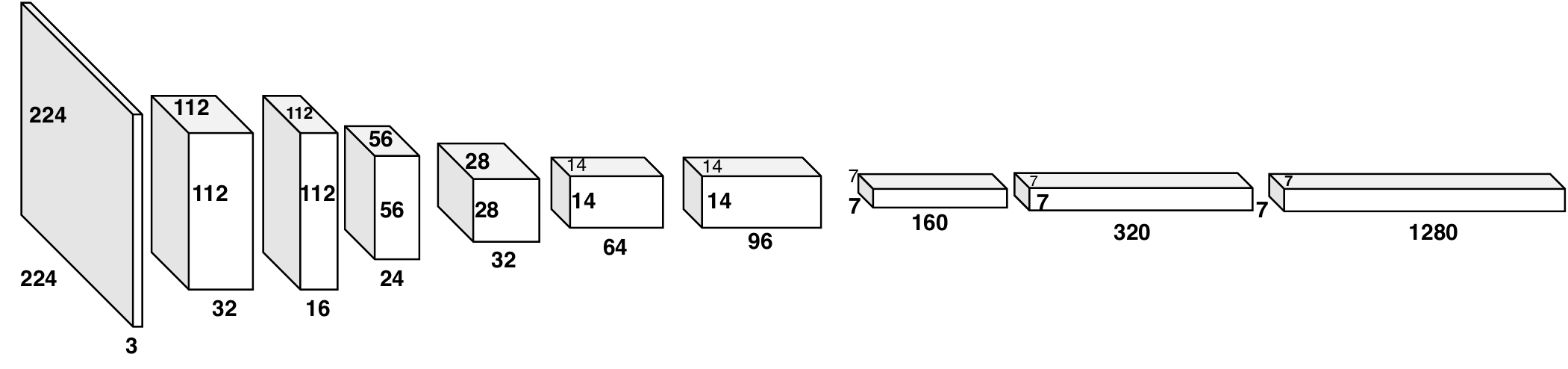}\vspace{-4mm}
    \caption{Feature matrix dimensions of Mobilenet}
    \label{fig:mobilenet_architecture}
    \vspace{-6mm}
\end{figure}
}

\Scut{
There appear to be two main reasons for why a DNN model has limit on amount of parallelism it exploits. First, some of the DNN's layers are unable to fully use all the SMs in the GPU due to small amount of computation they require. Second, some of the DNN kernels are memory bound, thus, providing more compute SMs do not improve their runtime.\Scut{ Combination of these memory limited and compute limited kernels in a DNN there exists a knee GPU\%. Providing a higher GPU\% will not significantly benefit either type of DNN, while reducing the GPU\% below that knee value rapidly increases the inference latency for both types of DNNs. } Moreover, there are the fixed costs of running DNN in GPU\Scut{ such as non-parallelizable (serialized) tasks such as launching kernel in GPU, DMA read/write to host memory \etc} that also contribute to limit the ability of DNN to fully utilize the GPU all the time.
}
\Scut{
There appear to be
two main reasons for why a DNN model has a limit on the amount of 
parallelism it exploits. First, DNN inference involves different layers of the DNN model, with different compute requirements, processing the input data. These layers \eg convolution, fully connected)  are often unable to fully use all the SMs in the GPU for their computation. 
For example, the feature matrix size of convolution layers of Mobilenet is shown in Fig.~\ref{fig:mobilenet_architecture}. With the variation in the dimensions of this matrix, the amount of computation required also varies, often not using all
of the available parallel processing capability of GPU. Compute heavy layers may utilize a lot of parallel GPU threads (using a lot of SMs). But the layers with light compute do not fully utilize the GPU.
Second, several operations performed during DNN execution,  such as launching kernels in the GPU, serialized global memory read/write do not benefit from the massive parallelism that the GPU offers. We also consider the issues with DNNs that may be GPU memory bandwidth limited. A DNN that has more number of memory reads/writes will often be limited by the memory bandwidth available. Reducing the number of SMs will reduce the bandwidth available. \Scut{Since there is also a limit in the physical memory bandwidth available, increasing the number of SMs beyond the memory bandwidth available will also not result in reduced latency for DNNs that are limited by memory bandwidth.} Thus, for both memory limited and compute limited DNNs\Scut{(DNN whose computation is more responsible for the latency)}, there exists a knee GPU\%. Providing a higher GPU\% will not significantly benefit either type of DNN, while reducing the GPU\% below that knee value rapidly increases the inference latency for both types of DNNs.
}

\Scut{Moreover, kernels in DNN may also be limited 
by GPU memory throughput rather than the SM's compute capability [cite google's TPU paper]. Therefore, adding more SMs by having a higher GPU\% beyond the knee does not result in a reduction of the execution latency for DNN inference. 
}

\vspace{-2mm}
\subsection{Modeling DNNs}
We now model an analytical DNN model that exhibits the characteristics of most actual DNN models, in terms of the variation in the compute workload across their different kernels.\Scut{, to illustrate the limits of a typical DNN's parallelism.} We model the DNN composed of multiple sequential \textit{kernels} executing in GPU (and other accelerators) instead of \textit{layers} as often used in other ML studies. We have observed using NVPROF profiling that each layer (\eg convolution layer) is often implemented as combination of multiple kernels in GPU, thus, we use kernel as basic component of DNN execution in this model. The model guides the determination of the best operating point (Knee) GPU\% for a DNN.  In our model, we breakdown the DNN workload into parallelizable operations (compute tasks), memory read/write as well as serialized (non-parallelizable) operations, and observe the effect of changing GPU resources. While our model is simple, it captures all the system level overheads that contributes to DNN latency, and provides us with good approximation of the Knee of each model. The simplicity of the model further aids in evaluating DNNs in different GPUs, with different numbers of SMs, as well as other accelerator hardware.\looseness-1

\Scut{In this section, we model a simple (hypothetical) DNN application that exhibits the characteristics of real-life DNN models and the variation in the workload across different layers of operation. We use this model to illustrate the limit 
on parallelism and 
to determine the amount of GPU resources that provide the best operating point\Scut{\sknote{define.}} for a model.}

Selected notation used in the analysis is shown in Table.~\ref{tab:dnn_limit_on_parallelism}.\Scut{We model a DNN as having $\mathbf{Lmax}$ number of layers to process an inference request. 
We also consider the execution of a batch \textit{\textbf{b}} of user requests.\Scut{ as batching is commonly supported by most ML frameworks to improve throughput and GPU utilization.}

We define the number of parallelizable operations per layer as $\mathbf{N_{i}}$, \ie it represents the degree of parallelism inherent in layer $\mathbf{L{_i}}$. $\mathbf{N_{i}}$ is analogous to the grid of cooperative thread groups (CTGs) used in GPU programming.} 
As in typical GPUs, each of the $\mathbf{S}$ SMs allocated to a DNN will process one parallel operation per $\mathbf{t_p}$ time. From a modeling perspective, we order the kernels by their amount of computation without losing generality. DNNs have an arbitrary order in kernel execution. However, the knee of the model is dependent on peak computation requirements of the kernels rather than the order of execution of each kernel.

\Scut{
Although the workload of the layers ordered by the actual sequence of  execution may be different, for our modeling we order the layers in decreasing order of the size of parallelizable computations.\looseness-1}
\Scut{
\begin{table}
    \centering
    \caption{Symbol Table for DNN Model}
    \vspace{-2mm}
    \resizebox{\columnwidth}{!}{
    \begin{tabular}{|c|c|}\hline
        \textbf{Symbol} & \textbf{Description} \\\hline
         $b$ & Batch Size\\
         $p$ &  1st layer's number of concurrent operations (tasks)\\
         $Lmax$ & Maximum number of layers\\
         $L_{i}$ & $i^{th}$ layer\\
         $N_{i}$ & Number of parallelizable operations for $L_{i}$\\
         $R_{i}$ & Number of repetition of $L_{i}$ layer in DNN\\
         $M$ & Memory Bandwidth per SM\\
         $d_{i}$ & Data for $i^{th}$ layer (parameters \& input)\\
         $S$ &  Number of allocated SMs  \\\hline
         
    \end{tabular}
    }
    \label{tab:dnn_limit_on_parallelism}
    \vspace{-3mm}
\end{table}
}
\Scut{In our simplified model}

We set the first kernel 
$\mathbf{K_{1}}$ as that with the greatest amount of parallelizable operations $\mathbf{N_{1}}$, which is selected as 
$N_1 = \mathbf{p}$ for modeling purposes. For subsequent kernels, the workload decreases by a fixed amount, 
so that $\mathbf{N_{i} > N_{i+1}}$.\Scut{Although the amount of reduction in the actual DNN may vary across layers, we consider a simplified model such that the reduction is constant to make our model simulation easier.}\Scut{ as the CNN processing progresses.} Eq.~\ref{eqn:layer_num_ops} specifies the amount of parallelizable operations for each kernel in the DNN.
\Scut{To accommodate batching, we increase the workload of the first layer by the factor of the batch size, and} We decrease the amount of parallelizable tasks by a fixed amount, $\frac{p\times b}{Kmax}$,
\vspace{-2mm}
\begin{equation}
    N_{i} =  \begin{cases}
    p\times b, & i=1\\
    \left \lfloor{N_{i-1}-\frac{p\times b}{Kmax}}\right \rfloor, & i\geq 2
    \end{cases}
    \label{eqn:layer_num_ops}
\end{equation}
\begin{figure*}
    \centering
    \subfloat[Inference Latency.]{
    \includegraphics[width=.24\linewidth]{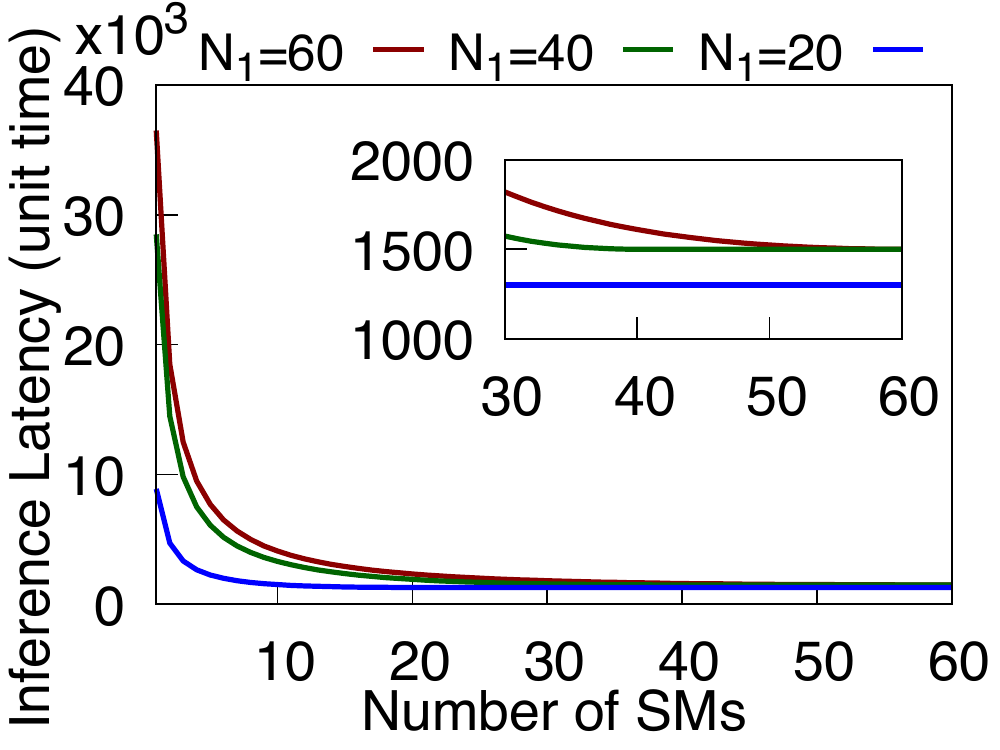}
    \label{fig:latency_varying_parallelism}}
    \subfloat[First order derivative(Eq.\ref{eq:first_order_derivative})]{
\includegraphics[width=.24\linewidth]{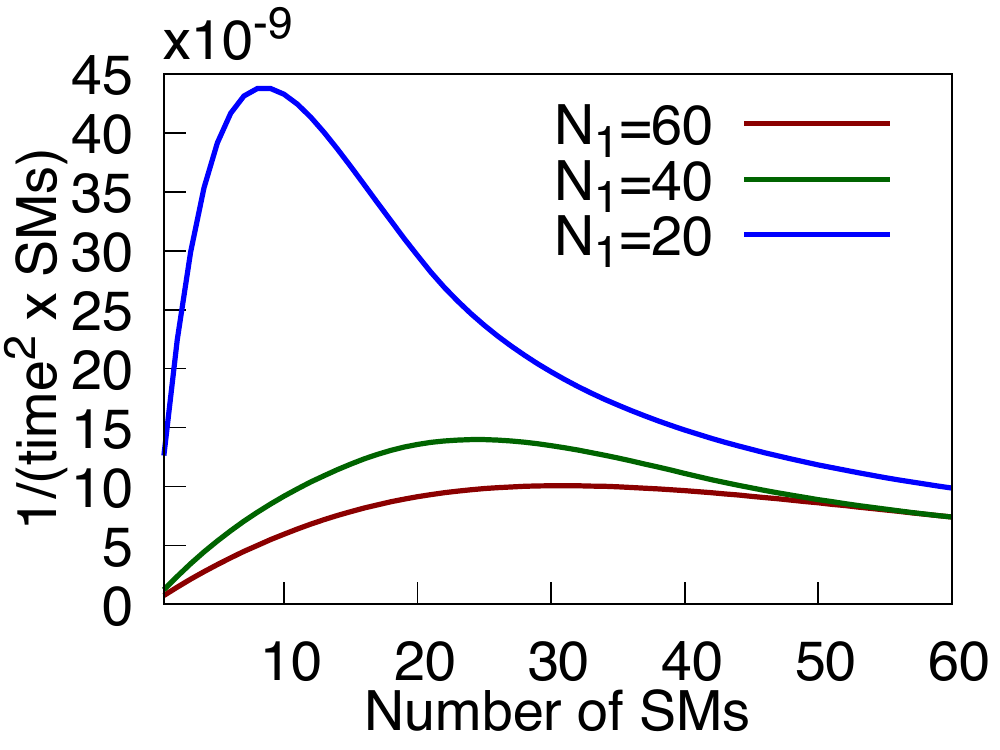}
    \label{fig:first_order_div_threads}
    }
    \subfloat[Mobilenet latency vary batch size]{
    \includegraphics[width=.24\linewidth]{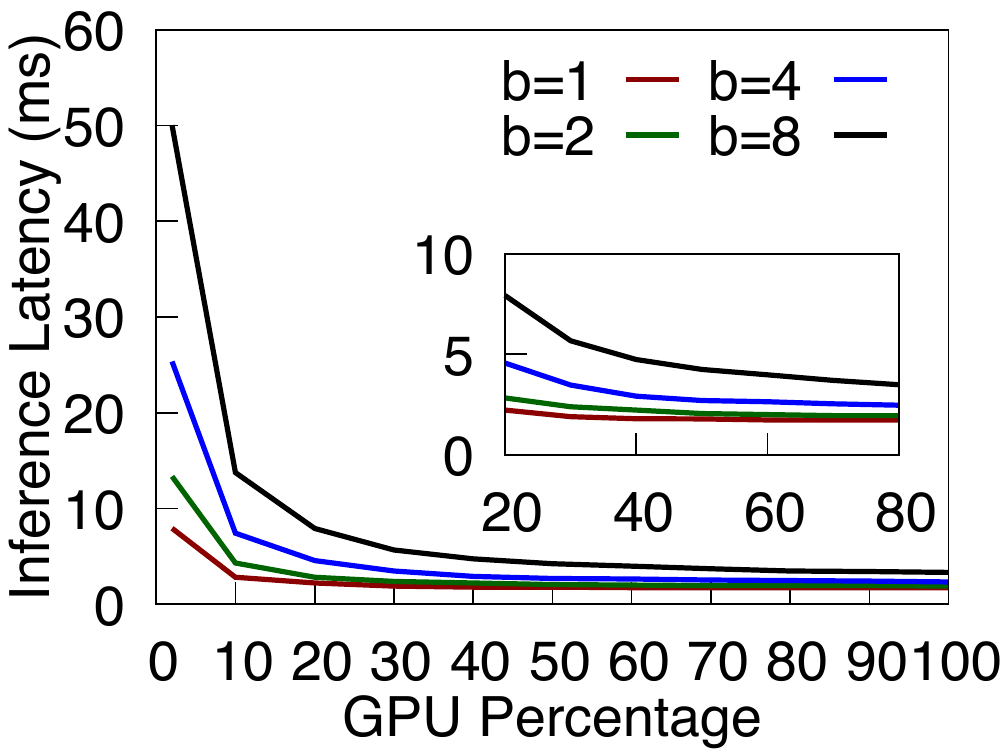}
        \label{fig:mobilenet_latency_multiple_batch}
    }
    \subfloat[First derivative (Eq.~\ref{eq:first_order_derivative})]{
    \includegraphics[width=.24\linewidth]{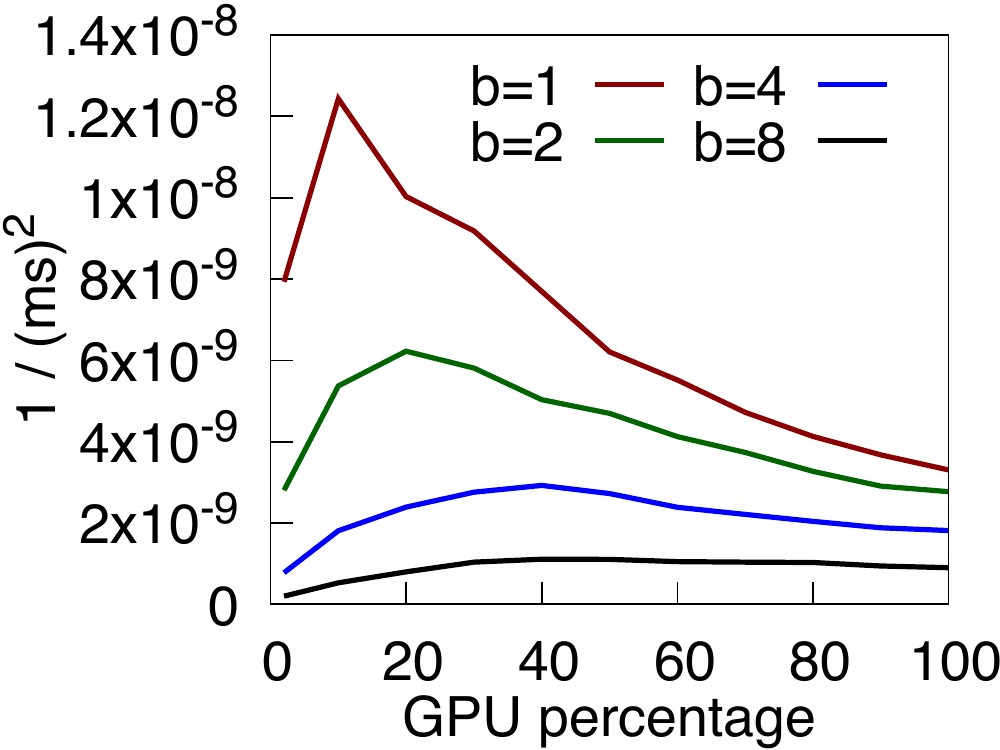}
    \label{fig:mobilenet_latency_derivative_batch}
    }
    \Scut{
\subfloat[Latency with higher batch size]{
     \includegraphics[width=.24\linewidth]{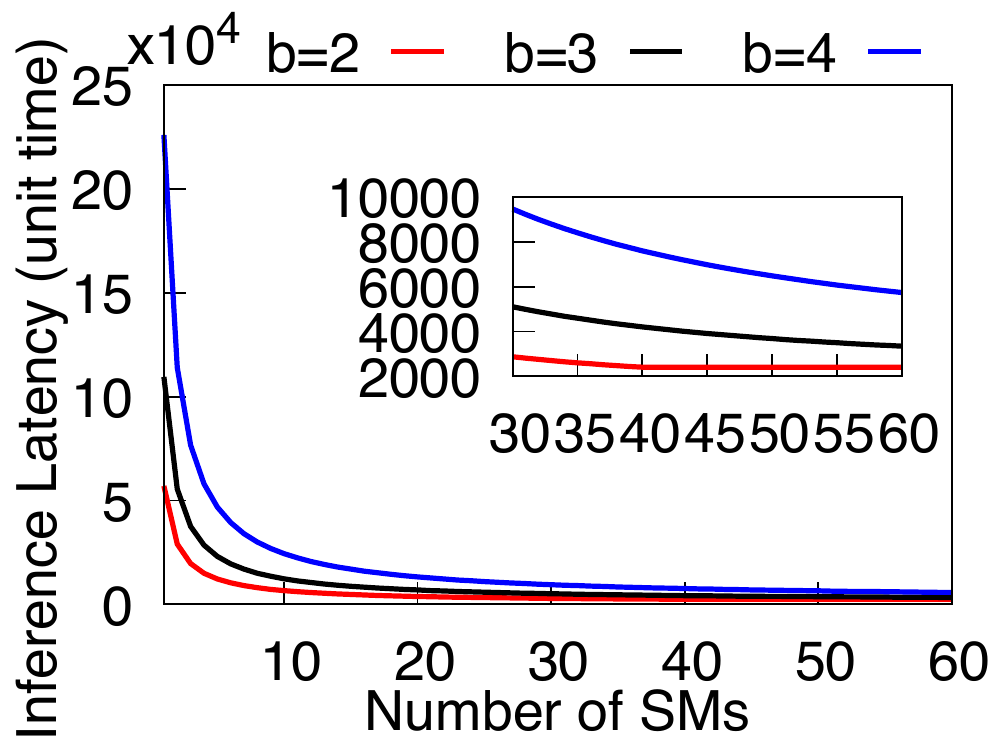}
     \label{fig:latency_different_batch_sizes}
     }
     \subfloat[Effect of higher batch size]{
     \includegraphics[width=.24\linewidth]{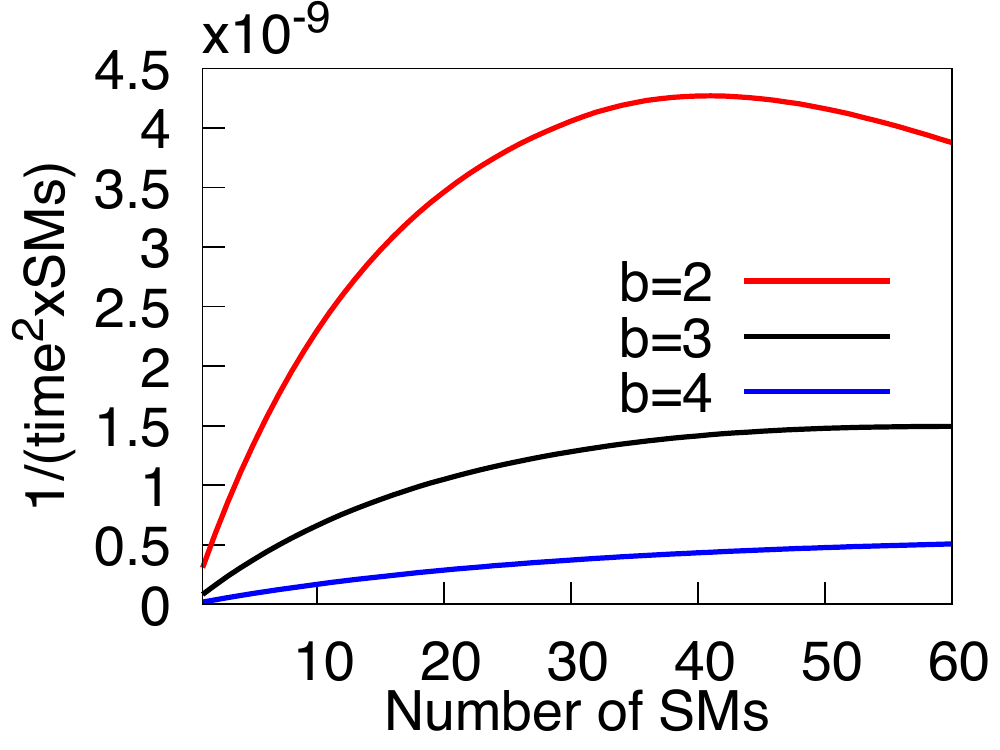}
    \label{fig:derivative_different_batch_sizes}
    }
    }
    \caption{(a), (b) Inference characteristics of analytical DNN models with varying amounts of parallelism and hardware resources.(c), (d) Demonstration of analytical model's understanding on real DNN Mobilenet}
    
\end{figure*}
for each subsequent kernel.
The number of concurrent operations decrease and reaches $\sim 0$ for the last ($K_{max}$) kernel. Correspondingly, we define the total execution time for each kernel's parallelizable tasks as $\mathbf{W_{i}} = N_{i} \times t_{p}$.\looseness-1 
\setlength{\belowdisplayskip}{0pt} \setlength{\belowdisplayshortskip}{0pt}
\setlength{\abovedisplayskip}{0pt} \setlength{\abovedisplayshortskip}{0pt}


Note: Ideally, $W_{i}$ can potentially be completed in $t_p$ units of time when we allocate greater than or equal to the $N_i$ SMs to execute $W_i$.  
If we consider that the GPU hardware is able to provide $S$ SMs to execute $K_i$, then, without loss of generality, we can show that the time taken to finish processing the kernel would depend on the minimum of the inherent parallelism, as defined by $N_i$, and the number of SMs allocated for executing the operation. 
Thus, the execution time for parallelizable operations at each kernel of the DNN can be computed using Eq.~\ref{eqn:execution_time_per_layer}. Individual kernels 
\begin{equation}
E_{i} = \frac{W_{i}}{max(1, min(S,N_{i}))}
\label{eqn:execution_time_per_layer}
\end{equation}
\Scut{DNNs are composed of several kernels with different thread count}
in the DNN often run repeatedly during a DNN inference. We define the number of repetitions of kernel $K_{i}$ as $\mathbf{R_{i}}$. \Scut{ can model the repetition of \Scut{working of a specific kernel with a desired thread count similar to the operation of a} a specific kernel in layer $L_{i}$ \Scut{of a CNN.} In addition, we define the number of repetitions of each layer $L_{i}$ as $R_{i}$ to factor the repetitive execution of the kernels.} 
We then factor the time taken to run all the serialized operations, including for kernel starting and kernel waiting for data.
The kernel starting time is considered
a constant, $\mathbf{t_{np}}$, per layer.  
The kernel's time waiting for data, however, depends on the kernel's input and parameters.
Each kernel of a DNN has a certain amount of data (model parameters, input data) that has to be fetched from GPU DRAM (main/global memory of GPU) to the CUDA cores in the SMs. We have observed that the total global memory read/write bandwidth increases with the proportion to the number of SMs allocated.
Other studies~\cite{zhang2020qosaware,micikevicius2012gpu} also point to a proportional increase. \Scut{ in memory read/write bandwidth with the increase in the number of SMs.}\Scut{Intuitively when more SMs, and more
active GPU threads running in an SM, read input or parameter data, and write the output, memory bandwidth requirements increase.}\looseness-1 
We define the latency per kernel, caused by kernel waiting for parameters, input, and other data to be loaded, as Eq.~\ref{eqn:memory_latency}. Thus, we can define the total time 
of non-parallelizable (sequential) operations $\mathbf{W_{se}}$ as Eq~\ref{eqn:total_non_paralleizable_time}. We use Eqs.~\ref{eqn:execution_time_per_layer} and~\ref{eqn:total_non_paralleizable_time} to compute DNN execution time, $\mathbf{E_{t}}$ as in Eq.~\ref{eqn:total_time_shrinking_kernel}.
\begin{equation}
     E_{m} = \frac{d_{i} \times S }{M}\label{eqn:memory_latency}
\end{equation}
\begin{equation}
  W_{se} = b\times\sum_{i=1}^{K_{max}}R_{i}\times\left( t_{np}+E_{m}\right)\label{eqn:total_non_paralleizable_time}   
\end{equation}
\begin{equation}
E_{t} = W_{se} + \sum_{i=1}^{K_{max}}R_{i}E_{i}\label{eqn:total_time_shrinking_kernel}
\end{equation}
\Scut{
\begin{equation}
 E_{st} = W_{se}+K_{max}\times p\times b
    \label{eqn:total_time_static_kernel}
\end{equation}
}
\Scut{
\begin{wrapfigure}[9]{l}{.4\textwidth}
\vspace{-4mm}
\begin{minipage}{0.4\textwidth}
\begin{align}
     E_{m} = \frac{d_{i} \times S }{M}\label{eqn:memory_latency}\\
     W_{se} = b\times\sum_{i=1}^{K_{max}}R_{i}\times\left( t_{np}+E_{m}\right)\label{eqn:total_non_paralleizable_time}\\
     E_{t} = W_{se} + \sum_{i=1}^{K_{max}}R_{i}E_{i}\label{eqn:total_time_shrinking_kernel}
\end{align}
\end{minipage}
\end{wrapfigure}
}
\Scut{
\useshortskip
\begin{equation}
    E_{m} = \frac{d_{i} \times S }{M}
    \label{eqn:memory_latency}
\end{equation}
\Scut{Where, $\mathbf{d_i}$ is data per layer (parameters \& input) and $\mathbf{M}$ is the memory bandwidth per SM. }
\begin{equation}
     W_{se} = b\times\sum_{i=1}^{K_{max}}R_{i}\times\left( t_{np}+E_{m}\right)
     \label{eqn:total_non_paralleizable_time}
\end{equation}
}
\Scut{
Therefore, the total overhead for these serialized tasks for the entire DNN, \ie including all layers, can be estimated using:  
\begin{equation}
\vspace{-1mm}
  W_{se} = \sum_{i=1}^{L_{max}}R_{i}\times t_{np}\times b 
  \label{eqn:total_non_paralleizable_time}
\end{equation}
}
\Scut{
\begin{equation}
    \vspace{-1mm}
    E_{t} = W_{se} + \sum_{i=1}^{K_{max}}R_{i}E_{i}
    \label{eqn:total_time_shrinking_kernel}
\end{equation}
}
\Scut{Unlike CNNs, other DNNs such as Long Short Term Memory (LSTM), transformers kernels do not change the dimensions of the matrix they operate with. Thus, their parallelizable workload remains relatively static.
\Scut{We considered Google's state-of-the-art Neural Machine Translation (GNMT~\cite{wu2016google}),  as a representative RNN model. 
To translate a sentence from one language to another, GNMT uses multiple LSTM kernels.
\Scut{ and an attention network to connect between the encoder and decoder.
These LSTM kernels benefit from parallelism offered by the GPU.Unlike convolution kernels whose compute requirement changes with change in input feature matrix, LSTM kernels execute with same input or output size, therefore, same number of GPU SMs each time.} We can define such kernels by using Eq.~\ref{eqn:layer_num_ops} such that the variable $i=1$ and $N_{i}$ remains the same during the DNN's execution. }
We define total execution time for RNNs, $\mathbf{E_{st}}$,\Scut{
\begin{wrapfigure}{l}{.4\textwidth}
\vspace{-2mm}
\begin{minipage}[h]{.4\textwidth}
\begin{equation}
    E_{st} = W_{se}+K_{max}\times p\times b
    \label{eqn:total_time_static_kernel}
\end{equation}
\end{minipage}
\end{wrapfigure}}
as a special case of Eq.~\ref{eqn:total_time_shrinking_kernel} (replacing the summation in Eq.~\ref{eqn:total_time_shrinking_kernel} by the execution time of first layer $(p\times b)$ repeated $K_{max}$ times), we get Eq.~\ref{eqn:total_time_static_kernel}. }


We now simulate the total time to execute a DNN under varying conditions \ie by varying the amount of parallelizable and non-parallelizable operations at each kernel and the number of SMs in the GPU.\Scut{ that can perform the computation for parallelizable operation.} As in typical GPUs, we assume the number of SMs\Scut{profile once} allocated for an DNN remains static. \Scut{ until execution of all layers of that DNN complete.}\looseness-1
Fig.~\ref{fig:latency_varying_parallelism} shows the impact on the DNN execution time when assigning  different numbers of SMs. First, we created a DNN with 50 kernels \ie $K_{max}=50$. We set the time taken for the parallel operation $t_{p}$ to be 40 units and for serialized operations $t_{np}$ to be 10 units. We repeat the simulation for 3 cases, varying the maximum amount of parallelization (concurrent operations at the first kernel) 
$N_{1}$  as 60, 40, and 20\Scut{concurrent operations }.\looseness-1


For all three cases, the execution time is very high when the number of SMs is small (1 to 5 SMs)\Scut{, especially for DNNs with a larger number of parallelizable operations}, reflecting the penalty of insufficient resources for the inherent degree of parallelism while executing the DNN kernel. 
However, as the number of SMs increases, the execution latency  decreases. Interestingly (see zoomed part of Fig.~\ref{fig:latency_varying_parallelism}), there occurs a point when giving more SMs beyond a point does not improve latency further, in each of the scenarios.
When the number of SMs provisioned exceeds the amount of parallelism inherent in the DNN kernel, there is no further reduction in the latency. 
Even before reaching this point, the latency improvements from having an increased number of SMs reaches a point of diminishing returns\footnote{ 
\ie showing marginal improvements.\Scut{We refer it as the point of diminishing returns, as adding more threads results. }
The DNN execution latency is impacted by both the number of parallelizable and non-parallelizable operations and it varies inversely with the number of allocated SMs, by Amdhal's law~\cite{amdahl1967validity}. Batching increases parallelizable work~\cite{gustaffson}.\looseness-1
}.
We seek to find the most efficient number of SMs ($S$) needed for executing a given DNN, so that the utilization of the allocated SMs is maximized. 
To compute this, we have to find the maximum of $\frac{1}{E_{t}*{S}}$, which 
represents the DNN work 
processed per unit time per SM. 
For this, we differentiate $\frac{1}{E_{t}*{S}}$ with respect to the time taken to execute the DNN.\looseness-1 
\begin{equation}
    \frac{d}{dE_{t}}\left(\frac{1}{E_{t}*S}\right) = -\frac{1}{\left(E_{t}\right)^2*S}
    \label{eq:first_order_derivative}
\end{equation}
Fig.~\ref{fig:first_order_div_threads} shows this first order derivative of the inverse of latency (Eq.\ref{eq:first_order_derivative}), showing 
that SMs for $N_{1} = 20, 40$ and $60$ reaches a maximum at $9$, $24$ and $31$  SMs respectively. 
Hence, operating at this derived `maximum' point for a DNN guarantees that there are sufficient number of SMs to provide low latency while achieving the most efficient use of the SMs.
Moreover, we can see from this that the `maximum' 
\Scut{that the amount of DNN work processed per SM in unit time, }peaks at a much lower SMs than the corresponding value of $N_{1}$. 
This is due to the impact of performing  serialized tasks adjacent to the parallelizable tasks. 
This results in lower (or no) utilization of many of the allocated SMs for the serialized tasks. Thus, further reduction in latency by increasing SMs is minimal. 

\Scut{20 and 40 threads will not improve the latency, therefore, exhibit limits to parallelism. For the simulation with $N_{i}=60$ we do not see the curve flattening with 50 or less than 50 threads. That points to DNN starting with $N_{1}=60$ can still utilize more than 50 hardware threads.  \knote{I am very confused by all of this, because it is not clear what you're referring to here. Which figure??} 
}

\Scut{
\noindent\textbf{Effect of Batching}:\Scut{ Batching multiple requests in a DNN 
is meant to correspondingly increase the degree of parallelism while executing the DNN, but also adds more non-parallelizable operations.} To show the effect of batching, we simulated a DNN with $K_{max}=50$, $N_{1}=20$, $t_p = 40$, and $t_{np}=10$. We varied the batch size from 2 to 4. 
From Fig.~\ref{fig:latency_different_batch_sizes}, we observe that the latency of execution increases with increasing batch sizes and consequently also shifts the point of diminishing returns (knee) for each case. We plot Eq.~\ref{eq:first_order_derivative} 
for different batch sizes  as shown in Fig.~\ref{fig:derivative_different_batch_sizes}.
Increasing the batch size also shifts the maximum point of the derivative\Scut{ $-\frac{1}{S*\left(E_{t}\right)^2}$,} to a higher number of SMs. For a batch $b=2$, the maximum is at 34 SMs. But, for 
$b = 3, 4$ that point exceeds 60. \Scut{
(not shown in  plot).} 
Thus, increasing the batch size increases DNN's \Scut{the degree of }parallelism.\looseness-1
}
\Scut{Thus, it used by the computation and the DNN can utilize more SMs.} 
\begin{figure*}
    \centering
    \includegraphics[width=\linewidth]{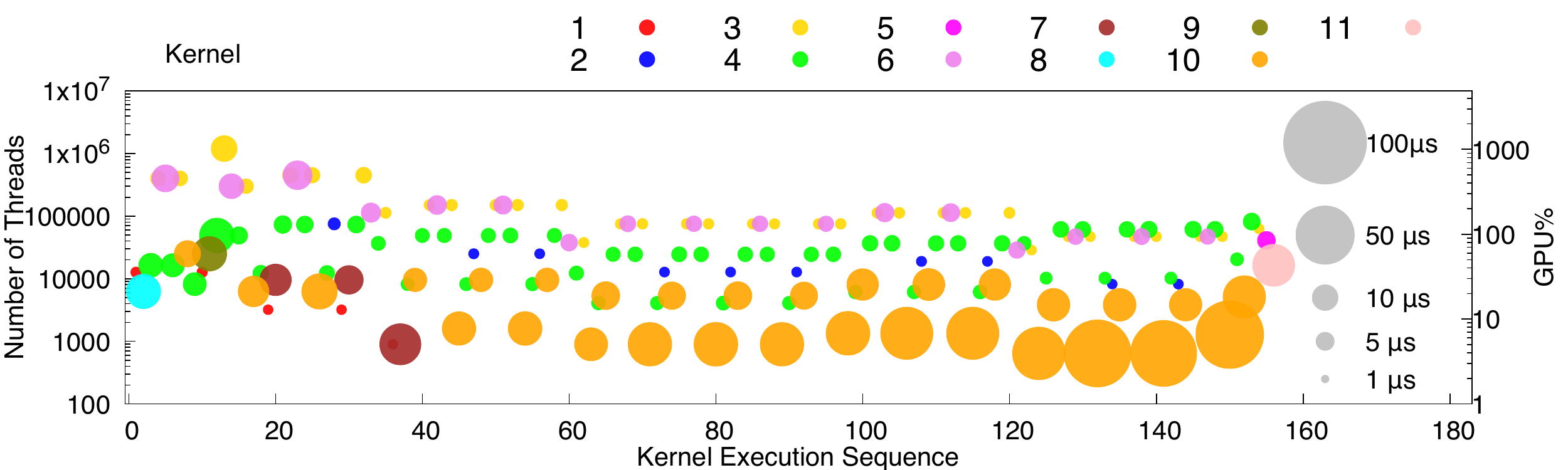}
   \caption{Thread count \& runtime (shown as area of circle) of 156 kernel of Mobilenet.}
   \label{fig:mobilenet_thread_count}
\end{figure*}
\Scut{
\begin{figure}[h]
\vspace{-6mm}
\Scut{
\subfloat[Thread count \& runtime (shown as area of circle\Scut{for each kernel}) of 156 kernel \Scut{(11 kernels, differently colored)} of Mobilenet.\Scut{ Convolution kernels are 9,10 and 11}]{
\includegraphics[width=.5\linewidth]{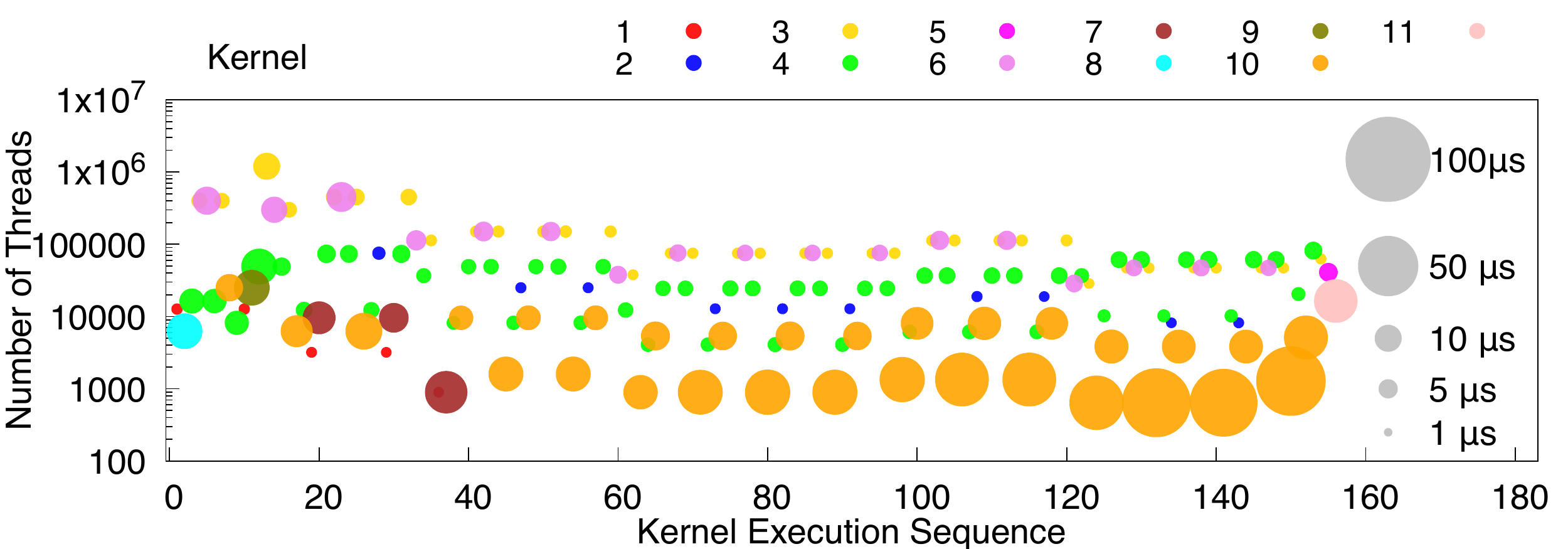}
\label{fig:mobilenet_thread_count}
}
}
\subfloat[ResNet-18 (82 kernels)\Scut{in PyTorch, 100\% GPU.}]{
\includegraphics[width=.5\linewidth]{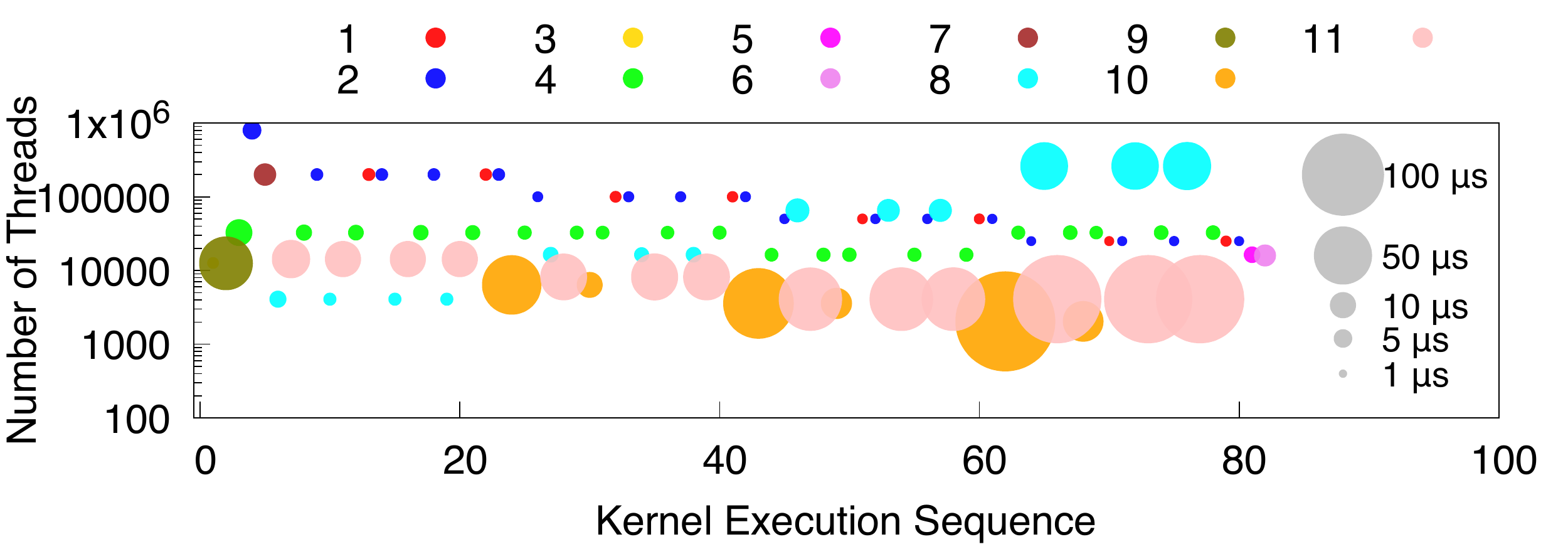}
\label{fig:Resnet18_thread_count}
}
\subfloat[GNMT (3579 kernels)]{
\includegraphics[width=.5\linewidth]{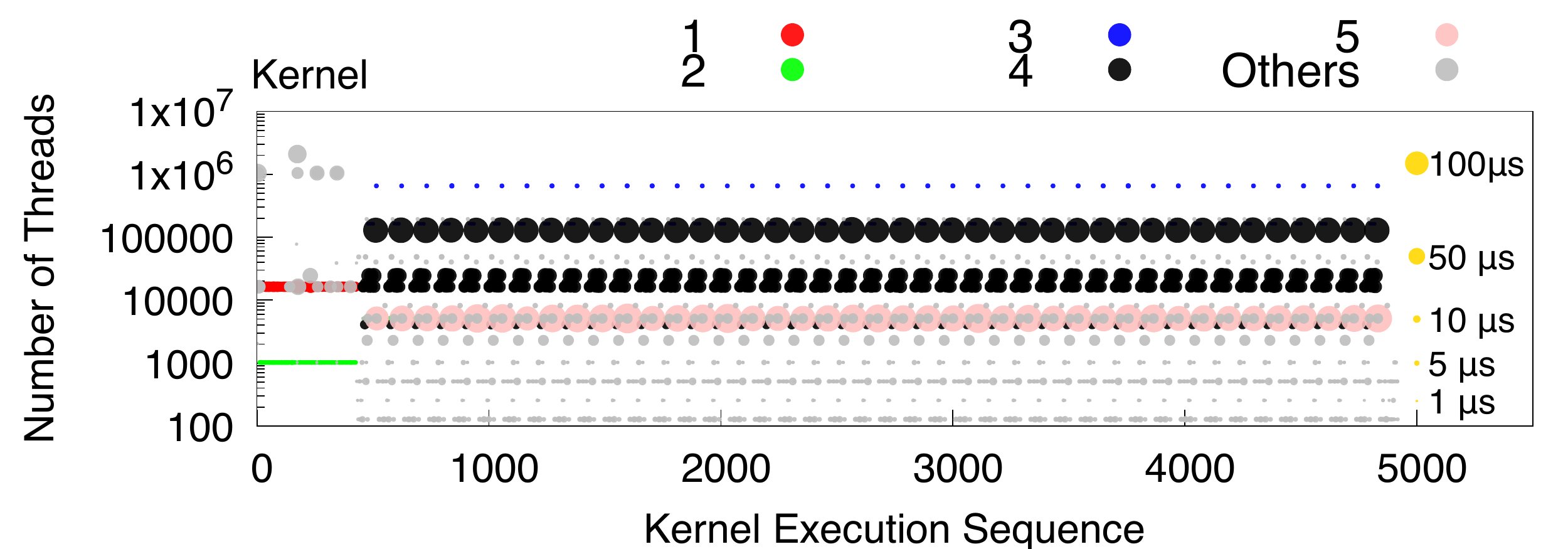}
\label{fig:GNMT_thread_count}
}
\vspace{-4mm}
\caption{Thread count and runtime of DNN kernels of ResNet-18 and GNMT in Pytorch with 100\% GPU}
\vspace{-4mm}
\end{figure}

\vspace{-1mm}
}
\subsection{Analyzing Execution of Typical DNNs}
\Scut{
\begin{figure}[]
 \includegraphics[width=\linewidth]{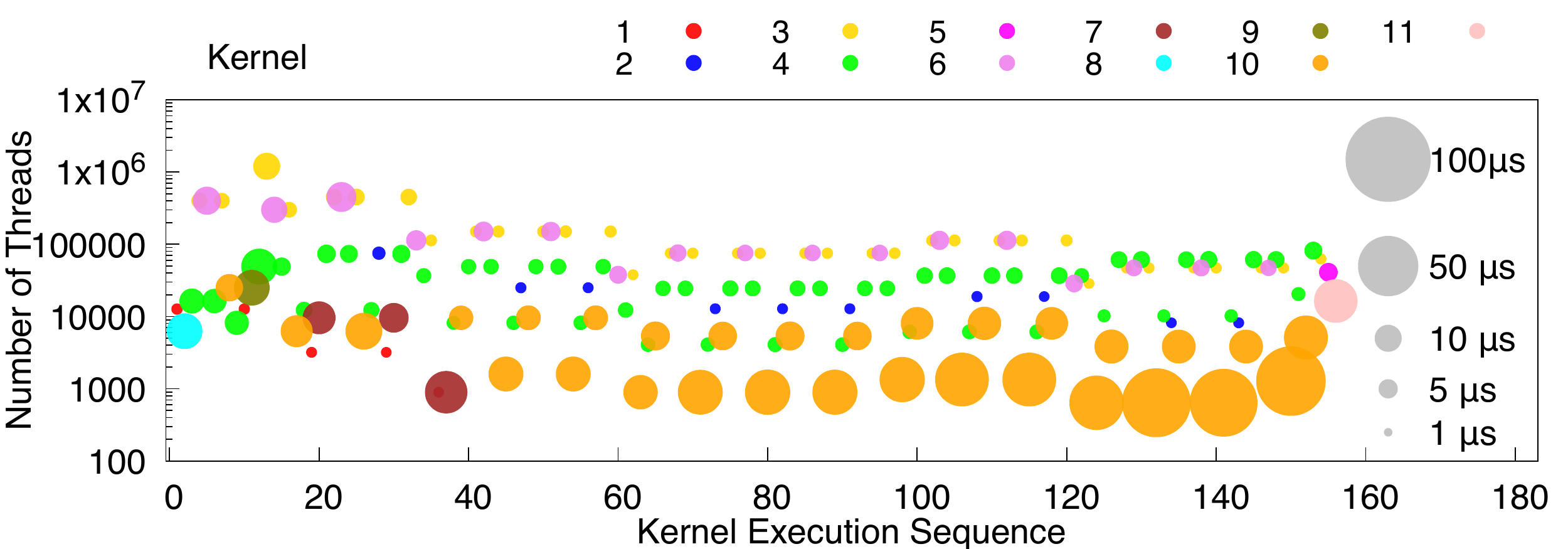}
\vspace{-3mm}
\caption{Thread count \& runtime (shown as area of circle\Scut{for each kernel}) (100\% GPU) of 156 kernel \Scut{(11 kernels, differently colored)} of Mobilenet in PyTorch.\Scut{ Convolution kernels are 9,10 and 11} 
}
\label{fig:mobilenet_thread_count}
\vspace{-2mm}
\end{figure}
}

\Scut{We examine the internal execution of the DNN kernels of several representative DNN models for both Convolutional Neural Network (CNN) and Recurrent Neural Network (RNN) models. }
We profiled and analyzed Mobilenet, ResNet and \Scut{Google's Neural Machine Translation System} GNMT DNNs using the NVPROF profiler~\cite{NVIDIAVisualProfiler} to capture the GPU resource usage and the execution time of the DNN kernels. 
\vspace{-1mm}
\subsubsection{CNN model: Mobilenet}
We profiled the inference of Mobilenet using 100\% of a V100 GPU. 
For each kernel, we show the GPU thread count on the y-axis (in log scale) and the corresponding runtime as the area of the bubble in Fig.~\ref{fig:mobilenet_thread_count}. The approximate GPU\% required for all the threads to run concurrently is on Y2-axis (log scale, on the right). We approximate this GPU\% by considering that only 2048 threads can run in an SM concurrently, due to limits on the number of concurrent blocks and warps~\cite{NVIDIAVolta}. The kernel's design and thread distribution across different threadblocks can lead to a higher SM demand than absolutely required.\looseness-1

We plot 11 distinct kernels of a Mobilenet model (each identified by a different color in Fig.~\ref{fig:mobilenet_thread_count}). These kernels are executed a total of 156 times per inference. We observe that few of the kernels (kernel 3, 4 and 6, in particular) require more than 100\% of the GPU to run. These kernels demand more threads than a GPU can run concurrently. However, these kernels run for a very short time and do not contribute significantly to the total inference latency. The kernels that contribute more to the total latency, such as kernels 10 and 7 utilize less than 10\% of the GPU. This is due to the fact that the DNN's inference feature matrix gets smaller, thus, resulting in limiting the inherent parallelism. Thus, these kernels use fewer parallel GPU threads and run for long time with low GPU\% demand. They contribute to lowering the Knee GPU\% of the entire DNN model.
\Scut{
Kernel 10 is a convolution kernel that contributes the most to the processing latency. We see that Kernel 10's thread count decreases as the inference operation progresses. This is due to the fact that the inference feature matrix gets smaller, resulting in that kernel using fewer parallel GPU threads. However, the execution time of kernel 10 increases as the inference progresses. 
This indicates that this convolution kernel is still compute-heavy, even though it uses fewer parallel threads. } 
From this understanding, 
when the amount of parallelism of a kernel is low, increasing the number of GPU SMs will not reduce the execution time of the kernel, since the additional SMs will not be utilized.\looseness-1 \Scut{In fact, we see a similar trend with}\Scut{ convolution kernels 9, 10 and 11 in the}\Scut{ ResNet-18 model (Convolution Kernels 10 and 11) in Fig.~\ref{fig:Resnet18_thread_count}. 
Similar to Mobilenet, ResNet-18's kernels also show the same characteristic of decreasing thread count and increased execution time as the inference execution progresses.\looseness-1 }
\color{black}

\begin{figure}
\vspace{-2mm}
    \centering
    \Scut{
    \subfloat[Latency, vary batch size]{
    \includegraphics[width=.32\linewidth]{figures/limits_to_parallelism/mobilenet_model/mobilenet_latency_plot.pdf}
        \label{fig:mobilenet_latency_multiple_batch}
    }
    \subfloat[First derivative (Eq.~\ref{eq:first_order_derivative})]{
    \includegraphics[width=.32\linewidth]{figures/limits_to_parallelism/mobilenet_model/mobilenet_derivative_latency_plot.pdf}
    \label{fig:mobilenet_latency_derivative_batch}
    }
    }
    \subfloat[First derivative (others)]{
    \includegraphics[width=.5\linewidth]{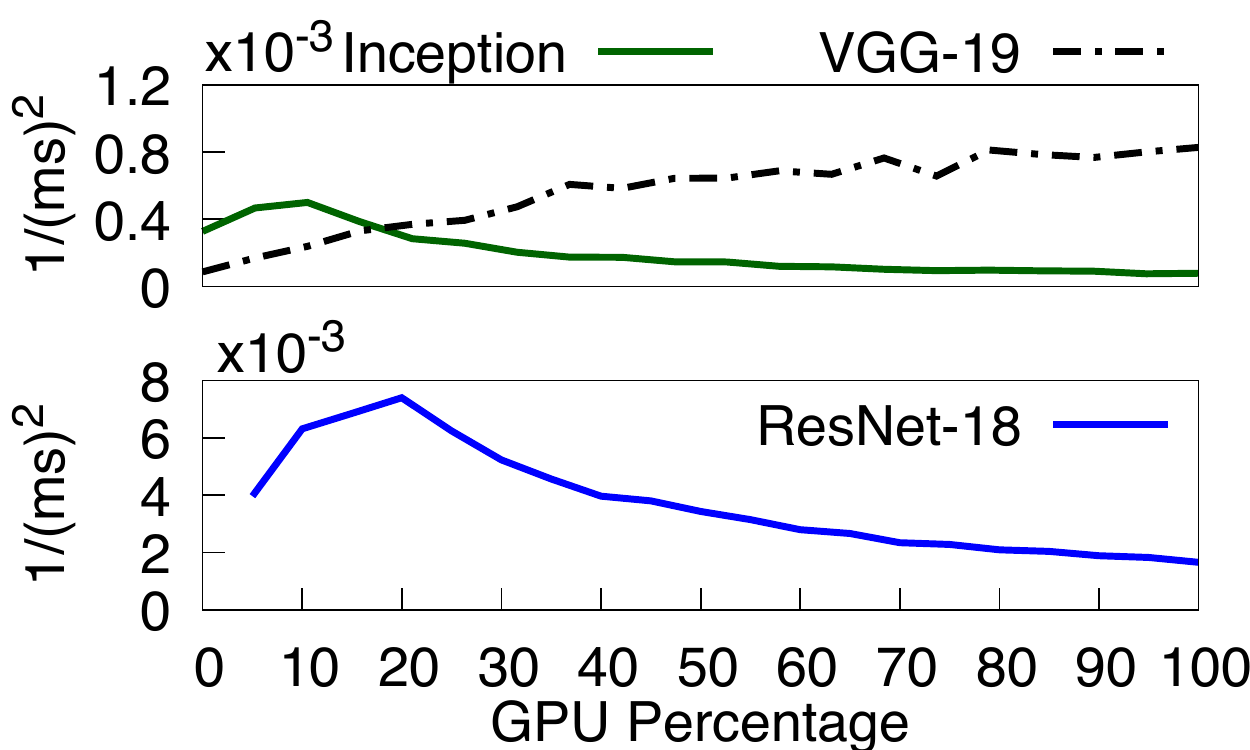}
    \label{fig:multiple_model_derivative_plot}
    }
    \subfloat[BERT]{
    \includegraphics[width=.5\linewidth]{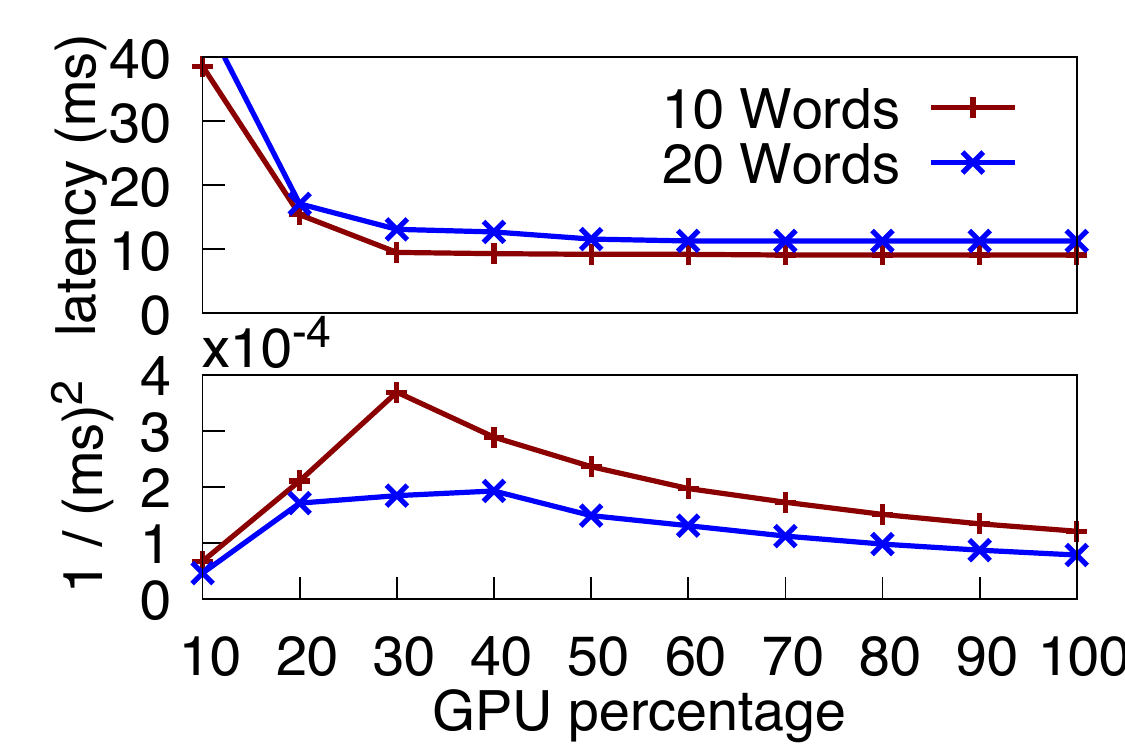}
    \label{fig:bert_first_derivative}
}
    \caption{DNN Latency, first derivative as in Eq.~\ref{eq:first_order_derivative}\Scut{ and Energy profile for different GPU \% with various DNN models. (a), (b) are for Mobilenet only.}}
\end{figure}

We also analyzed the inference time with different batch sizes of Mobilenet (Fig.~\ref{fig:mobilenet_latency_multiple_batch}). In all the cases, for a given batch size, the latency reduces with an increase in GPU\%. But, across all evaluated GPU percentages, the latency \emph{increases} with increasing batch sizes. 
Fig.~\ref{fig:mobilenet_latency_derivative_batch} shows the first derivative of the inverse of Mobilenet's latency obtained using Eq.~\ref{eq:first_order_derivative}. 
The maximum of the derivative, \ie the most efficient point for DNN operation, for batch sizes of 1, 2, 4 and 8 occurs at GPU\% of $\sim$ 10, 20, 40, and 50 respectively. 
This shows that with increasing batch size, \ie increased parallelism, the GPU\% at which the maximum utilization point occurs, based on Eq.~\ref{eq:first_order_derivative}, also increases. 
Fig.~\ref{fig:multiple_model_derivative_plot} shows the different maximum utilization points for the different models. 
Lightweight models such as Inception and ResNet-18 have a maximum at a lower GPU\%, while compute-heavy VGG-19 does not see an inflection point up to 
100\% GPU.\Scut{ The lightweight models show limits to their parallelism, while the compute-heavy VGG-19 has enough parallel work that it can exploit all of the parallelism that this V100 GPU has to offer.}  
These characteristics of the individual DNN's execution strongly correlate and match with the theoretical DNN model we presented.\looseness-1 

\subsubsection{Transformer Model BERT}
\Scut{
\Scut{
We profiled the working of GNMT to translate 25-word sentences from English to German. Fig.~\ref{fig:GNMT_thread_count} shows the thread count and runtime of all the GPU kernels used in GNMT.
\Scut{ The GNMT DNN uses a lot more kernels (4919) compared to CNNs.} 
For simplicity, we show the 5 most-repetitive, distinct kernel types and group the rest  (less-repetitive kernels) as `Others'. Most of GNMT's kernels run for less than 100$\mu$s.
\Scut{, and many kernels have runtime lower than 50$\mu$s.} 
A large number of short-running LSTM kernels indicates that GNMT occupies a considerable part of the GPU memory and may be limited by memory bandwidth~\cite{jouppi2017datacenterlong}.}
\Scut{ which is a relatively small time compared to overall inference time of \textasciitilde120 ms. }
\Scut{
We observe that 
\Scut{just like Mobilenet and ResNet-18 (CNN models) the thread count for different kinds of kernels vary (order of $10^2$  to $10^6$), but} 
a majority of the kernels that repeat show the same thread count and execution time throughout the model's execution (except for kernel 4). This is because LSTM's GPU kernels 
do not change the dimension of the feature matrix. However, variations in thread count and execution time between different kernels means that the GPU utilization changes dynamically while inferring a request.}\Scut{Since GNMT's GPU-thread demand is steady and modest, and is composed of a lot of memory-bound LSTM kernels, it has a small knee\%.\looseness-1\Scut{ and relatively modest over time, it gives us the opportunity to multiplex, at the complete model level, across multiple models that can spatially share the GPU.}
}\Scut{\knote{the next two sentences are a repeat and need to be commented out.}
We evaluated a RNN model, GNMT, a DNN that translates sentences into a target language. GNMT is composed of a large number of short-running LSTM kernels and is\Scut{indicates that GNMT occupies a considerable part of the GPU memory and may be} limited by memory bandwidth~\cite{jouppi2017datacenterlong}.
\knote{up to here should be commented out.}}
We computed the 99th-percentile inference latency of GNMT with different sentence size at different GPU\%,
\Scut{computed the first derivative of the inverse of the measured latency. The 99th-percentile of latency of GNMT} 
as shown in Fig.~\ref{fig:gnmt_latency}.
\Scut{, based on inferring/translating sentences of different lengths 
as the GPU resource percentage changes. We also increased batch size to 10 for inference}\Scut{ 25-word sentences and evaluated the latency for different GPU\%} 
The first order derivative of these latencies is shown in Fig.~\ref{fig:gnmt_first_derivative}. The latency for inference of 8, 12 and 25 word sentences (batch size of 1) improves with upto 15-20\% of GPU and then shows diminishing returns. \Scut{This observation is similar to what we see with CNN models, as in Fig.~\ref{fig:inference_latency}.}  Even with longer sentences, this knee for GNMT's does not change, indicating that increasing the 
total amount of computation by a factor of 2-3$\times$ does not increase inference latency substantially. \Scut{ 
GNMT model also shows the limit of parallelism it can use.} 
But, just like CNN models, with
\Scut{ a 25-word sentence and} 
a batch size of 10, \ie when presented with a higher (instantaneous) compute demand with batching, the inference latency improves with a higher GPU\%, upto about 25\% GPU (i.e., an additional 5-10\% GPU), but shows diminishing returns after that. }
We also present the evaluation of the inference latency for the transformer-based natural language processing DNN, BERT, as well as the first order derivative, 
per GPU\% in Fig.~\ref{fig:bert_first_derivative}. We evaluated sentences with 10 and 20 words.\Scut{These sentences resulted in about 20 and 40 tokens respectively. A sentence with a higher number of tokens, \ie} We can observe that longer sentences results in higher inference latency. But again, we see that the inference latency does not improve after a point. The first order derivative of the latency for 10 and 20 word sentences shows a peak at around 30\% and 40\% GPU respectively.\Scut{This is also confirmed with the theoretical DNN model presented in $\S \ref{sec:modelling_dnn_parallelism}$, with the peak value of first order derivative being $\sim$ 15\% GPU for inferring 8 and 12-word sentences, and about 10\% for 25-word sentence (and 25\% GPU for the higher batch size using 25-word sentences).} 
Thus, both our model prediction and our evaluation of representative compute-heavy CNN and memory-bound Transformer models show that there is indeed a limit to\Scut{the amount of} parallelism utilized by DNNs. This motivates our approach to further examine improving GPU utilization with spatio-temporal scheduling.\looseness-1


\Scut{
\subsection{Impact on GPU Utilization and Energy}
\cut{
\vspace{-1mm}
\Scut{
\knote{I am wondering if we want to use this to justify the limits of parallelism instead, and move this also to the place that shows the validation of our theoretical model, and yet another part of the evidence that the GPU usage levels off for different ML models when going beyond a certain percentage GPU.}}
\Scut{
We have measured GPU power consumption while it is inferring an image using different DNN models in PyTorch. We used NVPROF's system profiling to get GPU power utilization figures. We can see from Fig.~\ref{fig:power_statistics} that GPU uses less power when inferring with DNN at less GPU\%. Furthermore, we also see that compute heavy model VGG-19 use lot more power than other models at the same GPU\%. 

Flip fig. 9 and fig. 8. The point of power is to say that as we add GPU\%, we see the power also flattens out - point of diminishing returns for problems of smaller dimensions (like the smaller image). This again reflects that we are only able to gainfully use the GPU up to a point, and beyond that we do not use the GPU, and the power usage doesn't go up (if the GPU power truly reflects how much of the GPU is being used). It is only if the problem becomes much much larger (1000x1000 image) then we use up more of the GPU and therefore more power is used also.
}
\Scut{
\knote{all we need to say is: since GPU utilization is not measured properly, we don't really know how well a GPU us used by a particular model. Yes, we consider it for relatively small batch sizes, because we know if you use a large batch size, unleasing many copies of the same model on the GPU to run concurrently on multiple data items can fully use up the GPU. We use energy consumption of the GPU (apparently measured more reliably in nVIDIA) as a good proxy for the utilization of the GPU. Many others have observed that this is a good proxy. So, we looked at the energy consumption in Joules (taking both the power and the time taken for inference (to make sure we reflect the fact that at very low GPU\%, we will see much longer execution times) into account) we see that when we increase the percentage of GPU that the energy consumption does not go up. It implies that the fraction of GPU corresponding to the increased percentage of the GPU is idle, confirming our hypothesis that the DNN model has a limited amount of parallelism. As the complexity of the model goes up (e.g., from Mobilenet, which flattens out early, to VGG-19 which appears to not flatten out at all), then more energy is consumed as we allocate more and more of the GPU to the DNAA model. }
}
\anote{Reevaluate if we need the follow subsection}

Existing tools do not measure the actual GPU utilization very accurately. For example, the NVIDIA profiler reports only the time the GPU is active as its measure of utilization, and does not reflect the extent to which the GPU resources are actually occupied at any instant. Thus, we can only estimate this when performing inference with a particular DNN model.
Previous studies have shown that with an increase in GPU computation \ie actual GPU utilization, the power consumption of the GPU increases~\cite{7457766}.
NVIDIA's nvprof system profiler provides the power consumption of the GPU during the time it is running, and we monitor this during the DNN execution period. Therefore, we use the power metric as a proxy to estimate GPU utilization. We compute energy used for inference of 4000 images of resolution $224\times224$ with a batch size of 4 (thus, 1000 batches), taking the average power and total inference time. 
The result in Fig.~\ref{fig:power_statistics} shows that the inference at very low GPU\% takes more energy, primarily contributed by the significant increase in inference time at this low GPU\% (even though the peak instantaneous power is lower).  
However, after an inflection point, increasing the GPU\% does not cause energy consumption to go up. This indicates that a fraction of the GPU corresponding to the increased percentage remains idle, confirming  our  hypothesis that the DNN model has a limited amount of parallelism. As the complexity of the DNN model goes up (e.g., from Mobilenet, whose energy flattens out early, at a low GPU\%, to VGG-19 which flattens out later), more energy is consumed as we allocate more of the GPU. 
}

}
\Scut{
\begin{figure}[h]
    \centering
    \subfloat[\Scut{Power Consumption of Different DNNs (batch size:16, image size 1000x1000) running in Pytorch with different GPU\%. Max power is 250 Watts for V100 GPU. Base power with nothing running in GPU was 48 watts.}]{\includegraphics[width=.5\linewidth]{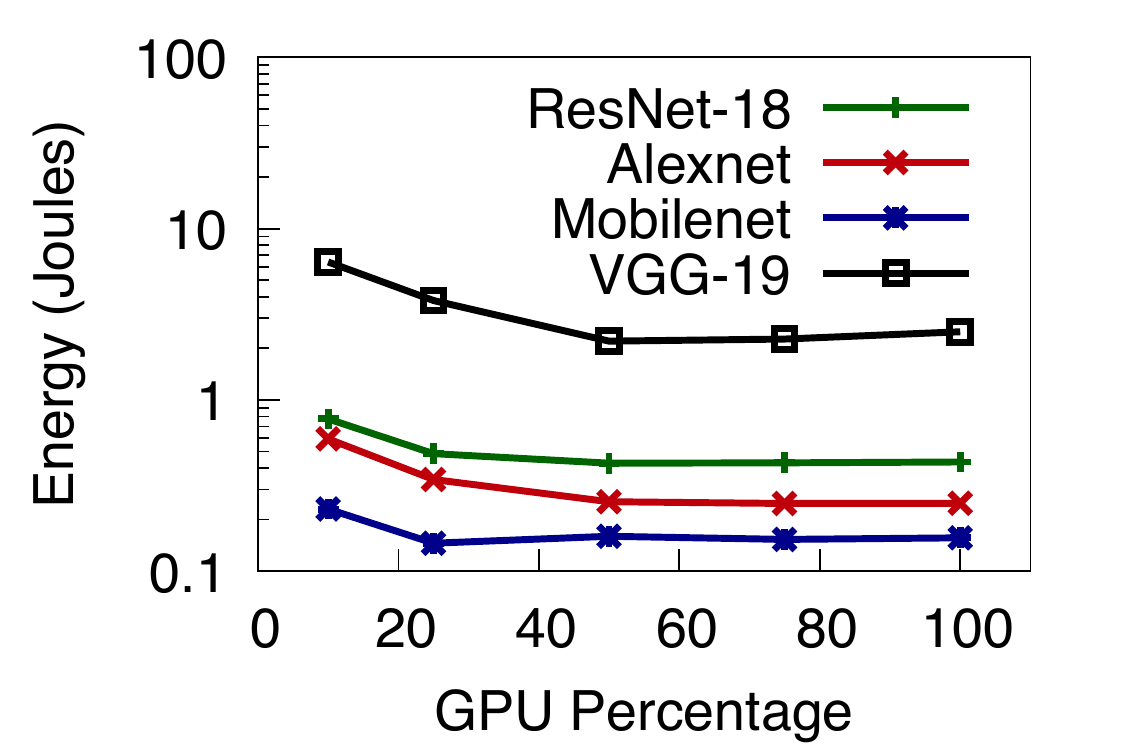}\label{fig:power_statistics}}
   \caption{Energy spent on inferring 4000 images (1000 inference, batch size of 4)}
\end{figure}

}

\Scut{
\begin{figure}
\vspace{-4mm}
   \subfloat[Latency of GNMT Models' inference vs. sentence length]{
    \includegraphics[width=.33\linewidth]{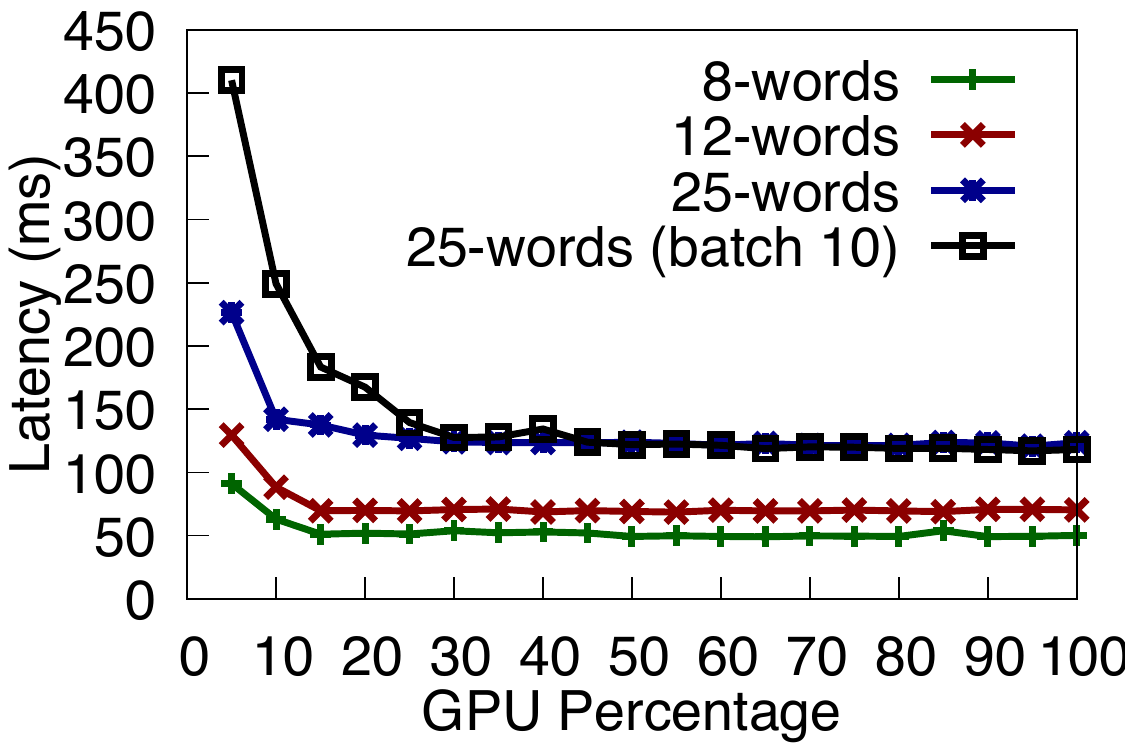}
    \vspace{-2mm}
     \label{fig:gnmt_latency}
     }
\subfloat[First Order derivative of GNMT Models' inference latency]{
     \includegraphics[width=.33\linewidth]{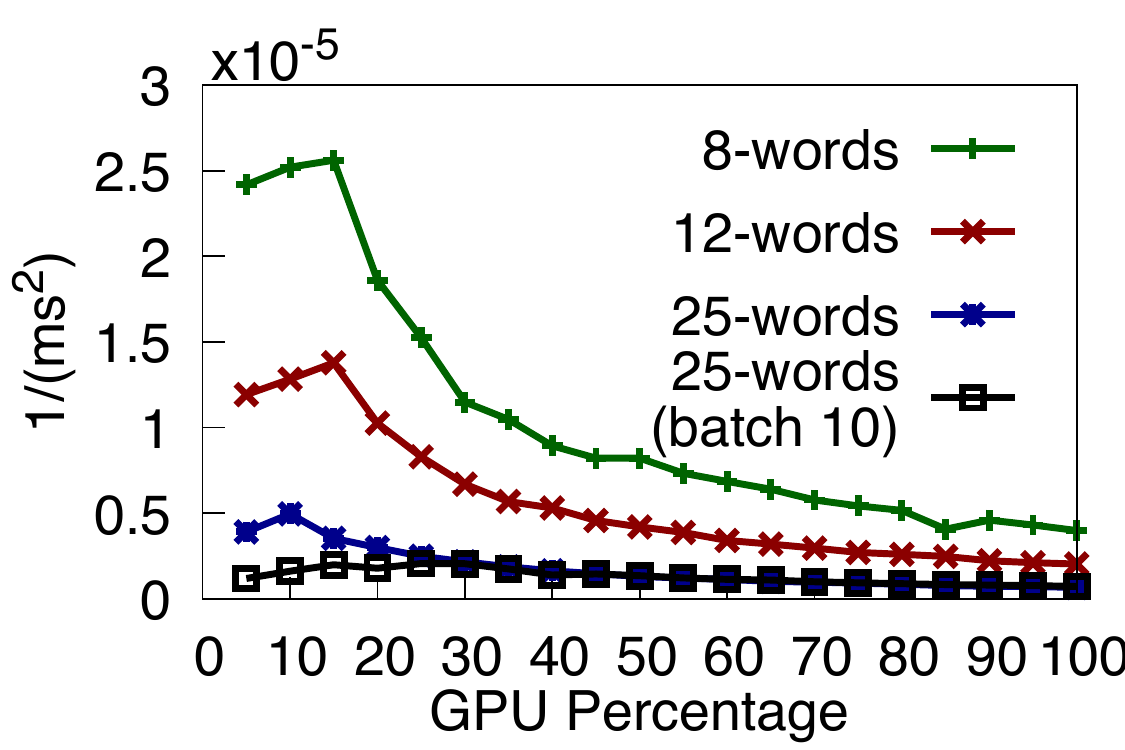}
     \vspace{-2mm}
    \label{fig:gnmt_first_derivative}
    }
\subfloat[Latency and First Order Derivative of latency of BERT]{
    \includegraphics[width=.33\linewidth]{figures/BERT_Data/bert_derivative_latency_plot.pdf}
    \vspace{-2mm}
    \label{fig:bert_first_derivative}
}
\Scut{
\begin{subfigure}{0.34\linewidth}
    \includegraphics[width=\linewidth]{figures/power_statistics/energy_stats.pdf}
    \caption{Energy to infer 1 image}
    \label{fig:power_statistics}
\end{subfigure}
}
\vspace{-2mm}
\caption{(a) (b) Evaluation of an RNN model: GNMT, (c) evaluation of Transfromer model: BERT \Scut{GNMT Latency and derivative of latency.(c) Energy profile for different DNN Models}}
\vspace{-6mm}
\end{figure}
}



\section{Optimal Batching for DNNs}\label{sec:batching_in_dnn}
\Scut{Batching multiple requests together during inference amortizes cost of transferring data, loading of kernel instructions for a DNN, and reading of weights and parameter. This leads to higher inference throughput. Increasing the batch of requests for inference usually increases the number of GPU threads, leading to higher GPU utilization.}\Scut{
Batching multiple inference requests helps improve throughput as it amortizes the cost of transferring data to the GPU. 
Since batching is performed of requests to be inferred by the same DNN, it amortizes loading of the kernel instructions to the GPU and the reading of weights and parameters during kernel execution. 
Note that a DNN application usually produces more threads, \ie batch of 2 requests would produce 2$\times$ the number of GPU threads.}
Batching is a trade-off between improving throughput at the cost of higher latency. Inferring a batch of requests requires more computation, thus increasing inference time. Preparing a bigger batch, \ie receiving and transferring data from the network to GPU also contributes additional latency.\Scut{
A higher batch size increases the overall inference latency. First, the latency for inferring a larger batch itself is higher than inferring a single request. Second, the time to prepare the batch, \ie waiting and receiving all the data from the network to prepare a batch of requests and transferring data from host memory to GPU contributes to higher latency for large batch sizes.}\Scut{Hence, batching implies a trade-off between higher throughput and higher latency.} 
Providing a higher GPU\% for a bigger batch can mitigate the inference latency increase. However, giving more than a certain GPU\% may be wasteful. We use the metric 
\begin{equation}
    \textit{Efficacy }(\eta) = \frac{\textit{Throughput}}{\textit{Latency}\times GPU\%}
    \label{eqn:efficiency}
\end{equation}
\Scut{ GPU resources if they may be better utilized multiplexing additional applications.} of \emph{Efficacy} ($\eta$) of using GPU resources as the basis to find a good operating point with respect to batch size and GPU\%. We define $\eta$ of a DNN at a certain batch size and GPU\% as Eq.~\ref{eqn:efficiency}.
Efficacy, $\eta$, lets us know how much throughput the GPU produces per unit time, per unit of GPU resource (GPU\%).\looseness-1 
\subsection{Optimum Batch Size for Inference}
\label{sec:efficient_batch_size}


We profiled the ResNet-50 model for inference at different batch sizes \& GPU\% configuration. Fig.~\ref{fig:resnet50_efficiency} shows that both very high and very low batch size leads to low Efficacy due to high latency and reduced throughput respectively, thus, an optimal batch size is desired.
We now develop an optimization formulation that can provide us with the right batch size and GPU\% for a model, given a deadline. First, we present the key notations used for the optimization in Table~\ref{tab:optimization_formulation}.\looseness-1
\begin{table}[h]
\centering
\captionof{table}{Notation for Optimization Formulation}
\vspace{-4mm}
\resizebox{\linewidth}{!}{%
\begin{tabular}{|cc|}\hline
    Notation & Description\\\hline
     $p_{i}$ & GPU\% for Session $i$\\
     $b_{i}$ & batch size for Session $i$\\
     $f_{L}(p_{i},b_{i})$ & inference latency of batch $b_{i}$ for model $M_{i}$ at GPU\% $p_{i}$\\
     $C_{i}$ & Request assembly time for Session $i$\\\hline
          \end{tabular}}
        \label{tab:optimization_formulation}
\end{table}

\Scut{
\begin{figure}
    \centering
    \includegraphics[width=.70\linewidth]{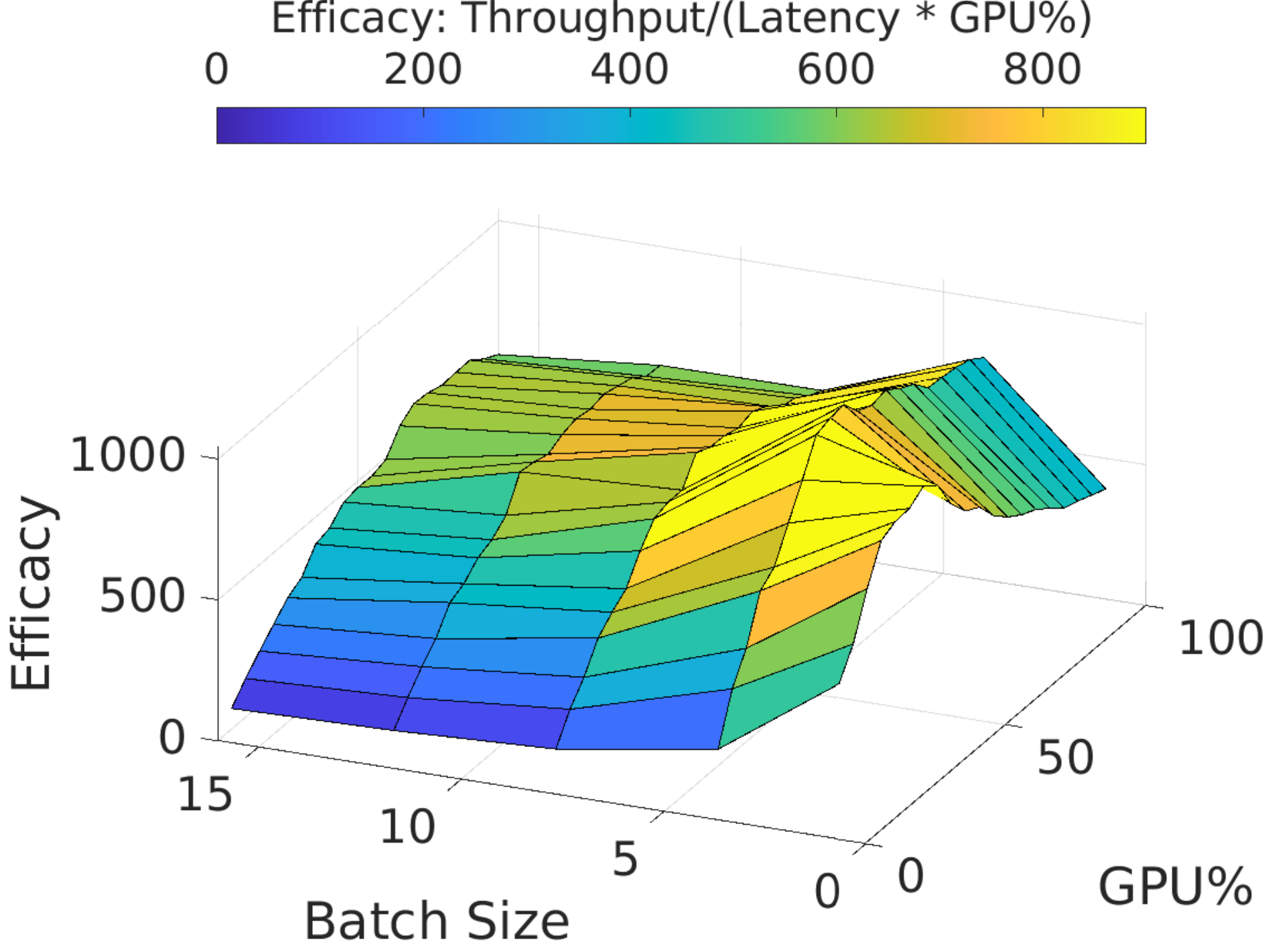}
    \caption{Caption}
    \knote{is this new? Is it supposed to be similar to 10a? removed it and put }
    \label{fig:my_label}
\end{figure}
}
\Scut{
\begin{figure*}[h!]
\vspace{-3mm}
    \centering
    \subfloat[\textit{Efficacy} of ResNet-50 ]{ 
    \includegraphics[width=.30\linewidth]{figures/optimization_plots/ResNet50/new_data/resnet50_boat.pdf}
    \label{fig:resnet50_efficiency}}\subfloat[Optimization profile for Mobilenet]{\includegraphics[width=.33\linewidth]{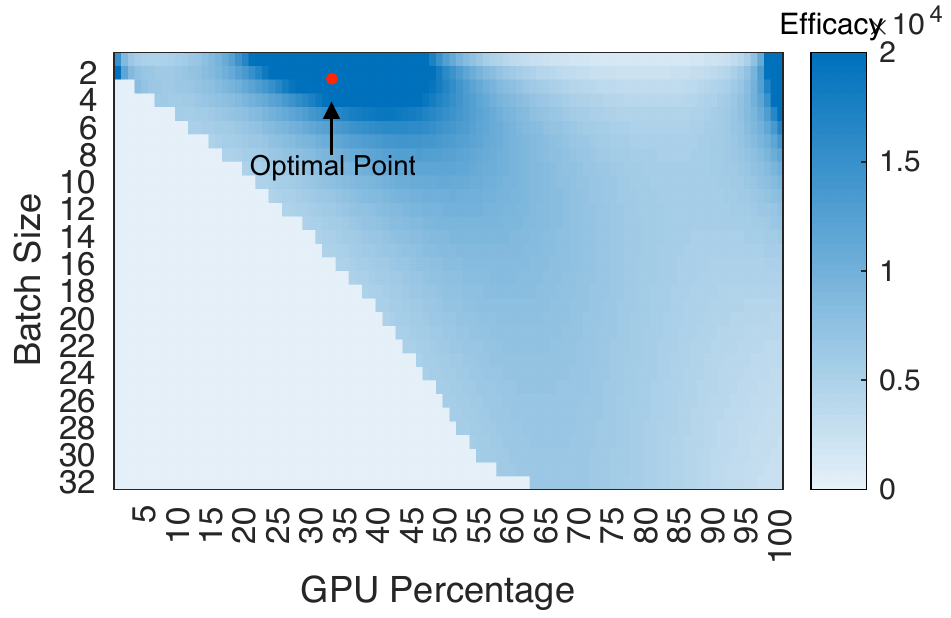}\label{fig:optimal_point_mobilenet}}
    \subfloat[Optimization profile for ResNet-50]{
    \includegraphics[width=.30\linewidth]{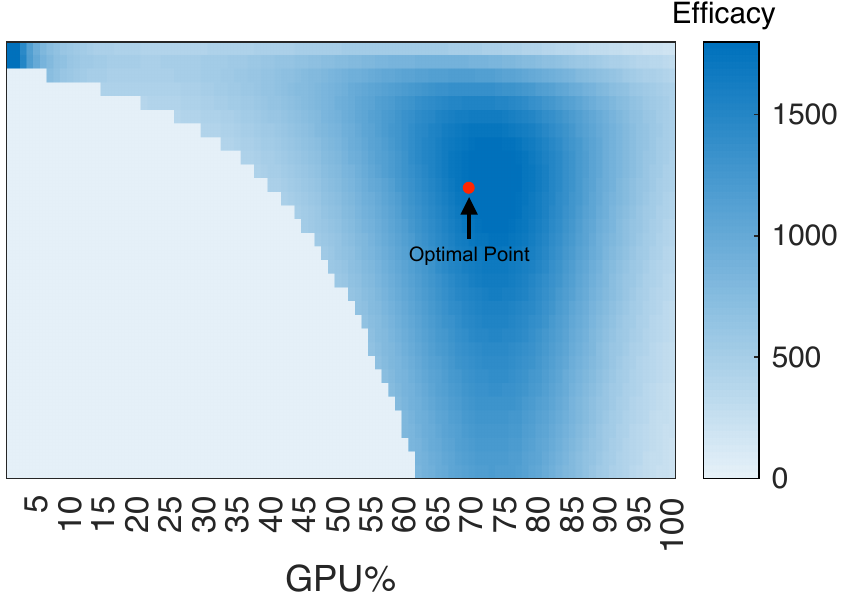}\label{fig:optimal_point_resnet50}}
    \vspace{-2mm}
    \caption{(a): Efficacy from measurements in V100 GPU. (b), (c): Feasibility region (darker shade) and optimal point.\Scut{ obtained from optimization.}}
    \vspace{-2mm}
\end{figure*}
}

\Scut{
\begin{table}
\vspace{-2mm}
\centering
\caption{Notation for Optimization Formulation}
\vspace{-1mm}
\resizebox{\linewidth}{!}{%
\begin{tabular}{|cc|}\hline
    Notation & Description\\\hline
     $p_{i}$ & GPU\% for Session $i$\\
     $b_{i}$ & batch size for Session $i$\\
     $f_{L}(p_{i},b_{i})$ & inference latency of batch $b_{i}$ for model $M_{i}$ at GPU\% $p_{i}$\\
     $C_{i}$ & Request assembly time for Session $i$\\\hline
          \end{tabular}}
        \label{tab:optimization_formulation}
\Scut{
\begin{minipage}[t]{.5\linewidth}
\caption{Avg. Throughput \& Latency in 3 GPU cluster}
\vspace{-2mm}
\resizebox{\columnwidth}{!}{
\begin{tabular}{|c|c|c|c|c|c|c|}\hline
         Model&\multicolumn{2}{|c|}{Default MPS}&\multicolumn{2}{|c|}{Fixed Batch (8)}  & \multicolumn{2}{|C|}{CSS}   \\\hline
         & Lat. (ms) & Thpt. (Ips) & Lat. & Thpt. &Lat. &Thpt.\\\hline
         Mobilenet&2.1&476 & 5.27&1702&5.23&1708\\
         ResNet-50&6.8&147&37.88&503&24.83&713\\
         VGG-19&31& 32& 133.5&60&133.1&60\\\hline
    \end{tabular}}
    \label{tab:css_all_together}
\end{minipage}
}
\vspace{-2mm}
\end{table}
}

The batch size is a product of the average incoming request rate and request assembly time. \Scut{The request assembly time is the time to aggregate the data from network packets to an application unit, such as an image. }Thus, $b_{i} =\texttt{Request-Rate}\times C_{i}$. 
Throughput $T_{i}$ is \Scut{defined as}number of images inferred per unit time (Eq.~\ref{eqn:throuhgput}. Knowing throughput (Eq.~\ref{eqn:throuhgput}) we can write $\eta$ (Eq.~\ref{eqn:efficiency}), as Eq.~\ref{eq:efficiencySquareLatency}. Eq.~\ref{eq:efficiencySquareLatency} is of the same form as the first derivative of inverse of latency, Eq.~\ref{eq:first_order_derivative},  \S\ref{sec:modelling_dnn_parallelism}. 

\begin{minipage}{\linewidth}
 \begin{minipage}{.4\linewidth}
  \begin{equation}
      T_{i} = \frac{b_{i}}{f_{L}({p_{i},b_{i})}}\label{eqn:throuhgput}
  \end{equation}
 \end{minipage}%
 \begin{minipage}{.59\linewidth}
  \begin{equation}
      \eta = \frac{b_{i}}{\left(f_{L}(p_{i},b_{i})\right)^{2}\times GPU\%}
    \label{eq:efficiencySquareLatency}
  \end{equation}
 \end{minipage}
\end{minipage}
\Scut{
\begin{wrapfigure}[5]{l}{.4\textwidth}
\vspace{-4mm}
\begin{minipage}{0.4\textwidth}
\begin{align}
T_{i} = \frac{b_{i}}{f_{L}({p_{i},b_{i})}}\label{eqn:throuhgput}\\
\eta = \frac{b_{i}}{\left(f_{L}(p_{i},b_{i})\right)^{2}\times GPU\%}
    \label{eq:efficiencySquareLatency}
\end{align}
\end{minipage}
\end{wrapfigure}
}
\Scut{
\begin{equation}
    \eta = \frac{b_{i}}{\left(f_{L}(p_{i},b_{i})\right)^{2}\times GPU\%}
    \label{eq:efficiencySquareLatency}
\end{equation}}
\Scut{Eq.~\ref{eq:efficiencySquareLatency} is of the same form as the first derivative} 

We seek to maximize Efficacy ($\eta$) to get the best balance in parameters based on the constraints~\ref{eq:max_batch_size},~\ref{eq:meet_slo}, and ~\ref{eq:meet_half_slo}. The constraints express following requirements: \textbf{Eq.~\ref{eq:max_batch_size}}: Batch size must be less than or equal to maximum batch size a 
\begin{align}
    1\leq b_{i}\leq \mathit{Max Batch Size}\label{eq:max_batch_size}\\
    f_{L}(p_{i},b_{i})+C_{i}\leq SLO_{i}\label{eq:meet_slo}\\
    f_{L}(p_{i},b_{i})\leq \frac{SLO_{i}}{2}\label{eq:meet_half_slo}
\end{align}
\Scut{
\begin{equation}1\leq b_{i}\leq \mathit{Max Batch Size}
\label{eq:max_batch_size}
\end{equation}
\begin{equation}f_{L}(p_{i},b_{i})+C_{i}\leq SLO_{i}
\label{eq:meet_slo}
\end{equation}
\begin{equation}
\vspace{-1mm}
    f_{L}(p_{i},b_{i})\leq \frac{SLO_{i}}{2}
    \label{eq:meet_half_slo}
\end{equation}
}
model can accept. \textbf{Eq.~\ref{eq:meet_slo}}: The sum of times taken for aggregation of batch via network\Scut{ data transfer to GPU }$(C_{i})$, and its inference execution, which has to satisfy the SLO. \textbf{Eq.~\ref{eq:meet_half_slo}}: When working with a high request rate, we can regularly gather large batch sizes for inference. However, a request that cannot be accommodated into the current batch due to constraint Eq.~\ref{eq:meet_slo}, has to be inferred in the next batch. Then the deadline for next batch is the deadline of the oldest pending request.\Scut{ Furthermore, The next batch cannot be executed until the current batch's inference is over. Due to high request rate, the next batch might be big batch too.} Therefore, we make sure that SLO is twice the time required to run a batch.\looseness-1

We computed the latency function $f_{L}(p_{i},b_{i})$, by fitting the latency observed while inferring DNN models with a batch size of 1,2,4,8,10,12,16 and GPU\% from 10-100 at 10\% intervals\Scut{, for ResNet-50} on our testbed\Scut{with the NVIDIA V100 GPU}. 
The optimization is solved using the non-linear programming solver 'fmincon' in MATLAB. 
Requests (images of resolution $224\times224$) arrive over a 10 Gbps link. 1 image is 
assembled every $\sim 481{} \mu{}s$. 
We use an SLO of 50 ms, \revise{allowing for an interactive system that can be used in safety critical environments such as autonomous driving\mbox{\cite{qiu2018avr}}}.\looseness-1 \Scut{ giving a chance to form a significant batch (e.g., a batch of 16 images takes \textasciitilde 25 ms to be formed and inferred in our system).}
We present the feasibility region (where the SLO constraints are fulfilled) and optimal point provided by the optimization formulation in Fig.~\ref{fig:optimal_point_mobilenet}\Scut{ and 
Fig.~\ref{fig:optimal_point_resnet50}}. The infeasible area is in a lighter shade. \Scut{Darker shading reflects higher Efficacy, with the optimal points provided by the solver also marked.}It is particularly revealing that Mobilenet has an optimal point close to 30\%.\looseness-1\Scut{ GPU, and ResNet-50 (a more complex model) has a much higher optimal GPU\% of $\sim$70\%.}
\Scut{
\begin{figure}[h]
\vspace{-2mm}
    \centering
    \subfloat[\textit{Efficacy} of ResNet-50 ]{ 
    \includegraphics[width=.5\linewidth]{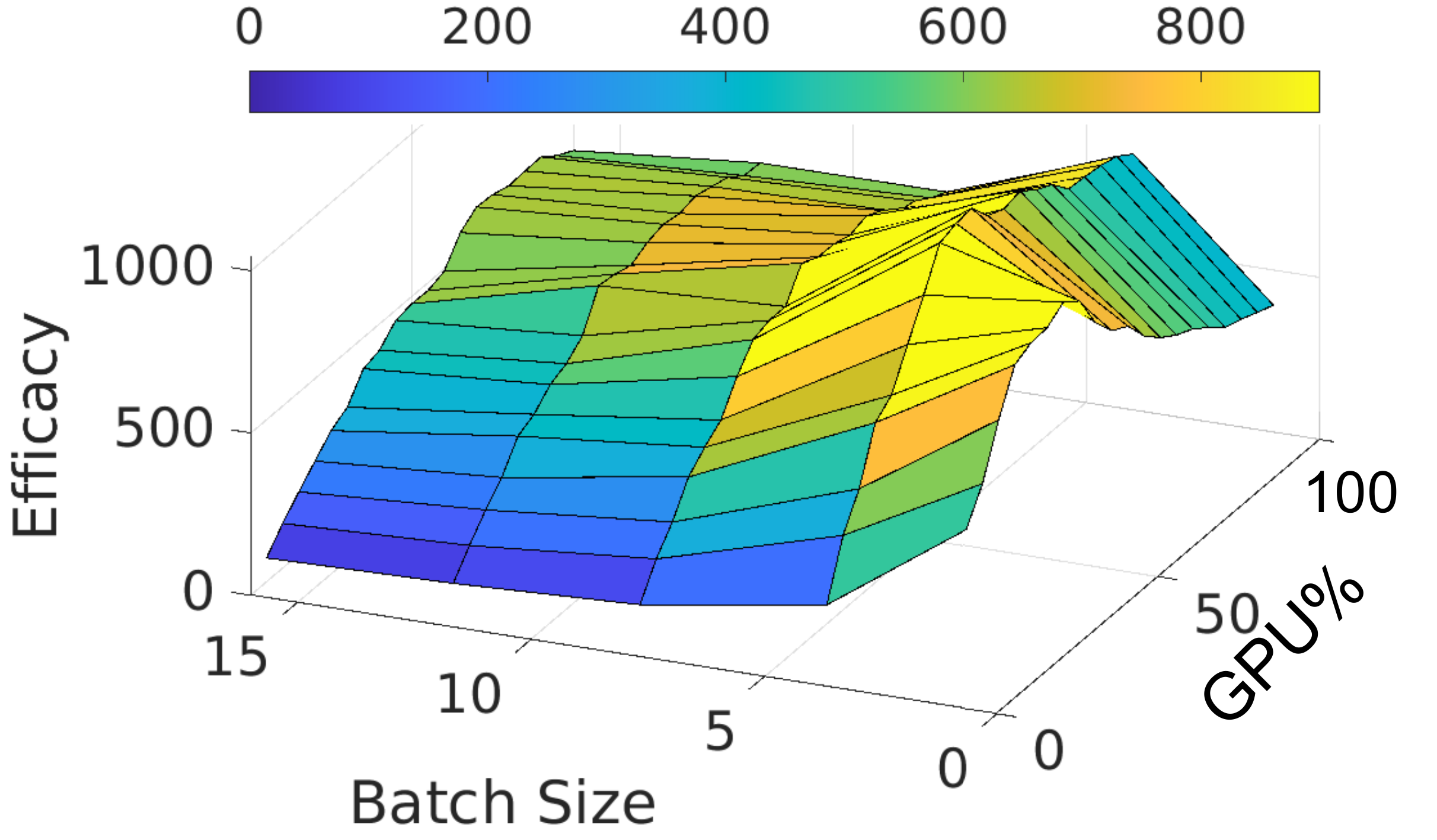}
    \label{fig:resnet50_efficiency}}
    \Scut{
    \subfloat[Optimization profile for Mobilenet]{\includegraphics[width=.5\linewidth]{figures/optimization_plots/mobilenet/optimization_profile_2.pdf}\label{fig:optimal_point_mobilenet}}
    }
    \Scut{
    \subfloat[Optimization profile for ResNet-50]{
    \includegraphics[width=.30\linewidth]{figures/optimization_plots/ResNet50/efficiency/optimization_results6_small.pdf}\label{fig:optimal_point_resnet50}}
    }
    \vspace{-2mm}
    \caption{(a): Efficacy from measurements in V100 GPU. (b): Feasibility region (darker shade) and optimal point.\Scut{ obtained from optimization.}}
    \vspace{-3mm}
\end{figure}
}
\begin{figure}
\begin{minipage}{\columnwidth}
    \centering
    \includegraphics[width=.95\linewidth]{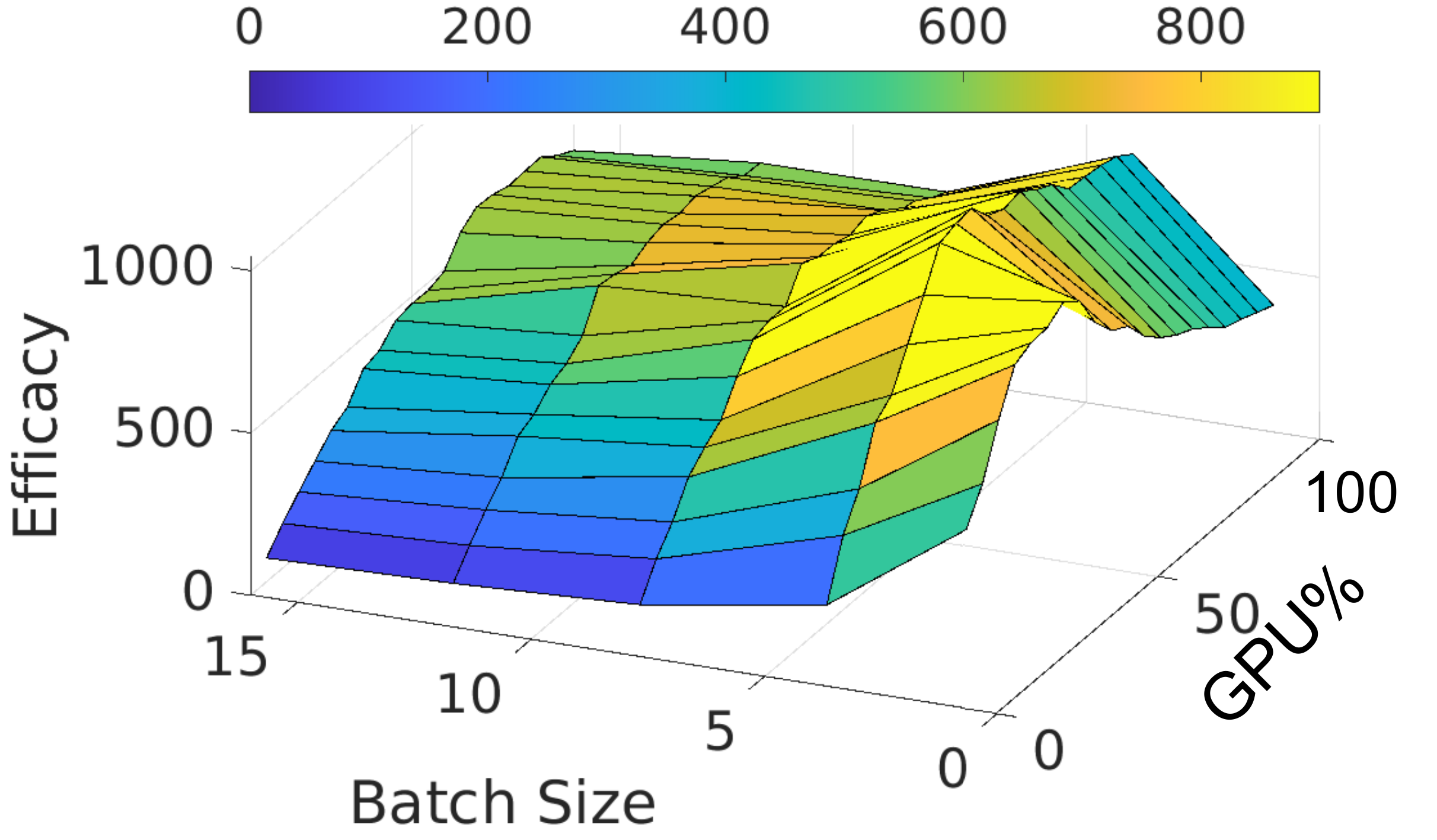}
    \vspace{-3mm}
    \caption{\textit{Efficacy} of ResNet-50 }
    \label{fig:resnet50_efficiency}
    \end{minipage}%
\end{figure}
\begin{figure}
    \centering
    \includegraphics[height=45mm,width=\linewidth, keepaspectratio=false]{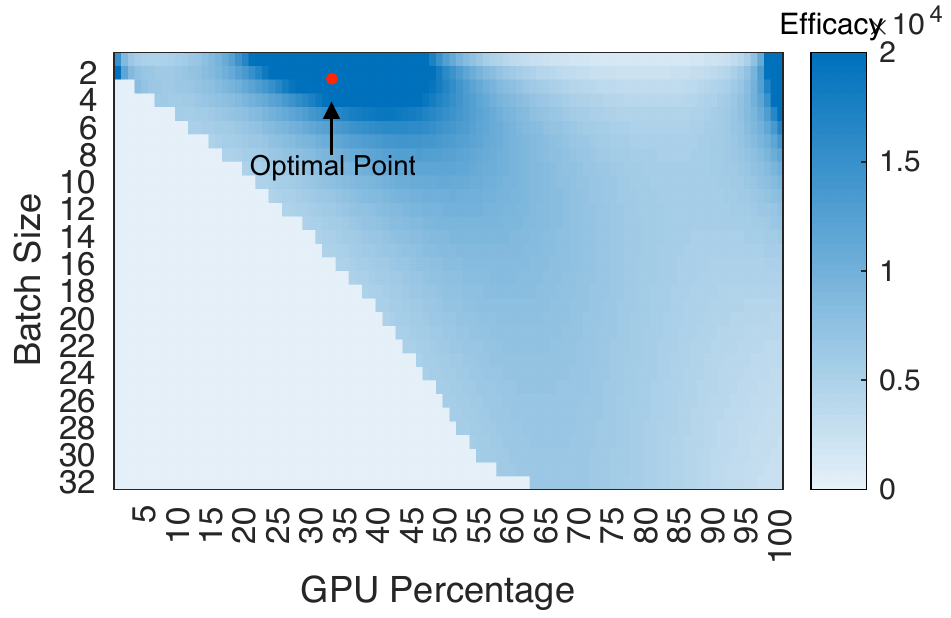}\label{fig:optimal_point_mobilenet}
    \caption{Mobilenet feasibility region (darker shade)}
    \label{fig:optimal_point_mobilenet}
    \end{figure}

\noindent\textbf{Estimation of the Knee for Real Systems:}
We view these optimal values in relative terms, representative of the limit to parallelism that the model exhibits,\Scut{ while not strictly interpreting it as being accurate in matching the knee of the latency curve observed empirically through measurements. This is  }
because the optimization does not necessarily factor all the aspects that influence the execution of the model in the real system. We, however, pick a batch size and GPU\% values\Scut{ that satisfy our deadline requirement} from the high efficacy region\Scut{pointed out by} in the optimization output in Fig.~\ref{fig:optimal_point_mobilenet} and over-provision the GPU\% by 5-10\% while deploying the model in a real system.\looseness-1

\section{GPU Scheduling of DNN models}\label{sec:scheduling}
We now discuss the Spatio-temporal scheduling with \name{}. We run the DNN models concurrently and meet their SLO while keeping the GPU from over-subscription. Over-subscription\Scut{`of the GPU} occurs when the aggregate GPU\% of concurrent models exceed 100\%.\looseness-1 
\subsection{Scheduling with varying SLO}
We schedule multiple models with different SLOs (deadlines), optimal batch sizes, and GPU\% with \name{}. Our scheduler considers two primary constraints. First, the DNN model must be scheduled at least once before an interval equal to its SLO, using an optimal batch size as predicted by the model in \S~\ref{sec:batching_in_dnn}. Second, the aggregate GPU demand at any point in the schedule\Scut{ (sum of GPU\% of all concurrently running models)} should not exceed 100\%.\Scut{We now define the terms used in the scheduling algorithm.} We choose a time period defined by the largest SLO to be a \textit{Session}.\Scut{ such that all the DNN models are scheduled at least once every \textit{Session}.}\Scut{We then and produce a schedule for all the DNN models for the interval of a session.} Models with an SLO smaller than a session will run multiple times in a session. \Scut{ We term the smaller time-slice defined by shorter SLO as a \textit{sub-session}. }\eg for a 100 ms session, a model with 25ms SLO will run at least 4 times. Our spatio-temporal scheduling also accommodates dynamic arrivals of requests by utilizing a Fair, Opportunistic and Dynamic scheduling module which dynamically recomputes the schedule, thus increasing the effective utilization of the GPU.
\begin{table}
    \centering
    \captionof{table}{Characteristics of different DNN models}
    \vspace{-3mm}
    \resizebox{\columnwidth}{!}{%
    \begin{tabular}{|c|c|c|c|c|}\hline
         Model & Knee\% & \makecell{SLO\\(ms)} & \makecell{Batch ($B_{i}$)\\Sentence len.} & \makecell{Runtime ($L_{i}$)\\(ms)} \\\hline
         Mobilenet & 20 & 25 & 16 & 10\\
         Alexnet & 30 & 25 & 16 & 8\\
         BERT&30&25&16 (10-words)& 9\\
         ResNet-50 & 40 & 50 & 16 & 28\\
         VGG-19 & 50 & 100 & 16 & 55\\
         ResNet-18 & 30 &25 &16 &12\\
         Inception &40 &50 &16 &25\\
         ResNeXt-50 &50 &100 &16 &40 \\\hline
    \end{tabular}
    }
    \label{tab:model_characteristics}
\end{table}

We use 8 different DNN models and\Scut{ Alexnet, Mobilenet, BERT, ResNet-50, VGG-19 for our experiments. We} present their optimal batch size, GPU\% and the latency of inference at that batch-size/GPU\% in Table~\ref{tab:model_characteristics}. We obtain the knee GPU\% and Batch Size from the model in \S~\ref{sec:batching_in_dnn}. We chose our SLO based on safety-critical work such as autonomous driving\mbox{\cite{qiu2018avr}}, where it is determined that less than 130ms processing is required to safely stop a car running at 80 miles/hr ($\sim$130 kmph). We choose a much more conservative 100 ms (effectively about 50 ms as rest is spent for preparing batch) for higher accuracy (VGG-19 and ResNext-50) and smaller SLOs (50 ms and 25 ms) for latency-optimized models (ResNet-50, Inception, Mobilenet, Alexnet and ResNet-18) aimed for application such as 30fps video stream. \looseness-1
\Scut{
Alexnet and Mobilenet are considered compute-optimized models, which are designed for low latency inference to be used in an interactive system. Models such as ResNet-50 and VGG-19 are optimized for higher accuracy. Therefore, they are typically expected to take slightly more time for execution. 
We set a short SLO of 25 ms for compute-optimized models (Alexnet and Mobilenet). We set SLOs of 50 ms and 100 ms for ResNet-50 and VGG-19, respectively, for these accuracy-optimized models\Scut{ that have higher runtimes}.\looseness-1
}
\Scut{
For each model, using the optimal batch ($B_i$) and GPU\% ($p_i$), 
we compute the inference latency $L_{i}$ 
as: $L_{i} = f_{L}(p_{i},B_{i})$
\Scut{ for the batch size $b_{i}$ where $b_{i}\leq B_i$ and GPU\% $p_{i}$ as:}
where $f_{L}$ is a 
function that provides latency for\Scut{ model's} batch size $B_{i}$\Scut{ for a $p_{i}$ based on the profiled data. }.
We then use GPU\% $p_{i}$ (spatial allocation) and batch inference latency $L_{i}$ (that guides the temporal schedule) to schedule a model over a session\Scut{$S_{i}$}. 
\Scut{Moreover, we assume that SLO of a model is larger than the batch inference latency we compute, $L_{i}$, for that model.} Furthermore, we assume, realistically, that a model's execution cannot be pre-empted during the execution of the inference of one batch of requests. 
}Unlike~\cite{zhang2019laius}, we realistically consider that a model's execution \Scut{on the GPU }cannot be preempted from GPU.\looseness-1\Scut{ during the execution of a given batch of requests.} 
\Scut{We now demonstrate temporal scheduling with several current, popular DNN models.\looseness-1}

\update{
We first examine a temporal schedule\Scut{of the models} with Alexnet, ResNet-50, and VGG-19. 
We provide time slices proportional to the model's SLOs.\Scut{Thus, a model with a bigger SLO gets a larger time slice.} We utilize an adaptive batching algorithm mentioned in clipper~\cite{crankshaw2017clipperlong} and 
Nexus~\cite{shen2019nexuslong} 
to obtain the batch size for each model's time slice. Fig.~\ref{fig:model_schedule} is the visualization of such a schedule. The SLOs are visualized as the vertical dotted lines.\Scut{We see that Each model gets a full-share (100\%) of the GPU but only use a part of that GPU.}\Scut{There are no methods or APIs to count how many SMs are used when a model runs in a GPU \Scut{without MPS, as} } We compute GPU utilization by using Knee\% for each model as shown in Table~\ref{tab:model_characteristics}. With temporal sharing, we achieve mean GPU utilization of 44\%.\looseness-1 
\Scut{We also see that 
all the models run at least once before their SLO (dotted line).}}

\Scut{We set the SLOs of the models as 25 ms for Alexnet, 50 ms for ResNet-50 and 100 ms for VGG-19. Therefore, picking the session length to be 100 ms. We assume there is always work available for the models. Therefore, to fulfill the SLO, Alexnet runs for at least 4 times, Inception-V3 for 2 times and VGG-19 once.  }

\cut{\ie a model begins its next batch of inference \knote{another what?} as soon as its SLO for previous round has passed \knote{after it was executed or if it missed its execution? This last sentence is too vague}.} 
\cut{
such a schedule may result in concurrent execution of a larger number of models, 
causing the GPU to be oversubscribed as shown by the red ares in the GPU utilization Fig.~\ref{fig:model_schedule}(Top plot). 
To avoid such an over-subscription\sknote{It is not clear as to what is the issue with over-subscription?} of the GPU, we develop an algorithm that ensures scheduling the appropriate number of models without the over-subscription of GPU. 
}
\subsubsection{\textbf{\name{}: Spatio-Temporal Scheduling}}
\Scut{Our scheduler seeks to maximize the gaps in the execution of different models such that the gaps can be used to accommodate additional models that can be concurrently scheduled. We take care that \Scut{GPU is never oversubscribed, \ie }demands of all concurrently running models do not exceed 100\%.}
Our \name{}'s scheduler aims to fit as many models as possible (potentially being different from each other) and run them concurrently in the GPU. We seek to be able to meet each model's (potentially different) SLO. We employ a simple version of the Earliest Deadline First Scheduling (EDF) algorithm to schedule all the models.\Scut{The pseudo-code for \name{}'s scheduler is listed in Algorithm~\ref{algo:spatio-temporal} (Appendix).} EDF schedules the model with the tightest deadline to run first.\Scut{We seek to schedule all the models we can fit, to be able to meet their individual (potentially different) SLOs. For that, we set a scheduling {\it Session} to be as long as the largest SLO, so that all the models, including one with largest SLO, get executed at least once in a Session. We also compute number of times each model can run in one session by dividing the session length by each model's individual SLO.} However, we should note that as a model's inference is not preempted, this simple schedule cannot guarantee that the GPU will not be oversubscribed at any moment in the schedule. To aid in fitting in as many models as possible, we schedule consecutive executions of any model with the shortest SLOs to be as far apart as possible. This allows us to fit longer running models in the GPU in the interim without oversubscribing it. We demonstrate a schedule generated by spatio-temporal only algorithm in Fig.~\ref{fig:model_schedule_scheduled}. We observe that the model with the smallest SLO, Alexnet (bottom), is scheduled to meet its SLO, but the time between the execution of the first instance and the second can be large because its execution time is short. This allows us to run ResNet-50 (second from the bottom) and VGG-19 (third) in between consecutive executions of Alexnet. 
Note that \name{}'s scheduler can also schedule a model with GPU\% lower than its Knee, albeit with high inference latency when necessary. \name{} also considers the additional latency of launching a new DNN model at lower GPU\% into the schedule. This latency-GPU\% trade-off has to be considered carefully before starting inference. Once a DNN process starts with its allocated GPU\%, it cannot be changed for that instance's execution lifetime.

\Scut{
In a subsequent step, if we encounter GPU being oversubscribed (GPU being utilized more than 100\%), we iteratively adjust the schedule of the models, starting with the one with the smallest runtime such that we do not oversubscribe the GPU.}


\Scut{
\setlength{\textfloatsep}{0pt}
\begin{algorithm}[t]
\caption{Spatio-Temporal Scheduling}
\label{algo:spatio-temporal}
\label{algo:dynamic_scheduling}
\begin{algorithmic}[1]
\Function{\texttt{ST-Schedule}$(\texttt{Models[],Length,schedule})$}{}
\State \texttt{repeat[]} $\gets$ \texttt{Length$\div$(Models[].SLO)} 
\State \texttt{\textbf{EDF}($\texttt{Schedule}$, Models[])}
\If{GPU\_Util($\texttt{Schedule}$)$<$100}
\State \textbf{return} $\texttt{Schedule}$
\EndIf
\For {\texttt{i:=0} \textit{to} \texttt{numModels-1}}
\For{\texttt{j:=0} \textit{to} \texttt{repeat[i], STEP 2}}
\State \textbf{Start-Late}($\texttt{Schedule}$, Model[j+1])
\EndFor
\If{GPU\_Util($\texttt{Schedule}$)$<$100}
\State \textbf{return} $\texttt{Schedule}$
\EndIf
\EndFor
\EndFunction
\Scut{
\State
\Function{\texttt{Dynamic-schedule}$(\texttt{Schedule, priority})$}{}
\For{\texttt{i:=Priority[max] to Priority[min]}}
\If {\texttt{Remaining-GPU > model[i].GPU\%}}
\State Slice $\gets$ time(\texttt{Schedule, model[i].GPU\%})
\EndIf
\State batch $\gets$ Batch-Size(\texttt{model[i], Slice})
\State \textbf{Run-Batch}(\texttt{model[i],batch})
\EndFor
\EndFunction
}
\end{algorithmic}
\end{algorithm}
}
\Scut{
Our spatio-temporal scheduling algorithm is presented in Algorithm~\ref{algo:spatio-temporal}\Scut{(Function \texttt{Schedule})}. Model characteristics obtained through profiling for models apriori (as in Table.~\ref{tab:model_characteristics}) are used as an input for scheduling.  We also compute the number of times each model can run in one session (line 2).\Scut{The algorithm picks the largest SLO as the session length (line 2 of Algorithm~\ref{algo:spatio-temporal}). We also compute the number of times each of the models has to run in one session (line 3). This is obtained by dividing the session length by SLO of each individual model. 
}
After using the EDF algorithm to schedule, we check if the GPU is oversubscribed (Algo.~\ref{algo:spatio-temporal}, line 4). If oversubscribed, we iterate through each model and increase the time between consecutive executions of a model.\Scut{, while ensuring at least one execution within each SLO interval.} We use a helper function \textbf{Start-Late} to schedule the model (line 8), such that the execution occurs 
as late as possible while still fulfilling its SLO and not oversubscribing the GPU. \Scut{It also finds a suitable time to schedule the model such that at any point in the schedule we do not exceed 100\% GPU.}We see in Fig.~\ref{fig:model_schedule_scheduled} that Alexnet (bottom) and ResNet-50 (middle) are scheduled with time in-between their executions so as to\Scut{ enable fitting} fit in an execution of VGG-19. We can further see from the top row that the total GPU\% does not exceed 100\%. We can use the remaining GPU\% to dynamically schedule an additional application to utilize the residual GPU\%.\looseness-1
}

\Scut{First, we fix the schedule of longest running model. Since this model has to run at least once within a session, we schedule it to start running at the last moment possible so that the model finishes its inference at the SLO expiration, as seen in line 6 of Algorithm~\ref{algo:spatio-temporal}.
We use a helper function \textbf{Start-Late} as described in Algorithm.~\ref{algo:spatio-temporal-helper2} to generate the schedule. This function helps schedule the model as late as possible (without oversubscribing GPU) and still make it to the deadline. We also update the GPU-utilization array to keep track of GPU resource available over the entire session length. We then place the model with the smallest execution time in the schedule with help of \textbf{Start-Early} function described in Algorithm.~\ref{algo:spatio-temporal-helper}, such that, the first occurrence of that model is placed as close as possible to the starting point of the models's sub-session. The next instance of the model is scheduled at the end of its second sub-session. This way, a gap is created in the schedule. We can see the schedule produced by this algorithm in Fig.~\ref{fig:model_schedule_scheduled}. }
\Scut{
We now demonstrate spatio-temporal scheduling with \Scut{with several current, popular DNN models,} Alexnet (30\%), ResNet-50 (40\%) and VGG-19 (50\%), each with their corresponding GPU knee \% (demand), based on our profiling and output from \S~\ref{sec:efficient_batch_size}. 
Table.~\ref{tab:model_characteristics} summarizes the characteristics of the models.

As shown in Fig.~\ref{fig:model_schedule_scheduled} (Top plot), the total GPU\% does not exceed 100\%, \ie GPU is never oversubscribed.We can further see \Scut{from Fig.~\ref{fig:model_schedule_scheduled} }that there are still GPU resources available during the session. We can use this remaining resource to dynamically schedule an application that can fit with the remaining GPU resources available.  
}
\Scut{We also computed the average GPU utilization of the schedules and average utilization of schedule in Fig.~\ref{fig:model_schedule} is 57\% while in output of our scheduler in  Fig.~\ref{fig:model_schedule_scheduled} is 74\%. Moreover, our solution does not oversubscribe the GPU for these models. \knote{this improvement could be considered trivial. Why would we say this?} \anote{please see now.}}\Scut{We further note that the scheduler generates a schedule where all the models run at least once within their SLO.} 
\Scut{
\begin{algorithm}[!htbp]
\caption{Fair, Opportunistic and Dynamic Scheduling}
\label{algo:dynamic_scheduling}
\begin{algorithmic}[1]
\Function{\texttt{Dynamic-schedule}$(\texttt{Schedule, priority})$}{}
\For{\texttt{i:=0 to len(priority)}}
\State \Comment{Highest Priority First}
\If {\texttt{Remaining-GPU > model[i].GPU\%}}
\State Slice $\gets$ time(\texttt{Schedule, model[i].GPU\%})
\EndIf
\State batch $\gets$ Batch-Size(\texttt{model[i], Slice})
\State \textbf{Run-Batch}(\texttt{model[i],batch})
\EndFor
\end{algorithmic}
\end{algorithm}
}

\subsubsection{\textbf{Fair, Opportunistic, Dynamic Scheduling}}
\begin{figure*}
    \centering
    \subfloat[Temporal Schedule]{
    \includegraphics[width=.5\columnwidth]{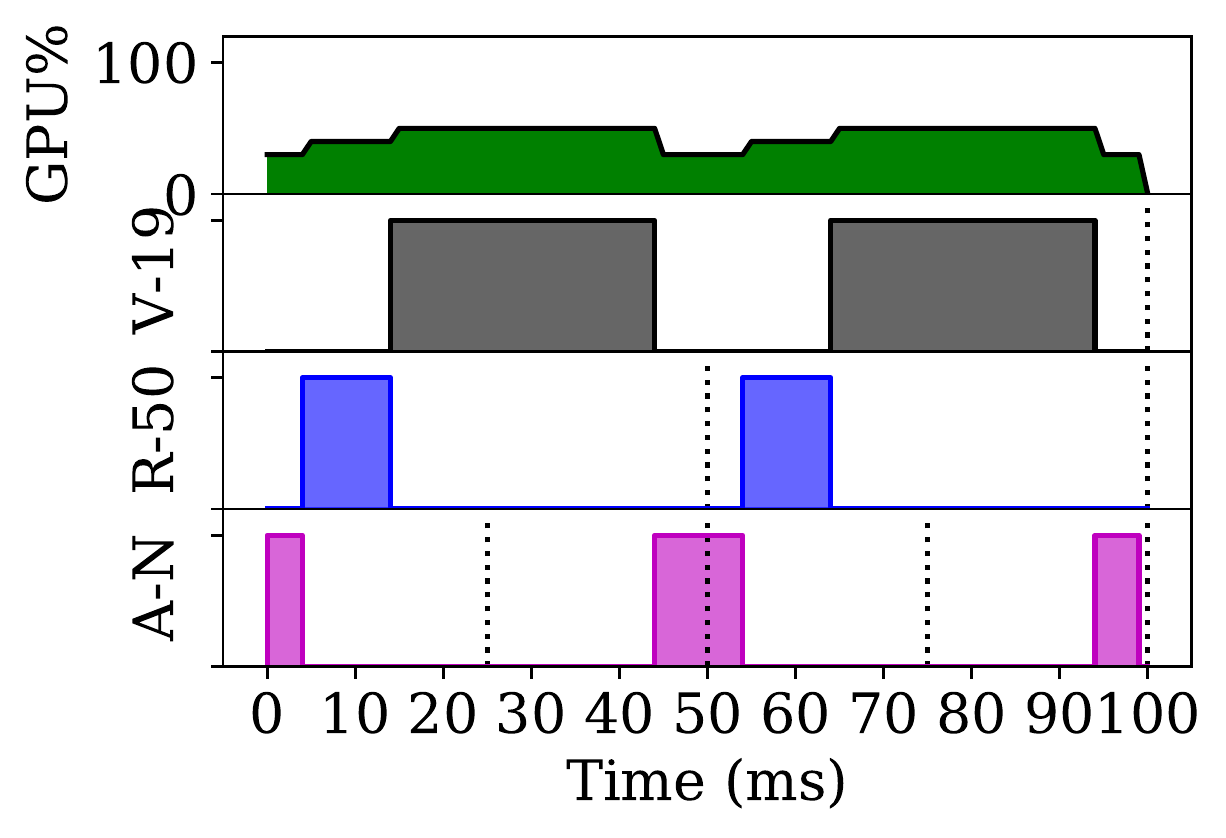}
    \vspace{-2mm}
    \label{fig:model_schedule}
    }
    \subfloat[ST-only Schedule]{
    \includegraphics[width=.5\columnwidth]{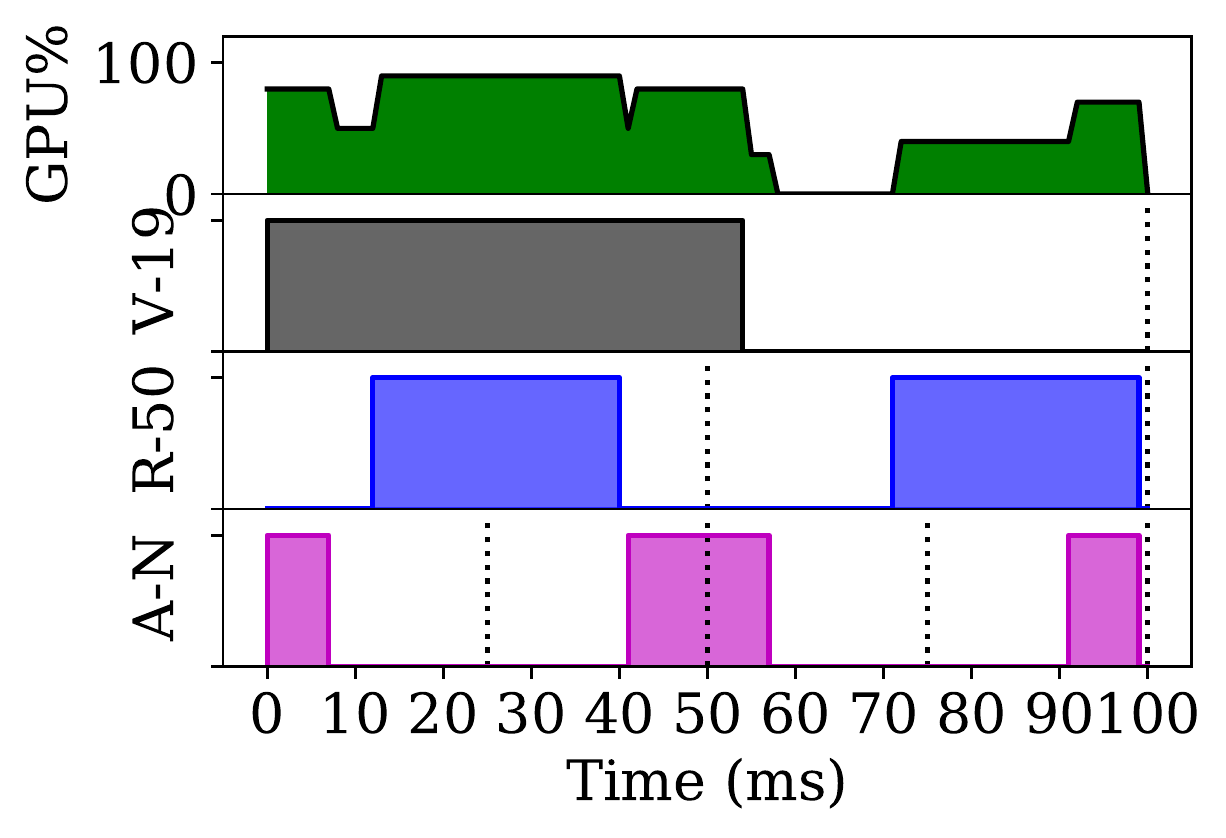}
    \vspace{-2mm}
    \label{fig:model_schedule_scheduled}
    }
    \centering
    \subfloat[\name{} schedule]{
        \includegraphics[width=.5\columnwidth]{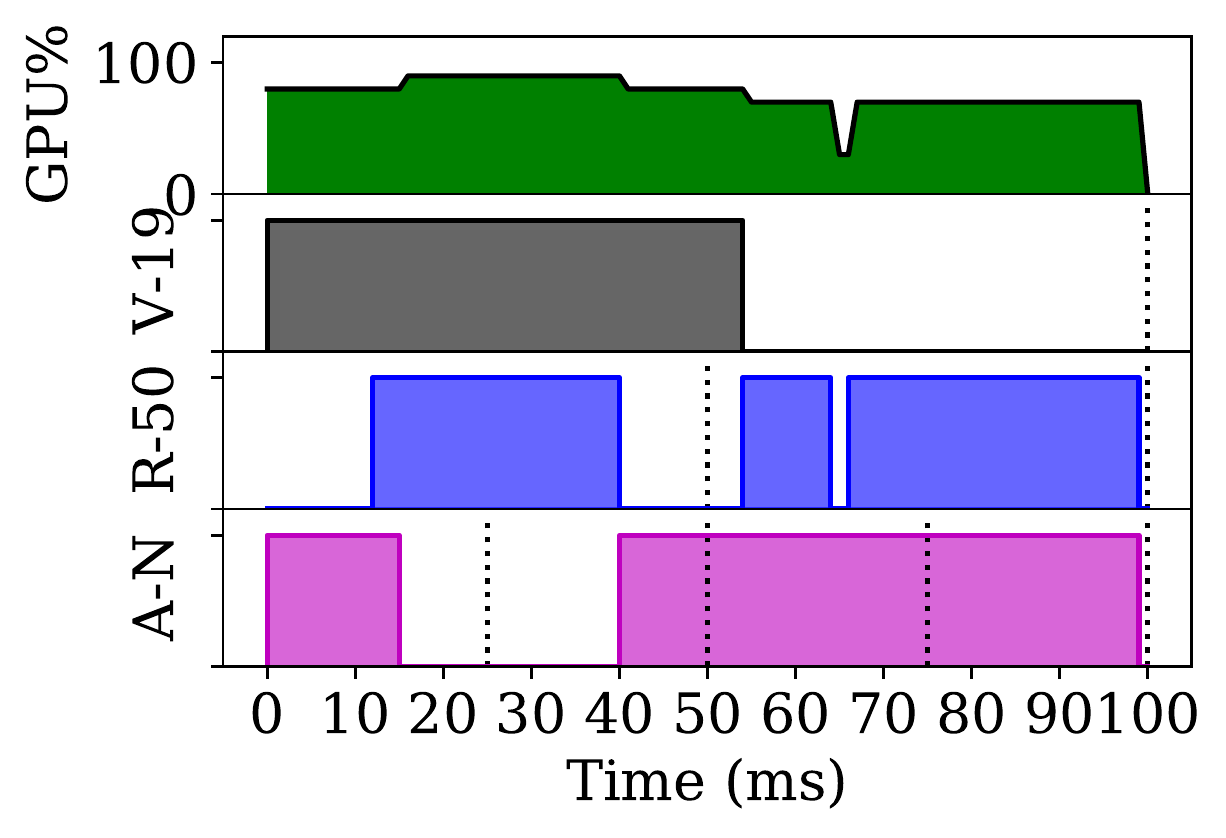}
        \vspace{-2mm}
    \label{fig:spatio-temporal-dynamic}
    }
    \subfloat[Throughput/GPU util.]{
    \includegraphics[width=.25\linewidth]{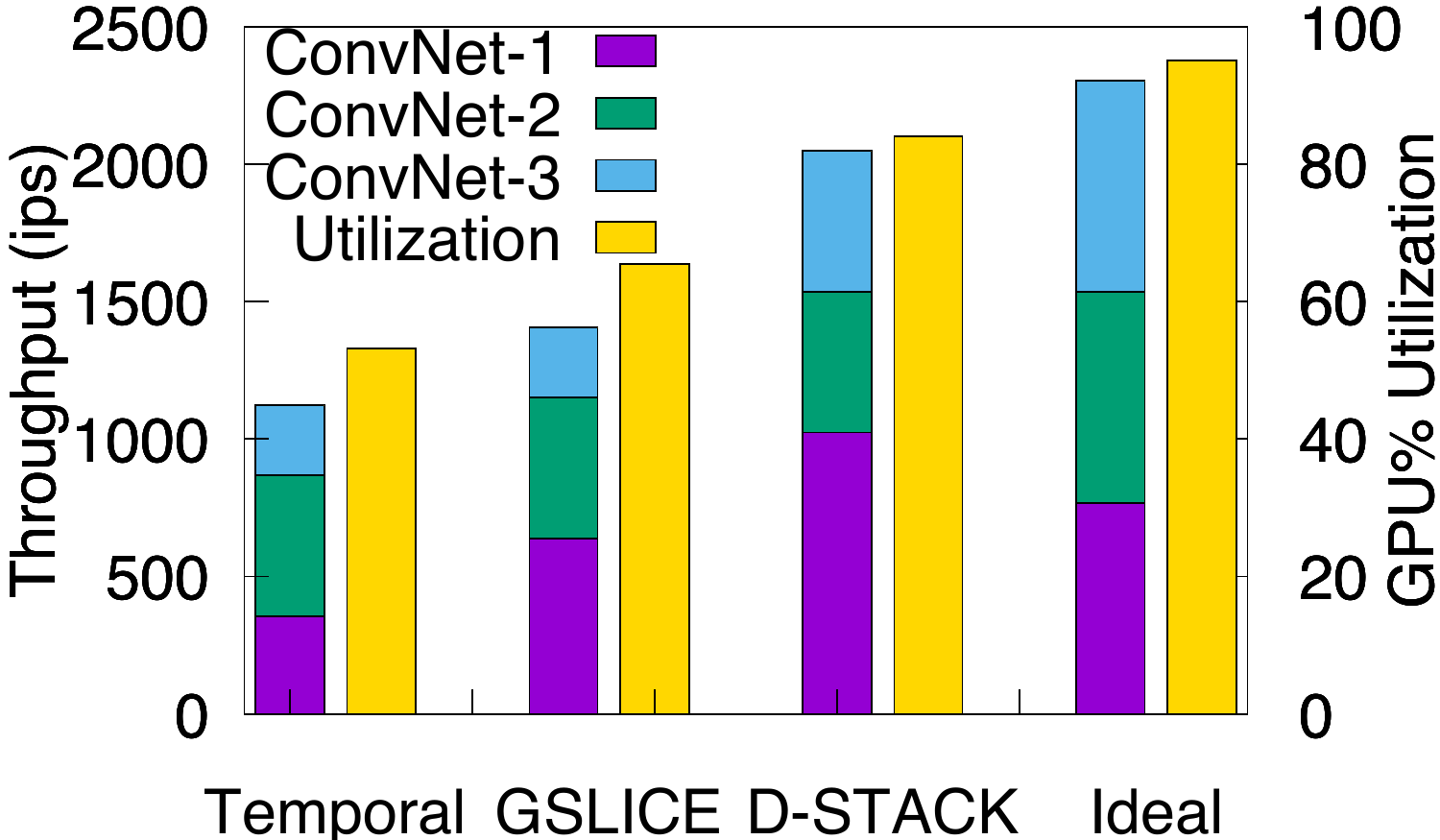}
\label{fig:ideal-throughput}
    }
    \Scut{
    \subfloat[Convolution Net Models]{
    \resizebox{.25\linewidth}{!}{%
    \begin{tabular}[b]{|c|c|c|}\hline
         Model & Knee\% & \makecell{Runtime\\(ms)}\\\hline
         \makecell{ConvNet-1} & 30\% & 10.3\\
         \makecell{ConvNet-2} & 40\% & 14.6\\
         \makecell{ConvNet-3} & 60\% & 15.4\\\hline
    \end{tabular}}
    \label{tab:ideal-schedule}
    }
    }
    \caption{(a, b, c) Scheduling Algorithms; (A-N=Alexnet, R-50=ResNet-50, V-19=VGG-19) \Scut{(d), Ideal scheduler DNN models} (d) Comparison with ideal scheduler}
    
\end{figure*}

\Scut{
\begin{figure}
\vspace{-4mm}
    \centering
    \subfloat[Temporal Schedule]{
    \includegraphics[width=.32\columnwidth]{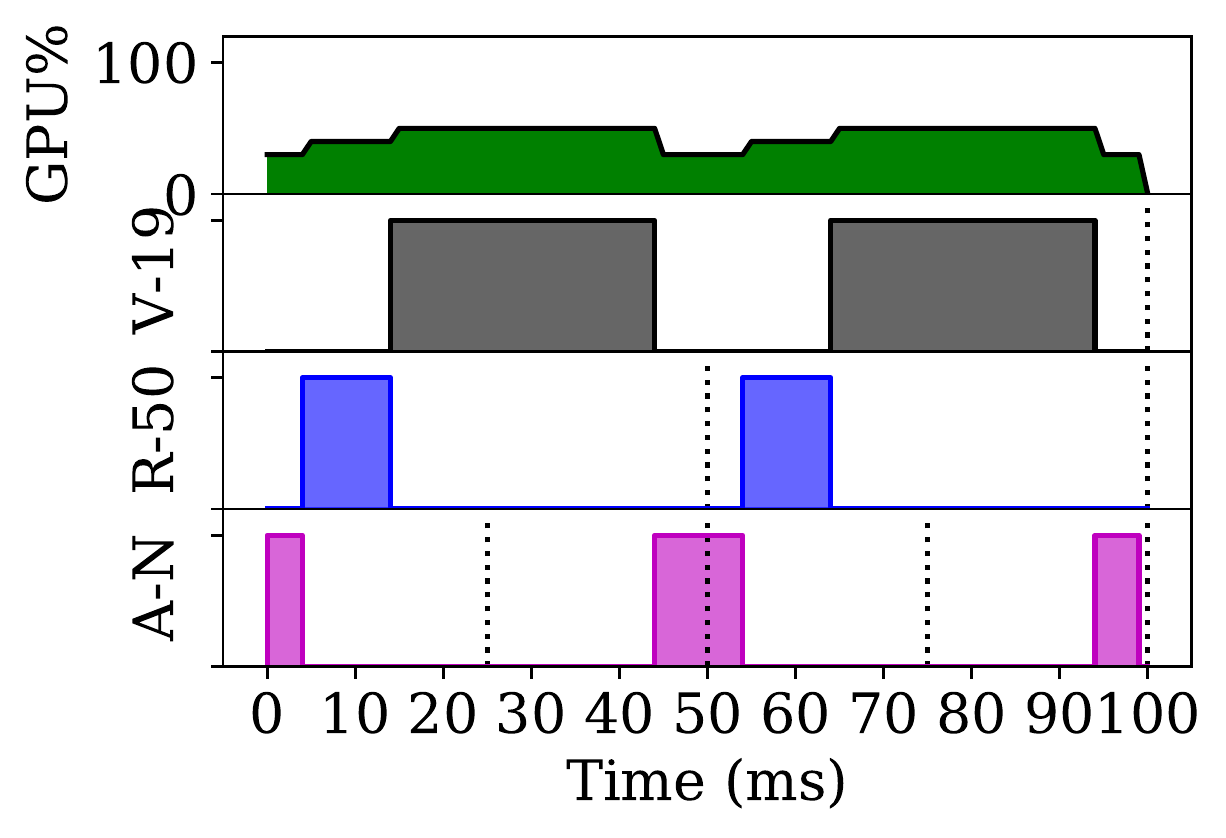}
    \vspace{-2mm}
    \label{fig:model_schedule}
    }
    \subfloat[ST-only Schedule]{
    \includegraphics[width=.32\columnwidth]{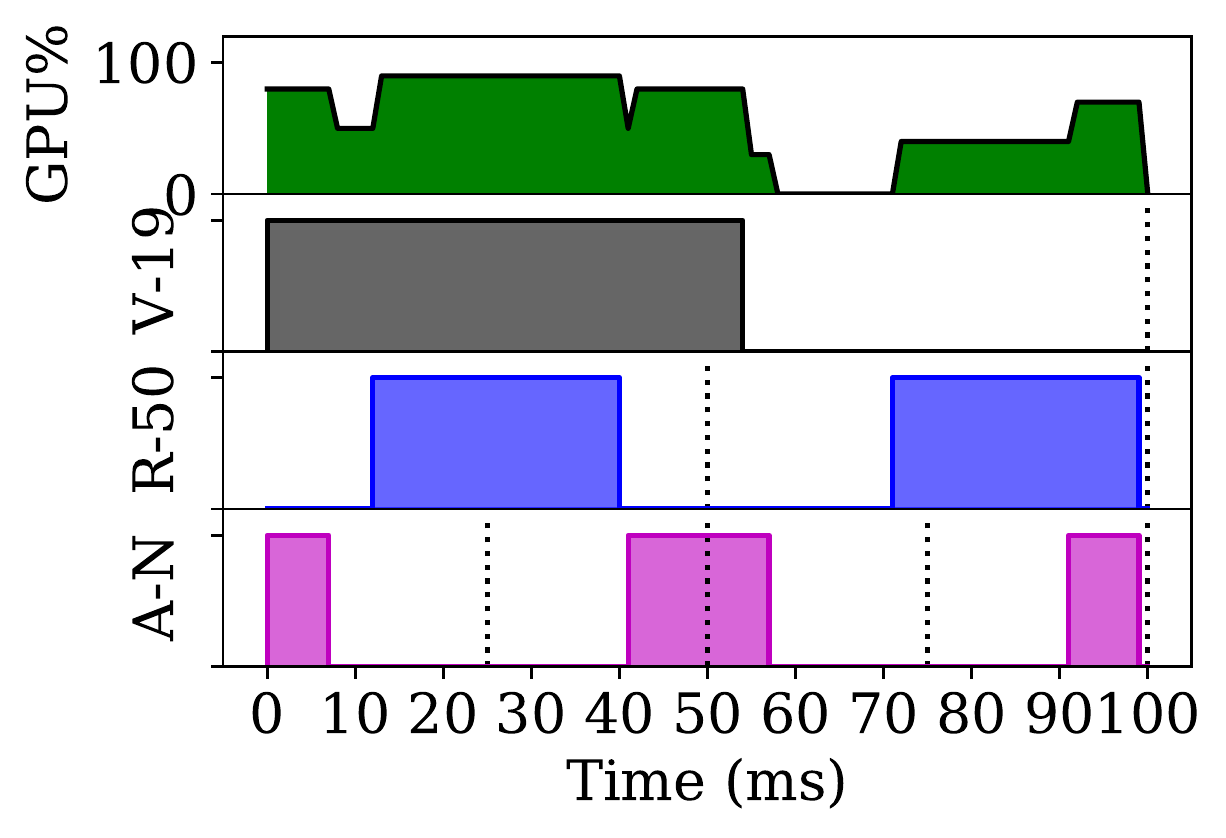}
    \vspace{-2mm}
    \label{fig:model_schedule_scheduled}
    }
    \centering
    \subfloat[\name{} schedule]{
        \includegraphics[width=.32\columnwidth]{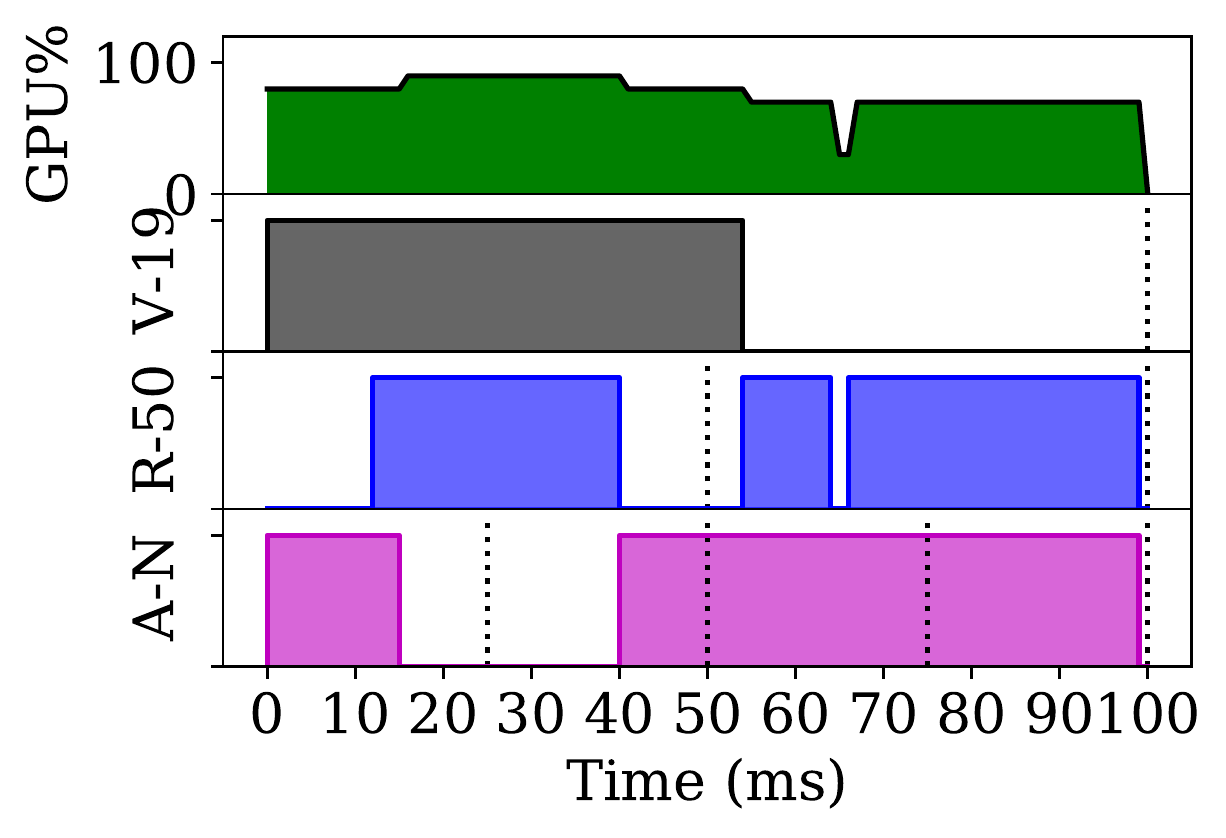}
        \vspace{-2mm}
    \label{fig:spatio-temporal-dynamic}
    }
    \vspace{-4mm}
    \caption{Behavior of Alternate Scheduling Algorithms; (A-N=Alexnet, R-50=ResNet-50, V-19=VGG-19)}
    \vspace{-6mm}
    
\end{figure}
}
To efficiently utilize the GPU resource while ensuring that the system meets SLO guarantees, we further propose an opportunistic dynamic scheduling enhancement. The dynamic scheduling\Scut{ in Algorithm~\ref{algo:dynamic_scheduling}} is 
triggered when a new request dynamically arrives for a model and when a model ends inference.\Scut{and when a scheduled model has no 
inference requests to process} The dynamic scheduler picks a 
model that is not \Scut{scheduled by the Spatio-Temporal scheduling Algorithm,~\ref{algo:spatio-temporal}}active.  This opportunistic addition is allowed as long as the GPU is not oversubscribed (so as to not interfere with the already scheduled models). 
To ensure fairness among available models, we use a scoreboard that tracks \Scut{each model's total execution time} how many times each model has run in the last few (e.g., ten) sessions and prioritizes the models that have run the fewest. 
\Scut{The model with the highest priority that can fit within the available GPU\% is chosen.}\Scut{, as in Algorithm~\ref{algo:dynamic_scheduling} (line 20).} The algorithm then finds a time slice for the model to finish inferring\Scut{without interfering with the other scheduled models (line 22)} and also determines a batch size 
that can complete within the time slice.\Scut{ (line 24) and runs the model with that batch} If the highest priority model cannot be run, the algorithm picks the model with the next higher priority. We show the output of the \Scut{spatio-temporal opportunistic dynamic scheduling}\name{} scheduling 
in Fig.~\ref{fig:spatio-temporal-dynamic}. With this dynamic scheduling packing more models to be scheduled opportunistically, the average GPU utilization increases from 60\% in the plain spatio-temporal 
schedule (Fig.~\ref{fig:model_schedule_scheduled}) to 74\% with the \Scut{spatio-temporal, dynamic scheduler}\name{} schedule (Fig.~\ref{fig:spatio-temporal-dynamic}).\looseness-1

\Scut{
\begin{figure}
    \centering
    \begin{subfigure}{.49\linewidth}
        \includegraphics[width=\linewidth]{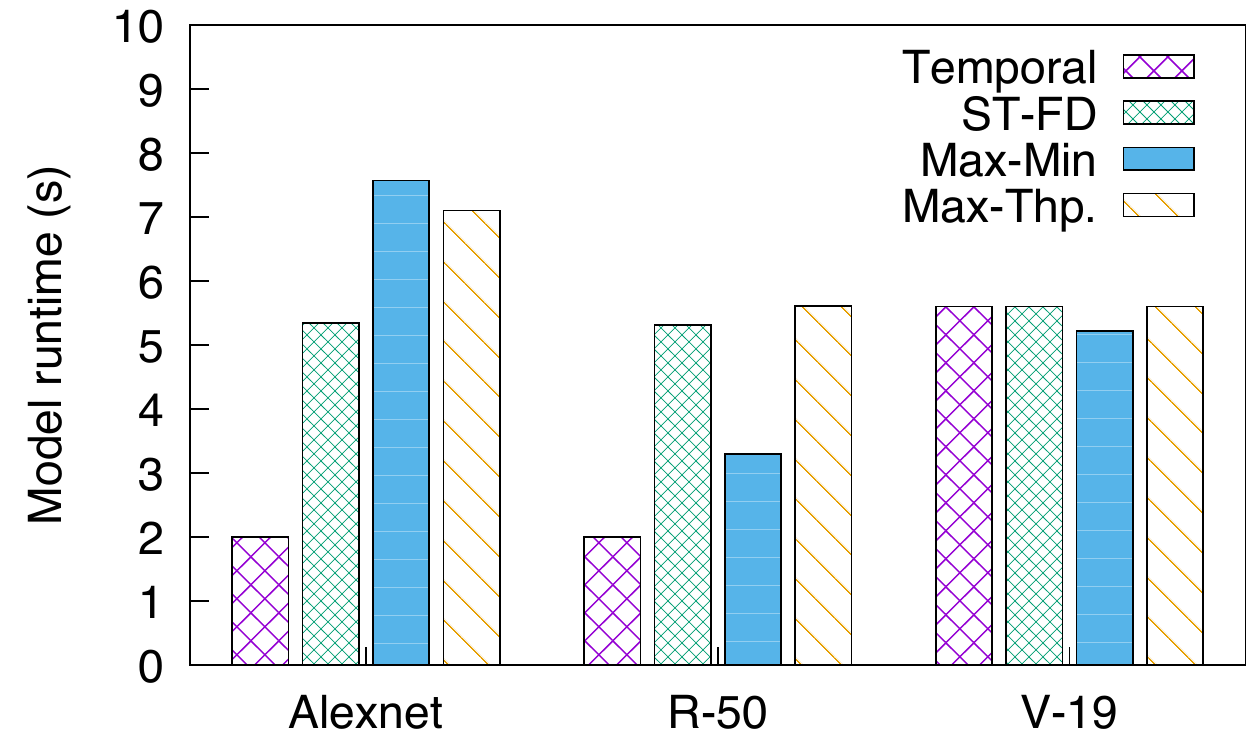}
            \caption{Total execution time (s)}
    \label{fig:time_max_min_vs_STFD}
    \end{subfigure}
    \begin{subfigure}{.49\linewidth}
        \includegraphics[width=\linewidth]{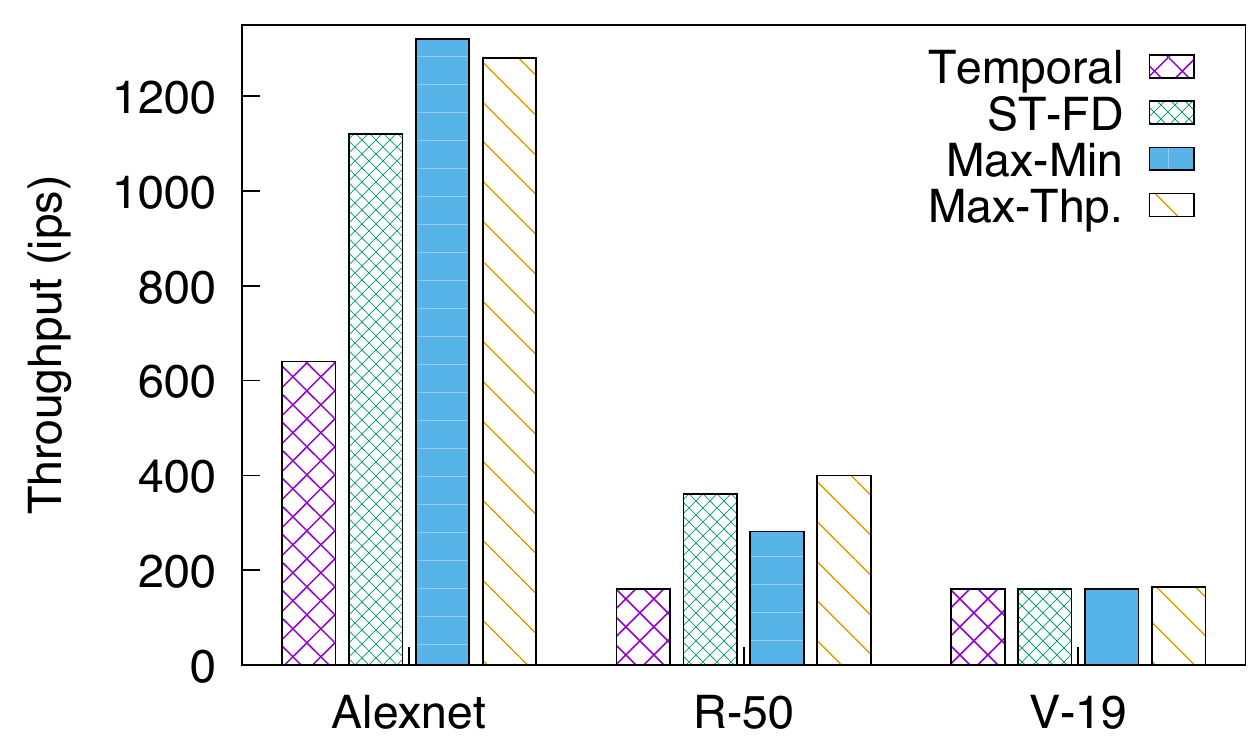}
            \caption{Throughput}
            \label{fig:throughput_max_min_vs_STFD}
    \end{subfigure}
    \caption{(a) Total execution time and (b) Throughput attained by models running with Max-Min vs. \name{} }
    \knote{be careful with the caption - opposite from the actual models placement.}
\end{figure}
}

\Scut{
\begin{algorithm}[!htbp]
\caption{Dynamic Scheduling}
\label{algo:dynamic_scheduling}
\begin{algorithmic}[1]
\Function{\texttt{Dynamic-schedule}$(\texttt{Schedule, priority-list})$}{}
\For{\texttt{i:=0 to len(priority-list)}}
\State \Comment{Highest Priority First}
\If {\texttt{Schedule.Remaining-GPU > model[i].GPU\%}}
\State Slice $\gets$ time-length(\texttt{Schedule, model[i].GPU\%})
\EndIf
\State batch $\gets$ Batch-Size(\texttt{model[i], Slice})
\State \textbf{Run-Batch}(\texttt{model[i],batch})
\EndFor
\end{algorithmic}
\end{algorithm}
}
\subsection{An Ideal Spatio-Temporal Schedule vs \name{}} 

We compare \name{} against an ideal scheduler,
\Scut{ that operates without the constraints of a practical GPU environment and software runtime,} 
which is a theoretical spatial and temporal schedule at the granularity of individual DNN kernels. For the ideal case, we assume
\Scut{that a particular DNN once scheduled can run on the GPU immediately (\ie}
GPU kernel preemption is allowed, a DNN's instantaneous GPU demand is known and the GPU's allocated resources are adjusted instantaneously. Any realistic system that does not preempt a currently running DNN model until its inference is completed, together with scheduling overheads to switch from one model to another inevitably under-utilizes the GPU.  \Scut{Thus, our ideal scheduler enables a very fine-grained GPU resource allocation to maximize GPU utilization and DNN inference throughput.} 
Thus, the ideal scheduler provides a theoretical 'optimal' performance achievable by \name{} or other schedulers.\Scut{By comparing with ideal scheduler, we seek to show how close a particular scheduling approach can come to the ideal. Closer the scheduling approach to ideal scheduling, the better it utilizes the GPU.}\looseness-1

\Scut{
We first present comparison of \name{} scheduling with an ideal scheduling mechanism. We consider ideal scheduling mechanism to be able to change the GPU\% of DNN instantly, in contrast to current GPU systems which do not allow to change GPU\% once a process starts. With ideal scheduling, we can easily provide right amount of GPU resource for each DNN kernel, and enable even fine-grained GPU resource allocation, potentially increasing the GPU utilization. We therefore, evaluate the GPU utilization of \name{} and an ideal scheduling mechanism.
}

We consider a time-slotted system (\eg 100$\mu$s for experiments with a small scale DNN), where $S_{i}$ represents $i^{th}$ time slot in the schedule. We schedule the kernel $k_{m}$ from DNN model $m$. We include as many model's kernels as will fit in the GPU at their Knee\%, ordered by their earliest deadline. We compute the aggregate GPU\% as $G_{ui}=\sum_{k\in S_{i}} GPU\%_{k}$  for each time slot $S_{i}$. We use an exhaustive search-based schedule to maximize the GPU utilization for every time slot (Eq.~\ref{eq:maximization}). The overall GPU utilization $G_{u}$ is maximized as:
\begin{equation}
    \max{G_{u}}, \quad \texttt{where } G_{u}=\sum_{i}G_{ui} = \sum_{i}\sum_{k\in S_{i}} GPU\%_{k} 
        \label{eq:maximization}
\end{equation}
\begin{align}
           \texttt{such that } G_{ui} \leq 100\% &\texttt{ and } k_i \in E \implies k_{i-1} \in E   
    \label{eq:constraints}
\end{align}

\Scut{
While performing ideal scheduling,  we first divide the schedule into fine grained time slots (\eg we used 100$\mu$s for our experiments), such that $S_{i}$ represents $i^{th}$ time slot in the schedule. We first schedule the kernel $k_{m}$ from the model $m$, which have the earliest deadline and compute GPU utilization GPU\% $GPU_{ui}$ such that $GPU_{ui}=\sum_{k\in S_{i}GPU\%_{k}}$ for each $S_{i}$ as sum of GPU\% of all kernels running on the session $S_{i}$. We then schedule kernels from other models on basis of minimizing the free GPU\% such that GPU utilization can be maximized. }

\Scut{
\begin{align}
\label{eq:gpu_ui}
    GPU_{ui} &= \sum_{k\in S_{i}} GPU\%_{k}\\
    GPU_{u} &= \sum_{i}GPU_{ui}\\\label{eq:eligible}
    k_i \in E &\implies k_{i-1} \in E\\\label{eq:hundred}
    GPU_{ui} &\leq 100\%\\\label{eq:maximization}
    \max_{GPU_{u}}\sum_{i}GPU_{ui} &\texttt{ w.r.t eq.} \ref{eq:gpu_ui},\ref{eq:eligible},\ref{eq:hundred}
\end{align}
}

\Scut{
}
The first constraint for scheduling kernels of different models (Eq.~\ref{eq:constraints}) is that the sum of the GPU\% of all concurrent kernels in a time slot should not exceed 100\%. Second, only eligible kernels (set $\left(E\right)$) can run concurrently in the time slot $S_{i}$ being scheduled.\Scut{We have to determine all the kernels that are eligible to run, represented by set $\left(E\right)$ for each scheduling instance.} 
DNN kernels are executed sequentially.\looseness-1

\Scut{We maximize (Eq.~\ref{eq:maximization}) the overall GPU utilization $G_{u}$ picking kernels
(using brute force search) for each time slot, by searching
\Scut{
\begin{table}
    \centering
    \vspace{-4mm}
    \captionof{table}{ConvNet Models}
    \vspace{-4mm}
    \resizebox{.5\linewidth}{!}{%
    \begin{tabular}{|c|c|c|}\hline
         Model & Knee\% & Runtime (ms)\\\hline
         ConvNet-1 & 30\% & 10.3\\
         ConvNet-2 & 40\% & 14.6\\
         ConvNet-3 & 60\% & 15.4\\\hline
    \end{tabular}
    }
    \label{tab:ideal-schedule}
    \vspace{-2mm}
\end{table}
}
exhaustively for eligible kernels from all the concurrently running models. Our ideal scheduler allows the preemption of a kernel unlike typical GPU systems and their software runtimes, allowing the 
selection of the most suitable set of kernels 
to maximize the GPU utilization for each time slot.
\looseness-1}

\Scut{
Unlike real GPU hardware, where Kernel pre-emption is not allowed when running multiple processes concurrently, ideal scheduling should allow to preempt the Kernel. We utilize kernel preemption to pick the kernels that will minimize the free GPU\%, thus, maximize the GPU utilization for every scheduling timeslice. We then utilize brute force search to find the kernels sequences that can lead to maximum GPU utilization for each timeslice in the schedule as shown in eq.~\ref{eq:maximization}.
}
\Scut{
As we run a kernel, we utilize a module to search the eligible kernels that can be run from other concurrently running DNNs. We should note that we search for eligible kernels with 2 constraints, first, it should be the kernel that runs next as DNN kernels run sequentially. Second, the sum of GPU\% should be less than 100\%. We then brute force the search among eligible kernels to find the sequence of kernels that utilizes the greatest amount of GPU. This way, ideal scheduling maximizes the GPU utilization, while, scheduling kernel wise.} 
\Scut{
\begin{minipage}{.49\textwidth}
\begin{algorithm}
\caption{Ideal Scheduling Algorithm}\label{algo:ideal}
\begin{algorithmic}
\Function{Schedule}{Model}
\State $Model \gets EDF(DNN-Models)$
\For{\texttt{each Kernel in Model}}
\State $\texttt{RemGPU\%} \gets 100 - \texttt{Kernel.GPU\%}$
\State $\texttt{Concur-Kernels} \gets \texttt{FindKernels(RemGPU\%)}$
\EndFor
\State $\texttt{RunKernels(Concur-Kernels)}$
\EndFunction
\end{algorithmic}
\end{algorithm}
\end{minipage}%
\begin{minipage}{.49\textwidth}
\begin{algorithm}
\caption{Ideal Scheduling Algorithm}\label{algo:ideal2}
\begin{algorithmic}
\Function{FindKernels}{RemGPU\%}
\For{\texttt{each Kernel in AllKernels}}
\State $\texttt{EKernels} \gets \texttt{isEligible(Kernel)}$
\EndFor
\For{\texttt{each eKernel in EKernels}}
\While{\texttt{eKernel.next != NULL}}
\State $\texttt{GPU-Util}\gets \texttt{eKernel.GPU\%}$
\State $Kernel-list.append(ekernel)$
\State $eKernel \gets eKernel.next$
\EndWhile
\EndFor
\State $return-kernels \gets max(GPU-util)$
\Return $return-Kernels$
\EndFunction
\end{algorithmic}
\end{algorithm}
\end{minipage}

}

We experimented by scheduling 3 convolution neural networks (ConvNet) based on LeNet~\cite{lecun1989backpropagation}. Each ConvNet has 3 convolution, 2 average-pool and 2 linear kernels. The dimensions of filters of the convolution layers are varied, varying the compute requirement for each ConvNet model. The inference image has a resolution of 224$\times$224. The knee-runtime combination for ConvNet-1, ConvNet-2 and ConvNet-3 are 30\%-10.3ms; 40\%-14.6ms, and 60\%-15.4ms, respectively.  
We computed the knee of each kernel of each model, for use by the ideal scheduling during inference. We present the GPU utilization and throughput in Fig.~\ref{fig:ideal-throughput}.
\Scut{
\begin{figure}
\begin{minipage}{.25\textwidth}
\includegraphics[width=\linewidth]{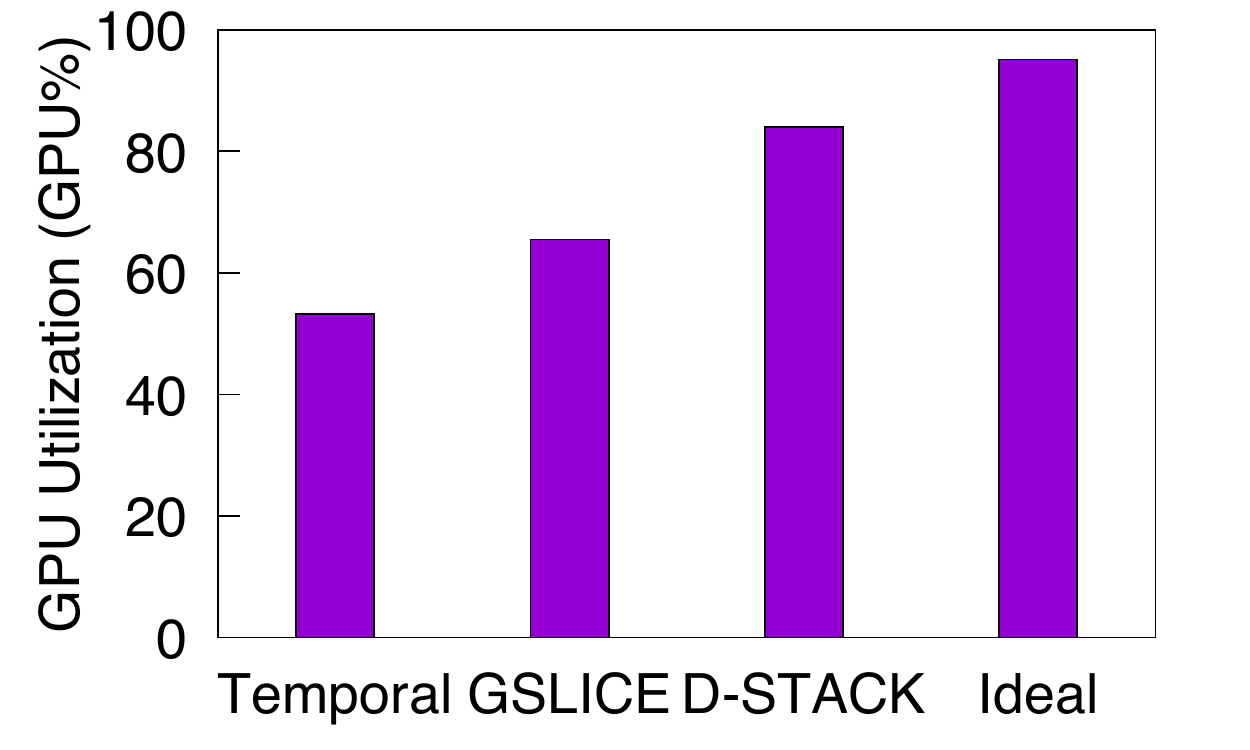}
\vspace{-8mm}
\caption{GPU utilization}
\label{fig:ideal-utilization}
\end{minipage}%
\begin{minipage}{.25\textwidth}
\includegraphics[width=\linewidth]{figures/ideal-schedule/ideal_throughput.pdf}
\vspace{-8mm}
\caption{Throughput (ip/s)}
\label{fig:ideal-throughput}
\end{minipage}%
\vspace{-6mm}
\end{figure}
}
Temporal scheduling has a much lower GPU utilization, as it runs a single kernel on the GPU at a time. GSLICE improves the GPU utilization, but its static schedule leads to lower utilization when not enough models are running on the GPU. Ideal scheduling attains almost 95\% GPU utilization, because it schedules kernels leveraging preemption. 
\name{} schedules without preemption of a kernel, 
runs a DNN kernel to completion even if a kernel that could utilize the GPU better is waiting.
Nonetheless, \name{} still achieves $\sim$86\% GPU utilization. The throughput attained by the three CNN models follows the same trend. 
\name's overall throughput is slightly higher than 90\% of the throughput of ideal scheduling - a measure of how close it comes to the ideal scheduler.\looseness-1  


\Scut{can clearly see Knees of ConvNet-1 being about 20\%, ConvNet-2 being 30\% GPU and ConvNet-3 being 40\% GPU. as knee information about each model's kernel in the Table~\ref{tab:LeNet}.}

\Scut{
\begin{figure}
    \centering
    \includegraphics[width=.33\linewidth]{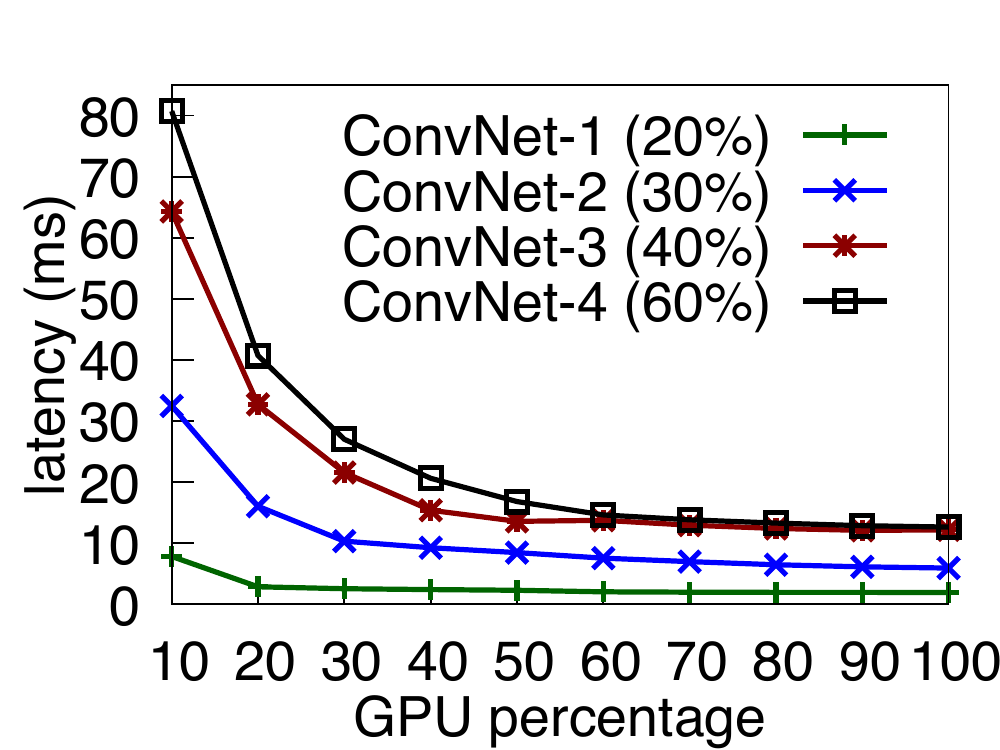}
    \caption{Latency (ms) vs GPU\% for different Convolution neural network Models\anote{Trying to get another Cov-4. Current looks liek it has knee at 60. .}}
    \label{fig:lenet_latencies}
    \begin{tabular}{|c|c|c|c|}\hline
         Model & Kernel & In Channels & Out Channels   \\\hline
         ConvNet-1 & Conv.1 & &\\\hline 
         ConvNet-1 & Conv.2 & &\\\hline 
         ConvNet-1 & Conv.3 & &\\\hline 
    \end{tabular}
\end{figure}
}
 \subsection{Evaluation of \name{} Scheduler}
We evaluate \name{} \Scut{spatio-temporal scheduler that includes the opportunistic and dynamic scheduling enhancement (hereafter called ???) }
using four popular DNN models (Alexnet, Mobilenet, ResNet-50, and VGG-19) that are run with fixed SLOs, GPU\%, and runtime as presented in Table~\ref{tab:model_characteristics}. We ran the models concurrently for 10 seconds.\Scut{ while providing a high rate (at 10 Gbps link rate) 
of inference requests.}\Scut{ that all the models can sustain while forming an optimal batch size for inference.} We took the workload mix from the Imagenet~\cite{imagenet_cvpr09} (vision DNNs), and IMDB dataset~\cite{imdb-data} (sentence classification with BERT). \Scut{(BERT \knote{I thought BERT was a model. Is it a database also? I think you should make sure what should be said here}\anote{updated with dataset citation for BERT}). }
\Scut{In the cloud, the IP 5-tuple of arriving inference requests is used to direct them to the right, concurrently running, application.} We introduce a random, uniformly distributed inter-arrival delay between requests destined for the same DNN model.\looseness-1

We compare the throughput, and GPU runtime of \name{} 
with the\Scut{ throughput achieved using the} baseline temporal sharing, and  
a schedule that maximizes the sum of the throughput across all the models ( \textit{max-throughput}).\Scut{by prioritizing scheduling the model with the least runtime.}\Scut{Finally we present throughput obtained by a brute force algorithm that that computes schdule that maximizes the throughput of each model.} 
We also evaluate the fairness of the schedulers, measured by the GPU runtime each model gets. For this, we compare \name{} against a Max-Min fair\Scut{fair share based, dynamic and opportunistic\knote{do we need to say dynamic and opportunistic?}} scheduler~\cite{bertsekas1992data}, which 
maximizes the placement of the minimum (smallest) demand (GPU\%). The throughput result is shown in Fig.~\ref{fig:throughput_max_min_vs_STFD}, and the GPU runtime each model gets is in Fig.~\ref{fig:time_max_min_vs_STFD}.\looseness-1

\Scut{For temporal sharing experiments, we made sure that each model gets 100\% of GPU resource for a time-slice to infer same batch size as spatio-temporal schedule. Any idle time remaining was then equally distributed to each of the model. This idle time was utilized by inferring an batch of request provided by an adaptive batching similar to the works~\cite{shen2019nexus,crankshaw2017clipper}, therefore, sharing all the time in the session with three DNN models.}\Scut{ We present the results in Fig.~\ref{fig:throughput_max_min_vs_STFD}.} 

\name{}
gets 2$\times$ the throughput of temporal sharing for the two compute-heavy models, ResNet-50 
and VGG-19 (Fig.~\ref{fig:throughput_max_min_vs_STFD}).
At the same time, the lighter-weight Alexnet and Mobilenet 
get 4$\times$ higher throughput.\Scut{Temporal scheduling\Scut{ with varying SLO causes models with a shorter deadline to be scheduled multiple times,} alone provides exclusive access of the GPU to a model.} In temporal scheduling,\Scut{As DNN inferences are not preempted,} running compute-heavy
\begin{figure}
     \subfloat[Throughput]{
        \includegraphics[width=.48\linewidth]{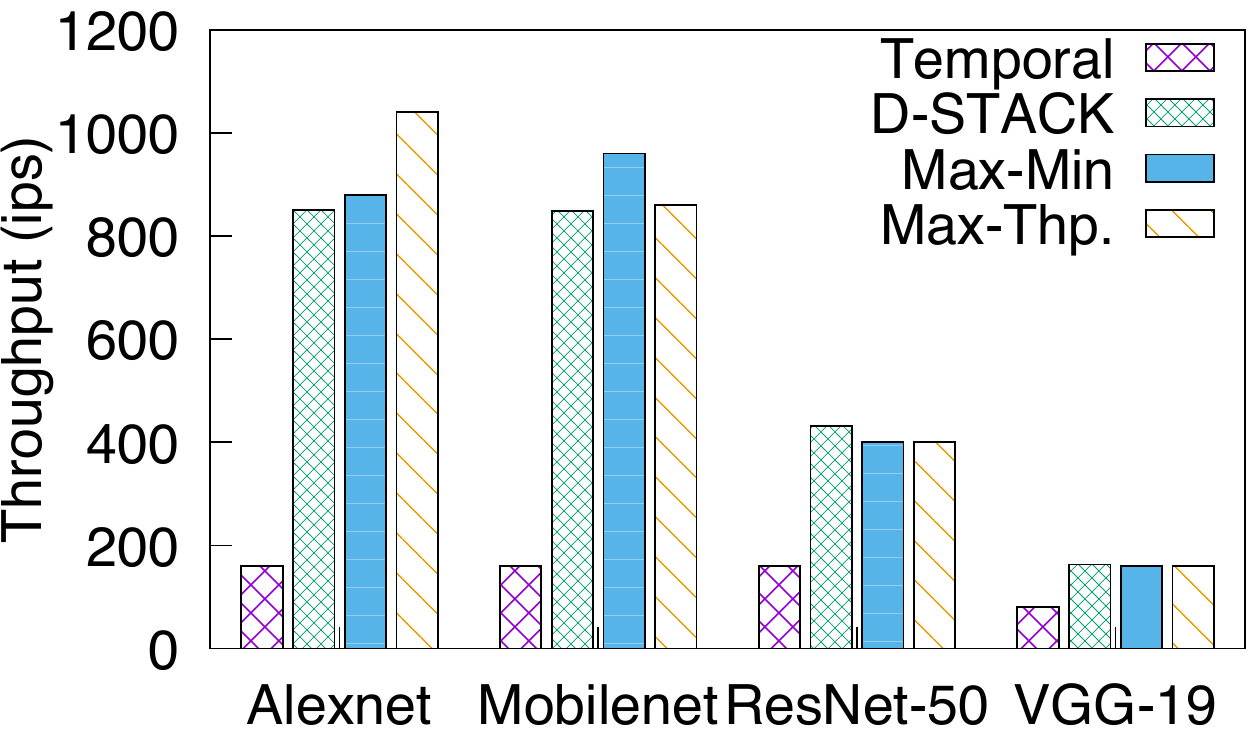}
        \vspace{-2mm}
            \label{fig:throughput_max_min_vs_STFD}
            }
    \subfloat[Runtime for each model (sec)]{
        \includegraphics[width=.48\linewidth]{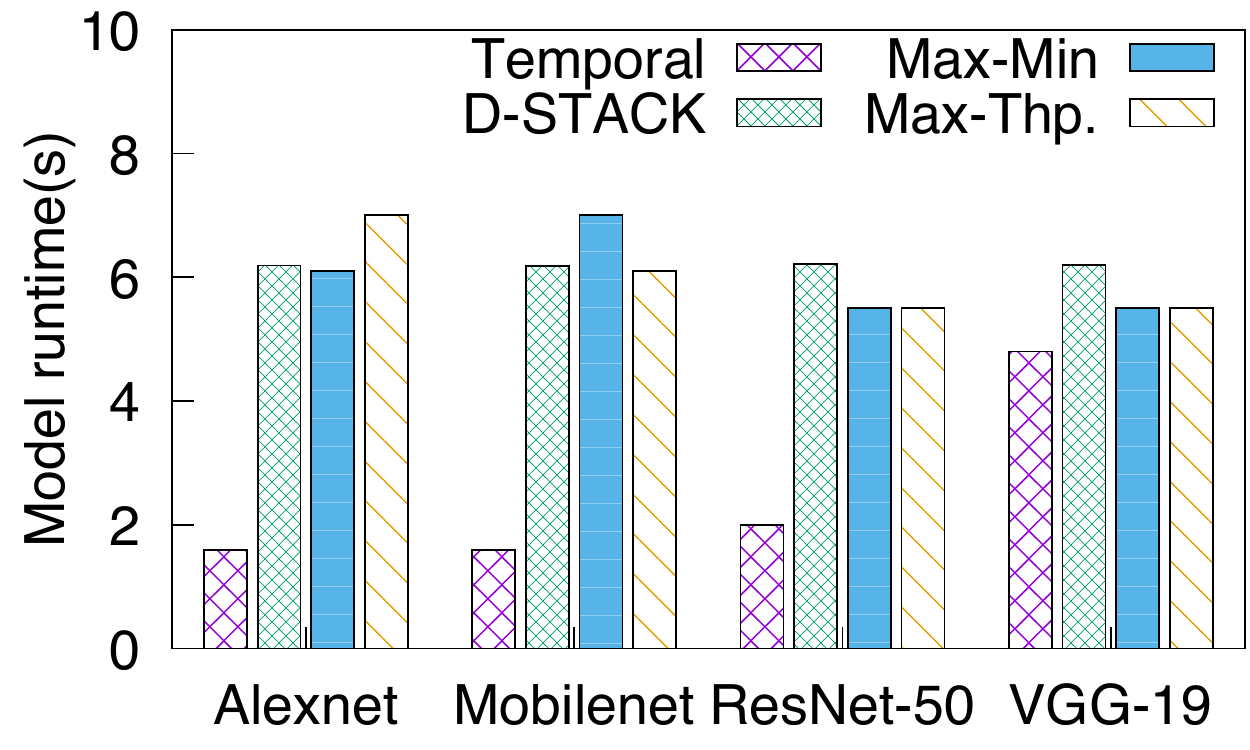}
        \vspace{-2mm}
    \label{fig:time_max_min_vs_STFD}
    }
    \caption{(a) Throughput of models running with different scheduling algo. and (b) Total runtime (s) per model  }
\end{figure}
DNNs with longer runtimes\Scut{cannot run with very large batch sizes to avoid their }
results in fewer opportunities for the other models, as there is no spatial sharing. 
Temporal scheduling runs models for only 1.6 sec. out of 10 secs. time, negatively impacting their throughput. \Scut{ this negatively impacts their throughput. }Fig.~\ref{fig:time_max_min_vs_STFD} shows that the \name{} runs all the models longer than temporal sharing \Scut{for more than 5 seconds each}. This is because \name{} can run multiple DNNs concurrently, \Scut{ 50\% of the experiment time}
\Scut{schedule permits running multiple models concurrently\Scut{ and allows models with a shorter deadline to run simultaneously with models having a long runtime. }
We can see from Fig.~\ref{fig:time_max_min_vs_STFD} that with \name{} all models run for more than 50\% of time we evaluated, showing they run concurrently with another model. }
providing higher throughput compared to temporal 
sharing (Fig.~\ref{fig:throughput_max_min_vs_STFD}).
\Scut{Thus, we can meet the SLO of models with a short deadline even while running other models with a larger batch size. Thus, we achieve a higher throughput.} 
We compare \name{}'s throughput with the 'max-throughput' schedule. 
\Scut{To estimate the ideal throughput, we generated a schedule by using brute force to maximize each individual model's throughput.} \name{} gets more than 80\% throughput of the max-throughput for the model with the lowest runtime (Alexnet) while providing better fairness as we see next.\looseness-1

The Max-Min fair schedule provides higher runtime for Mobilenet (Fig.~\ref{fig:time_max_min_vs_STFD}) 
than \name{} since Mobilenet has the minimum demand (25\% knee\%). 
However, \name{} achieves higher throughput than Max-Min for the medium runtime ResNet-50 (Fig.~\ref{fig:throughput_max_min_vs_STFD}). \name{}'s fairness measure picks the model that has run for the least time in the GPU over past sessions to schedule. Thus, \name{} seeks to act like a proportional fair scheduler, as with the Completely Fair Scheduler (CFS) in Linux~\cite{cfs_linux}. 
The fairness of \name{} is shown in Fig.~\ref{fig:time_max_min_vs_STFD}. Max-Min gives more time to a low-demand model like Mobilenet. With \name{}, all the models get similar GPU time, thus boosting the total throughput of higher demand models like ResNet-50. Overall, the \name{} 
scheduling beats temporal sharing's throughput by 4$\times$, \Scut{achieving close to} gets more than 80\% of the max-throughput scheduler and fairly shares GPU execution time while meeting SLOs.\looseness-1

 \Scut{increasing their throughout without making other models violate their SLO. }
 
 \Scut{Furthermore, the \name{} scheduler uses the optimal GPU\% and batch size, so that we achieve a throughput that is close to the model executing exclusively 100\% GPU, (thus achieving a higher throughput than temporal sharing.)  }

 \Scut{
\color{blue}
We further evaluated the fairness of \name{} scheduler. We compare \name{} with a Max-Min ~\cite{max-min} fair scheduler. A Max-Min fair scheduler would maximize the allocation to the DNN with the minimum demand. \name{} prioritizes fitting in a DNN which has run the least amount of time. We ran the 3 models (Alexnet, ResNet-50 and VGG-19) for 10 seconds to see how much time each model gets to run. We present the time received by each model in~\ref{fig:time_max_min_vs_STFD}. The Max-Min fair share algorithm prioritizes placing model with minimum demand (Alexnet), providing it with more runtime than other models. On the other hand, \name{} prioritizes all the models to get a fair allocation of time in the GPU while opportunistically and dynamically placing models. We also show the throughput of these allocation algorithms in Fig.~\ref{fig:throughput_max_min_vs_STFD}. We observe that our algorithm prevents the reduction of the throughput of ResNet-50 by dynamically providing more additional opportunities to run. We see in both Fig~\ref{fig:time_max_min_vs_STFD} and Fig~\ref{fig:throughput_max_min_vs_STFD} that VGG-19's execution time and throughput increases only by a very small amount. As VGG-19 is an accuracy-optimized, compute heavy model, inferring just a single image takes more time (10s of milliseconds), while the model itself requires high GPU\%. Therefore, it is more difficult to opportunistically place the model. Nonetheless, we see a slight increase in its throughput.
\color{black}
}
\Scut{
We prioritize the model that has run less times to be opportunistically placed whenever possible in the schedule. We measure the 
}
\Scut{
We also define a \textit{duty-cycle} as time taken by the slowest DNN model to execute a substantial batch, \eg a batch of 16 requests. Our intention to pick a \textit{duty-cycle} is to make sure that the slowest model could be scheduled at least once with big enough batch size to be useful. We then produce the schedule for each \textit{duty-cycle}.
\subsubsection{A 2-D Bin Packing problem and Solution}
We can now formulate a scheduling problem as a two dimensional bin packing problem, where we can see GPU spatio-temporal resource as a rectangle bin with dimensions of $D_{j}\times 100\%$ where 100\% represent all the GPU resources available for that duty-cycle. Similarly, each $i^{th}$ DNN model running in the GPU can be represented by a rectangle of dimensions $L_i\times p_i$. The constraints for packing are:
\begin{itemize}
    \item Rectangles for the models cannot overlap
\item    \[
    \sum_{i=1}^{k} L_i \times p_i \leq D_j \times 100 
\]
\end{itemize}
Where, $k$ is the number of models that can be scheduled.

We use Hybrid-First method\cite{chung1982packing} for 2-D bin packing. Hybrid-First (HF) 2-D bin packing algorithm is a two-phase method where first step uses a level-oriented algorithm to fit the models' rectangles into bins in non-increasing height order. In the second phase, the HF algorithm fits the bins according to first fit decreasing algorithm to minimize the number of bins. 

However, we should consider that hybrid-first algorithm is an approximation and might not lead to optimal packing. Furthermore, hybrid-first algorithm only fits one instance of the model. We can utilize the gaps in schedule by running DNN models that have requests available and can execute a batch on those gaps. 

\subsubsection{Algorithm}
}
\Scut{
\vspace{-2mm}
\subsection{GPU Memory Utilization During GPU\% Reconfiguration}\label{sec:dynamic_gpu_recon2}
\update{
\cut{
ML platforms today are designed with DNN training in mind. They cater to increase training throughput with higher batch sizes, however, an inference platform that multiplexes GPU and receives inference request from streaming data requires efficient data transfer to GPU, ability to adapt GPU resources of models according to request rate and efficient sharing of other hardware resources of GPU such as memory. 

\newline\noindent\textbf{Delivering data to GPU:} \Scut{DNN inference platforms that serve a huge amount of requests need to transfer the incoming streaming data to the GPU.} Usually, inference platforms transfer data to the GPU using data copies by the CPU into a contiguous region and then invoking GPU runtime API calls initiated by the CPU. This leads to the CPU becoming the bottleneck. Utilizing the GPU-resident DMA engine to scatter gather data has been shown to be effective ~\cite{dhakal2019netml,UVA}. We have adopted this GPU-DMA technique to transfer batches of DNN requests to the GPU. With our optimization, we attain higher throughput compared to the traditional data copy based  techniques using the CPU to push data to the GPU.
}
\Scut{
The GPU resources provided to an application will likely need to change due to the variations in the workload, \ie the change in the arrival rate of the tasks (especially with streaming data).
\Scut{ or the variations in the number of concurrently executing applications. Further, when an application terminates, it is necessary to reclaim and redistribute the GPU resources among the active applications to avoid any resource fragmentation and under-utilization of the GPU.} 
The latest CUDA MPS version, R450+~\cite{nvidiamps2019} strictly limits the configuration of the GPU resource partition (GPU\%) to a one-time (static) allocation that is set before the GPU initialization for the application. 
This limitation poses a major challenge, as any GPU resource readjustment would require us to spin up a new CPU process with an updated GPU\%, which can take several seconds (depending on the ML framework initialization). 
Some works~\cite{zhang2019laius} maintain a pool 
of (several) instances for each application with different statically assigned GPU\%. This allows the new incoming work to be directed towards an instance with desired GPU\%.}

\Scut{
Although effective for smaller GPU applications, we do not believe it is usable for large DNN models which often occupy gigabytes of GPU memory.\Scut{To avoid this idle time of several seconds, we utilize an \textit{overlapped execution} mechanism explained in ~\cite{GSLICE}.} Alternately,~\cite{GSLICE}, maintains a single standby instance of a DNN model that can be loaded into the GPU with dynamically computed GPU\%, while the original instance continues processing the requests.
Once the standby instance is ready, it seamlessly takes over the inference processing 
and terminates the original instance.}
\Scut{
While changing the GPU percentage two instances of the same model, the original and a new model, loaded with a new GPU\%, are running concurrently. This can double the amount of memory required by the model to run in the GPU.} 
To counter the increased GPU memory requirement while loading a new model with a different GPU\% \Scut{increase in memory required, }
we share DNN parameters and other buffers between the original and new model. As the model parameters remain same in both models, we create cudaIPC handles for all the parameter buffers of the original model and pass them to the new model instance that is loaded with new GPU\%, which, then uses those parameters for inference.  
Parameter sharing reduces the amount of memory needed by the new DNN model being  load by upto 40\% depending on size of DNN model's parameters. We adopt this method as it not only avoids idling the GPU, but also maintains a very low memory profile in the GPU and allows us to run larger DNN models. 
\cut{We further lower the GPU memory overhead by sharing the model parameters and GPU buffers between original model and its replica using the \textit{cudaIPC} mechanism. Reusing the original model's parameters reduces the GPU's memory footprint, especially while the GPU resources are being changed. Moreover, it does not have any effect on the latency and throughput of the currently active model.
}

\Scut{
We have observed that the replica's loading temporarily consumes additional memory as the original model and replica use GPU memory. To reduce the memory load, we use \textbf{parameter sharing} as explained in~\cite{GSLICE} to share the DNN model parameters between original and replica applications. \ie re-using the active model's parameters reduces the GPU's memory load while the GPU resources are being changed. Moreover, it does not have any effect on the latency and throughput of the currently running active model.}
}
}
\section{Validating Our Overall Approach}
\label{sec:validation}
\begin{figure*}
\subfloat[Throughput \& SLO violations]{
\includegraphics[width=.5\linewidth]{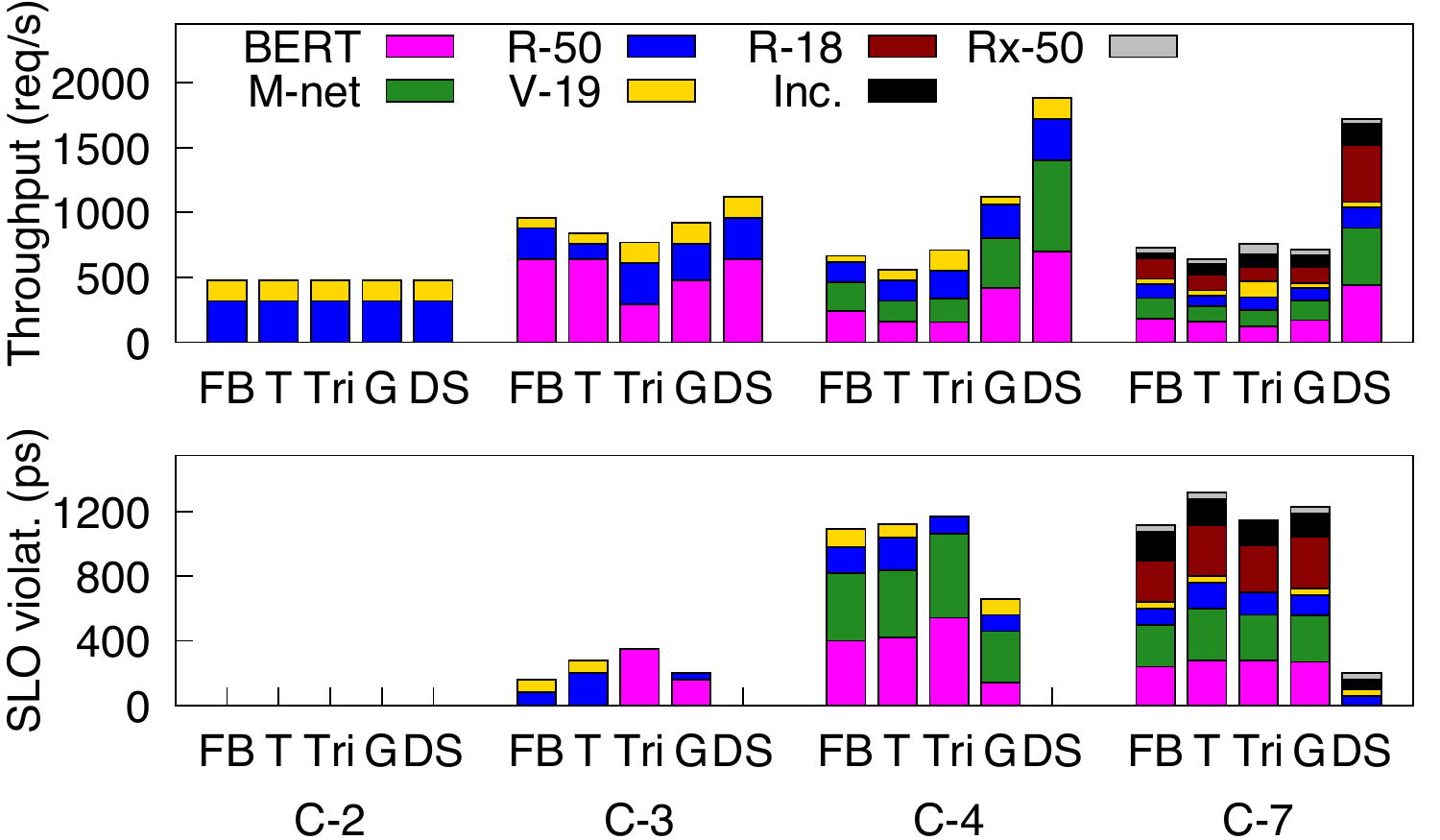}
    \label{fig:css_fig}
}
\subfloat[Baseline throughputs are shown in session $T_{0}$.]{
\includegraphics[width=.5\linewidth]{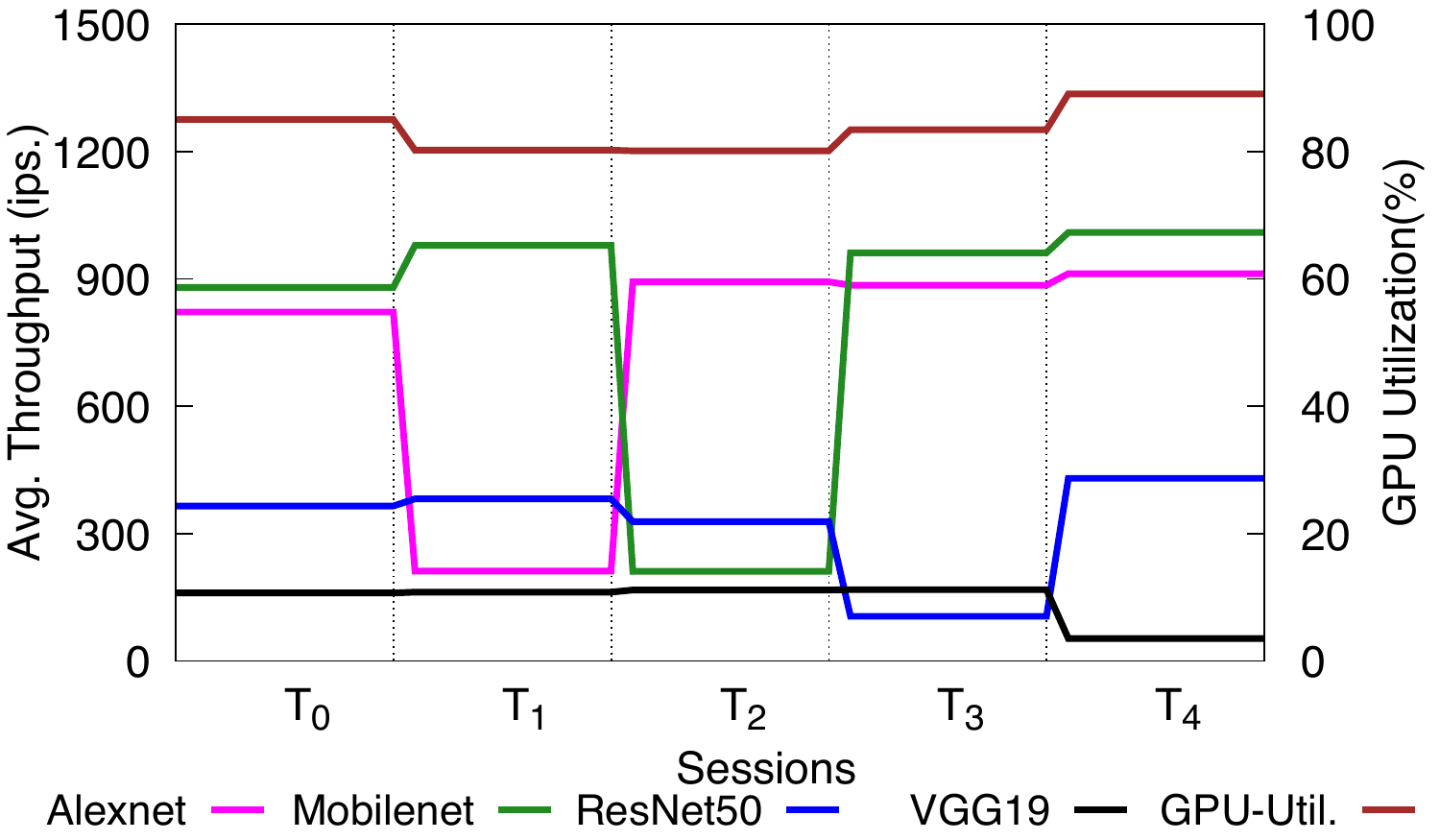}
    \label{fig:variable-rate}
}
\caption{(a) C-2 = ResNet-50 + VGG-19, C-3 = C-2 + BERT, C-4 = C-3 + Mobilenet, C-7 = C-4 + ResNet-18 + Inception + ResNeXt-50. (b) Throughput adjustment in \name{} with varying request rate }
\end{figure*}

\Scut{
\begin{figure}
   \includegraphics[width=\linewidth]{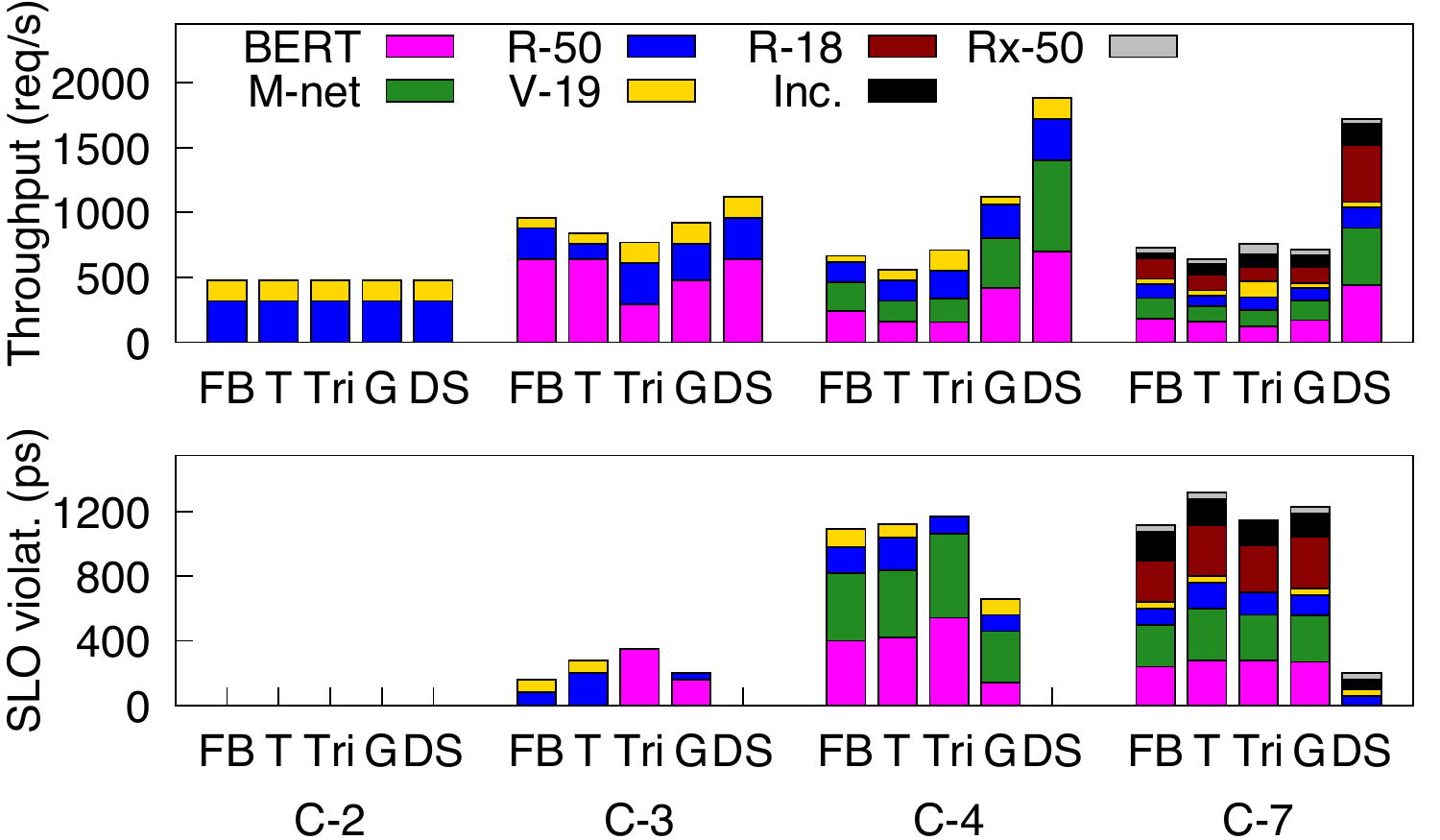}
    \caption{Throughput and SLO violations. M = Default MPS, FB = Fixed Batch,T = Temporal Sharing, ST-F = ST-FOOD w/GPU\% and opt. batch. C-2 = ResNet-50 \& VGG-19, C-3 = C-2 $+$ Alexnet, C-4 = C-3 $+$ Mobilenet. }
    \label{fig:css_fig}
    \vspace{-6mm}    
\end{figure}%
}

\Scut{
\begin{figure*}
   \subfloat[C-2=ResNet-50$+$VGG-19,C-3=C-2$+$Alexnet, C-4=C-3$+$Mobilenet,C-7=C-4$+$ResNet-18 $+$ Inception $+$ResNeXt-50 ]{
   \includegraphics[width=.5\linewidth]{figures/all_results/Multiplexing_chain_models_sigmetrics.pdf}
   \vspace{-2mm}
    \label{fig:css_fig}
    }
\Scut{
\begin{minipage}[]{.5\linewidth}
\captionof{table}{Avg. Throughput \& Latency in 3 GPU cluster}
\vspace{-2mm}
\resizebox{\columnwidth}{!}{
\begin{tabular}{|c|c|c|c|c|}\hline
         Model&\multicolumn{2}{|c|}{Default MPS}  & \multicolumn{2}{|C|}{CSS}   \\\hline
         & Latency (ms) & Throughput(Ips) & Latency & Throughput\\\hline
         Mobilenet&5.27&1702&5.23&1708\\
         ResNet-50&37.88&503&24.83&713\\
         VGG-19&24.80&40&24.80&40\\\hline
    \end{tabular}
    }
    \label{tab:css_all_together}
\end{minipage}
}
\Scut{
\begin{subfigure}{.5\linewidth}
   \includegraphics[width=\linewidth]{figures/all_results/Multiplexing_chain_models_sigmetrics.pdf}
    \caption{Throughput and SLO violations. M = Default MPS, FB = Fixed Batch,T = Temporal Sharing, ST-FD w/GPU\% and opt. batch. C-2 = ResNet-50 \& VGG-19, C-3 = C-2 $+$ Alexnet, C-4 = C-3 $+$ Mobilenet. }
    \label{fig:css_fig}
    \end{subfigure}
    }
\Scut{
\subfloat[ResNet-50 DNN]{
    \includegraphics[width=.33\linewidth]{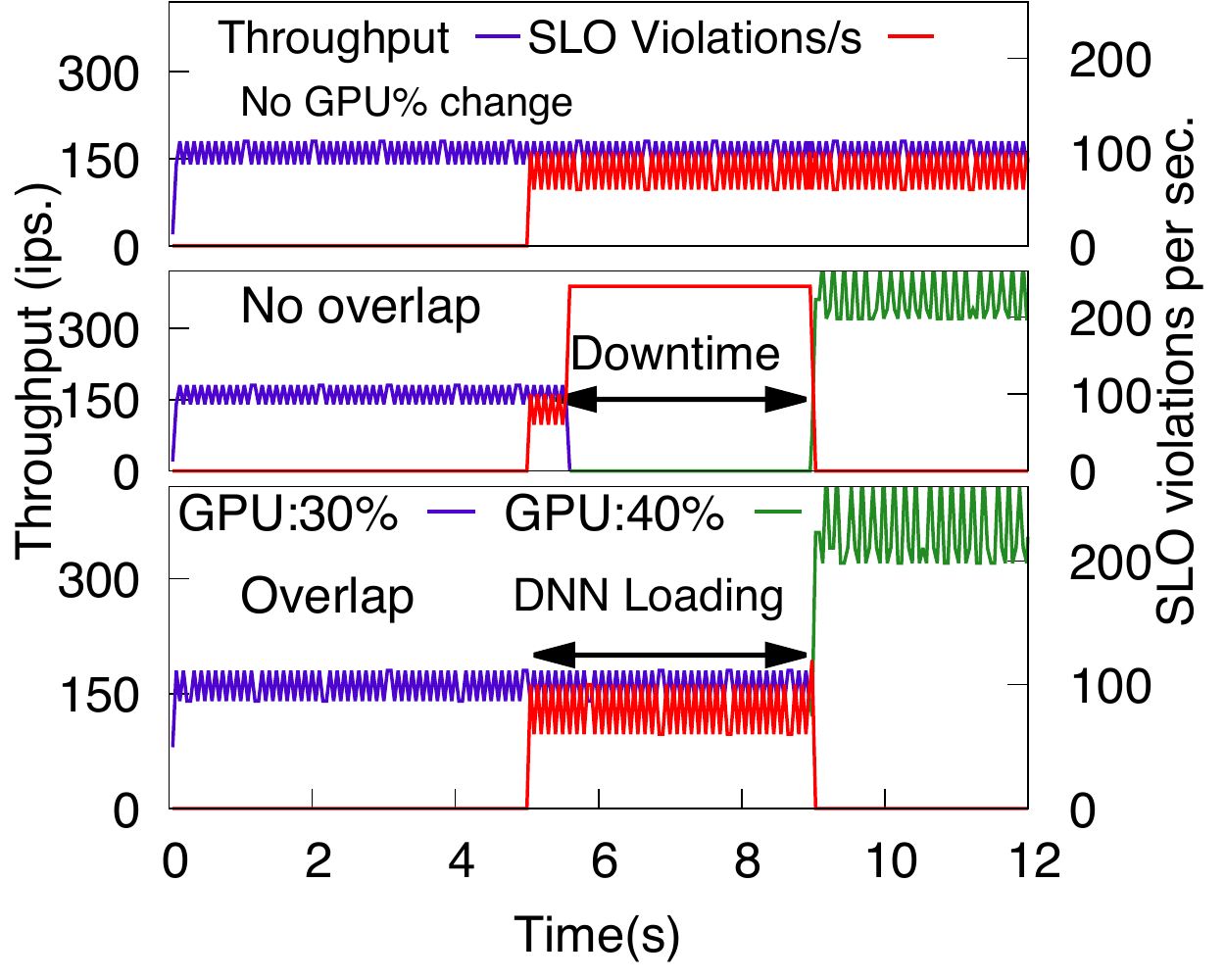}
    \vspace{-6mm}
    \label{fig:active-standby}
    }
}
\subfloat[Baseline throughputs are shown in session $T_{0}$.]{
    \centering
    \includegraphics[width=.33\linewidth]{figures/scheduling_variable/variable_rate_timeline.pdf}
    \vspace{-2mm}
    \label{fig:variable-rate}
    }

\vspace{-2mm}
\caption{(a) Throughput \& SLO violations.\Scut{M=Default-MPS,} FB=Fixed-Batch+MPS, T=Temporal Sharing, Tri=Triton Server, ST-FD w/GPU\%,optimal batch. (b) Dynamic GPU resource allocation and SLO counts.(c)\name{} with varying request rate\looseness-1 }
\vspace{-2mm}
\end{figure*}
}

We compare \name{} with other multiplexing methods.

\Scut{
\noindent\newline\textbf{Cluster-Wide GPU Multiplexing}: 
\knote{do we want to start with this experiment when all of our work is on a single GPU? Maybe we want to take it out of here and possibly use it in the future. Don't put in cluster-wide anything here, as it will confuse the reader. AD: I will cut it out.We will exapand on the multiplexing experiment Fig 15 (a) with temporal and spatio-temporal options.}
We ran an experiment with 5 DNN application instances on 3 of the GPUs (we were limited by network link bandwidth of 40 Gbps). 
For the `default MPS (M)' and 'Fixed batch (FB) cases, our setup includes two instances each of Mobilenet on one GPU, and two instances of ResNet-50 on another GPU and single instance of VGG-19 on a third GPU, and the batch size is set to 1 in the `default MPS' and to 8 for the 'Fixed Batch' case respectively. 
For CSS case, we multiplexed one instance of ResNet-50 with 70\% of GPU and Mobilenet  with remaining 30\% (close to the percentages obtained from the optimization above) of the GPU. We instantiate the same setup on two different GPUs. Since VGG-19 is compute heavy, we gave an entire instance of GPU (with its 'knee' at 100\%). 
CSS uses self-learning adaptive batching as described in~\cite{GSLICE}, ensuring latency remains below the specified SLO. In CSS, we also implement offloading data movement to the GPU DMA, and the task completion notification.
In all the cases we used the same workload (input traffic using 40Gbps link). We compute the throughput achieved by the system, with a nominal SLO of 50 ms. Table~\ref{tab:css_all_together} shows the results of `default MPS (M)', `default MPS with Fixed batch (FB)' and `CSS' modes.
\looseness-1

\Scut{
\begin{table}[h]
\vspace{-4mm}
    \centering
    \caption{Avg. Throughput \& Latency in 3 GPU cluster}
    \vspace{-3mm}
    \resizebox{\columnwidth}{!}{
    \begin{tabular}{|c|c|c|c|c|}\hline
         Model&\multicolumn{2}{|c|}{Default MPS}  & \multicolumn{2}{|C|}{CSS}   \\\hline
         & Latency (ms) & Throughput(Ips) & Latency & Throughput\\\hline
         Mobilenet&5.27&1702&5.23&1708\\
         ResNet-50&37.88&503&24.83&713\\
         VGG-19&24.80&40&24.80&40\\\hline
    \end{tabular}}
    \label{tab:css_all_together}
    \vspace{-2mm}
\end{table}
}
With CSS, we achieve much higher throughput 
compared to `M' and `FB' cases. 
ResNet-50 sees almost a {\bf 40\%}  throughput improvement over `FB', without sacrificing the performance for other models. CSS has the same average latency ($\sim$25 ms) for Mobilenet and VGG-19. Further, CSS has much better latency for Resnet-50, while `M' and `FB' show higher latency due to interference.\looseness-1 
}
\textbf{Multiplexing DNN models on the GPU}: \Scut{We present the case where our framework contributes to better performance with spatial multiplexing in the GPU. }\Scut{In Fig.~\ref{fig:css_fig}} We evaluate three different cases of multiplexing by running 2, 3, 4 and \revise{7} DNNs, respectively.\revise{ By multiplexing 7 different DNNs, we demonstrate how \mbox{\name{}} is still successful in scheduling a number of 
models with tight latency constraints, even if the sum-total of their demand (\mbox{\ie} knee-capacity) is substantially higher than 100\% GPU. We show \mbox{\name{}} can improve throughput and utilize the GPU better while reducing the SLO violations compared to the other approaches, with all, including \mbox{\name{}} having to compromise by missing the deadline on some inference requests.}  \Scut{We multiplex the GPU 4 different ways, Default MPS (M), Fixed Batching (FB), Temporal Sharing (T) and our Spatio-temporal sharing (ST).} 
We compare our approach, including \name{}, with four other methods of GPU multiplexing, namely,\Scut{default MPS (M),} Fixed batching with Default CUDA MPS (FB), and temporal sharing (T), Triton Inference Server (Tri) and GSLICE (G).\Scut{Default MPS spatial sharing (M) infers DNN requests with a batch size of 1, and the multiplexing models share the GPU with MPS without an explicit GPU\% allocation.}\Scut{ We also compare with multiplexing each model with} In Fixed batching with CUDA MPS (FB),\Scut{ the GPU is divided into equal proportions for all the models being multiplexed, and} the largest batch size of 16 is picked for inference every time and the multiplexing models share the GPU with MPS without an explicit GPU\%.\Scut{ fixed batch size is set to be as large as possible for each DNN while ensuring that the DNN can meet their SLO. For the fixed batching (FB) case, we multiplex the GPU among multiple DNNs using the default MPS} In temporal sharing (T), time slices are set in the proportion of the models' SLO length. With Triton server (Tri), we request the inference with multiple clients concurrently, allowing Triton server to dynamically batch and infer our requests. With GSLICE (G), we use all GSLICE's features, including adaptive batching and spatial sharing of the GPU at each DNN's knee. \Scut{ for all the multiplexing models.} Finally, in \name{}, we use the batch size and GPU\% from our optimization formulation and utilize \name{} scheduling to schedule the models.\looseness-1

We evaluate the throughput and the SLO violations per second for each model in Fig.~\ref{fig:css_fig}. We measure SLO violations per second as the sum of all the inference requests that violate the SLO and all the unserved requests. Inference requests are generated at the rate of $\sim$1920 images/sec (max. request rate limited by the 10 Gbps link in testbed). 
Requests are divided into the multiplexed models in proportion to their SLOs. Thus, for the experiments C-2, C-3 and C4, Alexnet and Mobilenet get 700 inference requests/sec, ResNet-50 gets 320 requests/sec and VGG-19 gets 160 requests/sec.\revise{For the experiment with 7 DNN models running concurrently (\mbox{\ie} C-7), Alexnet, Mobilenet and ResNet-18 receive 440 inference requests/sec, ResNet-50 and Inception receive 220 requests/sec while ResNeXt-50 and VGG-19 get 80 requests/sec.}\Scut{ We present the throughput and the SLO violations per second for each model in Fig.~\ref{fig:css_fig}. We measure SLO violations per second as sum of all the inference requests that violate the SLO and all the unserved request.}
We observe from Fig.~\ref{fig:css_fig} that our framework provides more than a 3$\times$ increase in aggregate throughput when multiplexing \revise{7} different models. \name{} achieves the highest throughput even when fewer models are running concurrently. 
For MPS, the lack of batching causes it to miss most of the SLOs for requests. Fixed batch, temporal sharing, GSLICE and Triton server provide good throughput while running just 2 models.\Scut{ Fixed batch, performs well while multiplexing fewer models.} However, as the number of models multiplexed increases, \Scut{ each model's throughput is reduced, as }each new added model contends for GPU resources in Fixed Batch, decreasing the\Scut{model's} throughput. Meanwhile, in temporal sharing, each model gets less and less GPU time, impacting  throughput.\looseness-1

Models hosted in Triton server too have to multiplex GPU temporally, thus, get lower throughput when more models are added. With GSLICE, multiplexing more models means some models get resources lower than knee GPU\%, exponentially increasing the inference latency.
\Scut{ resources and having high latency due to big batch size.Temporal sharing provides good throughput while running just 2 models. However, as more models are added, each model gets less and less GPU time, thus, impacting  throughput.} \name{} provides both the right amount of GPU resources and \Scut{chooses }the appropriate batch size.\Scut{ GPU resource isolation in \name{} helps prevent bigger models taking away all resources, thus, gets 3-4$\times$ higher throughput than fixed batch, temporal sharing and Triton server approach.} Furthermore, there are no SLO violations in \name{} when multiplexing 2-4 models. However, when overloading the GPU by multiplexing 7 DNNs, we see a few SLO violations for the models with longer runtime (Inception, Resnet-50, ResNeXt-50 and VGG-19). \mbox{\name{}} misses SLOs for 10\% of all requests, compared to more than 68\% for the alternatives. SLO misses for \name{} are from the smaller fraction of requests sent to compute heavy models such as ResNet-50, ResNext-50 and VGG-19. Even with some of the medium-to-large sized models with longer runtimes, such as  ResNet-50 and Inception, only 13\% of requests see a SLO violation. \Scut{ 13\% of all requests for ResNet-50 and Inception and 50\% of all requests for ResNeXt-50 and VGG-19. We meet the SLO for all other models.} This is due to the fact that 
running 7 models concurrently exceeds the capacity of GPU even with \name{}.\Scut{Nonetheless, \revise{the overall SLO misses for \mbox{\name{}} are low, 10\% compared to more than 68\% for the other alternatives.} Overall, \mbox{\name{}} provides very high throughput while keeping the SLO violations low.} 
With \mbox{\name{}} the average GPU utilization is 92\% while multiplexing with 7 models. 
With all the models having a knee greater than 10\%, this is close to fully utilizing the GPU.\looseness-1

\Scut{
\begin{figure}
    \centering
    \includegraphics[width=\linewidth]{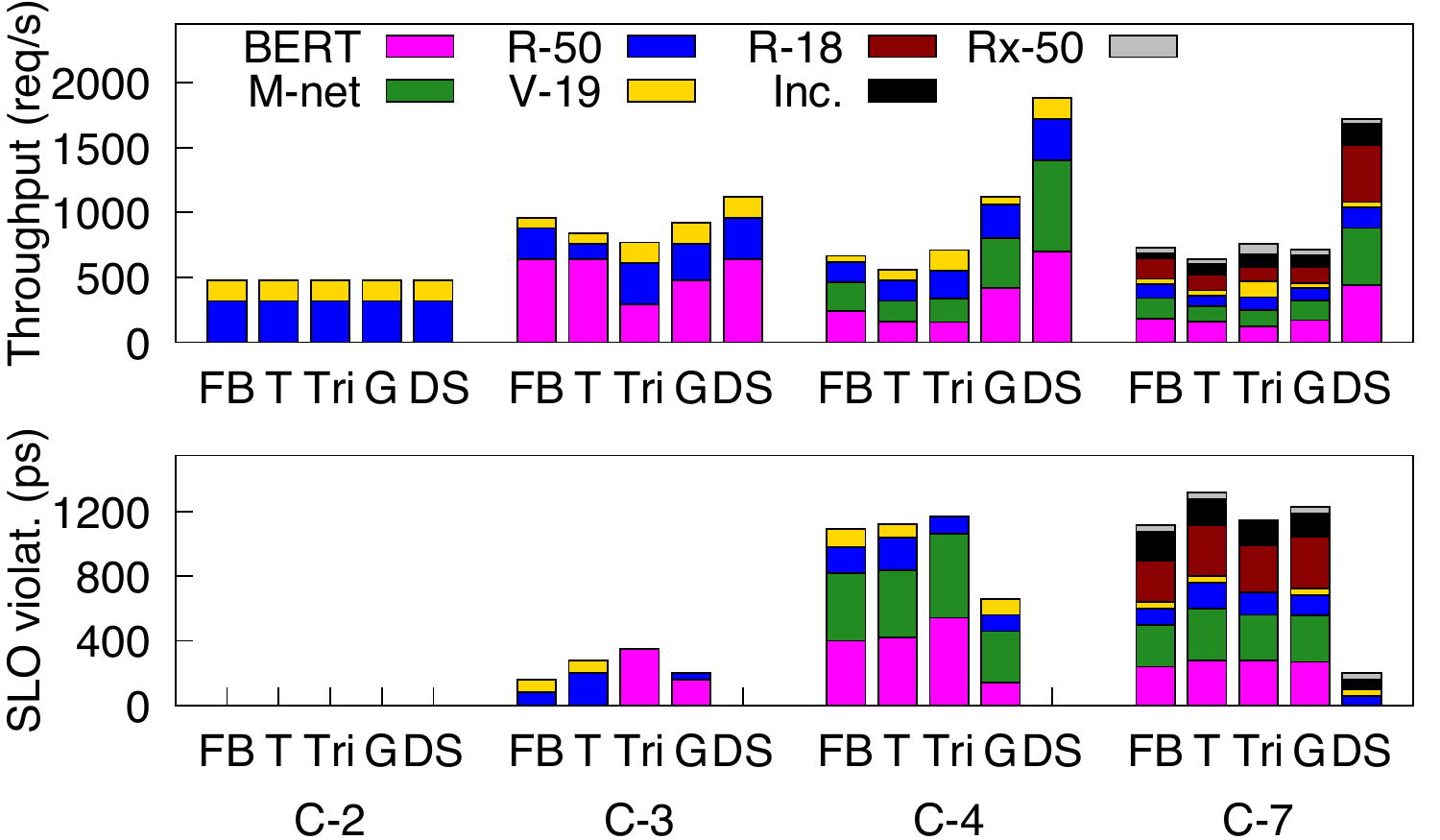}
    \caption{Throughput \& SLO violations. C-2=ResNet-50$+$VGG-19, C-3=C-2$+$Alexnet, C-4=C-3$+$Mobilenet, C-7=C-4$+$ResNet-18$+$Inception$+$ResNeXt-50}
    \label{fig:css_fig}
    \vspace{-4mm}
\end{figure}
\begin{figure}
    \centering
    \includegraphics[width=\linewidth]{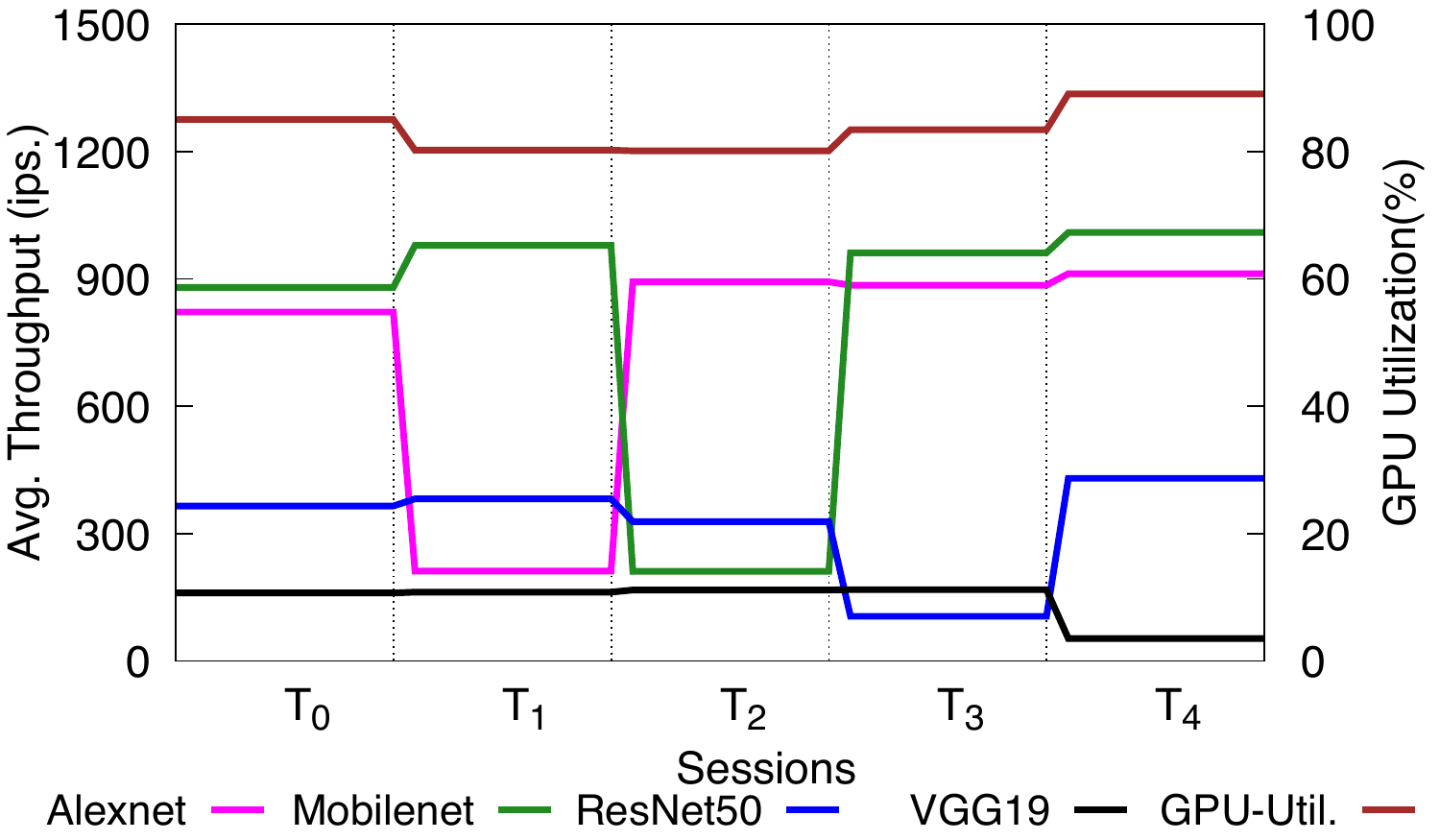}
    \caption{\name{} with varying request rate. Baseline throughputs are shown in session $T_{0}$.}
    \label{fig:variable-rate}
    \vspace{-2mm}
\end{figure}
}

\Scut{
\noindent\textbf{Resource re-provisioning with Overlapped Execution:}
We present the results for our interruption-free dynamic GPU resource allocation using the active and standby instances described in \S~\ref{sec:dynamic_gpu_recon1}\Scut{ and \S~\ref{sec:dynamic_gpu_recon2} through overlapped execution based on Algorithm~\ref{algo:activetostandby}.}. We show the timeline for three different scenarios, presenting the instantaneous (computed every 100 ms) inference throughput with ResNet-50 in Fig.~\ref{fig:active-standby}.\Scut{ whose request rate increases at\Scut{some point in the middle} 5 sec. mark.} In all three experiments, ResNet-50 starts with 30\% GPU that is below its "knee" value of 40\% GPU. The top plot, in Fig.~\ref{fig:active-standby}, shows the baseline case (when GPU\% remains fixed at 30\% till the end of the experiment). The middle plot, in Fig.~\ref{fig:active-standby}, shows the throughput when the GPU\% is re-adjusted from 30\% to 40\% without overlapped execution, resulting in downtime. The bottom plot, Fig.~\ref{fig:active-standby}), shows the throughput when the GPU resource is re-adjusted with the active-standby overlapped execution approach. 
At the $\sim$5 second mark we increase the request rate. This increase causes SLO violations for all 3 approaches. The baseline (top) violates the SLO for many requests as the low GPU\% cannot keep up with the higher request rate. Our framework quickly detects the SLO violations (after 5 violations) and triggers the change in GPU\% to 40\%, as recommended by the optimization framework to meet the request demand. The DNN model load time into the GPU is shown by the black arrow. In the middle plot, when the GPU\% is reset without overlap, there is a downtime of $\sim$4 seconds while resetting the GPU\%. This is the time taken by the standby DNN to load the model onto GPU and be ready for inference. In the middle case, all the arrivals during this time (and many subsequent arrivals because they are queued) will experience SLO violations, as no model serves any requests during the downtime. On the other hand, with overlap (bottom plot), the active DNN continues to provide useful inference service, albeit with some SLO violations, while the standby is in the process of loading the model, thus marking the DNN loading time\Scut{ in the background}. After the GPU\% increase is complete, the higher GPU\% model will continue the inference with higher throughput and meeting the SLO of the subsequent incoming requests.
The actual switchover (true downtime) happens in just about 100$\mu$s.
\Scut{The DNN model is aided by spatio-temporal scheduler which allocates it extended timeslots to meet the increased demand, bringing the SLO violations to zero.} 
The key is that the task of loading the DNN does not affect the throughput of active DNN\Scut{ and cause minimal SLO violations ($5$) for requests}. The overlap approach reduces the SLO violation by more than 65\% by changing the GPU\% allocation. 
\looseness-1
}

\textbf{Benefit of \name{} Scheduler}: Wherever possible, \name{} tries to 
opportunistically schedule additional model instances during the session, possibly with a smaller batch size to utilize the available GPU. To show the effectiveness of the \name{}, we present a scenario where the request rate of the multiplexed DNN models varies dynamically. To start with, in session $T_{0}$, we have 4 models, Alexnet, Mobilenet, ResNet-50 and, VGG-19, same as in 'C-4' in Fig.~\ref{fig:css_fig} running with their request rates high enough to support the optimal batch size, as determined in Table~\ref{tab:model_characteristics}. The GPU utilization we achieve is $\sim 85\%$. 
We then change the request rate of one model (Alexnet in session $T_{1}$) by a random amount. We still allow for the optimal batch to form for each model. 
The throughput of the models dynamically adjust with the throughput of other models increasing due to use of the un-utilized resources left by Alexnet (see $T_{1}$). Since these three models have a high GPU\% requirement, there is not enough GPU to accommodate an instance of another model. Thus, the GPU utilization drops very slightly. At $T_{2}$, Alexnet's request rate goes back up, while Mobilenet request rate lowers, once again by a random amount. Alexnet opportunistically uses the GPU to achieve a throughput higher than what it achieved in the baseline session $T_{0}$. Similarly, when ResNet-50 and, VGG-19's arrival rates drop at $T_{3}$ and $T_{4}$, respectively, \Scut{both Alexnet and ResNet-50} the other models increase their throughput.
We also see that across these sessions, the GPU utilization is nearly unchanged, remaining high, 
indicating that the \name{} effectively uses the GPU.\looseness-1

\subsection{\name{} in Multi-GPU Clusters}
\label{sec:multi-gpu-cluster}

We evaluated \name{} in a multiple GPU cluster of 4 NVIDIA T4 GPUs, each having 40 SMs
(fewer than a V100) 
and 16 GB of memory.
We utilized 4 different vision models, Mobilenet, Alexnet, ResNet-50 and VGG-19
(knee GPU\% is different for T4 GPU vs. V100). 
We compare throughput of 3 different
multiplexing and scheduling scenarios. First, we provide one T4 GPU for each DNN model exclusively. In the second scenario, we place all 4 models in each GPU, temporally sharing the GPU.
Finally, we evaluate \name{} with the 4 DNN models.\looseness-1

Fig.~\ref{fig:multi-gpu} shows temporal scheduling has almost the same
throughput as each model having an exclusive GPU.
\begin{figure}
\vspace{-3mm}
\centering
\includegraphics[width=\linewidth]{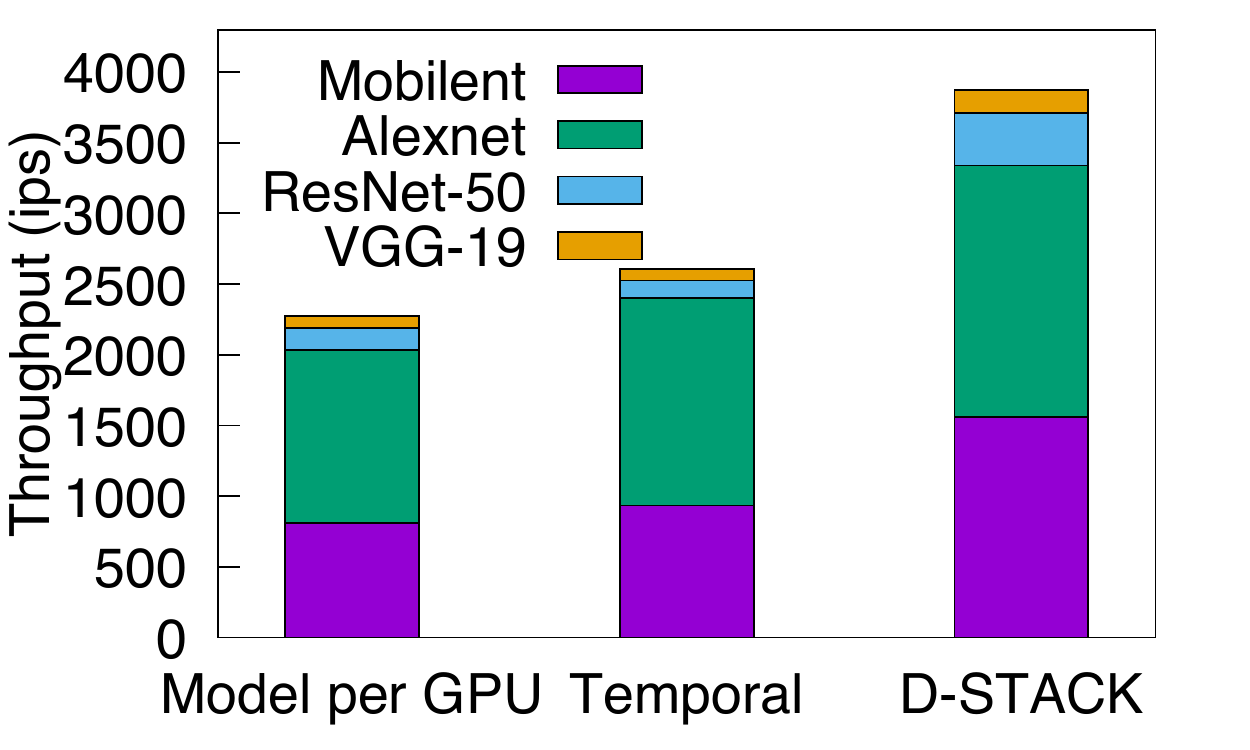}
\caption{GPU cluster throughput}
\label{fig:multi-gpu}
\vspace{-4mm}
\end{figure}
This is because of the under-utilization of the GPU by the DNN models. 
\name{} has much higher throughput for every model, with 160\% overall higher throughput than temporal sharing.  The overall inference throughput increases substantially as the multi-GPU cluster is better utilized by \name{}.\looseness-1 
\looseness-1
\Scut{
\begin{figure}
\centering
\includegraphics[width=.33\linewidth]{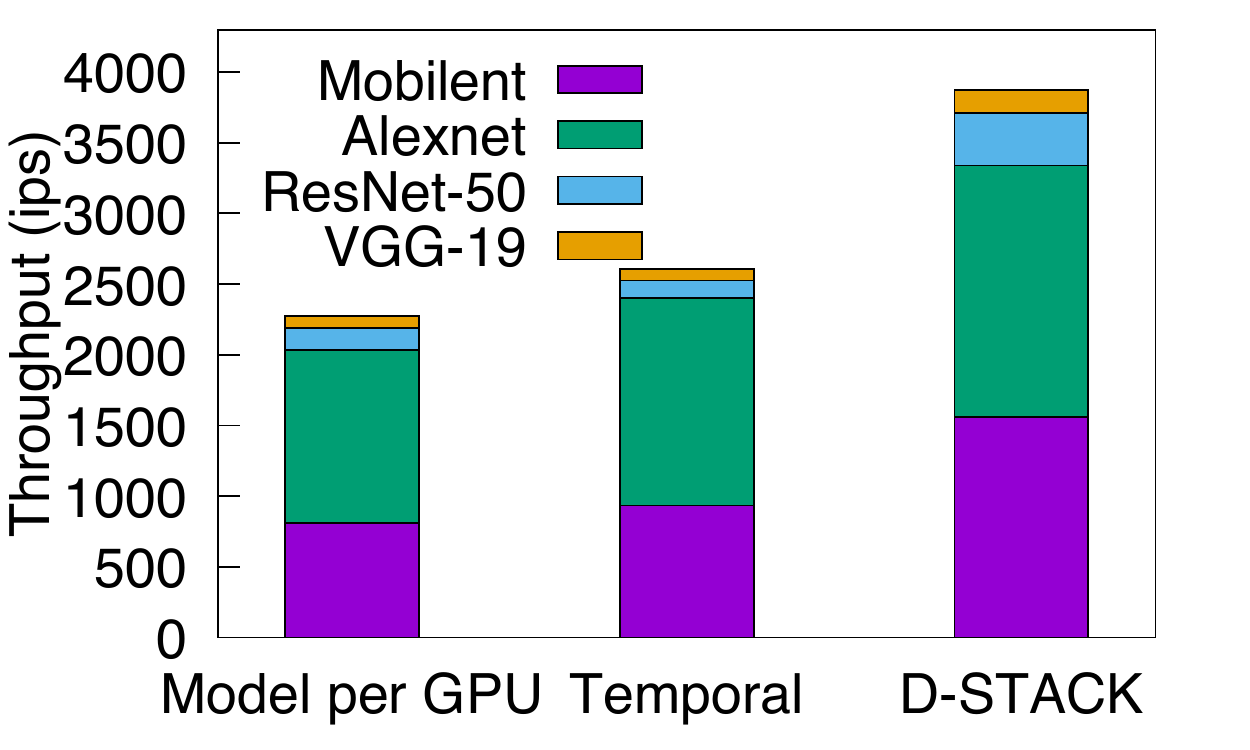}
\caption{Throughput in 4 T4 GPU cluster}
\label{fig:multi-gpu}
\end{figure}
}
\Scut{
\begin{figure}
    \centering
    \includegraphics[width=\linewidth]{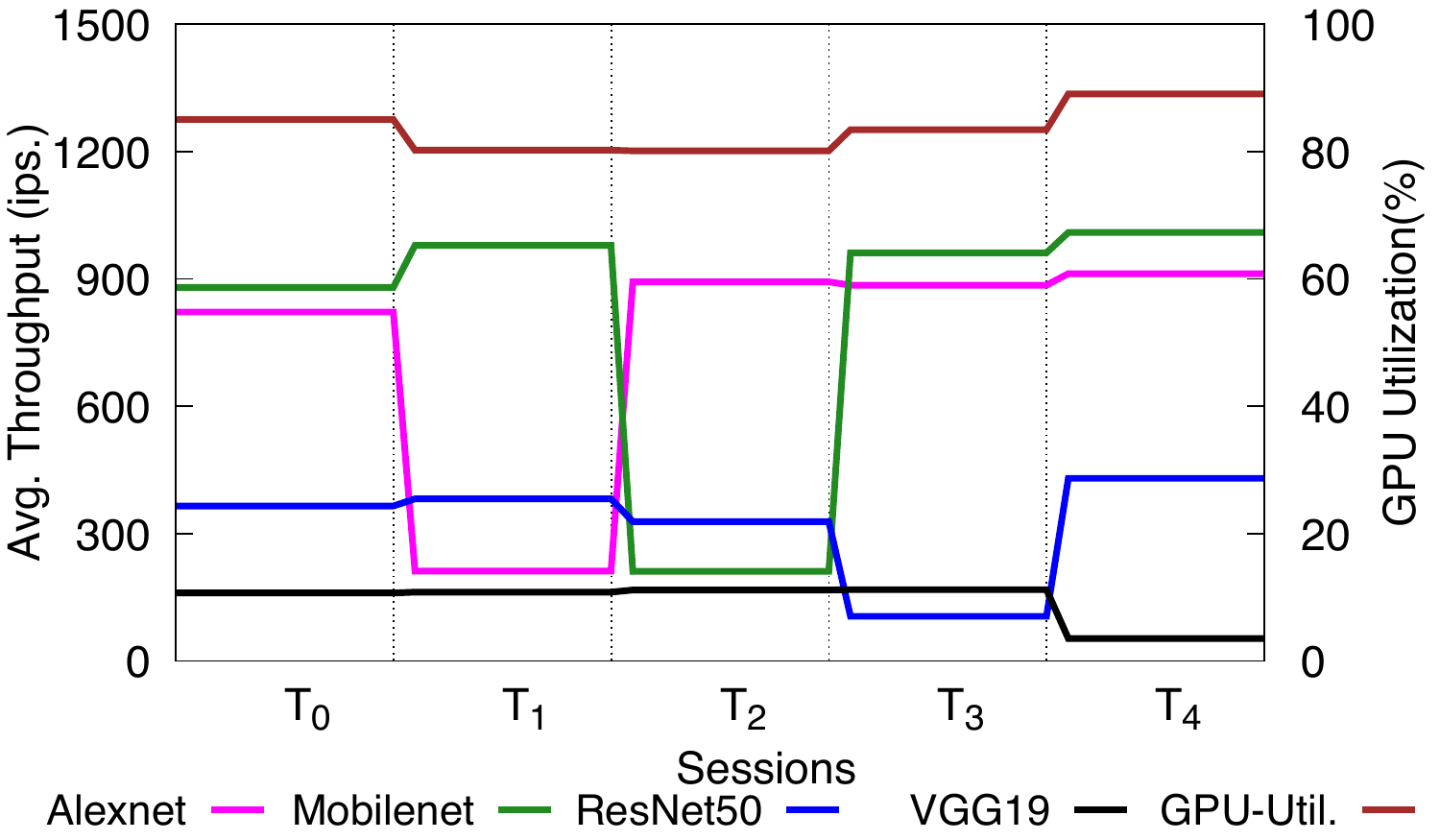}
    \caption{Spatio-Temporal scheduler's response to variable rates. Baseline throughput are shown in $T_{0}$ }
    \label{fig:variable-rate}
\end{figure}
}

\Scut{
\section{temporary for algorithm}
\begin{algorithm}[H]
\caption{Active to Standby Takeover: Manager Events}
\label{algo:activetostandby}
\begin{algorithmic}[1]
\Statex{{\bf Event:} \texttt{ResourceReadjust}}
\State $<\mathit{standbyDNN, GPUFlag}> = 1$
\State $<\mathit{standbyDNN, GPU\%}> = NewGPU\%$
\State sendMessage$(\mathit{standbyDNN,} \texttt{GetReady})$
\Statex{{\bf End Event}}
\State
\Statex{{\bf Event:} \texttt{ReceiveMessage}}
\If{msg = $<\mathit{standbyDNN,} \texttt{modelLoaded}>$}
    \State sendMessage$(\mathit{primaryDNN,} \texttt{CompleteWork})$
    \EndIf
    \If{msg$ =<\mathit{primaryDNN,\texttt{workCompleted}}>$}
    \State $< \mathit{primaryDNN, GPUFlag} >$ = 0
    \State wakeUp$\mathit{(standbyDNN})$
    \State shutdown$(\mathit{primaryDNN})$
    \State $\mathit{primaryDNN \gets standbyDNN}$
    \State startStandbyInstance$()$
    \EndIf
\Statex{{\bf End Event}}
\end{algorithmic}
\end{algorithm}

\begin{algorithm}[H]
\caption{Active to Standby Takeover: Active DNN}
\label{algo:activeDNN}
\begin{algorithmic}[1]
\Statex{{\bf Event:} \texttt{ReceiveMessage}}
\If{msg = $<\mathit{Manager,} \texttt{CompleteWork}>$}
\While{$\mathit{DNNInferencePending} > 0$}
\State wait() \Comment{Waits for inference results}
\EndWhile
\State sendMessage$(\mathit{Manager,} \texttt{workCompleted})$
\State wait() \Comment{Wait for Shutdown}
\EndIf
\Statex{{\bf End Event}}
\end{algorithmic}
\end{algorithm}

\begin{algorithm}[H]
\caption{Active to Standby Takeover: Standby DNN}
\label{algo:standbyDNN}
\begin{algorithmic}[1]
\Statex{{\bf Event:} \texttt{ReceiveMessage}}
\If{msg = $<\mathit{Manager,} \texttt{GetReady}>$}
\If{$GPUFlag == 1$}
\State setEnvVar($\mathit{newGPU\%}$)
\State LoadModel() \Comment{Load DNN model in GPU}
\State sendMessage$(\mathit{Manager,} \texttt{modelLoaded})$
\State sleep() \Comment{Sleep till woken by the manager}
\EndIf
\EndIf
\Statex{{\bf End Event}}
\Statex{{\bf Event:} \texttt{Wakeup}}
\State pendingInference() \Comment{Finish pending requests from active}
\State inferRequests() \Comment{Infer new requests}
\Statex{{\bf End Event}}
\end{algorithmic}
\end{algorithm}
}
\Scut{
\subsection{Multi-Instance GPUs}
Multi-Instance GPUs (MIGs) such as the NVIDIA A100 are hardware based approaches for coarser grained, spatial multiplexing. However, it is static. MIGs allow partitioning of a GPU into multiple smaller GPU instances (up to 7 instances with the A100) with a desired number of SMs and dedicated memory. These smaller GPU instances can be further spatially multiplexed. As we saw above,  
with multiple GPUs, \name{} improves inference throughput by more than 1.6$\times$ while using a cluster of multiple small GPUs. 
The case is similar with the MIG GPU design. 
We believe \name{} can improve utilization of each smaller MIG instances, improving utilization and throughput of the system. 
Moreover, note that MIG GPUs are also able to run as a single GPU (similar to V100). Thus, they can even benefit from \name{} without any modification.\Scut{\name{}'s effort is complementary to new technologies such as NVLINK~\cite{foley2017ultra} and faster GPU-GPU interconnection in MIGs. They 
greatly reduce inter-GPU data access time, thus lowering GPU idle time while fetching the data from CPU memory and network. Systems with NVLINK implementing \name{} improve GPU utilization by using the GPU's SMs better and taking advantage of the lower interconnect latency between GPUs.}\looseness-1
}
\section{Conclusions}
DNNs critically depend on GPUs and other accelerators, but 
often under-utilize the parallel computing capability of current high-performance accelerators. 
Due to uneven workloads of different DNN kernels, a DNN as a whole is unable to fully utilize all the parallelism of the GPU (i.e., all SMs). 
Furthermore, there are non-parallelizable tasks while executing a DNN on a GPU-based system limiting the effective use of a GPU's parallelism. We validated these conclusions from our model of a DNN through measurements of different types of DNNs (CNNs, and Transformers) on an V100 GPU.\Scut{ 
We developed a simple model to explain the inherent limits of parallelism a DNN is able to exploit and demonstrated through numerical simulations and measurement of different DNN models that GPUs are under-utilized by the models available in popular ML frameworks.
We presented the key reasons that limit how effectively the GPU is utilized by DNN models.
}
Since batching DNN requests improves inference throughput and GPU utilization, we develop an optimization framework to establish an optimal operating point (GPU\%, Batch Size) for a DNN utilizing the GPU at the highest efficacy.\Scut{ balances the batch size and the share of GPU resources to be provided for a DNN model using CSS.
and how 
different GPU multiplexing modes are inadequate, especially to meet SLO requirements.
Based on these insights, we develop an optimization framework to find an optimal operating point for a DNN that balances the batch size and the share of GPU resources to be provided for a DNN model using CSS.} 
We bring the optimal batch size and GPU\% together in \name{} to develop a spatio-temporal, fair, opportunistic, and dynamic scheduler to create an inference framework that effectively virtualizes the GPU. \name{} accounts for a DNN model’s SLO, GPU resource allocation, and batch size, 
to provide a schedule that maximizes meeting SLOs, across multiple DNN models while seeking to utilize the GPU fully.\Scut{Moreover, \name{} schedules DNNs fairly.} \name{} benefits both single GPUs and multi-GPU clusters. Our enhancements in \name{} do not require modifications to the GPU architecture, the runtime, or the DNN models themselves. \name's features can easily help improve existing DNN inference platforms (e.g., Triton server) as well. We show that \name{} can attain higher than 90\% throughput of an ideal scheduler, which we speculate can switch tasks instantaneously at a very fine time granularity, ignoring practical limitations.
Our controlled testbed experiments with 4 T4 GPU clusters show the throughput improvement of 160\%-180\% with \name{} compared to providing an entire GPU to each individual DNN model. With an NVIDIA V100 GPU, \name{} shows benefit in the range of \textasciitilde$1.6\times$ improvement in GPU utilization and 3$\times$ to 4$\times$ increase in throughput with no impact in latency compared to the baseline temporal sharing.\looseness-1 
\bibliographystyle{ACM-Reference-Format}
\bibliography{sigmetrics.bib}

\end{document}